\documentclass[11pt]{article}

\usepackage{graphicx,graphics}
\usepackage{amsmath,amssymb,slashed,dcolumn,mathcomp,amsfonts,cite,slashed,dsfont}

\allowdisplaybreaks[1]

\newcommand{\beq}{\begin{equation}}
\newcommand{\eeq}{\end{equation}}
\newcommand{\beqa}{\begin{eqnarray}}
\newcommand{\eeqa}{\end{eqnarray}}
\newcommand{\bal}{\begin{align}}
\newcommand{\eal}{\end{align}}

\newcommand{\Order}{\mathcal{O}}
\newcommand{\Lagr}{\mathcal{L}}
\newcommand{\F}{\mathcal{F}}
\newcommand{\M}{\mathcal{M}}
\newcommand{\Tr}{{\rm Tr}}
\newcommand{\mpp}{m_{\rm p}}
\newcommand{\mn}{m_{\rm n}}
\newcommand{\mpi}{M_{\pi}}
\newcommand{\mpii}{M_{\pi^0}}
\newcommand{\muu}{m_{\rm u}}
\newcommand{\md}{m_{\rm d}}
\newcommand{\unity}{\mathds{1}}
\newcommand{\Fpi}{F_\pi}
\newcommand{\ga}{g_{\rm A}}

\newcommand{\pp}{\mathbf{p}}
\newcommand{\diff}{\text{d}}
\newcommand{\Deltax}{\Delta_{\rm cex}}
\newcommand{\mx}{M_{\rm cex}}
\newcommand{\npi}{{\pi^{0}}}
\newcommand{\pt}{\hspace{-1pt}}
\newcommand{\Deltam}{\Delta \hspace{-1.3pt}m}
\DeclareMathOperator{\arsinh}{arsinh}
\DeclareMathOperator{\arcosh}{arcosh}
\DeclareMathOperator{\Li}{Li_2}

\newcommand{\settocdepth}[1]{%
\addtocontents{toc}{\protect\setcounter{tocdepth}{#1}}}

\def\Xint#1{\mathchoice
   {\XXint\displaystyle\textstyle{#1}}%
   {\XXint\textstyle\scriptstyle{#1}}%
   {\XXint\scriptstyle\scriptscriptstyle{#1}}%
   {\XXint\scriptscriptstyle\scriptscriptstyle{#1}}%
   \!\int}
\def\XXint#1#2#3{{\setbox0=\hbox{$#1{#2#3}{\int}$}
     \vcenter{\hbox{$#2#3$}}\kern-0.5\wd0}}
\newcommand{\dashint}[1]{\Xint{\hspace{#1}-}}

\begin{document}

\begin{flushright}
{\tiny{HISKP-TH-09/29}
{\tiny{FZJ-IKP-TH-2009-27}}
}
\end{flushright}

\vspace{.6in}

\begin{center}

\bigskip

{{\Large\bf Isospin violation in low-energy pion--nucleon scattering revisited
}}

\end{center}

\vspace{.3in}

\begin{center}
{\large 
Martin Hoferichter$^a$,
Bastian Kubis$^a$,
Ulf-G. Mei{\ss}ner$^{a,b}$
}

\vspace{1cm}

$^a${\it Helmholtz--Institut f\"ur Strahlen- und Kernphysik (Theorie) 
   and Bethe Center for Theoretical Physics, Universit\"at Bonn, D-53115 Bonn, Germany}

\bigskip

$^b${\it Institut f\"ur Kernphysik (IKP-3), Institute for Advanced Simulation, 
   and J\"ulich Center for Hadron Physics, Forschungszentrum J\"ulich, D-52425  J\"ulich, Germany}

\bigskip

\bigskip

\end{center}

\vspace{.4in}

\thispagestyle{empty} 

\begin{abstract}\noindent 
We calculate isospin breaking in pion--nucleon scattering in the threshold region in the framework of covariant baryon chiral perturbation theory.  All effects due to quark mass differences as well as real and virtual photons are consistently included. As an application, 
we discuss the energy dependence of the triangle relation that connects elastic
scattering on the proton $\pi^\pm p \to \pi^\pm p$ with the charge exchange
reaction $\pi^- p \to \pi^0 n$. 
\end{abstract}
\vspace{3cm}
{\it Key words:}
Pion--baryon interactions, Chiral Lagrangians,
Electromagnetic corrections to  strong-interaction processes \\[2mm]
{\it PACS:} 13.75.Gx, 12.39.Fe, 13.40.Ks

\vfill

\pagebreak

\tableofcontents

\newpage

\section{Introduction and summary}
\def\theequation{\arabic{section}.\arabic{equation}}
\setcounter{equation}{0}

In the Standard Model isospin violation (IV) is driven by strong and 
electromagnetic interactions, that is by the differences in the
light quark masses $\md-\muu$ and charges $Q_{\rm u}-Q_{\rm d}=e$,
\beq
\Lagr_{\rm QCD}^{\rm IV}=\frac{\md-\muu}{2}(\bar{u}u-\bar{d}d),\quad 
\Lagr_{\rm QED}^{\rm IV}=\frac{i e}{2}\left(\bar{u}\slashed{A} u-\bar{d}\slashed{A} d\right).
\eeq
Pion--nucleon scattering in the threshold region is particularly well-suited to 
test our understanding of isospin violation because it can be analyzed within
chiral perturbation theory (ChPT), which is the effective field theory of the Standard
Model, and in contrast to pion--pion interactions the quark mass difference $m_{\rm d}-m_{\rm u}$ already contributes at leading order in
the symmetry breaking \cite{Weinberg77}. This problem was addressed in the framework of
heavy-baryon chiral perturbation theory in a series of papers about a decade ago 
\cite{MS97,FMS99,MM99,FM00_above,FM01}. In \cite{HKM09} we
presented a novel analysis of all $\pi N$ scattering lengths based on a 
covariant formulation of baryon chiral perturbation theory. In this work,
the analysis of isospin violation in the $\pi N$ system will be extended
beyond threshold. 

The size of the isospin-breaking effects extracted from low-energy data in early
model-based analyses \cite{Gibbs95,Matsinos97} are in conflict with those found in heavy-baryon
ChPT \cite{FM01}, while more recently the ChPT picture seemed to be confirmed by a tree-level coupled-channel $K$-matrix approach \cite{GHBS06}. 
The aim of this work is to revisit isospin violation in $\pi N$ scattering in the framework of covariant baryon ChPT \cite{BL99}, where the following differences to previous analyses within ChPT should be noted: in \cite{FM01} the
full $\Order(p^3)$-amplitude was fitted to data in order to determine the 
low-energy constants (LECs), which are then used to switch off electromagnetic 
effects to isolate strong isospin breaking. This procedure is questionable in
the sense that strong forth-order terms are known to be important
\cite{FM00,BL01}, such that electromagnetic LECs might get contaminated
accounting for these contributions. For this reason, we restrict ourselves to 
isospin-breaking shifts in the amplitude quantified in terms of the so-called 
triangle relation. Hence, our results will comprise both strong and
electromagnetic isospin-violating effects. In addition, we include bremsstrahlung (in the soft-photon approximation), such that finite terms neglected in \cite{FM01} are taken into account consistently. 
Note that here we do not attempt to determine the appearing LECs by a fit to
the pertinent partial waves, but rather follow the path laid out in  
\cite{HKM09} and collect these from other determinations,
concentrating on the isospin-breaking shifts induced in the various partial
waves. Finally, we hope that the amplitude documented in the present work will prove valuable for a consistent incorporation of isospin breaking in forthcoming analyses of $\pi
N$ scattering based on Roy--Steiner equations.

\medskip

The main findings and conclusions of the present investigation can be summarized
as follows:
\begin{itemize}
\item[i)]
In this work, we have systematically analyzed isospin breaking in $\pi N$
scattering  in all measured channels ($\pi^\pm p\to \pi^\pm p$, $\pi^-p\to\pi^0n$)
and the energy dependence of the  triangle relation in $S$- and $P$-wave
projections of the scattering amplitude in the framework of covariant baryon 
chiral perturbation theory, accounting for all sources of isospin violation 
including virtual photons and bremsstrahlung. In all cases, we have provided a 
detailed estimate of the theoretical uncertainties.

\item[ii)]
Above threshold, isospin violation in terms of the triangle relation amounts
to about $(2.5-4.0)\,\%$ between $\sqrt{s}=(1.08-1.14)\,{\rm GeV}$ in the
$S$-wave, the bulk of which is due to virtual photons, whereas isospin
breaking in the $P$-waves is very small, at the order of $1\,\%$ at most. We
have compared our work to former studies in heavy-baryon ChPT and analyses 
based on phenomenological models.
 
\end{itemize}

\newpage
Our work is organized as follows. Section~\ref{chap:ampl} contains the calculation
of the isospin-breaking corrections to third order in the chiral expansion.
After some preliminary chapters on the kinematics and isospin structure of the
scattering amplitude, we systematically display the calculation of the tree
and loop graphs, including the effects of virtual photons and soft
bremsstrahlung. Many details of these calculations are given in the various
appendices. In Sect.~\ref{chap:above} we analyze isospin violation above
threshold by concentrating on the triangle relation in the $S$- and
$P$-waves. It vanishes in the isospin limit and is thus an excellent
measure of the effects of isospin breaking. We give a thorough discussion of
the theoretical uncertainty of our calculations. We also compare to earlier
work based on heavy-baryon ChPT and more model-dependent analyses.

\section{Scattering amplitude} 
\def\theequation{\arabic{section}.\arabic{equation}}
\setcounter{equation}{0}
\label{chap:ampl}

\subsection{Kinematics and isospin structure of $\pi N$ scattering}
\label{sec:kinematics}

In this section, we consider various formal aspects of the reaction
\beq
N(p)+\pi^b(q)\rightarrow N(p')+\pi^a(q').
\eeq
The masses of the initial (final) nucleon and pion are denoted by $m_{\rm i}$
($m_{\rm f}$) and $M_{\rm i}$ ($M_{\rm f}$), respectively; $\mpp$, $\mn$,
$\mpi$ and $\mpii$ are the masses of proton, neutron, and charged and neutral
pion, in order.
We define the isospin limit by the charged particle masses $\mpp$ and $\mpi$. The deviations can be expressed in terms of 
\beq
\Delta_\pi=\mpi^2-\mpii^2, \quad \Delta_{\rm N}=\mn-\mpp.
\eeq
Furthermore, we will use 
\beq
\Sigma=p+q=p'+q', \quad \Lambda=p-q'=p'-q, \quad \Delta=p-p'=q'-q, \quad Q=p+p',
\eeq
which are related to the usual Mandelstam variables by
\beq
\Sigma^2=s, \quad \Lambda^2=u, \quad \Delta^2=t.
\eeq
For convenience, we will often refer to these quantities in the center-of-mass system (CMS) defined by $\mathbf{p}+\mathbf{q}=\mathbf{p'}+\mathbf{q'}=\mathbf{0}$ and $\mathbf{p}\cdot \mathbf{p'}\equiv|\mathbf{p}| |\mathbf{p'}|\cos \theta_{\rm CMS}\equiv |\mathbf{p}| |\mathbf{p'}|z$, such that
\begin{align}
E_p&=\frac{s+m_{\rm i}^2-M_{\rm i}^2}{2\sqrt{s}},\quad E_q=\frac{s-m_{\rm i}^2+M_{\rm i}^2}{2\sqrt{s}},\quad E_{p'}=\frac{s+m_{\rm f}^2-M_{\rm f}^2}{2\sqrt{s}},\quad E_{q'}=\frac{s-m_{\rm f}^2+M_{\rm f}^2}{2\sqrt{s}},\notag\\
|\mathbf{p}|&=|\mathbf{q}|=\frac{\lambda^{1/2}\left(s,m_{\rm i}^2,M_{\rm i}^2\right)}{2\sqrt{s}},\quad |\mathbf{p'}|=|\mathbf{q'}|=\frac{\lambda^{1/2}\left(s,m_{\rm f}^2,M_{\rm f}^2\right)}{2\sqrt{s}},\notag\\
t&=-\left(M_{\rm i}^2-M_{\rm f}^2\right)-\frac{\lambda\left(s,m_{\rm i}^2,M_{\rm i}^2\right)}{2s}
+\frac{s-m_{\rm i}^2+M_{\rm i}^2}{2s}\left(m_{\rm f}^2-m_{\rm i}^2+M_{\rm i}^2-M_{\rm f}^2\right)\notag\\
&+\frac{\lambda^{1/2}\left(s,m_{\rm i}^2,M_{\rm i}^2\right)\lambda^{1/2}\left(s,m_{\rm f}^2,M_{\rm f}^2\right)}{2s}z\overset{m_{\rm i}=m_{\rm f},\ M_{\rm i}=M_{\rm f} }{=}-\frac{\lambda\left(s,m_{\rm i}^2,M_{\rm i}^2\right)}{2s}(1-z)\label{kin_CMS},
\end{align}
where the K\"{a}ll\'{e}n function is given by
\beq
\lambda(a,b,c)=a^2+b^2+c^2-2a b- 2 b c-2 a c.
\eeq
$u$ follows from the Mandelstam relation
\beq
s+t+u=m_{\rm i}^2+m_{\rm f}^2+M_{\rm i}^2+M_{\rm f}^2 \label{Mandelstam}.
\eeq
Finally, we will need
\beq
\nu=\frac{s-u}{2\left(m_{\rm i}+m_{\rm f}\right)}.
\eeq
Simplifications of the kinematics in loops will be addressed in Sect.~\ref{sec:loop_ampl}. In particular, it is convenient to introduce a generic nucleon mass $m$ for the $\Order(p^3)$ contributions, which may  be identified with $\mpp$ in the end (cf.\ Sect.~\ref{sec:tree}).

As pointed out in \cite{BL01}, the standard decomposition of the pion--nucleon scattering amplitude
\beq
T_{\pi N}=\bar{u}(p')\bigg\{A(s,t)+\frac{1}{2}\left(\slashed{q}+\slashed{q}'\right)B(s,t)\bigg\}u(p),\quad \bar{u}(p')u(p')=2 m_{\rm f},\quad\bar{u}(p)u(p)=2m_{\rm i}\label{ampl_AB},
\eeq
is not the appropriate form for a low-energy expansion, since $A(s,t)$ does not fulfil chiral power counting due to a cancelation between the leading pieces of $A(s,t)$ and $B(s,t)$. 
Therefore,  $T_{\pi N}$ is parameterized in terms of the two amplitudes
$D(s,t)$ and $B(s,t)$ according to 
\beq
T_{\pi N}=\bar{u}(p')\bigg\{D(s,t)-\frac{1}{2(m_{\rm i}+m_{\rm f})}[\slashed{q}',\slashed{q}]B(s,t)\bigg\}u(p)
,\quad D(s,t)=A(s,t)+\nu B(s,t) \label{ampl_DB}.
\eeq
The low-energy expansion of $D(s,t)$ and $B(s,t)$ starts at $\Order(p)$ and $\Order(p^{-1})$ and is determined up to $\Order(p^{3})$ and $\Order(p)$ by our leading-loop calculation, respectively. 

In the isospin limit,  $T_{\pi N}$ may be decomposed as
\beq
T^{ab}=\chi_{N'}^\dagger \left(T^{+}\delta^{ab}+T^{-}\frac{1}{2}[\tau^a,\tau^b]\right)\chi_{N},
\eeq
where $\tau^a$ are the Pauli-matrices and $\chi_{N}$ ($\chi_{N'}$) is the isospinor for the incoming (outgoing) nucleon.
In the Condon--Shortley phase convention, the physical amplitudes are related
to $T^\pm$ by (we only display the channels to be considered later)
\begin{align}
T_{\pi^- p}&\equiv T_{\pi^- p\rightarrow \pi^- p} = T^++T^-,\notag\\
T_{\pi^+ p}&\equiv T_{\pi^+ p\rightarrow \pi^+ p} = T^+-T^-,\notag\\
T_{\pi^- p}^{\rm cex}&\equiv T_{\pi^- p\rightarrow \pi^0 n} = -\sqrt{2}\,T^-.
\end{align}
\subsection{Partial wave expansion of the amplitude}

Above threshold, the partial wave decomposition of the $\pi N$ scattering amplitude takes the form \cite{Goldberger}
\begin{align}
T_{\pi N}(s,t)&=\sqrt{2m_{\rm i}}\sqrt{2m_{\rm f}}\chi^\dagger_{\rm f}\sum\limits_{l=0}^{\infty}\Big\{\left((l+1)T_{l+}(s)+l T_{l-}(s)\right)P_l(z)\notag\\
&-\left(T_{l+}(s)-T_{l-}(s)\right)i\boldsymbol\sigma\cdot\left(\mathbf{\hat{q}'}\times\mathbf{\hat{q}}\right)P_l'(z)\Big\}\chi_{\rm i},\quad P_l'(z)=\frac{\diff}{\diff z}P_l(z),\quad \mathbf{\hat{q}}=\frac{\mathbf{q}}{|\mathbf{q}|}\label{Goldberger},
\end{align}
where the $T_{l\pm}$ have definite orbital momentum $l$ and total angular momentum $j=l\pm 1/2$.
In the following, we consider the partial waves $f_{l\pm}$ defined by
\beq
T_{l\pm}(s)=\frac{8\pi\sqrt{s}}{\sqrt{|\mathbf{q}||\mathbf{q'}|}}\frac{1}{\cot\delta_{l\pm}-i}=
\frac{8\pi\sqrt{s}f_{l\pm}(s)}{\sqrt{2m_{\rm i}}\sqrt{2m_{\rm f}}},
\eeq
choosing the normalization such that $f_{0+}$ corresponds to the $S$-wave scattering length. The alternative representation of the scattering amplitude
\begin{align}
T_{\pi N}(s,t)&=\sqrt{2m_{\rm i}}\sqrt{2m_{\rm f}}\mathcal{N}_{\rm i}\mathcal{N}_{\rm f}\chi^\dagger_{\rm f}\left\{g(s,t)+i\boldsymbol\sigma\cdot\left(\mathbf{q'}\times\mathbf{q}\right)h(s,t)\right\}\chi_{\rm i},\notag\\
 \mathcal{N}_{\rm i}&=\sqrt{\frac{E_p+m_{\rm i}}{2m_{\rm i}}},\quad \mathcal{N}_{\rm f}=\sqrt{\frac{E_{p'}+m_{\rm f}}{2m_{\rm f}}},\label{HBCHPT}
\end{align}
is especially suitable for a calculation in heavy-baryon ChPT. In a non-relativistic expansion, $h(s,t)$ and $g(s,t)$ can be interpreted as the spin-flip and spin-non-flip part of the amplitude, respectively. 
The inversion of \eqref{Goldberger} and \eqref{HBCHPT} reads
\beq
f_{l\pm}(s)=\frac{\sqrt{E_p+m_{\rm i}}\sqrt{E_{p'}+m_{\rm f}}}{16\pi\sqrt{s}}\int\limits_{-1}^1\diff z\left\{g(s,t)P_l(z)
+h(s,t)|\mathbf{q}||\mathbf{q'}|\left(P_{l\pm 1}(z)-z P_l(z)\right)\right\},
\eeq
which can be rewritten in terms of $A(s,t)$ and $B(s,t)$ as
\begin{align}
f_{l\pm}(s)&=\frac{\sqrt{E_p+m_{\rm i}}\sqrt{E_{p'}+m_{\rm f}}}{16\pi\sqrt{s}}\left\{A_l(s)+\left(\sqrt{s}-\frac{m_{\rm i}+m_{\rm f}}{2}\right)B_l(s)\right\}\notag\\
&+\frac{\sqrt{E_p-m_{\rm i}}\sqrt{E_{p'}-m_{\rm f}}}{16\pi\sqrt{s}}\left\{-A_{l\pm 1}(s)+\left(\sqrt{s}+\frac{m_{\rm i}+m_{\rm f}}{2}\right)B_{l\pm 1}(s)\right\}\label{f_CHPT},
\end{align}
where we have defined the partial wave projections
\beq
A_l(s)=\int\limits_{-1}^1\diff z A(s,t)P_l(z),\quad B_l(s)=\int\limits_{-1}^1\diff z B(s,t)P_l(z).
\eeq
In the isospin limit, \eqref{f_CHPT} coincides with the expressions given in~\cite{Hoehler}. 
As pointed out in \cite{FM00_above}, it is favorable not to work with the $P$-waves $P_1(s)=f_{1-}(s)$ and $P_3(s)=f_{1+}(s)$ directly,
since in the $P_1$-projection the isospin conserving amplitude itself becomes very small, such that the expected large (relative) isospin breaking effect is experimentally not amenable. As we will see in Sect.~\ref{sec:triangle_res}, the results in this projection are very sensitive to the precise values of the LECs and therefore meaningful quantitative statements cannot be made.\footnote{This is similar to isospin-breaking corrections to the $\pi^0 N$ scattering lengths and their normalization with respect to the isoscalar scattering length $a^+$, cf.~\cite{HKM09}.}  Instead, we use $\mathcal{G}(s)$ and $\mathcal{H}(s)$ defined by
\beq
P_3(s)=\mathcal{G}(s)+\mathcal{H}(s),\quad P_1(s)=\mathcal{G}(s)-2\mathcal{H}(s).
\eeq
Actually, these projections exactly correspond to the spin-flip and spin-non-flip amplitudes
\begin{align}
\mathcal{G}(s)&=\frac{\sqrt{E_p+m_{\rm i}}\sqrt{E_{p'}+m_{\rm f}}}{16\pi\sqrt{s}}\int\limits_{-1}^1\diff z \,g(s,t)P_1(z),\notag\\
\mathcal{H}(s)&=\frac{\sqrt{E_p+m_{\rm i}}\sqrt{E_{p'}+m_{\rm f}}}{16\pi\sqrt{s}}\int\limits_{-1}^1\diff z\,
h(s,t)|\mathbf{q}||\mathbf{q'}|\left(P_2(z)-z P_1(z)\right).
\end{align}

\subsection{Effective Lagrangian and chiral counting}
\label{sec:Leff}

The framework to systematically analyze isospin breaking in pion--nucleon
scattering is baryon chiral perturbation theory. In this section, we
briefly lay out the framework. Our starting point is the chiral
effective Lagrangian,
\beq\label{eq:Leff}
{\cal L}_{\rm eff} = {\cal L}_{\pi N}^{(1)} +
{\cal L}_{\pi N}^{(2)} + {\cal L}_{\pi N}^{(3)} 
+ {\cal L}_{\pi}^{(2)} + {\cal L}_{\pi}^{(4)} 
+ \cdots \, , 
\eeq
formulated in terms of the asymptotically observable pion and nucleon fields,
and including also scalar sources and virtual photons to incorporate the
effects of strong and electromagnetic isospin breaking. The superscript in 
(\ref{eq:Leff}) denotes the chiral dimension, the ellipsis represents terms 
of higher order not needed in our investigation.  The couplings
appearing in ${\cal L}_{\rm eff}$ are called low-energy constants (LECs),
they encode information about the higher mass states of QCD that are
not active degrees of freedom of the effective field theory.

All monomials in the pertinent
fields and sources that make up the terms of the effective Lagrangian
are constructed according to the following counting rules:
\beq\label{eq:rules}
\mpi\sim\Order(p),~~~
m\sim\Order(1),~~~  e\sim\Order(p),~~~ 
m_u,m_d \sim\Order(p^2),~~~
 t\sim\Order(p^2),~~~
s-s_{\rm thr}\sim\Order(p),
\eeq
where $p$ is a genuine small parameter and $s_{\rm thr}$ denotes the threshold energy squared 
of the corresponding channel. Note that pion four- and
nucleon three-momenta are of $\Order(p)$. In what follows, we work to leading
order in isospin breaking, collected in the small  parameter $\delta$,
\beq
\delta = \{ m_{\rm d}-m_{\rm u}, e^2\}~.
\eeq
The precise form of the effective Lagrangian employed is given in 
App.~\ref{app:Leff}.

Based on the effective Lagrangian, the pion--nucleon scattering amplitude
acquires a chiral expansion in terms of tree and loop graphs. In this paper,
we work to third order in the chiral expansion.  All the various contributions 
will be developed in detail in the following sections, here we
only make a few general remarks. Tree diagrams start at $\Order(p)$ (including
one-photon-exchange graphs) and obtain strong and electromagnetic corrections
at orders  $\Order(p^2)$ and $\Order(p^3)$. One-loop graphs that
perturbatively restore unitarity start contributing at  $\Order(p^3)$.
Since we work in a covariant formulation of baryon chiral perturbation theory,
we utilize the infrared regularization method developed in~\cite{BL99} to deal
with the power counting violations generated by the nucleon mass, $m \sim
\Lambda_\chi$ with  $\Lambda_\chi = 4\pi F_\pi \simeq 1\,$GeV the scale of
chiral symmetry breaking. This
method allows for a consistent power counting by uniquely separating the soft
from the hard parts in any given one-loop diagram. In a nutshell, all one has
to do is a rearrangement of the Feynman parameter integration corresponding to the combination of light and heavy degrees of freedom,
\beq
\int_0^1 \diff z (\ldots) \to \int_0^\infty \diff z  (\ldots)
-  \int_1^\infty \diff z  (\ldots)
= I + R~,
\eeq
where the infrared singular part $I$ contains the soft (chiral) physics and
obeys power counting while the regular part, generated from the loop momenta
of the order of the nucleon mass, leads to power counting violations. However,
it can be shown that the contributions to $R$ are polynomials in external 
momenta and the pion mass squared. They may therefore be absorbed in the low-energy
constants of the effective Lagrangian and need not to be considered
explicitly. Moreover, the infrared-regularization prescription amounts to resumming $1/m$ insertions in certain heavy-baryon diagrams. As a consequence of these higher-order contributions, exact renormalization no longer works: a residual dependence on the renormalization scale $\mu$ is generated. Finally, the central values of our results will be given for a scale $\mu=1\,{\rm GeV}$, while the sensitivity to the variation of $\mu$ can be regarded as an indication how large the impact of higher orders will at least be. 
For more details on the various formulations and facets of baryon 
chiral perturbation theory, we refer to the recent review \cite{Bernard:2007zu}.

\subsection{Leading-order result (tree graphs)}
\label{sec:lead_ord}

\begin{figure}
\begin{center}
\includegraphics[width=\linewidth]{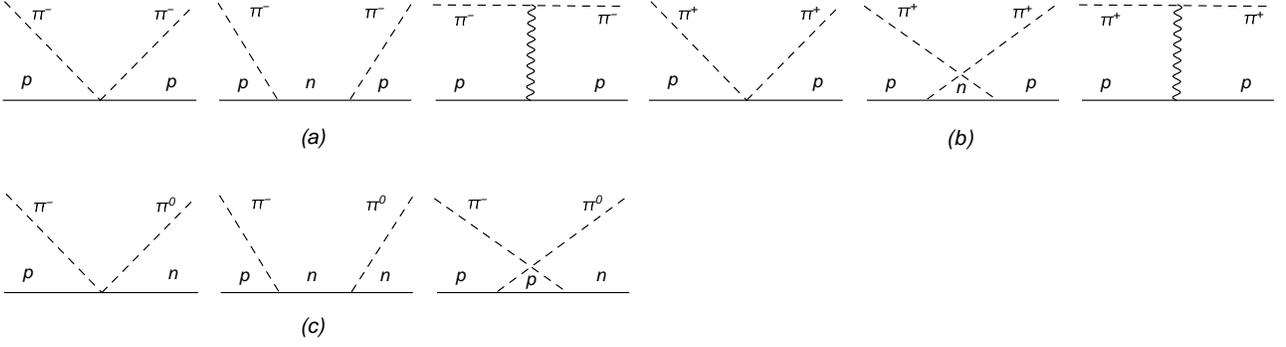}
\end{center}
\caption{Feynman diagrams for $\pi N$ scattering at $\Order(p)$. Solid, dashed, and wiggly lines, denote nucleons, pions, and photons, respectively.}
\label{fig:leading_order}
\end{figure}

In this work, we derive the full amplitude for the channels $\pi^\pm p\rightarrow \pi^\pm p$ and $\pi^-p\rightarrow \pi^0 n$ (referred to as the charge exchange reaction $T_{\rm cex}\equiv T_{\pi^- p}^{\rm cex}$), which are the only ones directly accessible in $\pi N$ scattering experiments.

First of all, we write down the leading-order amplitudes, which will be needed for future reference. For this purpose, all nucleon masses may be replaced by $m$. The diagrams contributing at $\Order(p)$ are depicted in Fig.~\ref{fig:leading_order} and yield within the given simplification
\begin{align}
B_{\pi^-p}^{\rm LO}(s,t)&=\frac{1}{2F^2}-\frac{g^2}{2F^2}\frac{s+3m^2}{s-m^2}-\frac{2e^2}{t},\quad 
B_{\pi^+p}^{\rm LO}(s,t)=-\frac{1}{2F^2}+\frac{g^2}{2F^2}\frac{u+3m^2}{u-m^2}+\frac{2e^2}{t},\notag\\
B_{\rm cex}^{\rm LO}(s,t)&=-\frac{\sqrt{2}}{2F^2}+\frac{\sqrt{2}\, g^2}{4F^2}\left(\frac{s+3m^2}{s-m^2}+\frac{u+3m^2}{u-m^2}\right),\quad 
D_{\rm cex}^{\rm LO}(s,t)=\nu B_{\rm cex}^{\rm LO}(s,t),\notag\\
D_{\pi^-p}^{\rm LO}(s,t)&=\frac{m g^2}{F^2}+\nu B_{\pi^-p}^{\rm LO}(s,t),\quad D_{\pi^+p}^{\rm LO}(s,t)=\frac{m g^2}{F^2}+\nu B_{\pi^+p}^{\rm LO}(s,t).\label{lead_order_ampl}
\end{align} 
In the context of bremsstrahlung and the cancelation of infrared divergences we will need the spin averaged squared matrix element 
($T_{\pi N}=\bar{u}\left(p'\right)\hat{T}_{\pi N}u(p)$)
\beq
|\M_{\pi N}|^2\equiv\frac{1}{2}\sum\limits_{\rm spins}|T_{\pi N}|^2=\frac{1}{2}\Tr\left\{\left(\slashed{p}'+m_{\rm f}\right)
\hat{T}_{\pi N}\left(\slashed{p}+m_{\rm i}\right)\gamma^0 \hat{T}_{\pi N}^\dagger \gamma^0\right\},
\eeq
for which we find at leading order
\begin{align}
|\M_{\pi^\pm p}^{\rm LO}|^2&=\left(4m^2-t\right)\left(D^{\rm LO}_{\pi^\pm p}\right)^2+2\nu t D^{\rm LO}_{\pi^\pm p} B^{\rm LO}_{\pi^\pm p}-\frac{t}{4}\left\{t-4\mpi^2+4\nu^2\right\}
\left(B^{\rm LO}_{\pi^\pm p}\right)^2,\notag\\
|\M_{\rm cex}^{\rm LO}|^2&=\left(4m^2-t\right)\left(D^{\rm LO}_{\rm cex}\right)^2+2\nu t D^{\rm LO}_{\rm cex} B^{\rm LO}_{\rm cex}
-\frac{1}{4}\left\{\lambda\left(t,\mpi^2,\mpii^2\right)+4t\nu^2\right\}\left(B^{\rm LO}_{\rm cex}\right)^2.\label{squ_matr_lead_ord}
\end{align}

\subsection{Tree diagrams beyond leading order}
\label{sec:tree}

\begin{figure}
\begin{center}
\includegraphics[width=0.9\linewidth]{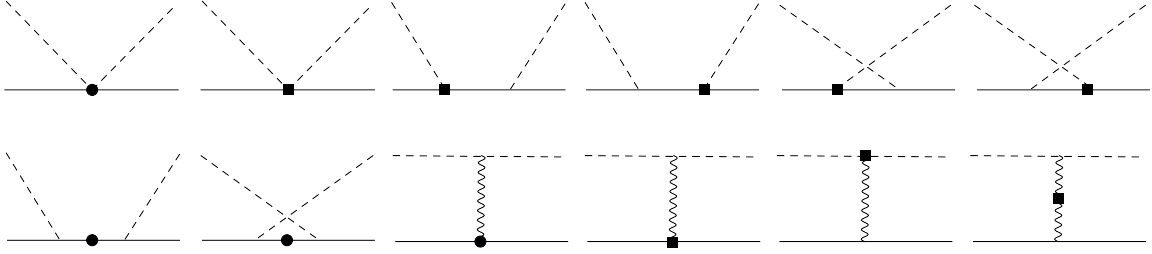}
\end{center}
\caption{Contact-term contributions. Heavy dots/squares refer to insertions suppressed by one/two chiral orders, respectively.}
\label{fig:counter_terms}
\end{figure}

As depicted in Fig.~\ref{fig:counter_terms}, there are four types of tree diagrams involving vertices beyond leading order: contact-term corrections to the Weinberg--Tomozawa (WT) term,  to the $\pi NN$ coupling and the one photon exchange, and mass insertions for the intermediate nucleon.
The effect of the latter can be accounted for by shifting the bare mass $m$ in the Lagrangian according to 
\begin{align}
m&\rightarrow \tilde{m}_{\rm p}=m-4c_1\mpii^2+2B c_5\left(m_{\rm d}-m_{\rm u}\right)-\frac{e^2F^2}{2}\left(f_1+f_2+f_3\right)\notag\\
&=\mpp-\frac{m g^2}{2F^2}\left(2\mpi^2\bar{I}_\pi\left(m^2 \right)+\mpii^2\bar{I}_\npi\left(m^2 \right)\right)=\mpp+\frac{g^2}{32\pi F^2}\left(2\mpi^3+\mpii^3\right)+\Order(p^4),\notag\\
m&\rightarrow\tilde{m}_{\rm n}=m-4c_1\mpii^2-2B c_5\left(m_{\rm d}-m_{\rm u}\right)-\frac{e^2F^2}{2}\left(f_1-f_2+f_3\right)\label{nucl_masses}\\
&=\mn-\frac{m g^2}{2F^2}\left(2\mpi^2\bar{I}_\pi\left(m^2 \right)+\mpii^2\bar{I}_\npi\left(m^2 \right)\right)=\mn+\frac{g^2}{32\pi F^2}\left(2\mpi^3+\mpii^3\right)+\Order(p^4),\notag
\end{align}
for the proton and neutron, respectively, where $\bar{I}_\pi$ is defined in App.~\ref{Scal_loop_func}. Actually, we must carefully distinguish between the bare mass $m$ in the Lagrangian, which is nothing but the mass of the nucleon in the chiral limit, and the masses of the proton and neutron entering the calculation by the requirement that the external particles are on their physical mass shell. However, in loops we can still use $m$ as a generic nucleon mass to be identified with $\mpp$ in the end, since the difference $\mpp-m$ only starts at second chiral order and is therefore beyond the accuracy we are working at. For the same reason, we may replace $m$ by $\mpp$ directly in the case of the contact-term contributions.
Nevertheless, once the shift \eqref{nucl_masses} in the mass parameters of the Lagrangian is performed, the error which is made when identifying $m\rightarrow \mpp$ is only of order $\Order(p^3)$. 

$\Lagr_{\rm N}^{(p^2)}$, $\Lagr_{\rm N}^{(p^3)}$, as well as $\Lagr_{\rm
  N}^{(e^2p)}$ (see App.~\ref{app:Leff} for precise definitions),  
yield corrections to the WT vertex. We will always consider $s$ and $t$ as the independent kinematical variables and $u$ to be fixed by the Mandelstam relation \eqref{Mandelstam}. Amplitudes originating from WT-type topologies and nucleon-pole diagrams will be referred to as vector- and axial-type contributions, respectively.  The full vector-type tree level amplitudes $D^{\rm v}(s,t)$ and $B^{\rm v}(s,t)$ read
\begin{align}
D^{\rm v}_{\pi^- p}(s,t)&=\frac{\nu}{2F^2}-\frac{4\mpii^2c_1}{F^2}+\frac{c_2}{8\mpp^2F^2}\left(16\mpp^2\nu^2-t^2\right)+\frac{c_3}{F^2}\left(2 \mpi^2-t\right)-\frac{e^2}{2}\left(4f_1+f_2\right)\notag\\
&+\frac{2\nu \left(d_1+d_2\right)}{F^2} \left(2\mpi^2-t\right)+\frac{d_3 \nu}{4\mpp^2F^2}\left(16\mpp^2\nu^2-t^2\right)
+\frac{8 \nu  \mpii^2 d_5}{F^2} +2e^2\nu \left(g_6+g_8\right),\notag\\
D^{\rm v}_{\pi^+ p}(s,t)&=-\frac{\nu}{2F^2}-\frac{4\mpii^2c_1}{F^2}+\frac{c_2}{8\mpp^2F^2}\left(16\mpp^2\nu^2-t^2\right)+\frac{c_3}{F^2}\left(2 \mpi^2-t\right)-\frac{e^2}{2}\left(4f_1+f_2\right)\notag\\
&-\frac{2\nu\left(d_1+d_2\right)}{F^2} \left(2\mpi^2-t\right)-\frac{d_3 \nu}{4\mpp^2F^2}\left(16\mpp^2\nu^2-t^2\right)-\frac{8 \nu  \mpii^2 d_5}{F^2} -2e^2\nu \left(g_6+g_8\right),\notag\\
\frac{1}{\sqrt{2}}D^{\rm v}_{\rm cex}(s,t)&=-\frac{\nu}{2F^2}+\frac{c_5B\left(m_{\rm d}-m_{\rm u}\right)}{F^2}+\frac{e^2f_2}{4}-\frac{\nu \left(d_1+d_2\right)}{F^2}\left(\mpi^2+\mpii^2-t\right)\frac{\mpp+\mn}{\mpp}\notag\\
 &-\frac{d_3}{32\mpp^3F^2}\Big\{\left(2(\mpp+\mn)\nu -\Delta_\pi\right)^2-\left(t+\mpp^2-\mn^2\right)^2\Big\}\left(2(\mpp+\mn)\nu -\Delta_\pi\right)\notag\\
 &-\frac{d_3}{32\mpp^3F^2}\Big\{\left(2(\mpp+\mn)\nu +\Delta_\pi\right)^2-\left(t-\mpp^2+\mn^2\right)^2\Big\}\left(2(\mpp+\mn)\nu +\Delta_\pi\right)\notag\\
 &-\frac{4\nu \mpii^2 d_5}{F^2}\frac{\mpp+\mn}{\mpp} +\frac{e^2 g_7}{8\mpp}\left(2(\mpp+\mn)\nu+\mpp^2-\mn^2\right),\notag\\ 
 B^{\rm v}_{\pi^- p}(s,t)&=\frac{1}{2F^2}+\frac{2\mpp c_4}{F^2}+\frac{4\mpp \nu \left(d_{14}-d_{15}\right)}{F^2},\quad 
  \frac{1}{\sqrt{2}}B^{\rm v}_{\rm cex}(s,t)=-\frac{1}{2F^2}-\frac{(\mpp+\mn) c_4}{F^2},\notag\\
 B^{\rm v}_{\pi^+ p}(s,t)&=-\frac{1}{2F^2}-\frac{2\mpp c_4}{F^2}+\frac{4\mpp \nu\left(d_{14}-d_{15}\right)}{F^2}.
\end{align}
Corrections to the $\pi NN$ coupling are generated by $\Lagr_{\rm N}^{(p^3)}$ and $\Lagr_{\rm N}^{(e^2p)}$. They may be incorporated replacing $g$ by
\begin{align}
\tilde{g}&=g+2\mpii^2\left(2d_{16} - d_{18}\right)+e^2F^2\left(g_1 +g_2 \right),\label{ax_coupl_shift}\\
\tilde{g}_{\rm p}&=g+2\mpii^2\left(2d_{16} -d_{18} \right)+e^2F^2\left(g_1 +g_2+g_3+g_4\right)
-2B\left(m_{\rm d}-m_{\rm u}\right)\left(2d_{17}-d_{18}-2d_{19}\right),\notag\\
\tilde{g}_{\rm n}&=g+2\mpii^2\left(2d_{16} -d_{18} \right)+e^2F^2\left(g_1 +g_2-g_3\right)+2B\left(m_{\rm d}-m_{\rm u}\right)\left(2d_{17}-d_{18}-2d_{19}\right),\notag
\end{align}
for $p n$, $p p$ and $n n$ couplings, respectively. In terms of these quantities, we find for the full axial-type tree level amplitudes $A^{\rm a}(s,t)$ and $B^{\rm a}(s,t)$:
\begin{align}
A^{\rm a}_{\pi^-p}(s)&=\frac{\tilde{g}^2}{2F^2}(\mpp+\mn)\frac{s-\mpp^2}{s-\mn^2}-\frac{\Deltam g^2}{2F^2}\frac{s-\mpp^2}{\left(s-\mn^2\right)^2}\left(s+\mn^2+2\mn \mpp\right),\notag\\
B^{\rm a}_{\pi^-p}(s)&=-\frac{\tilde{g}^2}{2F^2}\frac{s+\mpp^2+2\mpp \mn}{s-\mn^2}+\frac{\Deltam g^2}{F^2}(\mpp+\mn)\frac{s+\mpp \mn}{\left(s-\mn^2\right)^2},\notag\\
A^{\rm a}_{\pi^+p}(s,t)&=A^{\rm a}_{\pi^-p}(u),\quad B^{\rm a}_{\pi^+p}(s,t)=-B^{\rm a}_{\pi^-p}(u),\notag\\
\frac{1}{\sqrt{2}}A^{\rm a}_{\rm cex}(s,t)&=-\frac{\tilde{g}\tilde{g}_{\rm n}}{4F^2}\frac{1}{s-\mn^2}\Big\{\left(s-\mn^2\right)(\mpp+\mn)+\frac{\Delta_{\rm N}}{2}\left(s+2\mn \mpp+\mn^2\right)\Big\}\notag\\
&+\frac{\Deltam g^2}{4F^2}\bigg\{\frac{s+2\mpp \mn +\mn^2}{s-\mn^2}+\frac{\Delta_{\rm N}}{2}\frac{\mn\left(3s+\mn^2\right)+\mpp \left(s+3\mn^2\right)}{\left(s-\mn^2\right)^2}\bigg\}\notag\\
&+\frac{\tilde{g}\tilde{g}_{\rm p}}{4F^2}\frac{1}{u-\mpp^2}\Big\{\left(u-\mpp^2\right)(\mpp+\mn)-\frac{\Delta_{\rm N}}{2}\left(u+2\mn \mpp+\mpp^2\right)\Big\}\notag\\
&-\frac{\Deltam g^2}{4F^2}\bigg\{\frac{u+2\mpp \mn +\mpp^2}{u-\mpp^2}-\frac{\Delta_{\rm N}}{2}\frac{\mpp\left(3u+\mpp^2\right)+\mn\left(u+3\mpp^2\right)}{\left(u-\mpp^2\right)^2}\bigg\},\notag\\
\frac{1}{\sqrt{2}}B^{\rm a}_{\rm cex}(s,t)&=\frac{\tilde{g}\tilde{g}_{\rm n}}{4F^2}\frac{s+\mn^2+2\mpp \mn}{s-\mn^2}-\frac{\Deltam g^2}{4F^2}\frac{s(3\mn+\mpp)+\mn^2(3\mpp+\mn)}{\left(s-\mn^2\right)^2}\notag\\
&+\frac{\tilde{g}\tilde{g}_{\rm p}}{4F^2}\frac{u+\mpp^2+2\mpp \mn}{u-\mpp^2}-\frac{\Deltam g^2}{4F^2}\frac{u(3\mpp+\mn)+\mpp^2(3\mn+\mpp)}{\left(u-\mpp^2\right)^2}\label{axial_tree},
\end{align}
where 
\beq
\Deltam=m_{\rm p/n}-\tilde{m}_{\rm p/n}=\frac{m g^2}{2F^2}\left(2\mpi^2\bar{I}_\pi\left(m^2\right)+\mpii^2\bar{I}_\npi\left(m^2\right)\right)=-\frac{g^2\left(2\mpi^3+\mpii^3\right)}{32\pi F^2}+\Order(p^4),
\eeq
and $\tilde{g}^2$, $\tilde{g}\tilde{g}_{\rm p}$, and $\tilde{g}\tilde{g}_{\rm n}$, are understood such that only terms linear in the LECs should be retained. We have checked explicitly that the double poles in \eqref{axial_tree} and the loop diagrams $(s_1)$, $(s_{14})$, and $(a_{10})$, indeed cancel, as required by the analytic properties of the $S$-matrix.

Finally, $\Lagr_{\rm N}^{(p^2)}$, $\Lagr_{\rm N}^{(p^3)}$, $\Lagr_{\pi}^{(p^4)}$, and $\Lagr_{\pi}^{(e^2p^2)}$, provide contact-term contributions to the one photon exchange. The full tree amplitudes $D^{\rm p}(s,t)$ and $B^{\rm p}(s,t)$ turn out to be
\begin{align}
D^{\rm p}_{\pi^- p}(t)&=-D^{\rm p}_{\pi^+ p}(t)=-\frac{2e^2\nu}{t}+\frac{2e^2\nu}{F^2}l_6+2e^2\nu \left(d_6+2d_7\right)-\frac{40e^4\nu }{9t}\left(k_1+ k_2 + l_5 - 2h_2 + h_4\right),\notag\\
B^{\rm p}_{\pi^- p}(t)&=-B^{\rm p}_{\pi^+ p}(t)=-\frac{2e^2}{t}\left(1+c_6+c_7\right)+\frac{2e^2}{F^2} l_6-\frac{40e^4}{9t}\left(k_1+ k_2 + l_5 - 2h_2 + h_4\right).
\end{align}

\subsection[Renormalization of $F$ and $g$]{Renormalization of \boldmath{$F$} and \boldmath{$g$}}

For the numerical evaluation, we aim to replace the pion decay constant $F$ and the axial charge of the nucleon $g$ in the chiral limit by the physical values $\Fpi$ and $\ga$, respectively. Since both obtain corrections suppressed by two chiral orders, this affects only the leading-order diagrams. The explicit expressions in the isospin limit read \cite{GL84}
\beq
\Fpi=F\left\{1+\frac{\mpi^2}{F^2}\left(l_4^{\rm r}-\frac{1}{16\pi^2}\log\frac{\mpi^2}{\mu^2}\right)\right\}+\Order(\mpi^4),
\eeq
and \cite{BKLM94,BL01}
\beq
\ga=g\left\{1+\frac{4d_{16}^{\rm r}\mpi^2}{g}-\frac{\left(2g^2+1\right)\mpi^2}{16\pi^2F^2}\log\frac{\mpi^2}{\mu^2}-\frac{g^2\mpi^2}{16\pi^2F^2}\right\}+\Order(\mpi^3)\label{ga_chiral}.
\eeq
Numerically, we use $\Fpi=\left(92.2\pm 0.2\right)\,{\rm MeV}$ and $|\ga|=1.2695\pm 0.0029$ as quoted in \cite{PDG08}. We do not consider isospin breaking in $F_\pi$ and $\ga$.\footnote{In fact, radiative corrections are taken into account when extracting $\Fpi$ and $\ga$ from experiment. Moreover,  the final accuracy of our results will be rather limited by unknown LECs than by the uncertainties in $\ga$ and $\Fpi$.} However, to ensure consistency with the loop contributions, the chirally expanded version \eqref{ga_chiral} is replaced by the corresponding expression calculated in infrared regularization ($\bar{I}_{\rm A}^\pi$ and $\bar{I}_\pi$ are defined in App.~\ref{Scal_loop_func})
\begin{align}
\ga&=g\bigg\{1+\frac{g^2}{8F^2}\left(\mpi^2\left(20m^2-13\mpi^2\right)\bar{I}_{\rm A}^\pi(0)-11\mpi^2\bar{I}_\pi\left(m^2\right)-\frac{3\mpi^2}{8\pi ^2}\log\frac{\mpi^2}{\mu^2}\right)\notag\\
&+\frac{4d_{16}^{\rm r}\mpi^2}{g}-\frac{2\mpi^2}{F^2} \bar{I}_\pi\left(m^2\right)+\frac{g^2\mpi^2\left(2m^2-\mpi^2\right)}{128\pi ^2m^2F^2}-\frac{\mpi^2}{16\pi ^2F^2}\log\frac{\mpi^2}{\mu^2}\bigg\}.
\end{align}
Finally, we remark that the dependence on $d_{16}$ cancels between the renormalization of $\ga$ and the corrections to the nucleon-pole diagrams.

\subsection{Wave function renormalization}
\label{sec:wavefunction}

As a consequence of the LSZ reduction formula, the full $\pi N$ scattering amplitude depends on the wave function renormalization of pion and nucleon fields according to
\beq
T_{\pi N}=\sqrt{Z^{\rm f}_{\rm N}}\sqrt{Z^{\rm i}_{\rm N}}\sqrt{Z^{\rm f}_{\pi}}\sqrt{Z^{\rm i}_{\pi}}\bar{u}(p')\hat{T}_{\pi N} u\left(p\right)\label{scatt_ampl_wave_renom},
\eeq
where $\hat{T}_{\pi N}$ is the amputated amplitude, while $Z^{\rm i}_{\rm N}$ and $Z^{\rm i}_{\pi}$ ($Z^{\rm f}_{\rm N}$ and $Z^{\rm f}_{\pi}$) denote the wave function renormalization of the initial (final) nucleon and pion, respectively.

\begin{figure}
\begin{center}
\includegraphics[width=0.82\linewidth]{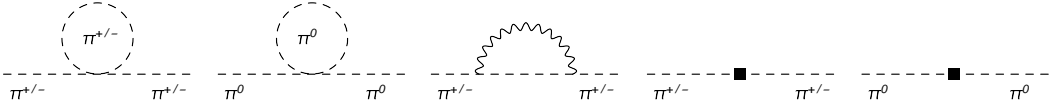}
\end{center}
\caption{Wave function renormalization of the charged and the neutral pion.}
\label{fig:z_pion}
\end{figure}

\begin{figure}
\begin{center}
\includegraphics[width=0.82\linewidth]{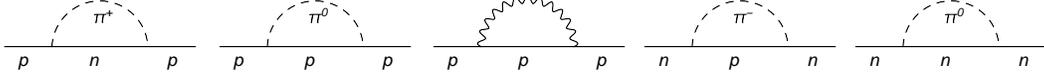}
\end{center}
\caption{Wave function renormalization of the nucleon.}
\label{fig:z_nucleon}
\end{figure}

The Feynman diagrams contributing to the wave function renormalization of pion and nucleon are shown in Figs.~\ref{fig:z_pion} and \ref{fig:z_nucleon}. The corresponding wave function renormalization factors can be deduced from the self energy $\Sigma(p)$ via 
\beq
Z_{\rm N}=1+\frac{\diff \Sigma\left(\slashed{p}\right)}{\diff \slashed{p}}\bigg|_{\slashed{p}=m}+\cdots,\quad 
Z_{\pi}=1+\frac{\diff \Sigma(p)}{\diff p^2}\bigg|_{p^2=\mpi^2}+\cdots,
\eeq
where the ellipsis indicates higher orders. Note that the wave function renormalization factors depend on the parametrization of the pion fields.\footnote{At $\Order(p^3)$, $Z_{\rm N}$ is accidentally independent of the parametrization, since no vertices involving more than two pions occur.} In the $\sigma$ parametrization \eqref{sigma_param} we obtain\footnote{$C_{\pi}$, $L_{\pi}$, and the loop functions $I_{A}^\pi$, $I_\pi$, are defined in Apps.~\ref{app:def} and \ref{Scal_loop_func}, respectively.}
\begin{align}
Z_\pi&=1-\frac{e^2}{8\pi^2}\left(L_{\pi}+1+C_{\pi}\right)-\frac{\mpi^2C_{\pi}}{16\pi ^2F^2}-\frac{20}{9} e^2\left(k_1+k_2\right),\notag\\
Z_\npi&=1-\frac{\mpii^2C_{\pi ^0}}{16\pi ^2F^2}-\frac{20}{9} e^2\left(k_1+k_2\right)-2e^2\left(k_4-2k_3\right),\notag\\
Z_{\rm n}&=1+\frac{g^2}{4F^2}\bigg\{2\left(4m^2-2\mpi^2\right)\mpi^2I_{A}^\pi(0)+\left(4m^2-2\mpii^2\right)\mpii^2I_{A}^\npi(0)-4 \mpi^2I_\pi\left(m^2\right)\notag\\
&-2 \mpii^2 I_\npi\left(m^2\right)-\frac{\mpi^2}{8\pi^2}C_{\pi}-\frac{\mpii^2}{16\pi^2} C_{\pi _0}\bigg\},\quad Z_{\rm p}=Z_{\rm n}-\frac{e^2}{8\pi ^2}\left(L_{\pi }+1+C_{\pi }\right).
\end{align}
Thus, the wave function renormalization factors are UV and (in the case of the charged particles) IR divergent and, with respect to the cancelation of divergences, constitute therefore an indispensable part of the amplitude. Their contributions $D^{\rm wf}(s,t)$ and $B^{\rm wf}(s,t)$ due to $\delta \pt Z=Z-1$ expressed in terms of the leading-order amplitudes \eqref{lead_order_ampl} read
\begin{align}
D^{\rm wf}_{\pi^\pm p}(s,t)&=\left(\delta\pt Z_\pi+\delta\pt Z_{\rm p}\right)D^{\rm LO}_{\pi^\pm p}(s,t),\quad B^{\rm wf}_{\pi^\pm p}(s,t)=\left(\delta\pt Z_\pi+\delta\pt Z_{\rm p}\right)B^{\rm LO}_{\pi^\pm p}(s,t),\notag\\
D^{\rm wf}_{\rm cex}(s,t)&=\frac{1}{2}\left(\delta\pt Z_\pi+\delta\pt Z_\npi+\delta\pt Z_{\rm p}+\delta\pt Z_{\rm n}\right)D^{\rm LO}_{\rm cex}(s,t),\notag\\
B^{\rm wf}_{\rm cex}(s,t)&=\frac{1}{2}\left(\delta\pt Z_\pi+\delta\pt Z_\npi+\delta\pt Z_{\rm p}+\delta\pt Z_{\rm n}\right)B^{\rm LO}_{\rm cex}(s,t)\label{wfr}.
\end{align}

\subsection{Loop diagrams}
\label{sec:loop_ampl}

\begin{figure}
\begin{center}
\includegraphics[width=\linewidth]{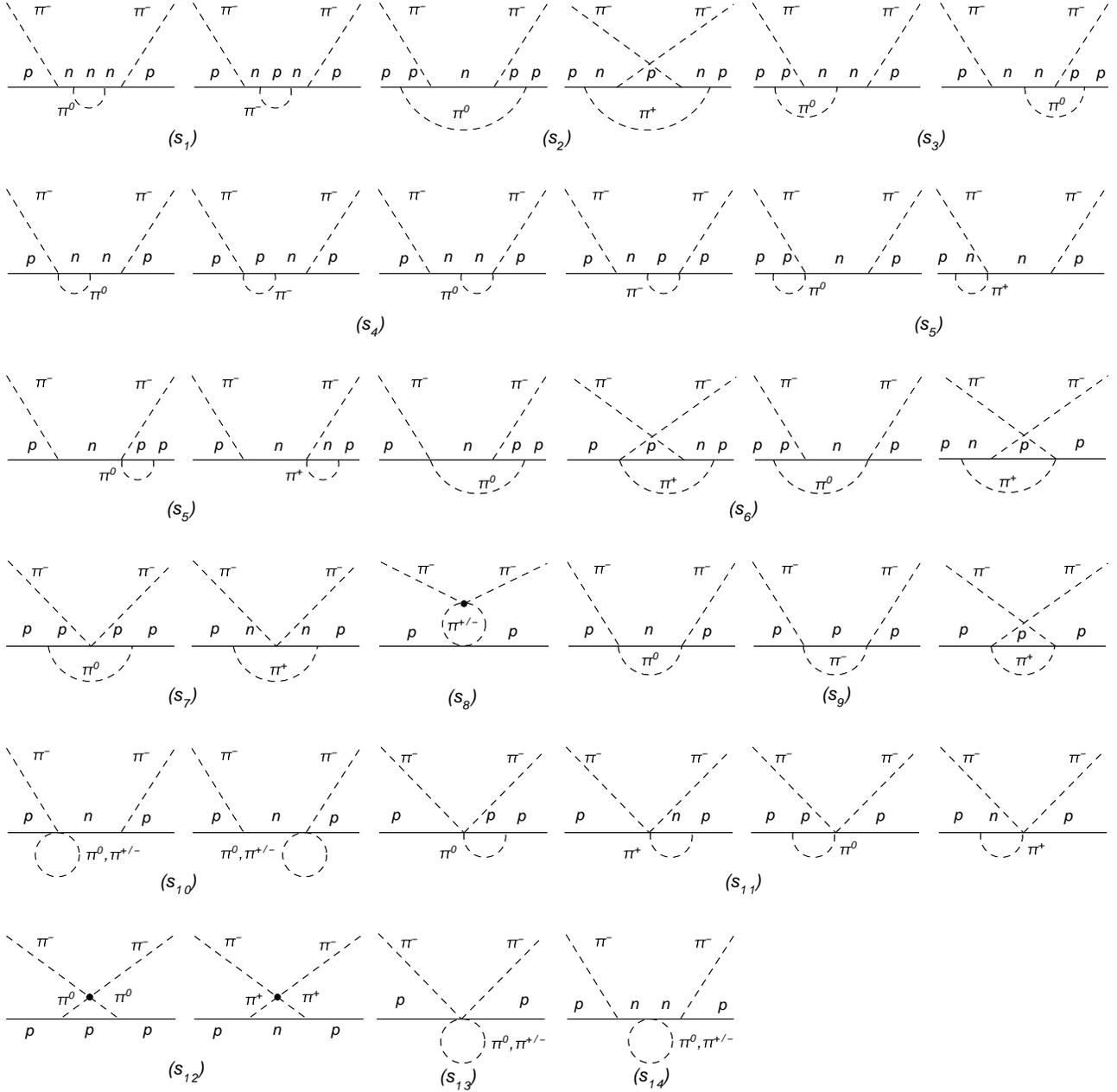}
\end{center}
\caption{Strong one-loop diagrams for $\pi^- p \rightarrow \pi^- p$.}
\label{fig:strong_a}
\end{figure}

\begin{figure}
\begin{center}
\includegraphics[width=\linewidth]{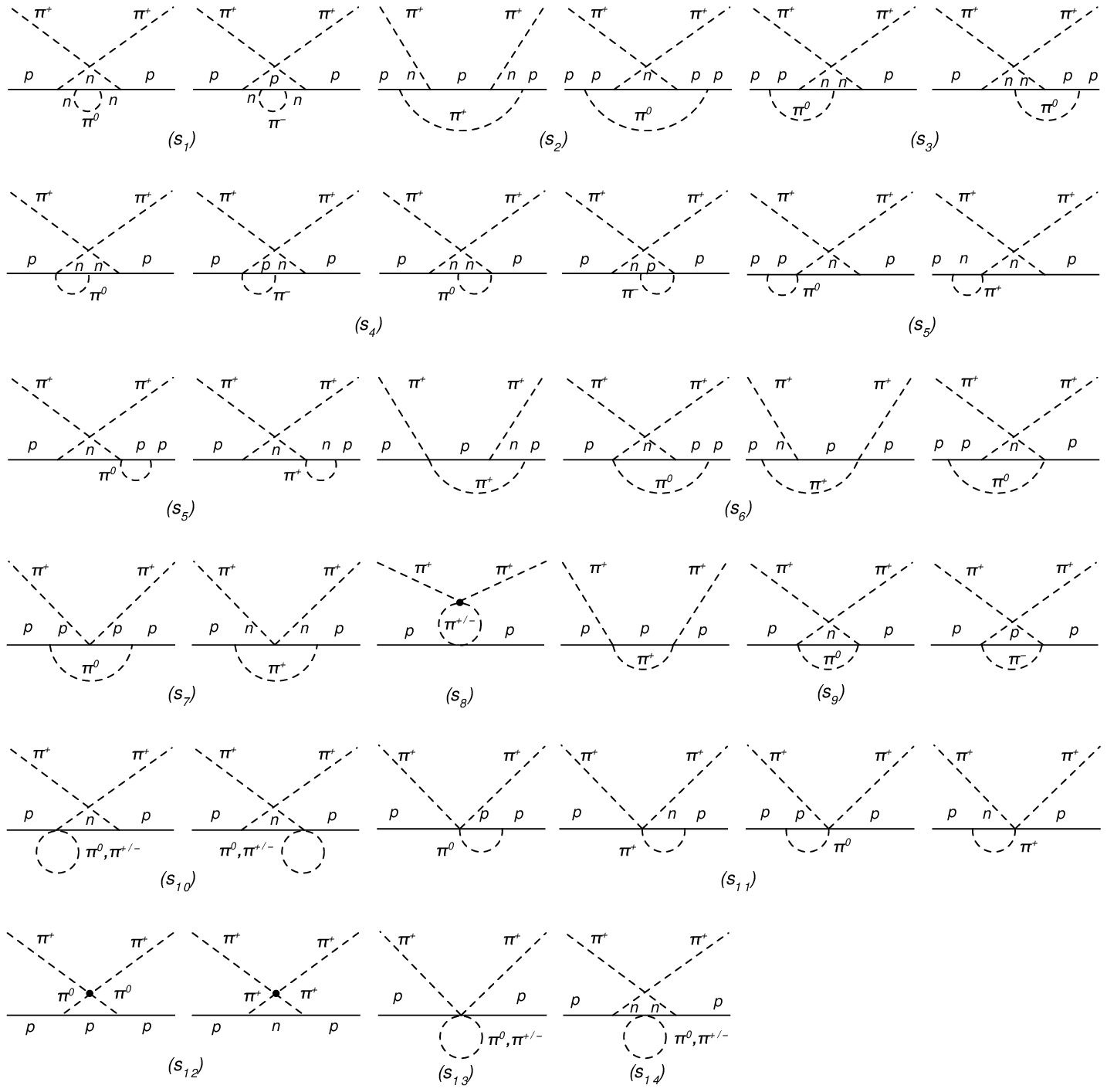}
\end{center}
\caption{Strong one-loop diagrams for $\pi^+ p \rightarrow \pi^+ p$.}
\label{fig:strong_b}
\end{figure}

\begin{figure}
\begin{center}
\includegraphics[width=0.968\linewidth]{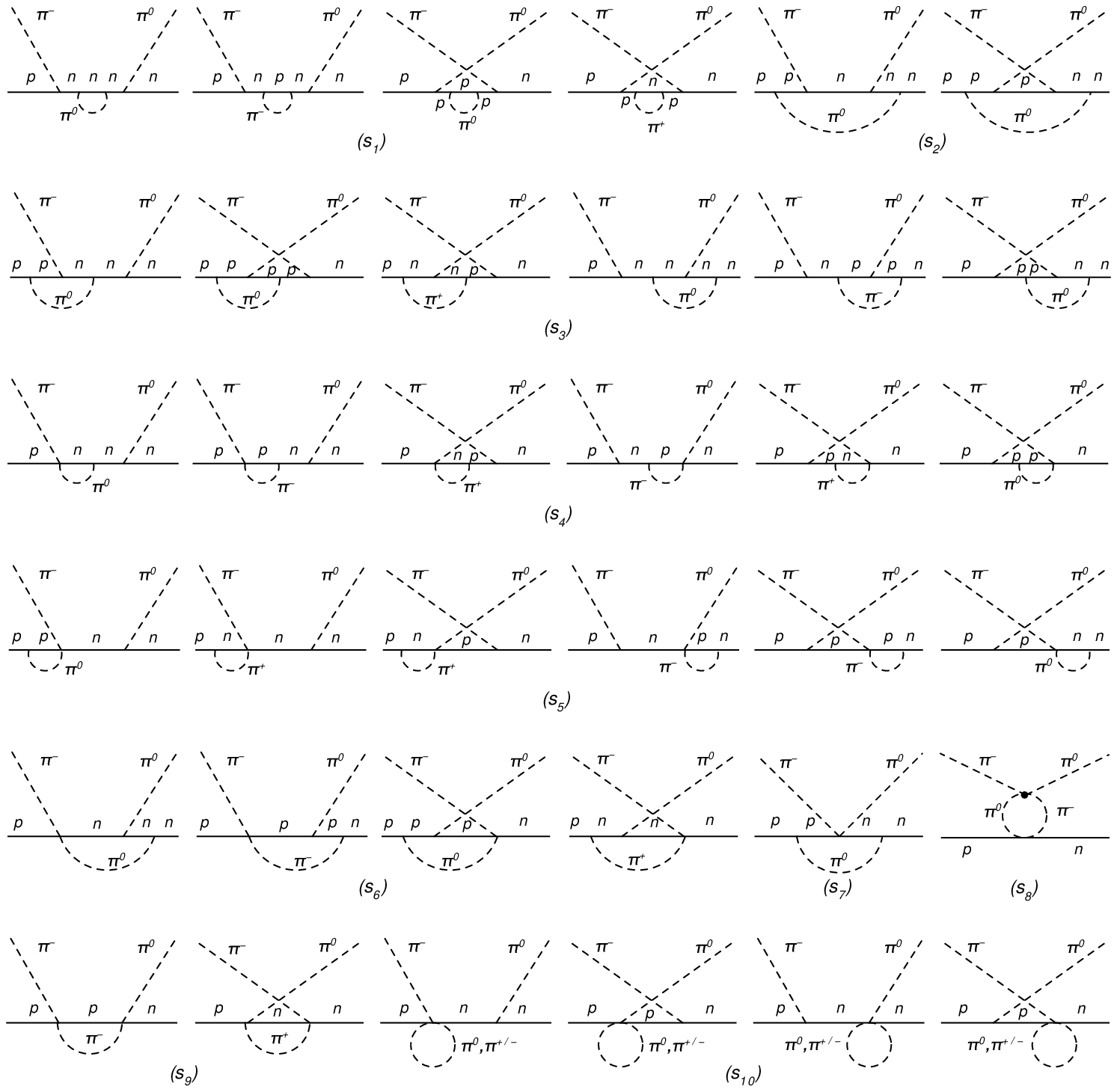}
\includegraphics[width=0.968\linewidth]{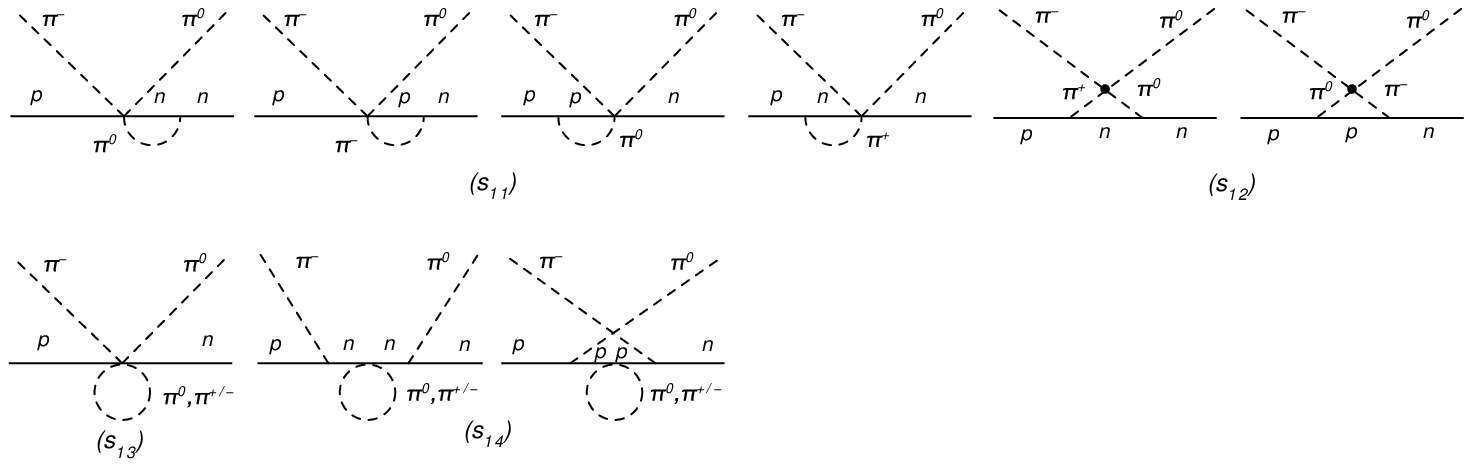}
\end{center}
\caption{Strong one-loop diagrams for $\pi^- p \rightarrow \pi^0 n$}
\label{fig:strong_c}
\end{figure}

\begin{figure}
\begin{center}
\includegraphics[width=\linewidth]{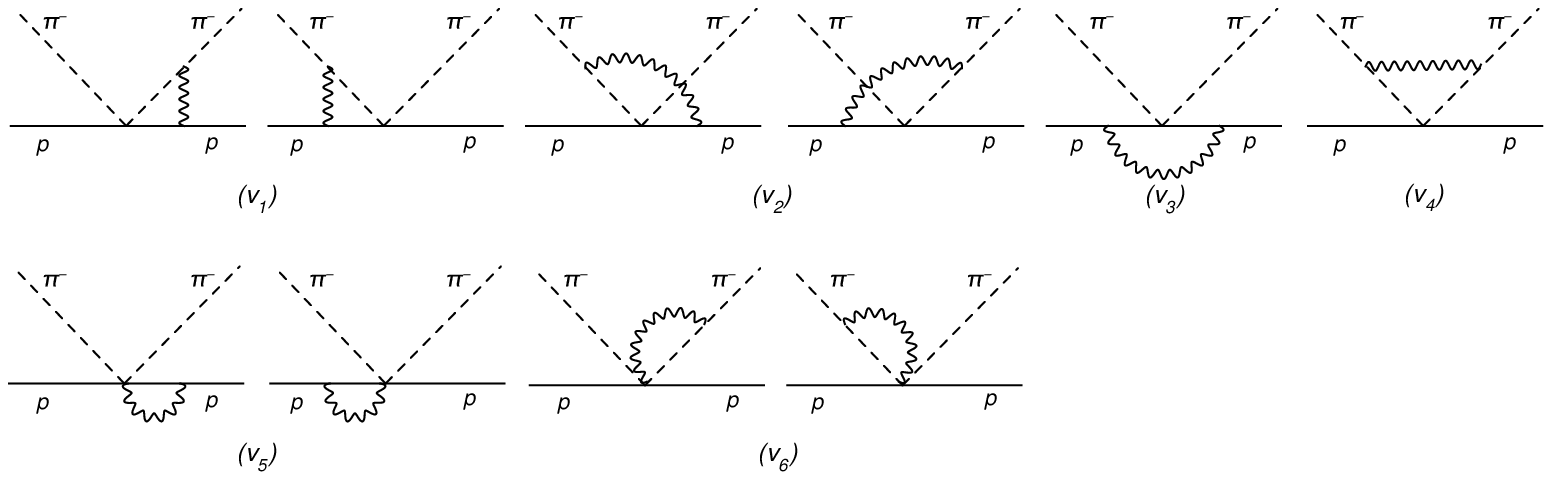}
\end{center}
\caption{Electromagnetic vector-type diagrams for $\pi^- p \rightarrow \pi^- p$.}
\label{fig:vector_a}
\end{figure}

\begin{figure}
\begin{center}
\includegraphics[width=\linewidth]{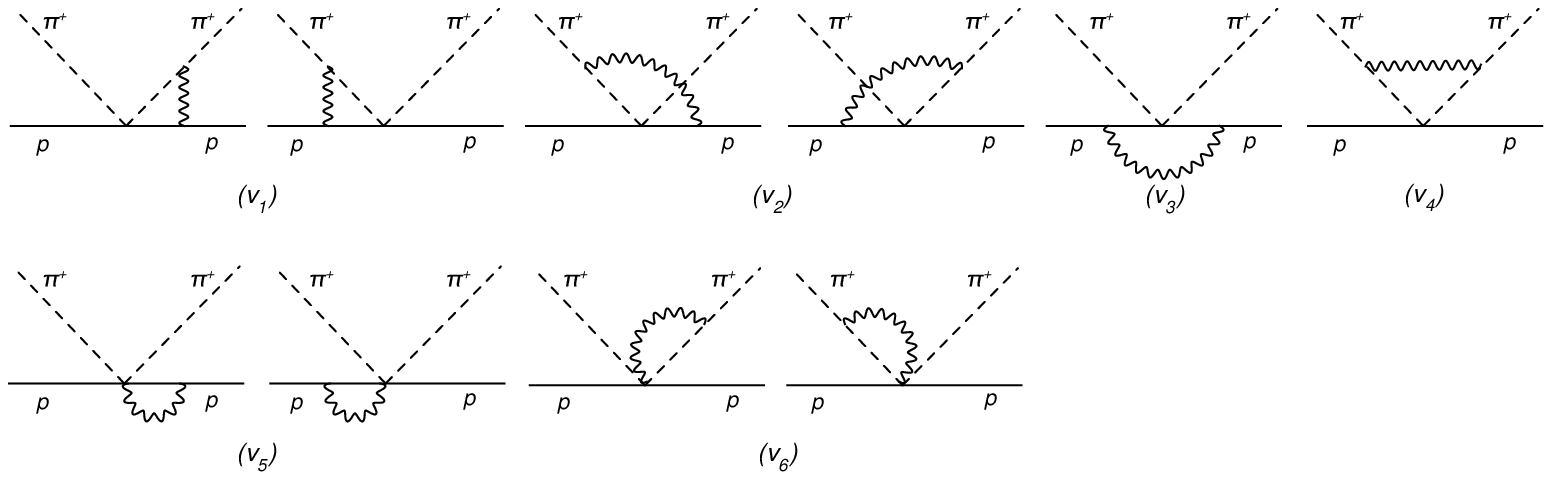}
\end{center}
\caption{Electromagnetic vector-type diagrams for $\pi^+ p \rightarrow \pi^+ p$.}
\label{fig:vector_b}
\end{figure}

\begin{figure}
\begin{center}
\includegraphics[width=0.5\linewidth]{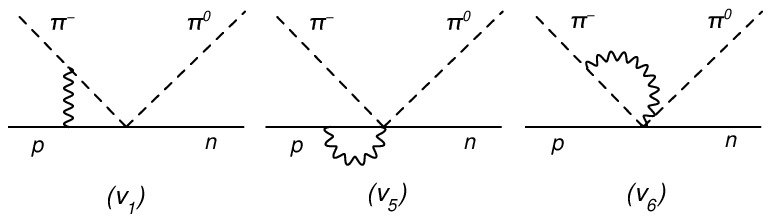}
\end{center}
\caption{Electromagnetic vector-type diagrams for $\pi^- p \rightarrow \pi^0 n$.}
\label{fig:vector_c}
\end{figure}

\begin{figure}
\begin{center}
\includegraphics[width=\linewidth]{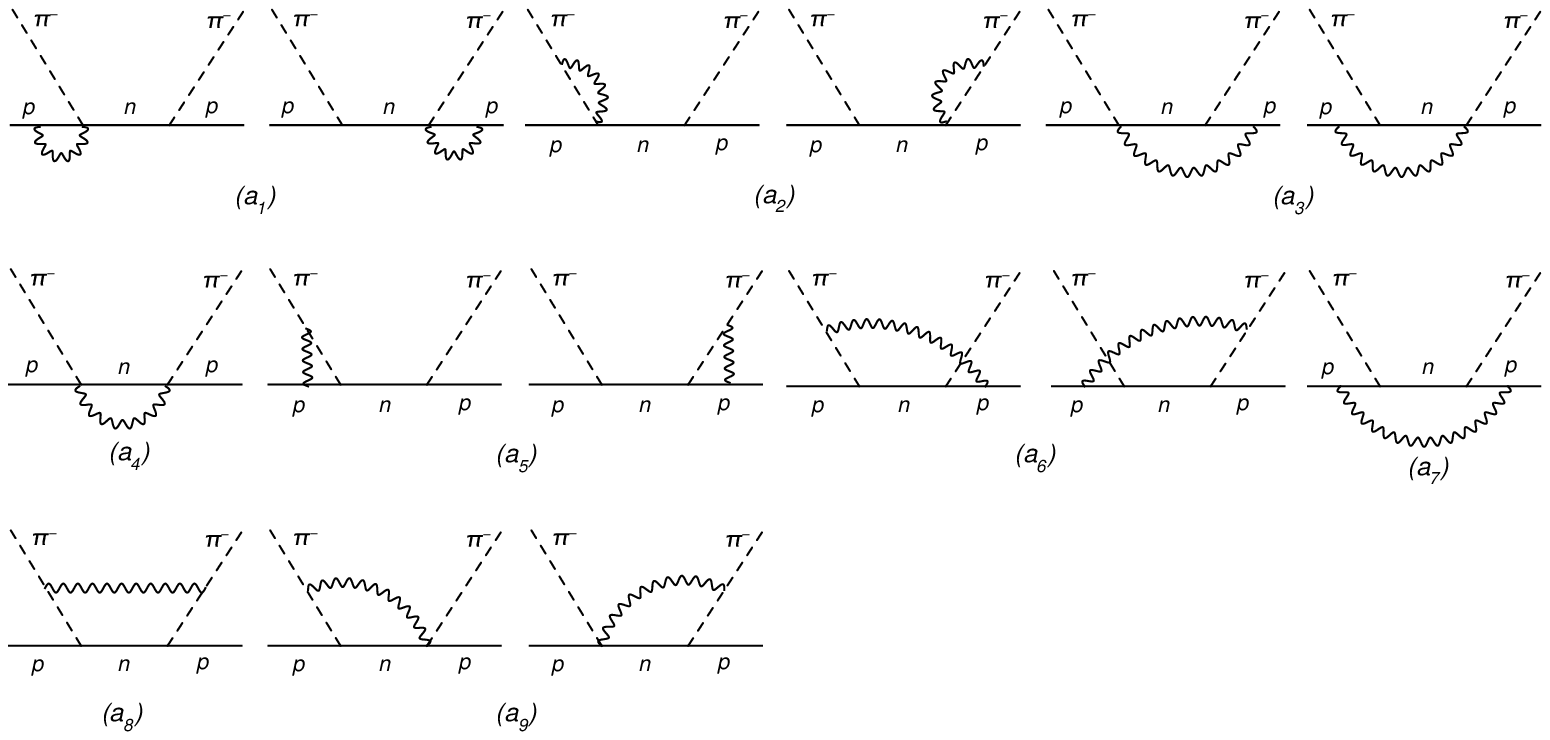}
\end{center}
\caption{Electromagnetic axial-type diagrams for $\pi^- p \rightarrow \pi^- p$.}
\label{fig:axial_a}
\end{figure}

\begin{figure}
\begin{center}
\includegraphics[width=\linewidth]{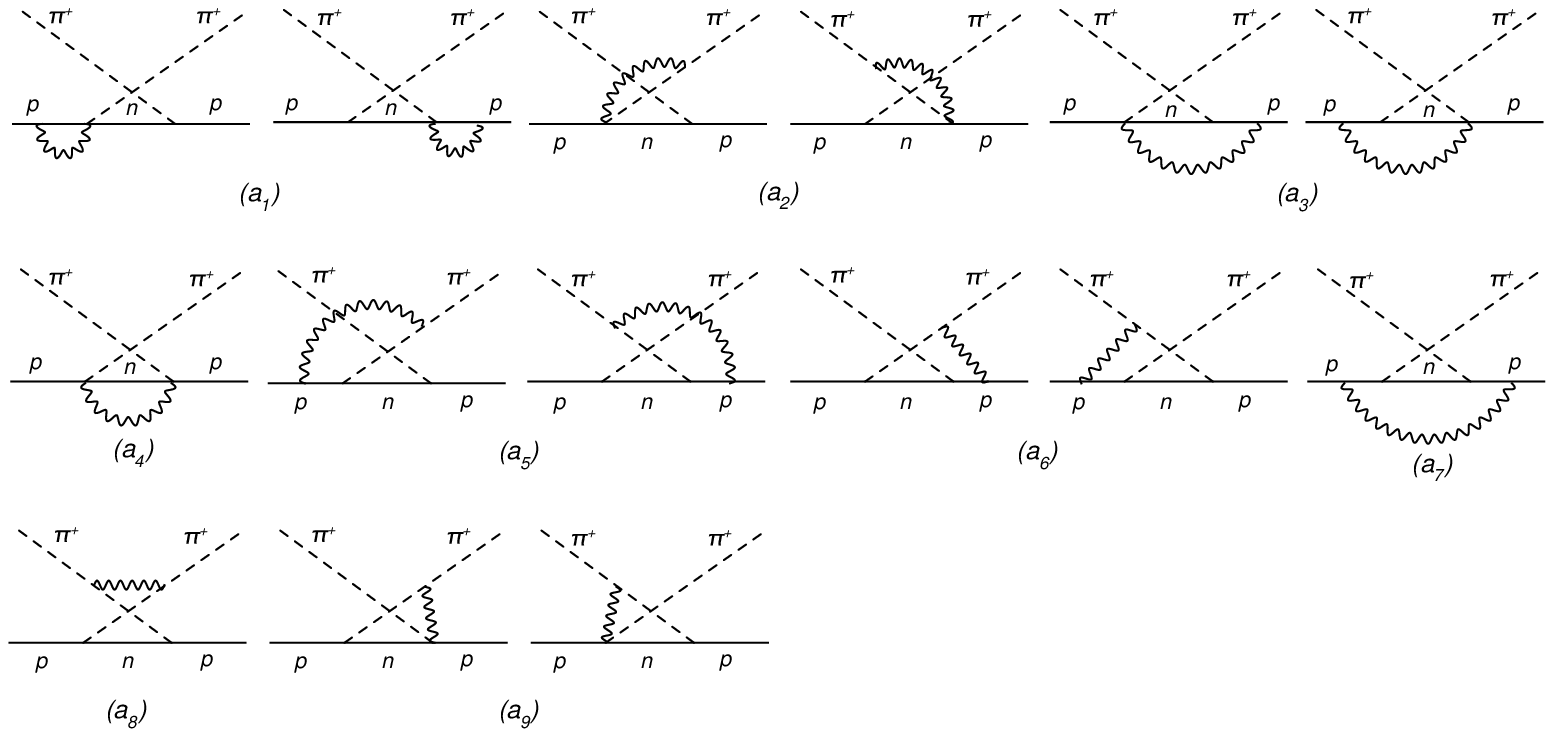}
\end{center}
\caption{Electromagnetic axial-type diagrams for $\pi^+ p \rightarrow \pi^+ p$.}
\label{fig:axial_b}
\end{figure}

\begin{figure}
\begin{center}
\includegraphics[width=\linewidth]{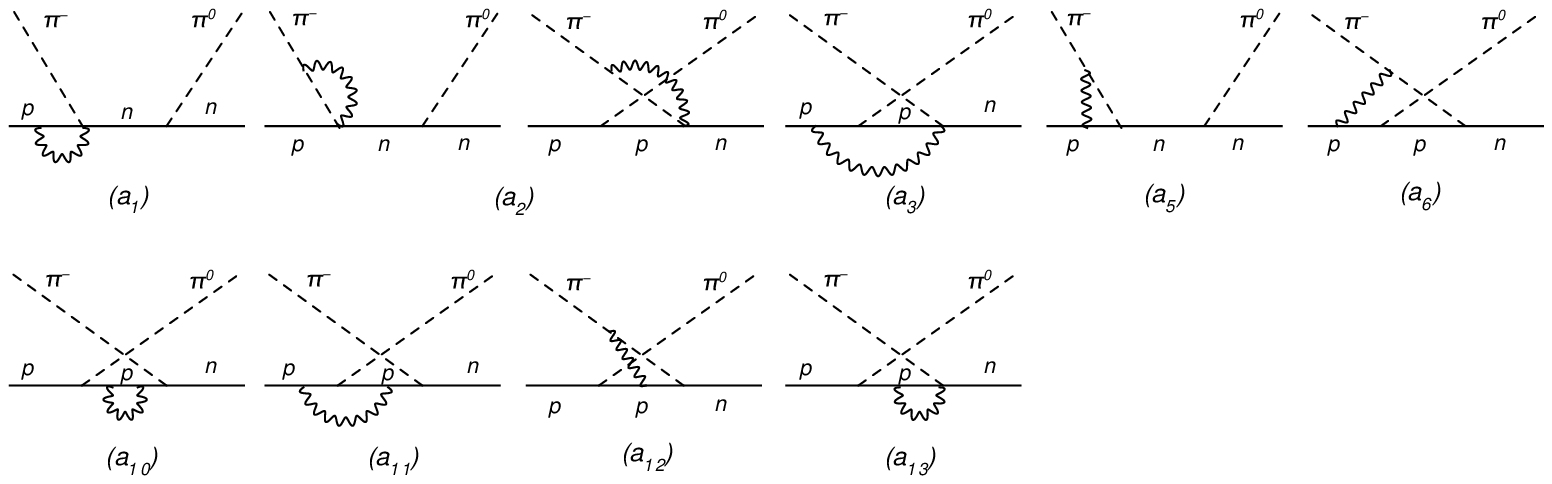}
\end{center}
\caption{Electromagnetic axial-type diagrams for $\pi^- p \rightarrow \pi^0 n$.}
\label{fig:axial_c}
\end{figure}

\begin{figure}
\begin{center}
\includegraphics[width=\linewidth]{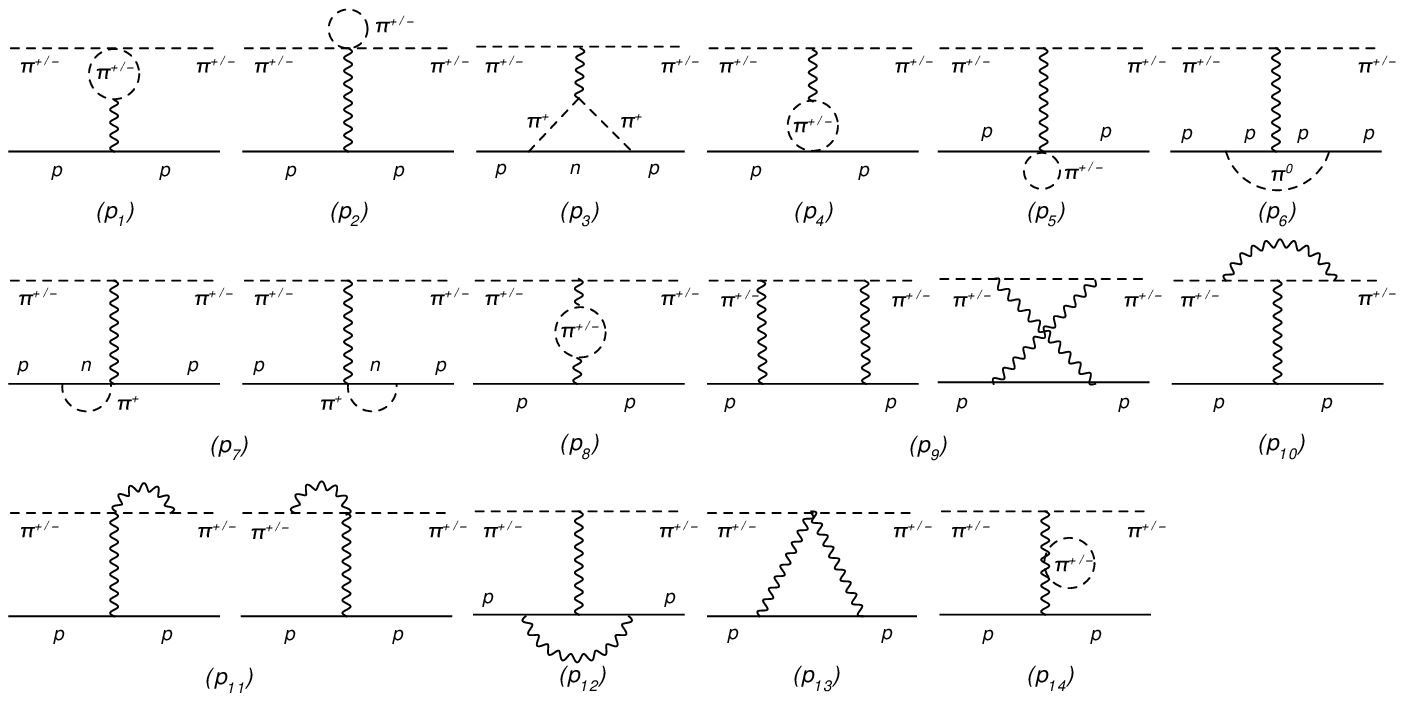}
\end{center}
\caption{Photon-exchange diagrams for $\pi^\pm p \rightarrow \pi^\pm p$ at $\Order(p^3)$.}
\label{fig:photon_ab}
\end{figure}

At order $\Order(p^3)$ loop graphs solely contain vertices from the leading-order Lagrangians \eqref{lead_lagr}. The list of diagrams depicted in Figs.~\ref{fig:strong_a}--\ref{fig:photon_ab} comprises contributions from purely strong diagrams, virtual photon corrections to vector-type and axial-type diagrams, and, for the elastic channels, loop corrections to the one photon exchange. At the order we are working all nucleon masses may be replaced by a generic nucleon mass $m$ (cf.\ Sect.~\ref{sec:tree}), which simplifies the calculation tremendously. However, counting $e\sim\Order(p)$ the pion mass difference is of the same chiral order as the pion masses themselves and therefore has to be kept. Apart from virtual photons, this is the second source through which isospin violation manifests itself in the loop contributions.

The evaluation of the diagrams proceeds as usual: at first, the Dirac structure is simplified by application of the Dirac equation, such that the power of the loop momentum $k^\mu$ in the numerator is reduced as far as possible. The remaining tensor integrals  are reduced to scalar loop functions by means of Lorentz invariance.  As for the strong diagrams, we follow the conventions in \cite{BL01} for the basis vectors to facilitate the comparison to this work. The details of the tensor decomposition can be found in App.~\ref{tensor_dec}, where in particular the question of how to properly incorporate the pion mass difference is addressed. 
Since $\bar{u}(p')\slashed{q}'u(p)=\bar{u}(p')\slashed{q}u(p)$ up to higher orders, $A(s,t)$ and $B(s,t)$ as defined in \eqref{ampl_AB} can be simply read off as prefactors of $\unity$ and $\slashed{q}$, respectively, and $D(s,t)$ then follows from \eqref{ampl_DB}. The results for the individual diagrams are given in App.~\ref{app:ind_diagrams}. To arrive at these expressions, we have made use of the relations between the basis functions, which in some cases simplifies the formulae considerably. 

The results for the scalar loop functions calculated in infrared regularization are presented in App.~\ref{Scal_loop_func}.
As most of the work to arrive at these expressions is rather straightforward but tedious, we will focus in the remaining part of this section on the issues which are more intricate, namely the calculation of the so-called box graphs, the extraction of infrared divergences  and the separation of the Coulomb pole generated by virtual photons in a threshold expansion of the amplitude.

\subsubsection{Box graphs}
\label{sec:box}

We will refer to any diagram whose loop contains four propagators as ``box graph''. These topologies lead to scalar loop integrals which are technically the most difficult ones we have to deal with, as three Feynman parameters are needed to perform the momentum integration in the usual manner. Having done the Feynman parameter integration corresponding to infrared regularization, we are still left with two Feynman parameter integrals. The analytic properties of these integrals are far from obvious, which in particular causes numerical integration to fail as soon as one goes above threshold. 

Let us first consider the box diagram consisting of three nucleon and one meson propagator
\beq
I_{13}(s,t)=\frac{1}{i}\int_{\rm I} \frac{\diff^d k}{(2\pi)^d}\frac{1}{\left(\mpi^2-k^2\right)\left(m^2-(p-k)^2\right)
\left(m^2-(\Sigma-k)^2\right)\left(m^2-(p'-k)^2\right)}.
\eeq
For simplicity, we will neglect the pion mass difference in this section (the general case is treated in App.~\ref{sec:scal_1m3n}). Below threshold, the representation
\begin{align}
I_{13}(s,t)&=\int\limits_0^1\diff x \int\limits_0^1\diff y\frac{y}{16\pi^2A^2}\left\{\frac{c_0+z_0^2}{2c_0^{3/2}}\arccos\bigg[-\frac{z_0}{\sqrt{c_0+z_0^2}}\bigg]+\frac{z_0}{2c_0}\right\},\notag\\
A&=(1-y)s+y m^2-x(1-x)y^2 t-y(1-y)\mpi^2, \notag\\
z_0=&\frac{\mpi^2+(1-y)(s-m^2)}{2A},\quad c_0=\frac{\mpi^2}{A}-z_0^2,\label{box_below_2int}
\end{align}
may be used. One way to obtain a representation valid above threshold is the application of dispersion relations, since the regular part is numerically harmless. However, $I_{13}(s,t)$ is also needed with the pion replaced by a photon, and the virtual photon leads to an IR divergence. The prescription how to deal with the vanishing photon mass is to send $m_\gamma\rightarrow 0$ at the very end of the calculation. Therefore, we first have to analytically continue above threshold and thereafter put $m_\gamma\rightarrow 0$, which cannot be done on the basis of \eqref{box_below_2int}. As shown explicitly in the appendix, dispersion relations are not appropriate either, such that it is inevitable to improve \eqref{box_below_2int}, i.e.\ the $y$ integration should be performed. The second term is a rational function in $y$ and can be integrated by means of a partial fraction expansion. After an integration by parts
\begin{align}
&\int\limits_0^1\diff y\frac{y\left(c_0+z_0^2\right)}{2A^2c_0^{3/2}}\arccos\bigg[-\frac{z_0}{\sqrt{c_0+z_0^2}}\bigg]=
\frac{4\mpi^2\left\{2y(y_1+y_2)-4y_1y_2\right\}\arccos\Big[-\frac{z_0}{\sqrt{c_0+z_0^2}}\Big]}{(c_2)^{3/2}(y_1-y_2)^2\sqrt{(y_1-y)(y-y_2)}}\bigg|^1_0\notag\\
&-\frac{4\mpi^2}{(c_2)^{3/2}}\int\limits_0^1\diff y\frac{2y(y_1+y_2)-4y_1y_2}{(y_1-y_2)^2\sqrt{(y_1-y)(y-y_2)}}\frac{\diff}{\diff y}\arccos\bigg[-\frac{z_0}{\sqrt{c_0+z_0^2}}\bigg],
\end{align}
where the notation of App.~\ref{sec:scal_1m3n} has been used, also the first term is reduced to an integral over a rational function in $y$, since as far as square roots are concerned
\beq
\frac{\diff}{\diff y}\arccos\bigg[-\frac{z_0}{\sqrt{c_0+z_0^2}}\bigg]\sim\frac{1}{\sqrt{(y_1-y)(y-y_2)}}.
\eeq
The $x$ integration turns out to be harmless, such that the continuation above threshold can be performed and the IR divergence for $m_\gamma\rightarrow 0$ extracted. The final results are again summarized in App.~\ref{sec:scal_1m3n}.

\subsubsection{Extraction of infrared divergences}

The following scalar loop functions possess an IR divergence: $I_A^\gamma(t)$ (cf.\ diagram $(v_3)$), $V_{20}(t)$ $(v_4)$, $V_{11}(s)$ $(v_1)$, $I_{13}^\gamma(s,t)$ $(a_7)$, $A_{21}(s,t)$ $(a_8)$, $A_{12}(s,t)$ $(a_6)$, and $P_{11}(s,t)$ $(p_9)$. Thereof, the first two integrals do not induce any problems and $I_{13}^\gamma(s,t)$ has already been treated in the previous section, but unfortunately this procedure cannot be directly adopted to the remaining cases. Let us first consider $V_{11}(s)$, which is the simplest remaining task, as $A_{21}(s,t)$, $A_{12}(s,t)$, as well as $P_{11}(s,t)$, are box graphs. The following procedure is inspired by \cite{GR02}.

We extract the IR divergence from the loop integral calculated in ordinary dimensional regularization\footnote{The regular part is finite in the limit of vanishing photon mass, such that $m_\gamma$ may be put equal to zero right from the start.} 
\beq
V_{11}^{\rm dim}(s)=\frac{1}{16\pi^2}\int\frac{\diff x_1\,\diff x_2\,\diff x_3\,\delta(1-x_1-x_2-x_3)}{x_1 m_\gamma^2+x_2^2\mpi^2+x_3^2m^2-x_2x_3\left(s-m^2-\mpi^2\right)}.
\eeq
The substitution
\beq
x_2\rightarrow (1-x_1)x_2,\quad x_3\rightarrow (1-x_1)x_3
\eeq
implies
\beq
\delta\left(1-x_1-x_2-x_3\right)\rightarrow \delta\left((1-x_1)(1-x_2-x_3)\right)=\frac{\delta\left(1-x_2-x_3\right)}{1-x_1},
\eeq
which leads to
\beq
V_{11}^{\rm dim}(s)=\frac{1}{16\pi^2}\int\limits_0^1\diff x \int\limits_0^1\diff y\frac{1-y}{y m_\gamma^2+(1-y)^2s(x)},\quad
s(x)=x m^2+(1-x)\mpi^2-x(1-x) s.
\eeq
Performing the $y$ integration, we finally arrive at
\beq
V_{11}^{\rm dim}(s)=\frac{1}{32\pi^2}\Bigg\{-L\int\limits_0^1\frac{\diff x}{s(x)}+\int\limits_0^1\frac{\diff x \log \frac{s(x)}{m^2}}{s(x)}\Bigg\}.
\eeq

For the remaining box graphs, this approach has to be generalized, which we will illustrate for $A_{12}(s,t)$ ($A_{21}(s,t)$ and $P_{11}(s,t)$ are treated analogously). The starting point is again 
\beq
A_{12}^{\rm dim}(s,t)=\frac{1}{16\pi^2}\int\frac{\diff x_1\,\diff x_2\,\diff x_3\,\diff x_4\,\delta(1-x_1-x_2-x_3-x_4)}{\left(x_1m_\gamma^2-x_3\left(s-m^2\right)+\left(x_2 q'+x_3\Sigma+x_4 p\right)^2\right)^2}.
\eeq
Separating (and performing) the $x_1$ integration with the help of the substitution
\beq
x_i\rightarrow (1-x_1)x_i, \quad i\in\{2,3,4\},\quad \delta(1-x_1-x_2-x_3-x_4)\rightarrow\frac{\delta(1-x_2-x_3-x_4)}{1-x_1}
\eeq
yields
\begin{align}
A_{12}^{\rm dim}(s,t)&=\frac{1}{16\pi^2}\int \diff x_2\,\diff x_3\,\diff x_4\bigg\{\frac{m_\gamma^2-x_3\left(s-m^2\right)}{\left\{f(x_2,x_3,x_4)-x_3\left(s-m^2\right)\right\}s^2(x_2,x_3,x_4)}\notag\\
&+\frac{4m_\gamma^2\left(\arctan\frac{m_\gamma^2+x_3\left(s-m^2\right)}{\sqrt{-s^2(x_2,x_3,x_4)}}-\arctan\frac{m_\gamma^2+x_3\left(s-m^2\right)-2f(x_2,x_3,x_4)}{\sqrt{-s^2(x_2,x_3,x_4)}}\right)}{(-s^2(x_2,x_3,x_4))^{3/2}}\bigg\}\delta(1-x_2-x_3-x_4),\notag\\
s^2(x_2,x_3,x_4)&=m_\gamma^4+x_3^2\left(s-m^2\right)^2-2m_\gamma^2\left\{2f(x_2,x_3,x_4)-x_3\left(s-m^2\right)\right\},\notag\\
f(x_2,x_3,x_4)&=\left(x_2 q'+x_3\Sigma+x_4 p\right)^2.
\end{align}
The structure of this formula suggests to repeat the same steps for $x_3$
\beq
x_i\rightarrow (1-x_3)x_i, \quad i\in\{2,4\},\quad \delta(1-x_2-x_3-x_4)\rightarrow\frac{\delta(1-x_2-x_4)}{1-x_3}.
\eeq
In this way, the $x_3$ integration factorizes and the $\delta$-function can be
applied. The result of this manipulation remarkably resembles
\eqref{box_below_2int}, and indeed the $x_3$ integration can be treated in the
same way as the $y$ integration in the previous section. Thereafter, the limit
$m_\gamma\rightarrow 0$ may finally be taken. In this way, we find the IR
divergence $\sim\log m_\gamma$ and the correct finite terms. As compared to
\eqref{box_one_int}, putting the photon mass to zero at the end simplifies the
result considerably, while the derivation requires about the same amount of work.

\subsubsection{Coulomb pole}
\label{Coulomb_pole_ampl}

The $\pi N$ scattering amplitude at threshold contains Coulomb singularities once virtual photons are included. In particular, the threshold expansion of loop diagrams which allow for an exchange of a virtual photon corresponding to a static Coulomb potential involves a term $\propto 1/|\mathbf{p}|$, where $\mathbf{p}$ is the CMS momentum of the incoming particles. The coefficient of the Coulomb pole can be associated with the $\pi N$ scattering amplitude itself. In this section, we gather the parts of the amplitude contributing to the Coulomb pole in order to confirm this statement.

The diagrams which give rise to the Coulomb pole are $(v_1)$ and $(a_5)$ for $\pi^-p\rightarrow \pi^- p$, $(v_1)$ and $(a_6)$ for $\pi^+p\rightarrow \pi^+ p$, and $(v_1)$, $(a_5)$, and $(a_6)$, for $\pi^-p\rightarrow \pi^0 n$, respectively. Based on the representations given in App.~\ref{sec:1m1n1p} and \ref{sec:1m2n1p}, one can show that for $\mathbf{p}\rightarrow 0$
\begin{align}
V_{11}(s)&=\frac{1}{32\left(m+\mpi\right)|\pp|}+\Order(1),\quad \left(u-m^2\right)A_{12}(u,t)=-\frac{1}{32\left(m+\mpi\right)|\pp|}+\Order(1),\notag\\
&\Rightarrow  \lim_{\pp\rightarrow 0}\left\{|\pp|\left(V_{11}(s)+\left(u-m^2\right)A_{12}(u,t)\right)\right\}=0\label{pole_scal_loop}.
\end{align}
For $A_{12}(u,t)$, this relies on the observation that
\beq
g_{12}^{\pi^\pm p}(u,0)=g_{12}^{\rm cex}(u,-\Delta_\pi)=-\frac{\pi^2}{\left(m+\mpi\right)|\pp|}+\Order(1).
\eeq
We denote the analogues of the tree amplitudes \eqref{lead_order_ampl} with the one-photon-exchange contributions $\propto 1/t$ subtracted by $\tilde{D}^{\rm LO}(s,t)$ and $\tilde{B}^{\rm LO}(s,t)$. The Coulomb pole in $V_{11}(s)$ may be written as 
\beq
V_{11}^{\rm pole}(s)=\frac{1}{16\lambda^{1/2}\left(s,m^2,\mpi^2\right)}=\frac{1}{32\left(m+\mpi\right)|\pp|}+\Order(|\mathbf{p}|).
\eeq
Expressing the pole in $A_{12}(u,t)$ by $V_{11}^{\rm pole}(s)$ via \eqref{pole_scal_loop}, we find that the Coulomb-pole contributions to the full amplitude can be summarized in the simple form
\begin{align}
B^{\rm pole}_i(s,t)&=-2(Q_{\rm in}+Q_{\rm out})_i\,e^2\left(s-m^2-\mpi^2\right)V_{11}^{\rm pole}(s)\tilde{B}^{\rm LO}_i(s,t),\quad i\in\left\{\pi^-p,\pi^+p,{\rm cex}\right\},\notag\\
D^{\rm pole}_i(s,t)&=-2(Q_{\rm in}+Q_{\rm out})_i\,e^2\left(s-m^2-\mpi^2\right)V_{11}^{\rm pole}(s)\tilde{D}^{\rm LO}_i(s,t)\label{Coulomb_pole_full_amp},
\end{align}
where
$Q_{\rm in/out}$ are given as the products of the charges of the particles emitting and absorbing the photon in the initial and final state, respectively, such that
\beq
(Q_{\rm in}+Q_{\rm out})_{\pi^- p}=-2,\quad (Q_{\rm in}+Q_{\rm out})_{\pi^+ p}=2,\quad  (Q_{\rm in}+Q_{\rm out})_{\rm cex}=-1.
\eeq

It is obvious from the above that the perturbative treatment of photon exchange breaks down for very small $|\mathbf{p}|$.  The Coulomb pole $\propto 1/|\mathbf{p}|$ discussed above is the leading, $\Order(e^2)$ approximation to the Gamow--Sommerfeld factor~\cite{Gamow28,Sommerfeld39}
\beq
G(\eta) = \frac{2\pi\eta}{\exp(2\pi\eta)-1} = 1-\pi\eta+\Order(\eta^2), \quad
\eta_{\rm in/out} = \frac{Q_{\rm in/out}e^2}{4\pi}\frac{s-m^2-\mpi^2}{\lambda^{1/2}(s,m^2,\mpi^2)} ,\label{Gamow}
\eeq 
which resums Coulomb photon exchange to all orders.  Obviously, the one-photon approximation is insufficient as soon as $\eta\approx 1$.  If one attempts to describe physical data also very close to threshold, one has to multiply the amplitude with a factor of $\sqrt{G(\eta_{\rm in})G(\eta_{\rm out})}$; in order to avoid double counting, we have to remove the approximation to the Coulomb pole as contained in our one-loop representation.  In Sect.~\ref{chap:above}, we will discuss results for amplitudes purified of Coulomb photon effects in this sense.

\subsection{Bremsstrahlung}

The appearance of infrared divergences in $T_{\pi N}$ renders the amplitude ill-defined if the photon mass is sent to zero. The solution to this problem is well known \cite{Bloch37,Yennie61,Weinberg65}: it is inconsistent to merely account for virtual photons, since real photons can be emitted with arbitrarily low energies in the final state as soon as charged particles are involved, which inevitably requires the inclusion of bremsstrahlung. One can never specify the number of escaped photons exactly, but has to decoherently sum over all processes integrating the photon energies up to some typical detector resolution $E_{\rm max}$ (and over the phase space available). The integration over propagators from which a photon is radiated generates another infrared divergence which cancels the singularities caused by virtual photons. In this way, the inclusive cross section becomes infrared finite. 

\begin{figure}
\begin{center}
\includegraphics[width=\linewidth]{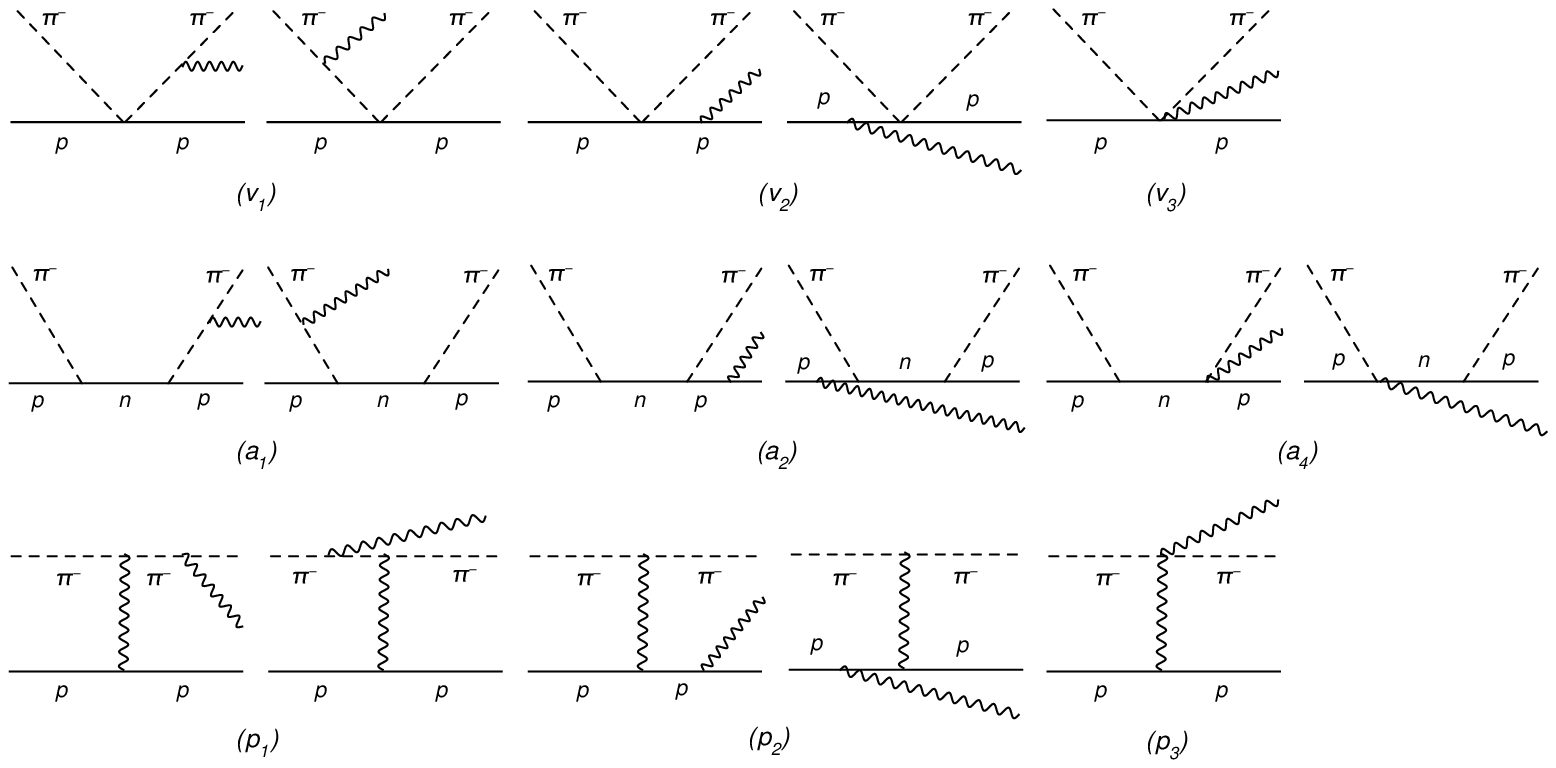}
\end{center}
\caption{Bremsstrahlung diagrams for $\pi^- p \rightarrow \pi^- p$.}
\label{fig:brems_a}
\end{figure}

\begin{figure}
\begin{center}
\includegraphics[width=\linewidth]{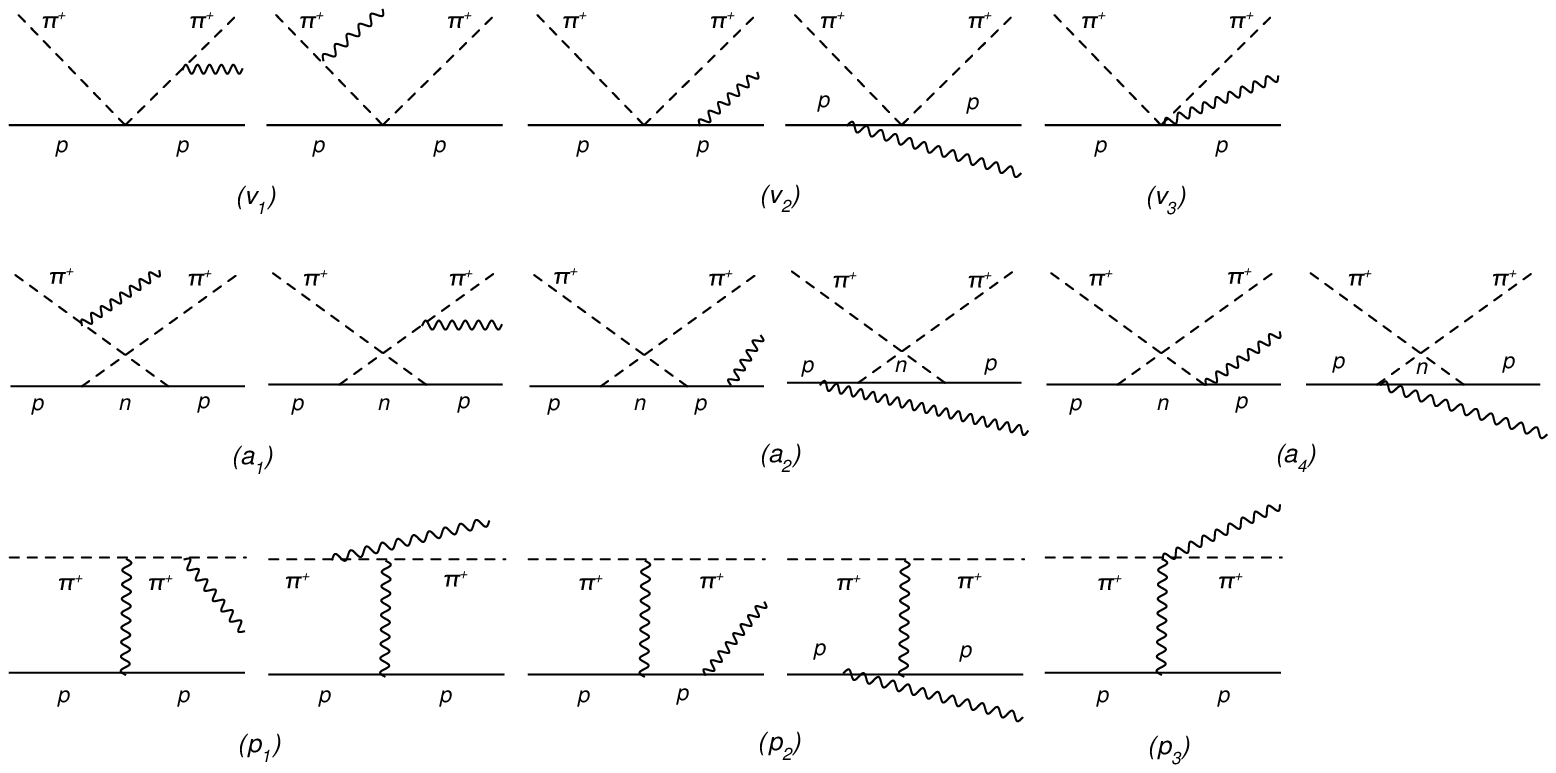}
\end{center}
\caption{Bremsstrahlung diagrams for $\pi^+ p \rightarrow \pi^+ p$.}
\label{fig:brems_b}
\end{figure}

\begin{figure}
\begin{center}
\includegraphics[width=0.82\linewidth]{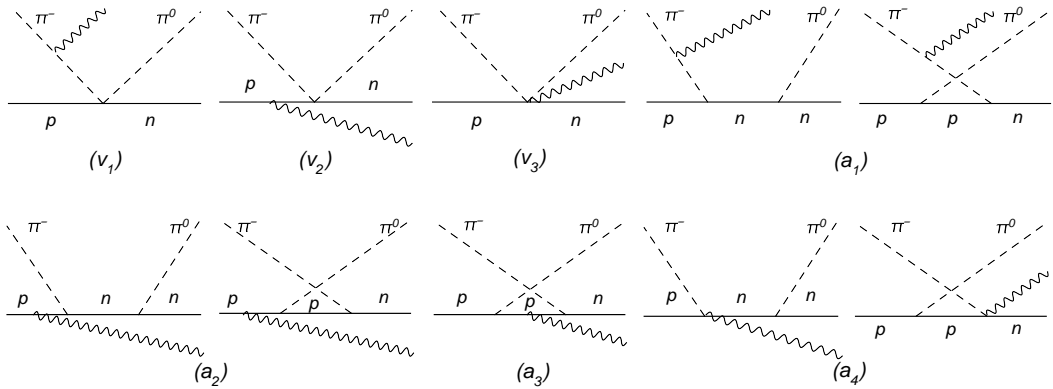}
\end{center}
\caption{Bremsstrahlung diagrams for $\pi^- p \rightarrow \pi^0 n$.}
\label{fig:brems_c}
\end{figure}

To the order we are working, it suffices to consider the emission of a single photon to achieve this cancelation of infrared divergences in the total cross section 
\beq
\sigma_{\rm tot}(s,t,E_{\rm max})=\sigma_{\pi N}(s,t)+\sigma_\gamma(s,t,E_{\rm max}).
\eeq
Denoting the photon momentum by $k=(E_\gamma,\mathbf{k})$ and the amplitude for bremsstrahlung by
\beq
T_\gamma=\bar{u}(p')\epsilon^*_\mu(k)T^\mu_\gamma(p,p',q,q',k)u(p)\label{brems_ampl},
\eeq
the cross section for the pure scattering process  $\sigma_{\pi N}(s,t)$ and for the bremsstrahlung correction $\sigma_{\gamma}(s,t,E_{\rm max})$ can be obtained from the spin and polarization averaged squared matrix elements
\beq
|\M_\gamma|^2\equiv\frac{1}{2}\sum\limits_{\rm spins, pol.}|T_\gamma|^2,\quad |\M_{\pi N}|^2\equiv\frac{1}{2}\sum\limits_{\rm spins}|T_{\pi N}|^2,
\eeq
via
\begin{align}
 \sigma_{\pi N}(s,t)&=\frac{1}{\F}\int \text{d}\phi_2|\M_{\pi N}|^2(2\pi)^4\delta^{(4)}(p+q-p'-q'),\notag\\
\sigma_\gamma(s,t,E_{\rm max})&=\frac{1}{\F}\int\limits_{E_\gamma<E_{\rm max}}  \frac{\text{d}\phi_2\text{d}^3\mathbf{k}}{(2\pi)^3 2E_\gamma}|\M_\gamma|^2 (2\pi)^4\delta^{(4)}(p+q-p'-q'-k),
\end{align}
where the flux factor $\F$ and the two-body phase space $\text{d}\phi_2$ are given by
\beq
\F=4\sqrt{(p\cdot q)^2-m^2q^2},\quad
\text{d}\phi_2=\frac{\text{d}^3\mathbf{p'}}{(2\pi)^3 2E_{p'}}\frac{\text{d}^3\mathbf{q'}}{(2\pi)^3 2E_{q'}}.
\eeq
Since the flux is determined solely by the incoming particles, $\F$ is the same for both processes.
The Feynman diagrams for bremsstrahlung are depicted in Figs.~\ref{fig:brems_a}--\ref{fig:brems_c}.

\subsubsection{Soft-photon approximation}

It is clear from the previous discussion that in general the infrared divergences only cancel at the level of cross sections. In particular, one has to perform the phase space integration of the amplitude, which is a considerable task if analytical expressions are aimed for. The amount of work can be reduced significantly be means of the so-called soft-photon approximation, in which already the sum of the squared amplitudes is infrared finite. It neglects terms of order $\Order(E_{\rm max})$, but reproduces singular structures and constant terms correctly.\footnote{A comparison of the soft-photon approximation and the exact result in the context of radiative corrections to kaon decays can be found in \cite{BFGKR09}.}

The crucial approximation is to drop the photon momentum in the delta function. In principle, this amounts to reducing the three-body phase space integration to a two-body one by setting $k=0$. In fact, every term in $|M_\gamma|^2$, which is not of order $\Order(k^{-2})$ may be neglected, such that effectively only expressions of the form 
\beq
\frac{1}{\left(2p_1\cdot k\pm m_\gamma^2\right)\left(2p_2\cdot k\pm m_\gamma^2\right)}\label{p1p2}
\eeq
need to be integrated over the photon momentum. That the error is indeed of order $\Order(E_{\rm max})$ follows from dimensional analysis
\begin{align}
 &\int\limits_{E_\gamma<E_{\rm max}} \frac{\text{d}^3k}{(2\pi)^3 2E_\gamma}\frac{1}{2p_1\cdot k \ 2p_2\cdot k}\sim
\int\limits_{E_\gamma<E_{\rm max}} \frac{\text{d}^3k}{k^3}\sim\log\left(\frac{E_{\rm max}}{m_\gamma}\right),\\
&\int\limits_{E_\gamma<E_{\rm max}} \frac{\text{d}^3k}{(2\pi)^3 2E_\gamma}\frac{1}{2p_1\cdot k}\sim
\int\limits_{E_\gamma<E_{\rm max}} \frac{\text{d}^3k}{k^2}\sim E_{\rm max}.\notag
\end{align}
In particular, it is clear that terms of order $\Order(1)$ can only originate from the phase space integration of \eqref{p1p2}. Moreover,  the $m_\gamma^2$ terms in the denominator of \eqref{p1p2} may be omitted, while the full energy-momentum relation
$E^2_\gamma=\mathbf{k}^2+m^2_\gamma$ is necessary to arrive at the correct constant terms. Neglecting $m_\gamma$ therein merely reproduces the infrared divergence. This result can be understood from the observation that $\mathbf{k}$ enters the energy-momentum relation quadratically, such that both terms are of the same size in the integration region where $|\mathbf{k}|\sim m_\gamma$, while the $m_\gamma^2$ terms in \eqref{p1p2} are still suppressed by one order of $m_\gamma$ in this domain and therefore constitute a higher-order effect. For $p_1=p_2\equiv p=(E_p,\pp)$, we obtain
\begin{align}
\int\limits_{E_\gamma<E_{\rm max}} \frac{\text{d}^3k}{(2\pi)^3 2E_\gamma}\frac{1}{(2p\cdot k)^2}=-\frac{1}{32\pi^2p^2}\bigg\{\log\frac{m_\gamma^2}{4E_{\rm max}^2}+\frac{E_p}{|\mathbf{p}|}\log\frac{E_p+|\mathbf{p}|}{E_p-|\mathbf{p}|}\bigg\}+\Order(m_\gamma),
\end{align}
while the formula for the general case reads
\begin{align}
\int\limits_{E_\gamma<E_{\rm max}}\frac{\text{d}^3k}{(2\pi)^3 2E_\gamma}\frac{1}{2p_1\cdot k \ 2p_2\cdot k}&=-\frac{1}{32\pi^2}\bigg\{\log\frac{m_\gamma^2}{4E_{\rm max}^2}\int\limits_0^1\frac{\diff x}{\tilde{p}^2}+\int\limits_0^1\frac{\diff x}{\tilde{p}^2}\frac{E_{\tilde{p}}}{|\mathbf{\tilde{p}}|}\log\frac{E_{\tilde{p}}+|\mathbf{\tilde{p}}|}{E_{\tilde{p}}-|\mathbf{\tilde{p}}|}\bigg\}+\Order(m_\gamma),\notag\\
\tilde{p}&=x p_1+(1-x) p_2=\left(E_{\tilde{p}},\mathbf{\tilde{p}}\right)\label{phase_int_brems}.
\end{align}

Finally, the soft-photon approximation allows for the application of the usual expressions for the Mandelstam variables, since the difference $(p'+q')^2-s=(p+q-k)^2-s=-2k\cdot(p+q)+m^2_\gamma\rightarrow -2k\cdot(p+q)$ yields only contributions of order $\Order(E_{\rm max})$ even when multiplied by $\Order(k^{-2})$ terms, which is beyond the accuracy of this approximation.

\subsubsection{Bremsstrahlung amplitude}
\label{sec:Bremsstrahlung_amplitude}

The contributions from the individual diagrams are given in App.~\ref{app:Bremsstrahlung}. Their sum can be brought into the following form
\begin{align}
T^{\mu}_{\pi^-p}&=a_1 \slashed{q}'\gamma^\mu\slashed{q}+a^\mu_2 \slashed{q}+a^\mu_3 \slashed{q}'+a_4(\slashed{q}-\slashed{q}')\gamma^\mu+a_5\gamma^\mu+a^\mu_6,\quad T^{\mu}_{\pi^+p}=T^{\mu}_{\pi^-p}\left(q\leftrightarrow -q'\right),\notag\\
T^{\mu}_{\rm cex}&=b_1 \slashed{q}\gamma^\mu\slashed{q}'+b_2\slashed{q}'\slashed{q}\gamma^\mu+b^\mu_3\slashed{q}'+b_4\gamma^\mu,\notag\\
a_1&=-\frac{e}{2F^2}\frac{1}{2p'\cdot k}\bigg\{1-g^2-\frac{4m^2g^2}{(p+q)^2-m^2}\bigg\}+\frac{e}{2F^2}\frac{1}{2p\cdot k}\bigg\{1-g^2-\frac{4m^2g^2}{(p'+q')^2-m^2}\bigg\},\notag\\
a^\mu_2&=\frac{e}{2F^2}\bigg\{1-g^2-\frac{4m^2g^2}{(p+q)^2-m^2}\bigg\}\bigg\{\frac{(2q'+k)^\mu}{2q'\cdot k}+\frac{2q'^\mu}{2p'\cdot k}\bigg\},\notag\\
a^\mu_3&=-\frac{e}{2F^2}\bigg\{1-g^2-\frac{4m^2g^2}{(p'+q')^2-m^2}\bigg\}\bigg\{\frac{(2q-k)^\mu}{2q\cdot k}+\frac{2q^\mu}{2p\cdot k}\bigg\},\notag\\
a_4&=\frac{m e g^2}{F^2}\bigg\{\frac{1}{2p'\cdot k}-\frac{1}{2p\cdot k}\bigg\},\quad
a_5=-\frac{e(1-g^2)}{2F^2}\bigg\{\frac{2p\cdot q+q^2}{2p'\cdot k}-\frac{2p'\cdot q'+q'^2}{2p\cdot k}\bigg\},\notag\\
a^\mu_6&=\frac{m e g^2}{F^2}\bigg\{\frac{(2q'+k)^\mu}{2q'\cdot k}-\frac{(2q-k)^\mu}{2q\cdot k}-\frac{2(q-q')^\mu}{2p'\cdot k}\bigg\},\notag\\
b_1&=\frac{\sqrt{2}\,m e g^2}{F^2((p'-q)^2-m^2)((p-q')^2-m^2)},\notag\\
b_2&=\frac{\sqrt{2}\,e}{2F^2}\frac{1}{2p\cdot k}\bigg\{1-g^2-\frac{2m^2g^2}{(p'+q')^2-m^2}-\frac{2m^2g^2}{(p'-q)^2-m^2}\bigg\},\notag\\
b^\mu_3&=\frac{\sqrt{2}\,e}{2F^2}\frac{(2q-k)^\mu}{2q\cdot k}\bigg\{1-g^2-\frac{2m^2g^2}{(p'+q')^2-m^2}-\frac{2m^2g^2}{(p-q')^2-m^2}\bigg\},\notag\\
b_4&=-\frac{\sqrt{2}\,e}{2F^2}\frac{1}{2p\cdot k}\bigg\{(1-g^2)(2p'\cdot q'+q'^2)-\frac{2m^2g^2}{(p'-q)^2-m^2}2q\cdot q'\bigg\},\label{brems_amplitude}
\end{align}
where terms generated by the one-photon-exchange diagrams are omitted. On the level of the soft-photon approximation they can be restored by the replacement
\beq
1-g^2\rightarrow 1-g^2-\frac{4e^2F^2}{t}
\eeq
in $a_1$, $a_2^\mu$, $a_3^\mu$ and $a_5$. As a check on \eqref{brems_amplitude}, we have verified explicitly that the Ward identity
\beq
k_\mu T^{\mu}_{\pi^-p}=k_\mu T^{\mu}_{\pi^+p}=k_\mu T^{\mu}_{\rm cex}=0 \label{ward}
\eeq
is fulfilled. Besides $m_\gamma^2$ terms in the denominators, \eqref{brems_amplitude} is still exact. The soft-photon approximation implies that all $k$'s in the numerators may be dropped and the usual definitions for the Mandelstam variables applied, since we are only interested in $\Order(k^{-2})$ terms in $|\M_\gamma|^2$ in the end. In particular, this means that $b_1$ does not contribute. In view of this procedure, one might worry what happens to gauge invariance. Indeed, within the above approximation \eqref{ward} is only valid up to terms of order $\Order(k)$, as the terms omitted in $T^\mu_i$ are by construction of order $\Order(1)$. However, using the photon polarization sum in the form \cite{Wein1}
\beq
\sum\limits_{\lambda=\pm 1}\epsilon_\mu^*(k,\lambda)\epsilon_\nu(k,\lambda)=-g_{\mu\nu}+k_\mu c_\nu+k_\nu c_\mu,\quad \mathbf{c}=-\frac{\mathbf{k}}{2\mathbf{k}^2},\quad c^0=\frac{1}{2|\mathbf{k}|}\label{pol_sum},
\eeq
it is clear that in
\beq
|\M_\gamma|^2=\frac{1}{2}\sum\limits_{\lambda=\pm 1}\epsilon_\mu^*(k,\lambda)\epsilon_\nu(k,\lambda)\Tr\left((\slashed{p}'+m)T^\mu(\slashed{p}+m)\gamma^0(T^\nu)^\dagger\gamma^0\right)\label{pol_sum_ampl}
\eeq
the additional terms due to $k_\mu T^\mu$ are at least of order $\Order(k^{-1})$, which goes beyond the soft-photon approximation. In this sense,  gauge invariance is only maintained up to the order one is working at. Nevertheless, it follows from \eqref{pol_sum} and \eqref{pol_sum_ampl} that the squared matrix element for the bremsstrahlung process is still given by 
\beq
|\M_\gamma|^2=-\frac{1}{2}\Tr\left((\slashed{p}'+m)T^\mu(\slashed{p}+m)\gamma^0T_\mu^\dagger\gamma^0\right)\label{M_gamma}.
\eeq
The evaluation of the traces in \eqref{M_gamma} is now straightforward, but tedious. Since we are eventually interested in
\beq
|\widetilde{\M}_\gamma|^2=\int \frac{\text{d}^3\mathbf{k}}{(2\pi)^3 2E_\gamma}|\M_\gamma|^2,
\eeq
the spin- and polarization-averaged squared matrix element integrated over the photon energies, all that remains to be done is the $\mathbf{k}$ integration, which may be performed on the basis of \eqref{phase_int_brems}. The final result 
can be expressed as a correction factor to the leading-order formula for $\pi N$ scattering \eqref{squ_matr_lead_ord}
\beq
|\widetilde{\M}_\gamma|^2=\frac{e^2}{8\pi^2}|\M^{\rm LO}_{\pi N}|^2C\left(s,t,E_{\rm max}\right)\label{brems_ampl_final},
\eeq
where 
\begin{align}
C_{\pi^-p}\left(s,t,E_{\rm max}\right)&=\log\frac{m_\gamma^2}{4E_{\rm max}^2}\bigg\{ 4-\frac{2m^2-t}{m^2}f(t,m)-\frac{2\mpi^2-t}{\mpi^2}f(t,\mpi)\notag\\
&+2\frac{s-m^2-\mpi^2}{m^2}g_{11}\left(s,m,\mpi\right)-2\frac{u-m^2-\mpi^2}{m^2}g_{11}\left(u,m,\mpi\right)\bigg\}\notag\\
&+2m^2g_{pp}(s)+2\mpi^2 g_{qq}(s)-2(s-m^2-\mpi^2)g_{pq}(s)\notag\\
&-(2m^2-t)g_{pp'}(s,t)-(2\mpi^2-t)g_{qq'}(s,t)-2(u-m^2-\mpi^2)g_{pq'}(s,t), \notag\\
C_{\pi^+p}\left(s,t,E_{\rm max}\right)&=\log\frac{m_\gamma^2}{4E_{\rm max}^2}\bigg\{ 4-\frac{2m^2-t}{m^2}f(t,m)-\frac{2\mpi^2-t}{\mpi^2}f(t,\mpi)\notag\\
&-2\frac{s-m^2-\mpi^2}{m^2}g_{11}\left(s,m,\mpi\right)+2\frac{u-m^2-\mpi^2}{m^2}g_{11}\left(u,m,\mpi\right)\bigg\}\notag\\
&+2m^2g_{pp}(s)+2\mpi^2 g_{qq}(s)+2(s-m^2-\mpi^2)g_{pq}(s)\notag\\
&-(2m^2-t)g_{pp'}(s,t)-(2\mpi^2-t)g_{qq'}(s,t)+2(u-m^2-\mpi^2)g_{pq'}(s,t), \notag\\
C_{\rm cex}\left(s,t,E_{\rm max}\right)&=\log\frac{m_\gamma^2}{4E_{\rm max}^2}\bigg\{ 2+\frac{s-m^2-\mpi^2}{m^2}g_{11}\left(s,m,\mpi\right)\bigg\}\notag\\
&+m^2g_{pp}(s)+\mpi^2 g_{qq}(s)-(s-m^2-\mpi^2)g_{pq}(s)\label{brems_ampl_result} .
\end{align}
$f(t,m)$ and $g_{11}\left(s,m,\mpi\right)$ are defined in App.~\ref{app:def}, while the remaining functions are relegated to App.~\ref{app:brems_func}. Note that the pion mass difference does not play a role for either of the three channels. For the charge exchange reaction this is true, as $q'$ does not enter. Furthermore, $C_{\rm cex}\left(s,t,E_{\rm max}\right)$ is in fact independent of $t$. Therefore
threshold kinematics are given by $s_{\rm thr}=(m+\mpi)^2$, $u_{\rm thr}=(m-\mpi)^2$ and $t=0$, since $s$ is the only Mandelstam variable needed for the charge exchange reaction. Using 
\begin{align}
f(0,m)&=f(0,\mpi)=1, \quad g_{11}(s_{\rm thr})=-\frac{m}{\mpi}, \quad g_{11}(u_{\rm thr})=\frac{m}{\mpi},\label{IR_div_threshold}\\
g_{pp}(s_{\rm thr})&=g_{pp'}(s_{\rm thr},0)=\frac{2}{m^2}, \quad  g_{qq}(s_{\rm thr})=g_{qq'}(s_{\rm thr},0)=\frac{2}{\mpi^2},
 \quad  g_{pq}(s_{\rm thr})=g_{pq'}(s_{\rm thr},0)=\frac{2}{m \mpi},\notag
\end{align}
one can easily show that $C(s_{\rm thr},0,E_{\rm max})$ vanishes as it should, since at threshold there is no phase space available for the emission of bremsstrahlung any more.

\subsection{Total amplitude}
\label{sec:total_ampl}

Eventually, we now have everything at hand to write down the full amplitude and show  how the cancelation of UV and IR divergences works. First, the results of Sects.~\ref{sec:tree}, \ref{sec:wavefunction}, and \ref{sec:loop_ampl}, are combined to
\begin{align}
B_i(s,t)&=B_i^{\rm tree}(s,t)+B_i^{\rm wf}(s,t)+B_i^{\rm loop}(s,t),\qquad i\in\left\{\pi^-p,\pi^+p,{\rm cex}\right\},\notag\\
D_i(s,t)&=D_i^{\rm tree}(s,t)+D_i^{\rm wf}(s,t)+D_i^{\rm loop}(s,t),\notag\\
B_i^{\rm tree}(s,t)&=B_i^{\rm v}(s,t)+B_i^{\rm a}(s,t)+B_i^{\rm p}(s,t)=B_i^{\rm LO}(s,t)+\Order(1),\notag\\
D_i^{\rm tree}(s,t)&=D_i^{\rm v}(s,t)+D_i^{\rm a}(s,t)+D_i^{\rm p}(s,t)=D_i^{\rm LO}(s,t)+\Order(p^2).
\end{align}
Second, the generalization of \eqref{squ_matr_lead_ord} beyond leading order reads
\begin{align}
|\M_{\pi^\pm p}|^2&=\left(4\mpp^2-t\right)D^{\rm tree}_{\pi^\pm p}D^{\rm LO}_{\pi^\pm p}+\nu t \left\{D^{\rm tree}_{\pi^\pm p} B^{\rm LO}_{\pi^\pm p}+D^{\rm LO}_{\pi^\pm p} B^{\rm tree}_{\pi^\pm p}\right\}-\frac{t}{4}\left\{t-4\mpi^2+4\nu^2\right\}
B^{\rm tree}_{\pi^\pm p}B^{\rm LO}_{\pi^\pm p}\notag\\
&+2\left(4m^2-t\right) D^{\rm LO}_{\pi^\pm p}\left(D^{\rm loop}_{\pi^\pm p}+D^{\rm wf}_{\pi^\pm p}\right)
+2\nu t \left\{ D^{\rm LO}_{\pi^\pm p}\left( B^{\rm loop}_{\pi^\pm p}+B^{\rm wf}_{\pi^\pm p}\right)+B^{\rm LO}_{\pi^\pm p} \left( D^{\rm loop}_{\pi^\pm p}+D^{\rm wf}_{\pi^\pm p}\right)\right\}\notag\\
&-\frac{t}{2}\left\{t-4\mpi^2+4\nu^2\right\}
B^{\rm LO}_{\pi^\pm p}\left(B^{\rm loop}_{\pi^\pm p}+B^{\rm wf}_{\pi^\pm p}\right),\notag\\
|\M_{\rm cex}|^2&=\left(\left(\mpp+\mn\right)^2-t\right)D^{\rm tree}_{\rm cex}D^{\rm LO}_{\rm cex}+\left(\nu t-\frac{\Delta_{\rm N}\Delta_\pi}{2}\right) \left\{D^{\rm tree}_{\rm cex} B^{\rm LO}_{\rm cex}+D^{\rm LO}_{\rm cex} B^{\rm tree}_{\rm cex}\right\}\notag\\
&-\frac{1}{4}\Big\{\lambda\left(t,\mpi^2,\mpii^2\right)+4t\nu^2+\left(2\mpi^2+2\mpii^2-t\right)\Delta_{\rm N}^2-4\nu\Delta_{\rm N}\Delta_\pi\Big\}
B^{\rm tree}_{\rm cex}B^{\rm LO}_{\rm cex}\notag\\
&+2\left(4m^2-t\right)D^{\rm LO}_{\rm cex}\left(D^{\rm loop}_{\rm cex}+D^{\rm wf}_{\rm cex}\right)
+2\nu t \left\{D^{\rm LO}_{\rm cex}\left( B^{\rm loop}_{\rm cex}+B^{\rm wf}_{\rm cex}\right)+B^{\rm LO}_{\rm cex} \left( D^{\rm loop}_{\rm cex}+D^{\rm wf}_{\rm cex}\right)\right\}\notag\\
&-\frac{1}{2}\Big\{\lambda\left(t,\mpi^2,\mpii^2\right)+4t\nu^2\Big\}
B^{\rm LO}_{\rm cex}\left(B^{\rm loop}_{\rm cex}+B^{\rm wf}_{\rm cex}\right)\label{squ_matr}.
\end{align}
To arrive at these expressions, we have used the following: the amplitude has been determined up to $\Order(p^3)$. Hence, the resulting squared amplitude can be described consistently up to $\Order(p^4)$ only, since, by interference with the leading order, $\Order(p^4)$ terms of the amplitude start contributing at $\Order(p^5)$. Therefore, terms of the form $B^{\rm loop}B^{\rm wf}$, etc., are of higher order in the chiral expansion. For the same reason, the imaginary parts of $D$ and $B$ were omitted. Strictly speaking, the squared amplitude looks like
\beq
|\M|^2=(4m^2-t)|D|^2+\nu t \left(B D^*+B^*D\right)-\frac{t}{4}\left\{t-4\mpi^2+4\nu^2\right\}|B|^2, \notag
\eeq
but the chiral order of the terms involving imaginary parts is at least $\Order(p^5)$.

\subsubsection{Cancelation of ultraviolet divergences}

We can only expect the UV divergences to cancel at $\Order(p^3)$, higher-order
singularities as introduced by infrared regularization have to be removed by
hand. Collecting the prefactors of the UV poles of $B(s,t)$ and
$D(s,t)$\footnote{Note that we cannot use $A(s,t)$ for this purpose, as it
  does not fulfil chiral power counting.} in $B^\lambda(s,t)$ and
$D^\lambda(s,t)$, the cancelation should work up to $\Order(p)$ and
$\Order(p^3)$, respectively.  The chiral expansion is performed in terms of
the rules given in (\ref{eq:rules}) and  
$\mpii^2=\mpi^2-2e^2F^2 Z$. In this way, we have checked explicitly that indeed
\beq
B^{\lambda}\sim\Order(p^2),\quad D^{\lambda}\sim\Order(p^4).
\eeq

\subsubsection{Cancelation of infrared divergences}
\label{sec:IR_cancel}

In contrast to the UV divergences, the IR singularities vanish exactly once bremsstrahlung is included. Inserting the IR divergent parts of $Z_{\rm p}$, $Z_\pi$, and the pertinent loop functions, into \eqref{squ_matr}, we obtain\footnote{The ellipsis denotes terms finite in the limit $m_\gamma\rightarrow 0$. Note that due to \eqref{IR_div_threshold} the coefficients of $\log m_\gamma$ vanish at threshold.}
\begin{align}
|\M_{\pi^- p}|^2&=-\frac{e^2}{8\pi^2}|\M^{\rm LO}_{\pi^- p}|^2\log\frac{m_\gamma^2}{m^2}\bigg\{ 4-\frac{2m^2-t}{m^2}f(t,m)-\frac{2\mpi^2-t}{\mpi^2}f(t,\mpi)\notag\\
&+2\frac{s-m^2-\mpi^2}{m^2}g_{11}\left(s,m,\mpi\right)-2\frac{u-m^2-\mpi^2}{m^2}g_{11}\left(u,m,\mpi\right)\bigg\}+\cdots\notag\\
|\M_{\pi^+ p}|^2&=-\frac{e^2}{8\pi^2}|\M^{\rm LO}_{\pi^+ p}|^2\log\frac{m_\gamma^2}{m^2}\bigg\{ 4-\frac{2m^2-t}{m^2}f(t,m)-\frac{2\mpi^2-t}{\mpi^2}f(t,\mpi)\notag\\
&-2\frac{s-m^2-\mpi^2}{m^2}g_{11}\left(s,m,\mpi\right)+2\frac{u-m^2-\mpi^2}{m^2}g_{11}\left(u,m,\mpi\right)\bigg\}+\cdots\notag\\
|\M_{\rm cex}|^2&=-\frac{e^2}{8\pi^2}|\M^{\rm LO}_{\rm cex}|^2\log\frac{m_\gamma^2}{m^2}\bigg\{ 2+\frac{s-m^2-\mpi^2}{m^2}g_{11}\left(s,m,\mpi\right)\bigg\}+\cdots,\label{IR_div_ampl}
\end{align}
which actually cancels the $\log m_\gamma$ terms in \eqref{brems_ampl_final}.  Thus, we have shown explicitly that 
\beq
|\M_{\pi N}|^2+|\widetilde{\M}_\gamma|^2
\eeq
is infrared finite.

\subsubsection{Checks on the amplitude}

First of all, the presence of UV and IR divergences in the theory provides a powerful consistency check  on the amplitude: in the final result both types of singularities must cancel. Beyond that, we have shown that the coefficient of the Coulomb pole is indeed related to the $\pi N$ scattering amplitude (Sect.~\ref{Coulomb_pole_ampl}) and that at threshold the IR divergences cancel among themselves, while bremsstrahlung is absent (Sects.~\ref{sec:Bremsstrahlung_amplitude} and \ref{sec:IR_cancel}).

In order to check the representations of the scalar loop functions based on Feynman parameter integrals we have also invoked dispersion relations (cf.\ Apps.~\ref{app:def} and \ref{Scal_loop_func}). If the dispersive representation does not converge due to an IR divergence, we have at least checked numerically that dispersion relations for finite $m_\gamma$ reproduce the Feynman representation with explicitly separated IR divergence when the photon mass is made sufficiently small. 

Finally, we have verified that in the isospin limit the result quoted in \cite{BL01} is reproduced.

\section{Isospin violation above threshold}
\def\theequation{\arabic{section}.\arabic{equation}}
\setcounter{equation}{0}
\label{chap:above}

First of all, one should note that the concept of a partial wave expansion becomes doubtful once electromagnetic interactions are included, since a partial wave projection of the $1/t$ terms generated by the one-photon-exchange topologies is not feasible. However, as isospin violation due to static Coulomb interaction is in principle well understood, we will focus on the real part of the scattering amplitude with the one-photon-exchange contributions amputated and the Coulomb pole as separated in \eqref{Coulomb_pole_full_amp} subtracted.\footnote{To avoid an unnecessarily clumsy notation, the resulting amplitude is again denoted by $T_{\pi N}$.} This procedure ensures that $T_{\pi N}$ is finite at threshold and guarantees consistency with the treatment of the scattering lengths in \cite{HKM09}.

\subsection{Triangle relation above threshold}
\label{sec:triangle_above}

We will quantify isospin violation above threshold in terms of the triangle relation
\beq
R_{l\pm}(s)=2\frac{f_{l\pm}^{\pi^+ p}(s)-f_{l\pm}^{\pi^- p}(s)-\sqrt{2}\, f_{l\pm}^{\rm cex}(s)}{f_{l\pm}^{\pi^+ p}(s)-f_{l\pm}^{\pi^- p}(s)+\sqrt{2}\, f_{l\pm}^{\rm cex}(s)}\label{triangle_above}.
\eeq
However, this expression involves several problems: first, the IR divergences due to virtual photons no longer cancel among themselves, such that the inclusion of bremsstrahlung is inevitable. Nevertheless, even in the soft-photon approximation this is in principle only possible at the level of squared amplitudes. Second, elastic and inelastic channels have a different threshold behavior in the $P$-wave projections, which renders \eqref{triangle_above} useless near threshold.

To remedy the bremsstrahlung problem, one might try to define an analogous triangle relation for the squared amplitudes, which may be regarded as infrared safe objects. Obviously, there is no way to combine
\beq
|T^{\pi^+ p}|^2=|T^+-T^-|^2, \quad |T^{\pi^- p}|^2=|T^++T^-|^2, \quad \text{and} \quad |T^{\rm cex}|^2=2|T^-|^2,
\eeq
in a non-trivial way such that they cancel in the isospin limit.\footnote{As discussed in Sect. \ref{sec:total_ampl}, we can even replace the amplitudes by their real part up to higher orders.} In \cite{FM01}, the infrared divergences are removed at the amplitude level just replacing $m_\gamma$ by the detector resolution $E_{\rm max}$. As we have seen in Sects. \ref{sec:Bremsstrahlung_amplitude} and \ref{sec:IR_cancel}, this admittedly catches the leading effect, but neglects finite terms, which e.g.\ affect the logarithm such that effectively $m_\gamma$ is replaced by $2E_{\rm max}$. Hence, the full outcome of the soft-photon approximation should somehow be accounted for. For that reason, we define the partial waves in \eqref{Goldberger} not for the pure scattering process $T_{\pi N}$, but include bremsstrahlung via 
\beq
T_{\pi N}\rightarrow T_{\pi N+\pi N\gamma}\equiv T_{\pi N}+T_{\pi N}^{\rm LO}\frac{e^2}{16\pi^2}C(s,t,E_{\max})\label{brems_above_thresh},
\eeq
where $T_{\pi N}^{\rm LO}$ is determined by \eqref{ampl_DB} and \eqref{lead_order_ampl} and $C(s,t,E_{\max})$ is defined in \eqref{brems_ampl_result}. The corresponding triangle relation 
is infrared finite, coincides with \cite{HKM09} at threshold and involves amplitudes satisfying
\beq
|\M_{\pi N+\pi N\gamma}|^2\equiv\frac{1}{2}\sum\limits_{\rm spins}|T_{\pi N+\pi N\gamma}|^2=|\M_{\pi N}|^2+|\widetilde{\M}_\gamma|^2+\Order(p^5),
\eeq
such that the required decoherent summation of the pure process and bremsstrahlung is maintained up to higher orders. 

The problem we encounter for the $P$-wave projections is due to the threshold behavior of the partial wave amplitudes $T_l$: for elastic scattering, their real part can be generically written as
\beq
\text{Re}\, T_l=|\pp|^{2l}\left(a_l+b_l|\pp|^2+\cdots\right),
\eeq
but once inelasticities arise, the prefactor has to be modified according to $|\pp|^{2l}\rightarrow |\pp|^l|\mathbf{p'}|^l$. While the $S$-wave is obviously not affected, the threshold behavior of the $P$-waves differs between the elastic channels and the charge exchange:  whereas the former vanish $\sim \lambda\left(s,\mpp^2,\mpi^2\right)$, the latter behaves as ${\lambda^{1/2}\left(s,\mpp^2,\mpi^2\right)\lambda^{1/2}\left(s,\mn^2,\mpii^2\right)}$. Hence, $R$ will be totally dominated by $f_{1\pm}^{\rm cex}(s)$ for $s\rightarrow(\mpp+\mpi)^2$. Even more, if we take the denominator in the isospin limit, $R$ will diverge as a square root at threshold. 

In order to circumvent this problem, we remove the known kinematical threshold behavior, i.e.\ instead of $f_{1\pm}(s)$ we consider the triangle relation for
\beq
 \frac{f_{1\pm}^{\pi\pm p}(s)}{|\pp|^2} \quad \text{and} \quad \frac{f_{1\pm}^{\rm cex}(s)}{|\pp||\mathbf{p'}|},
\eeq
respectively. In this way, the quantity of interest for the $P$-wave projections becomes
\beq
R_{1\pm}(s)=2\frac{f_{1\pm}^{\pi^+ p}(s)-f_{1\pm}^{\pi^- p}(s)-\sqrt{2}\, \frac{|\pp|}{|\mathbf{p'}|}f_{1\pm}^{\rm cex}(s)}{f_{1\pm}^{\pi^+ p}(s)-f_{1\pm}^{\pi^- p}(s)+\sqrt{2}\, \frac{|\pp|}{|\mathbf{p'}|}f_{1\pm}^{\rm cex}(s)}\label{triangle_pwave},
\eeq
which is not only regular at threshold,\footnote{In the actual evaluation, isospin violation in $R$ will be restricted to first order in $\delta$, such that the correction factor in the denominator disappears.} but reduces just to a triangle relation for the $P$-wave scattering lengths, completely analogous to the $S$-wave projection. If we incorporate bremsstrahlung according to \eqref{brems_above_thresh}, \eqref{triangle_above} (for $l=0$) and \eqref{triangle_pwave} should therefore yield a consistent framework to study isospin violation above threshold.

\subsection{Numerical results}

\subsubsection{Low-energy constants}
\label{sec:LEC_triangle}

In this section, we specify the required low-energy constants not yet collected in \cite{HKM09}.  In the strong sector, these are $c_4$,  $d_1^{\rm r}(\mu)+d_2^{\rm r}(\mu)$, $d_3^{\rm r}(\mu)$, and  $d_{18}^{\rm r}(\mu)$. The $d_i$ are often given in terms of the scale-independent quantities $\bar{d}_i$, which are related to  $d_i^{\rm r}(\mu)$ by 
\beq
\Fpi^2d_i^{\rm r}(\mu)=\Fpi^2\bar{d}_i+\frac{\beta_i}{16\pi^2}\log\frac{\mpi}{\mu},\quad \text{for} \quad i\in\{1,2,3\},\quad
\Fpi^2d_{18}^{\rm r}(\mu)=\Fpi^2\bar{d}_{18}+\frac{\ga l_4^{\rm r}(\mu)}{2}.
\eeq
$c_4$ will be taken from~\cite{M05}, where various previous analyses are briefly reviewed yielding
\beq
c_4=3.5^{+0.5}_{-0.2} \,{\rm GeV}^{-1}\label{c4}.
\eeq
$\bar{d}_{18}$ is determined by the Goldberger-Treiman discrepancy 
\beq
\Delta_{\rm GT}=1-\frac{\mpp \ga}{\Fpi g_{\pi NN}}=-\frac{2\mpi^2\bar{d}_{18}}{\ga}.
\eeq
There are many values for $\Delta_{\rm GT}$ available in the literature \cite{
BL01,BEM01,Nagy2004,Nasrallah99,deSwart97}, ranging from $(1.4\pm 0.9)\%$  up to a few percent. However, small values of $\Delta_{\rm GT}$ seem to be favored: first, the experimental value of $\ga$ has increased over the years, whereas $\Fpi$ decreased due to radiative corrections.\footnote{Recently, $\Fpi$ has further decreased from $92.4\, {\rm MeV}$ to $92.2\,{\rm MeV}$, which is not included in the quoted literature.} Both effects tend to reduce $\Delta_{\rm GT}$. Second, it is shown in \cite{BEM01} that only rather small values of $\Delta_{\rm GT}$ are consistent with the standard picture of chiral symmetry breaking. We conclude that values of $\Delta_{\rm GT}$ much larger than about $2\,\%$ seem very unlikely. Consequently, we will use \cite{Nasrallah99}
\beq
\Delta_{\rm GT}=(1.4\pm 0.9)\%
\eeq
in the numerical work, which corresponds to ($\mu=1\,{\rm GeV}$)
\beq
\Fpi^2d_{18}^{\rm r}=(-2.1\pm 2.6)\cdot 10^{-3}.
\eeq
The remaining $d_i^{\rm r}$ are fixed based on $\pi N$ threshold parameters (for details of this method we refer to \cite{Eil}): as far as LECs are concerned, the isovector $S$-wave scattering length $a^-$, the isovector $P$-wave scattering lengths $a_{1 \pm}^-$, and the isovector $S$-wave effective range $b_{0+}^-$, depend at $\Order(p^3)$ only on $c_4$, $\bar{d}_1+\bar{d}_2$, $\bar{d}_3$, and $\bar{d}_5$. Inverting this linear system yields e.g.\      
\begin{align}
\bar{d}_1+\bar{d}_2&=\frac{\ga \bar{d}_{18}}{2}-\frac{3+21\ga^2+2\ga^4}{1152\pi ^2\Fpi^2}+\frac{\ga^2\left(3+2\ga^2\right)\mpi}{192\pi  \Fpi^2(\mpp+\mpi)}-\frac{\ga^2\left(4\mpp^2+4\mpp \mpi+3\mpi^2\right)}{32\mpp^2\mpi^2}\notag\\
&+\frac{2\mpp+\mpi}{32\mpp \mpi(\mpp+\mpi)}-\frac{\pi  \Fpi^2}{\mpp \mpi}(\mpp a_{1-}^-+2\mpp a_{1+}^-+3\mpi a_{1+}^-). 
\end{align}
For $a^-$ and $a_{1\pm}^-$ we use the same values as in \cite{HKM09}. Together with
$
b_{0+}^-=(8\pm 8)\cdot 10^{-3}\mpi^{-3}
$
this leads to
\begin{align}
\Fpi^2(d_1^{\rm r}+d_2^{\rm r})&=(38\pm 8)\cdot 10^{-3},\quad \Fpi^2d_3^{\rm r}=(-38\pm 10)\cdot 10^{-3}, \quad
\Fpi^2d_5^{\rm r}=(-2.4\pm 3.0)\cdot 10^{-3},\notag\\
c_4&=(3.4\pm 0.3)\,{\rm GeV}^{-1}.
\end{align}
Note that \eqref{c4} is reproduced quite well, while this value for $d_5^{\rm r}$ is at least compatible with \cite{HKM09} within the errors. For consistency reasons the new value for $d_5^{\rm r}$ is used in order to treat all $d_i$ in question in the same manner. Note that the errors of the $d_i$ obtained in this way are strongly correlated.

In general, the remaining electromagnetic LECs are treated as in \cite{HKM09}. However, there is one linear combination of LECs, for which this method fails, since the $\beta$-function vanishes:
\beq
e^2\Fpi^2(2g_3^{\rm r}+g_4^{\rm r})-4B(\md-\muu)(2d_{17}^{\rm r}-d_{18}^{\rm r}-2d_{19}^{\rm r})\label{axial_LECs}.
\eeq
As except for $d_{18}^{\rm r}$ none of these LECs is known, the central value of the whole combination is set to zero in the numerical analysis. One possible way to estimate the uncertainty might be based on the observation that this combination of LECs is nothing but $\tilde{g}_{\rm p}-\tilde{g}_{\rm n}$ (cf.~\eqref{ax_coupl_shift}). In view of the finding for $\Delta_{\rm GT}$, we regard $0.5\%$ as a reasonable upper limit for the contribution of \eqref{axial_LECs} to isospin breaking in $\ga$. The same effect would be generated by a $\beta$-function of the magnitude
\beq
0.5\cdot 10^{-2}\ga \frac{16\pi^2}{e^2}\sim 11.
\eeq
On the other hand, the vanishing of the $\beta$-function means that this combination is particularly stable against variation of the renormalization scale, which hints at a value of the total counterterm not being exceptionally large. This is the reason why in \cite{DKM} in such cases the $\beta$-function is set to $1$, which just reproduces the naive order-of-magnitude estimate $1/16\pi^2$. Since the $\beta$-function of both $2g_3^{\rm r}+g_4^{\rm r}$ and $2d_{17}^{\rm r}-d_{18}^{\rm r}-2d_{19}^{\rm r}$ vanishes, one could also argue that these two combinations should be treated separately. Counting $e^2\Fpi^2\sim B(\md-\muu)$ and adding the errors linearly, this would correspond to an ``effective'' $\beta$-function of $\sim 5$. All in all, we conclude that it is very difficult to obtain a reliable error for \eqref{axial_LECs}. What we will eventually use is
\beq
\Big|e^2\Fpi^2(2g_3^{\rm r}+g_4^{\rm r})-4B(\md-\muu)(2d_{17}^{\rm r}-d_{18}^{\rm r}-2d_{19}^{\rm r})\Big|\sim \frac{5 e^2}{16\pi^2} \sim 3\cdot 10^{-3},
\eeq
which should give a reasonable estimate, though not the most conservative possible. Note that this value is somewhat smaller than the estimate given in~\cite{vanKolck96} in the context of the two-nucleon system.\footnote{The low-energy constants used in~\cite{vanKolck96} can be most easily compared to the present framework by matching to the charge-symmetry breaking corrections to the pion-nucleon coupling constant given in~\cite{EM05}.}

\subsubsection{Threshold divergences}
\label{sec:thres_div}

In the numerical analysis, there are several obstacles to be overcome, which are related to the behavior of the amplitude at threshold. First of all, the basis functions of the tensor decomposition associated with $I_B^{ij}(s)$ (cf.\ diagram $(s_6)$), $I_B^{\gamma j}(s)$ $(a_3)$, $V_{11}(s)$ $(v_1)$, and $A_{11}(s)$ $(a_9)$, and the box-graph topologies $I_{13}^{i}(s,t)$ $(s_2)$, $I_{13}^{\gamma}(s,t)$ $(a_7)$, $A_{12}(s,t)$ $(a_6)$, and $A_{21}(s,t)$ $(a_8)$, involve prefactors of the type\footnote{For the sake of simplicity, the following discussion refers to the isospin limit. However, we have checked that the conclusions persist if $\Delta_\pi \neq 0$ is taken into account.}
\beq
\frac{1}{\lambda\left(s,m^2,\mpi^2\right)} \qquad \text{and} \qquad \frac{1}{\lambda\left(s,m^2,\mpi^2\right)+s t},
\eeq
respectively, which suggest threshold divergences even worse than the Coulomb pole. In fact, these singularities are canceled by the numerators, such that the loop functions are actually finite at threshold once the Coulomb pole is subtracted. To avoid any complications inherent to this (numerical) cancelation we start the analysis not directly at threshold $s_{\rm thr}=(\mpp+\mpi)^2\sim 1.162 \,{\rm GeV}^2$  but at $s=1.17\,{\rm GeV}^2$. As the final results are smooth for $s\rightarrow s_{\rm thr}$ this implies no severe limitation. In addition, the tensor decomposition of the box graphs causes difficulties also above threshold: it follows from \eqref{kin_CMS} that $\lambda(s,m^2,\mpi^2)+s t=0$ for $z=-1$ independent of the actual value of $s$, which induces numerical problems in the partial wave projection of the corresponding amplitudes. Fortunately, it turns out that the $z$-dependence of these loop functions is very weak, and, moreover, can be well approximated by a straight line. For that reason, we replace the full $z$ dependence by the interpolation between $z_1=0.8$ and $z_2=0.9$. The precise choice of these points is of course rather arbitrary, but accounts for the fact that difficulties arise near $z=-1$.

Especially for the $P$-wave projections, another difficulty concerns the consistent treatment of the kinematics. Let us rewrite the relevant part of the numerator of \eqref{triangle_pwave} as
\begin{align}
\left(E_p+\mpp\right)&\Big\{A_1^{\pi^+ p}(s)-A_1^{\pi^- p}(s)-\sqrt{2}\,C_{A+}(s)A_1^{\rm cex}\notag\\
&+\left(\sqrt{s}-\mpp\right)\left(B_1^{\pi^+ p}(s)-B_1^{\pi^- p}(s)-\sqrt{2}\,C_{B+}(s)B_1^{\rm cex}\right)\Big\}\notag\\
+\left(E_p-\mpp\right)&\Big\{-A_{1\pm 1}^{\pi^+ p}(s)+A_{1\pm 1}^{\pi^- p}(s)+\sqrt{2}\,C_{A-}(s)A_{1\pm 1}^{\rm cex}\notag\\
&+\left(\sqrt{s}+\mpp\right)\left(B_{1\pm 1}^{\pi^+ p}(s)-B_{1\pm 1}^{\pi^- p}(s)-\sqrt{2}\,C_{B-}(s)B_{1\pm 1}^{\rm cex}\right)\Big\},
\end{align}
where we have defined the correction factors
\begin{align}
C_{A\pm}(s)&=\sqrt{\frac{\left(\sqrt{s}\pm\mn\right)^2-\mpii^2}{\left(\sqrt{s}\pm\mpp\right)^2-\mpi^2}}\sqrt{\frac{\lambda\left(s,\mpp^2,\mpi^2\right)}{\lambda\left(s,\mn^2,\mpii^2\right)}},\notag\\
C_{B\pm}(s)&=\Bigg\{\sqrt{\frac{\left(\sqrt{s}\pm\mn\right)^2-\mpii^2}{\left(\sqrt{s}\pm\mpp\right)^2-\mpi^2}}
\mp\frac{\Delta_{\rm N}}{2\left(\sqrt{s}\mp \mpp\right)}\Bigg\}\sqrt{\frac{\lambda\left(s,\mpp^2,\mpi^2\right)}{\lambda\left(s,\mn^2,\mpii^2\right)}}\label{corr_fact}.
\end{align}
To avoid artificial threshold divergences, it is important to treat these correction factors in the same way as the corresponding part of the amplitude they are multiplied with. Therefore, we adopt the following conventions:
for tree contributions (including the wave function renormalization) the full kinematics with $\Delta_{\rm N}\neq 0$ and $\Delta_\pi\neq 0$ are used. In particular, the nucleon masses in the axial parts of $A^{\rm LO}$ and $B^{\rm LO}$ in \eqref{wfr} are restored according to \eqref{axial_tree} with $\Deltam=0$. Consequently, \eqref{corr_fact} is applied as it stands. The sole exception is made for the counterterms involving $e^2 k_i$ or $e^2 g_i$, for which the correction factors \eqref{corr_fact} are set to $1$, since the effects of the mass differences are anyway outweighed by the uncertainty in these LECs. In contrast, we have neglected the nucleon mass difference as far as loops and bremsstrahlung are concerned. Thus, for these contributions $\mn$ should be replaced by $\mpp$ in \eqref{corr_fact}. Note that to ensure consistency $\Delta_\pi$ is kept in the electromagnetic loops, although this is formally an $\Order(e^4)$ effect. We conclude that the consistent treatment of the mass differences is crucial to prevent remnants of the different threshold behavior of the elastic channels and the charge exchange reaction as described in Sect. \ref{sec:triangle_above} from spoiling the threshold properties of $R_{1\pm}$.

\subsubsection{Triangle relation}
\label{sec:triangle_res}

\begin{figure}
\begin{center}

\vspace{-23pt}

\includegraphics[width=0.55\linewidth,clip]{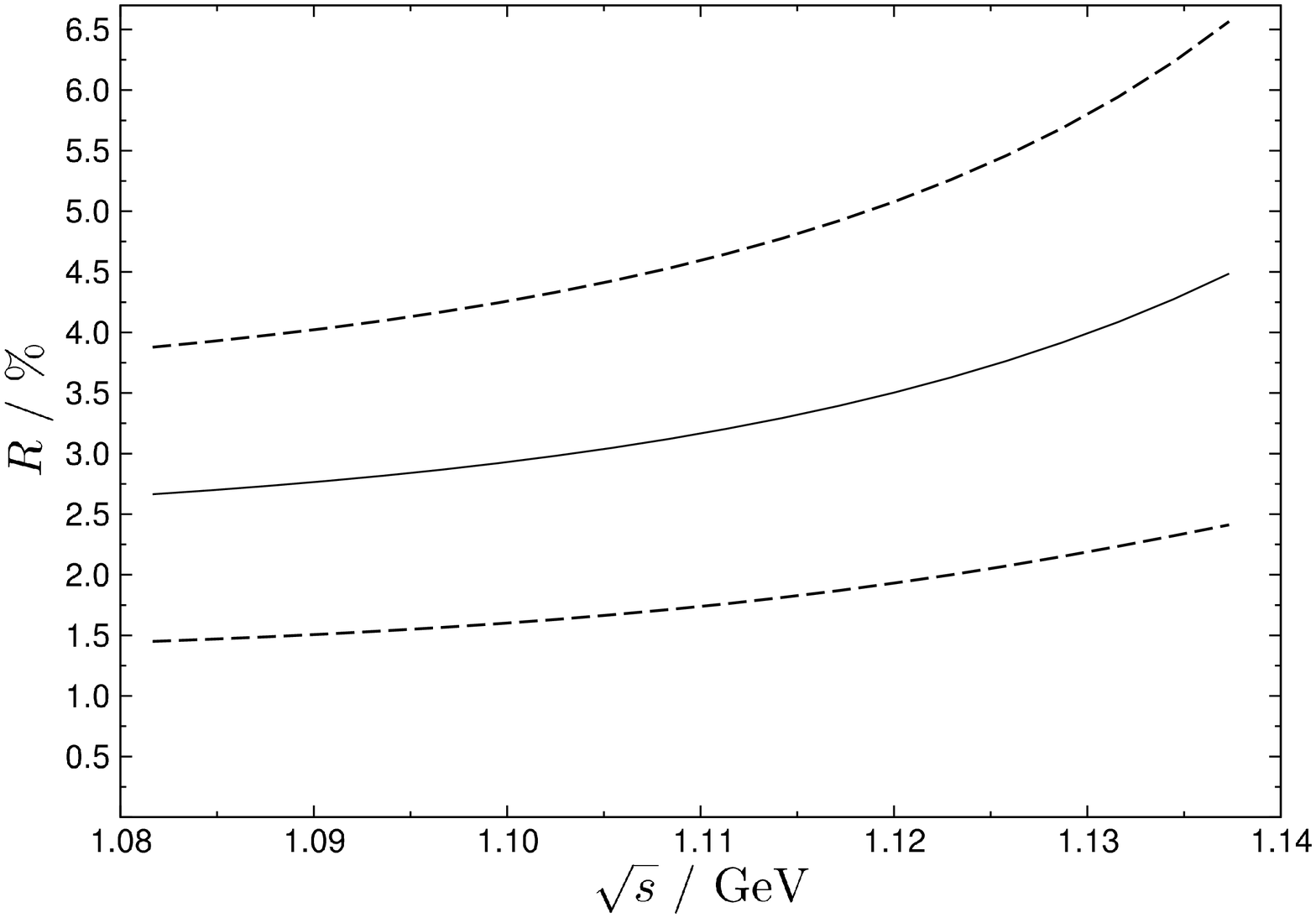}

\vspace{0.3cm}

\includegraphics[width=0.55\linewidth,clip]{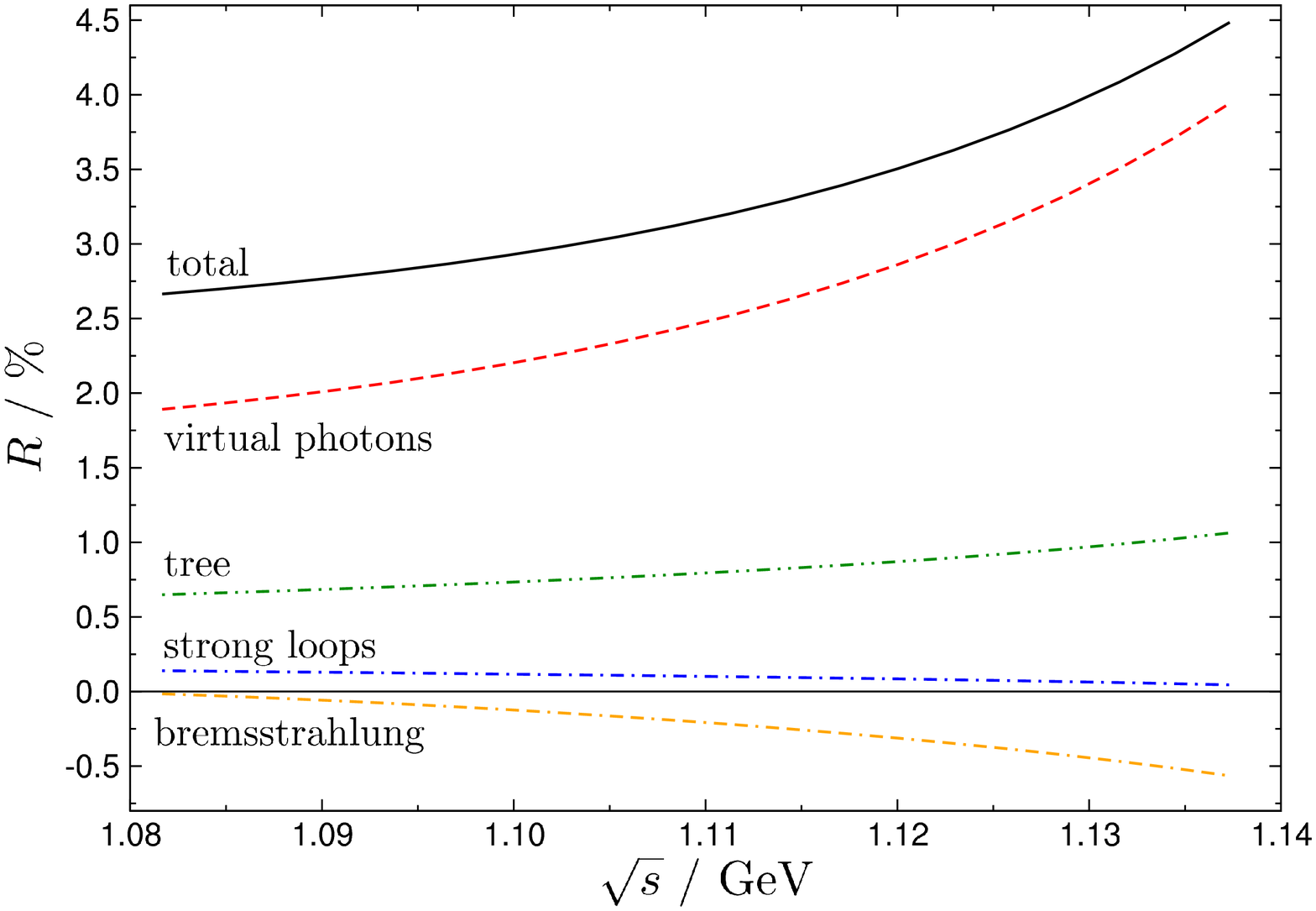}

\vspace{0.3cm}

\includegraphics[width=0.55\linewidth,clip]{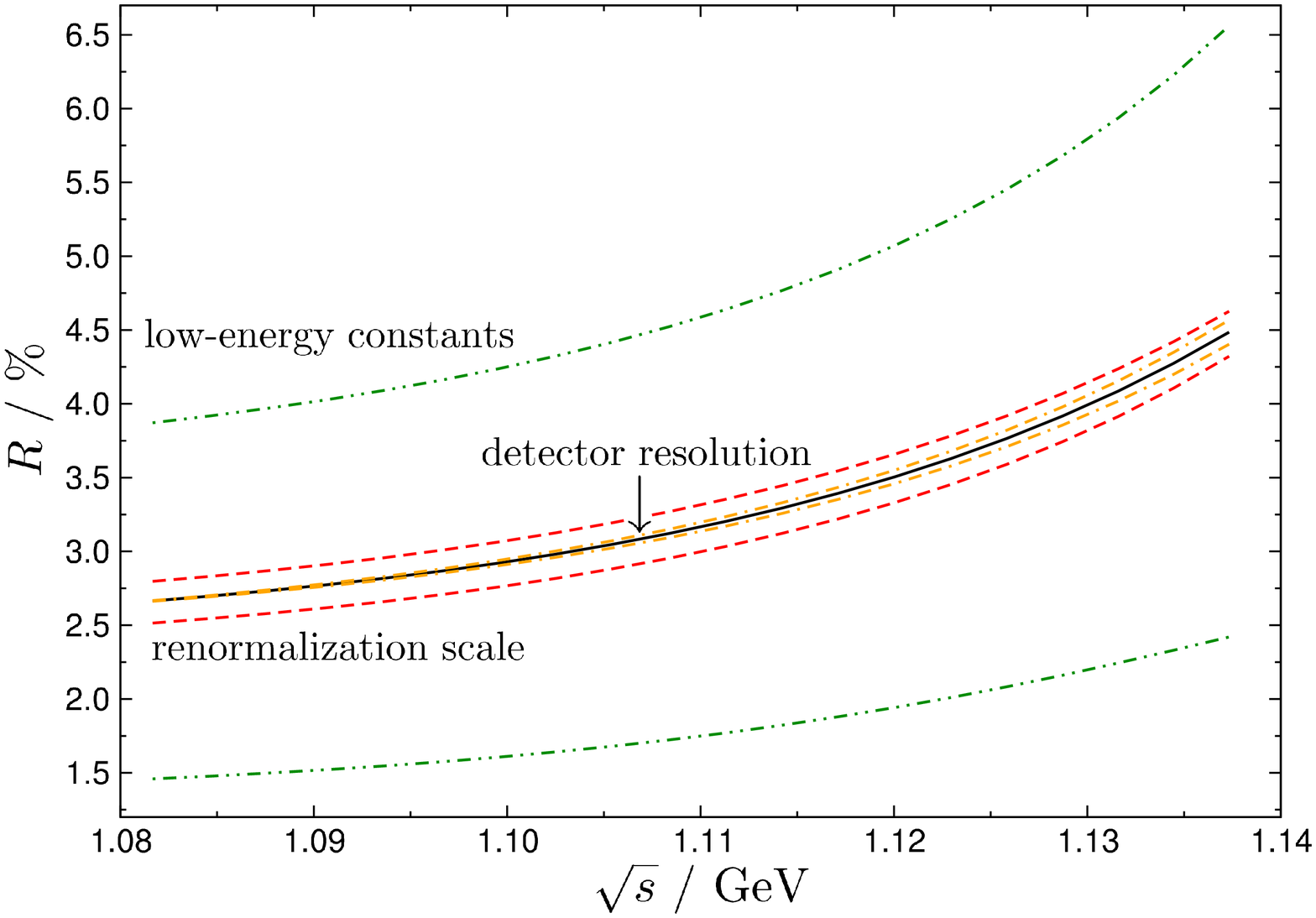}
\end{center}
\caption{Triangle relation in the $S$-wave. Upper panel: central result (solid
  line) with uncertainties (dashed). Middle panel: individual contributions to
the central result. Lower panel:  individual contributions to the uncertainty.}
\label{fig:swave}
\end{figure}

\begin{figure}
\begin{center}

\vspace{-23pt}

\includegraphics[width=0.55\linewidth,clip]{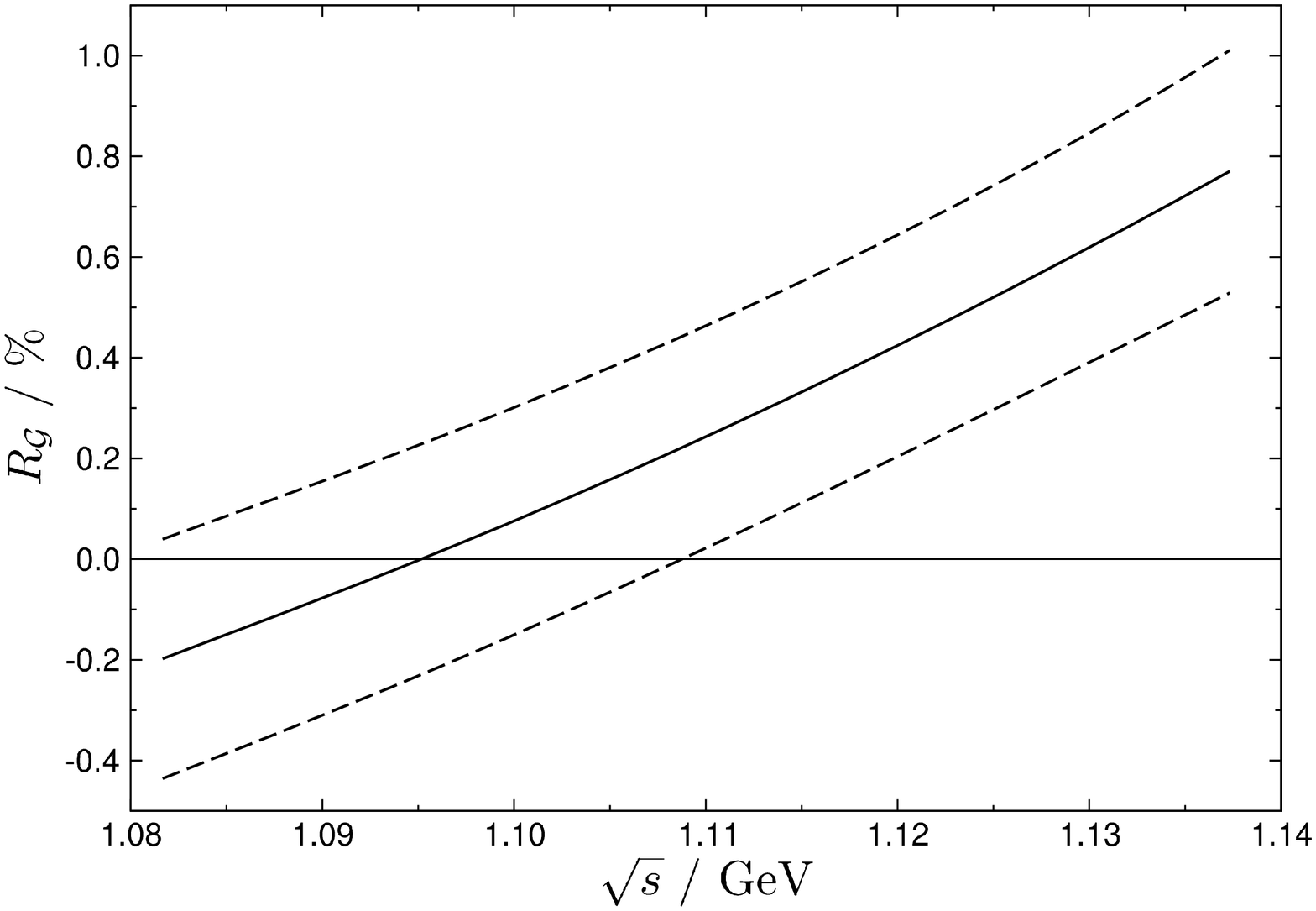}

\vspace{0.3cm}

\includegraphics[width=0.55\linewidth,clip]{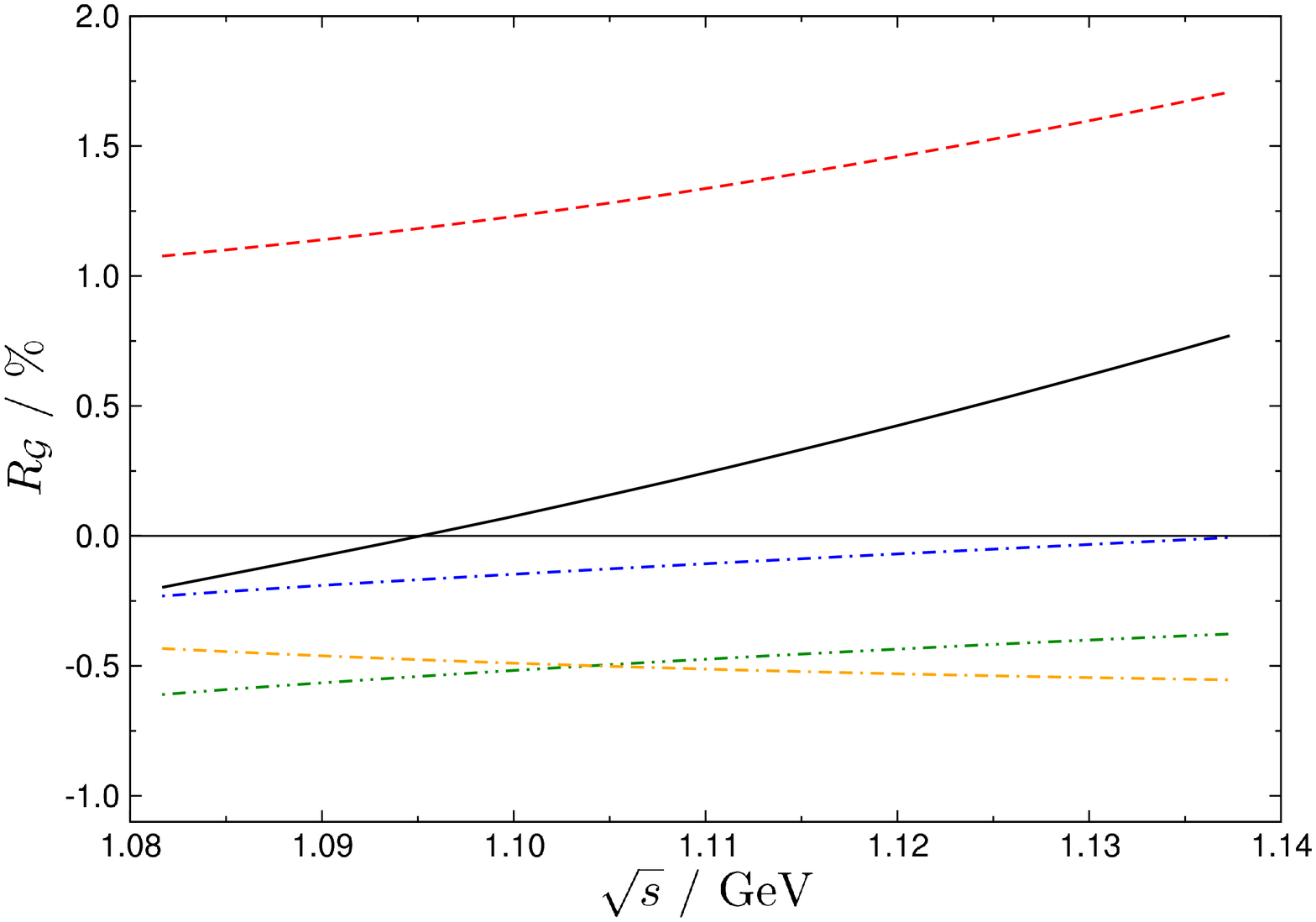}

\vspace{0.3cm}

\includegraphics[width=0.55\linewidth,clip]{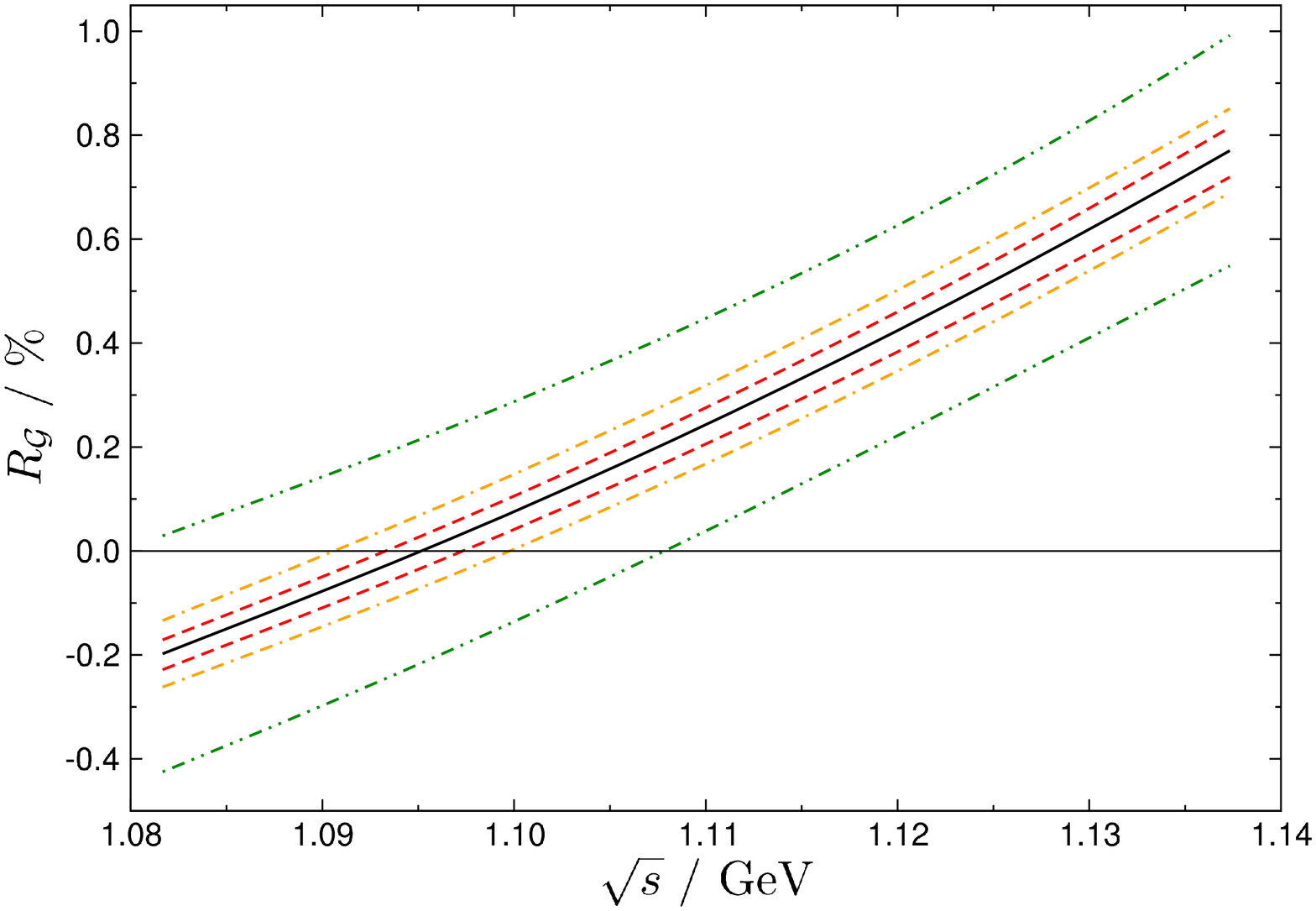}
\end{center}
\caption{Triangle relation in the $\mathcal{G}$-projection.
For notations, see Fig.~\ref{fig:swave}.
}
\label{fig:gwave}
\end{figure}

\begin{figure}
\begin{center}

\vspace{-23pt}

\includegraphics[width=0.55\linewidth,clip]{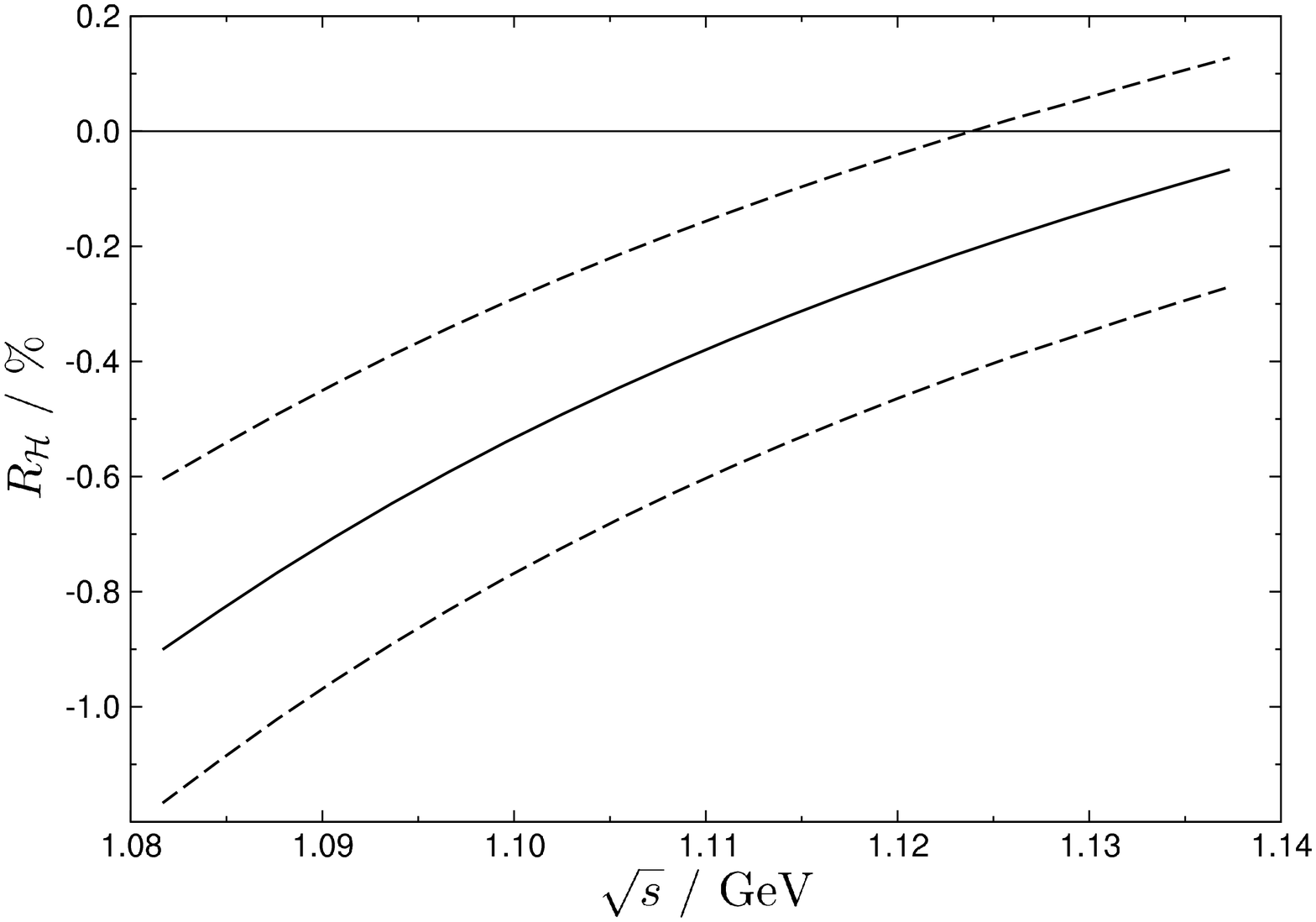}

\vspace{0.3cm}

\includegraphics[width=0.55\linewidth,clip]{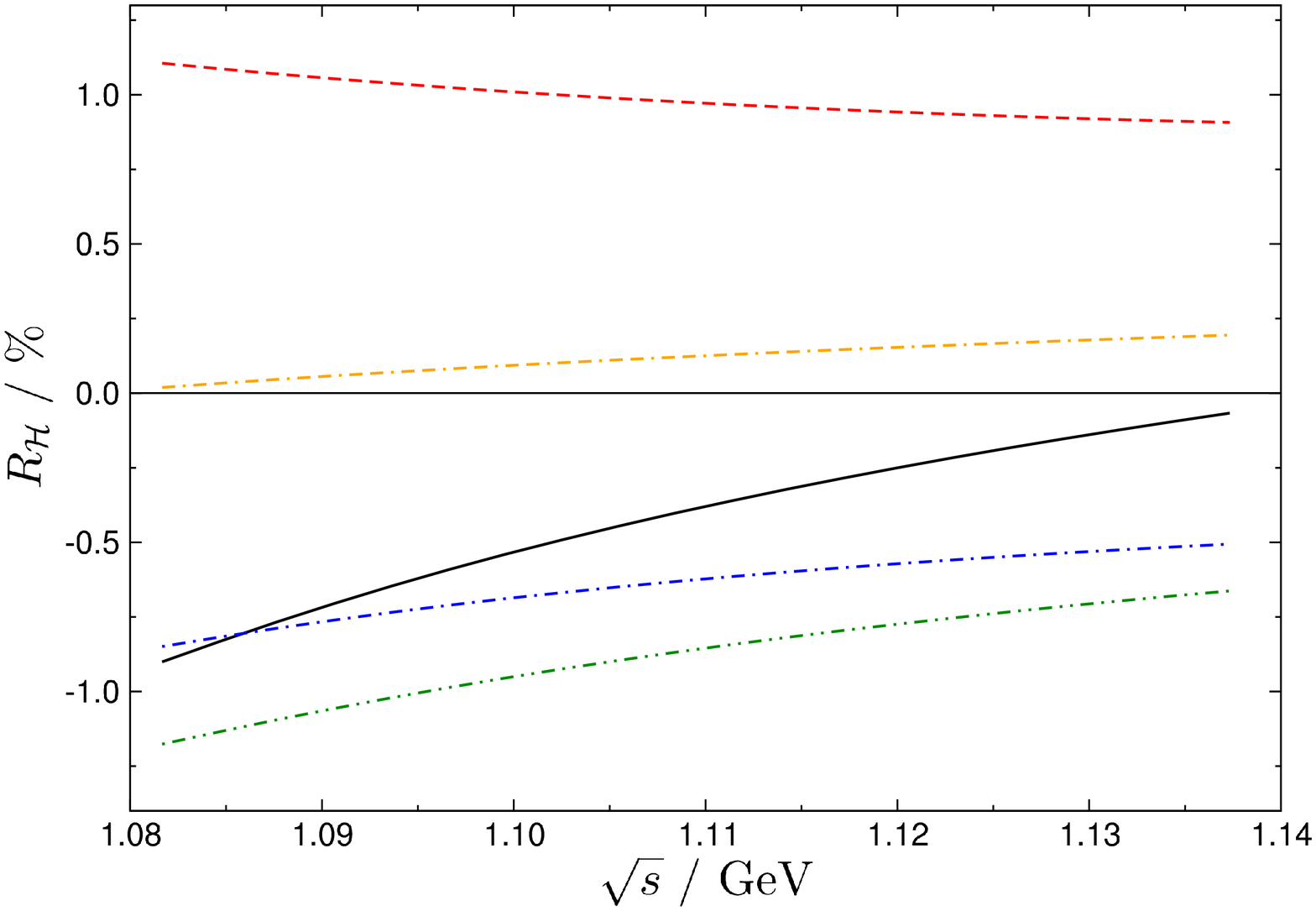}

\vspace{0.3cm}

\includegraphics[width=0.55\linewidth,clip]{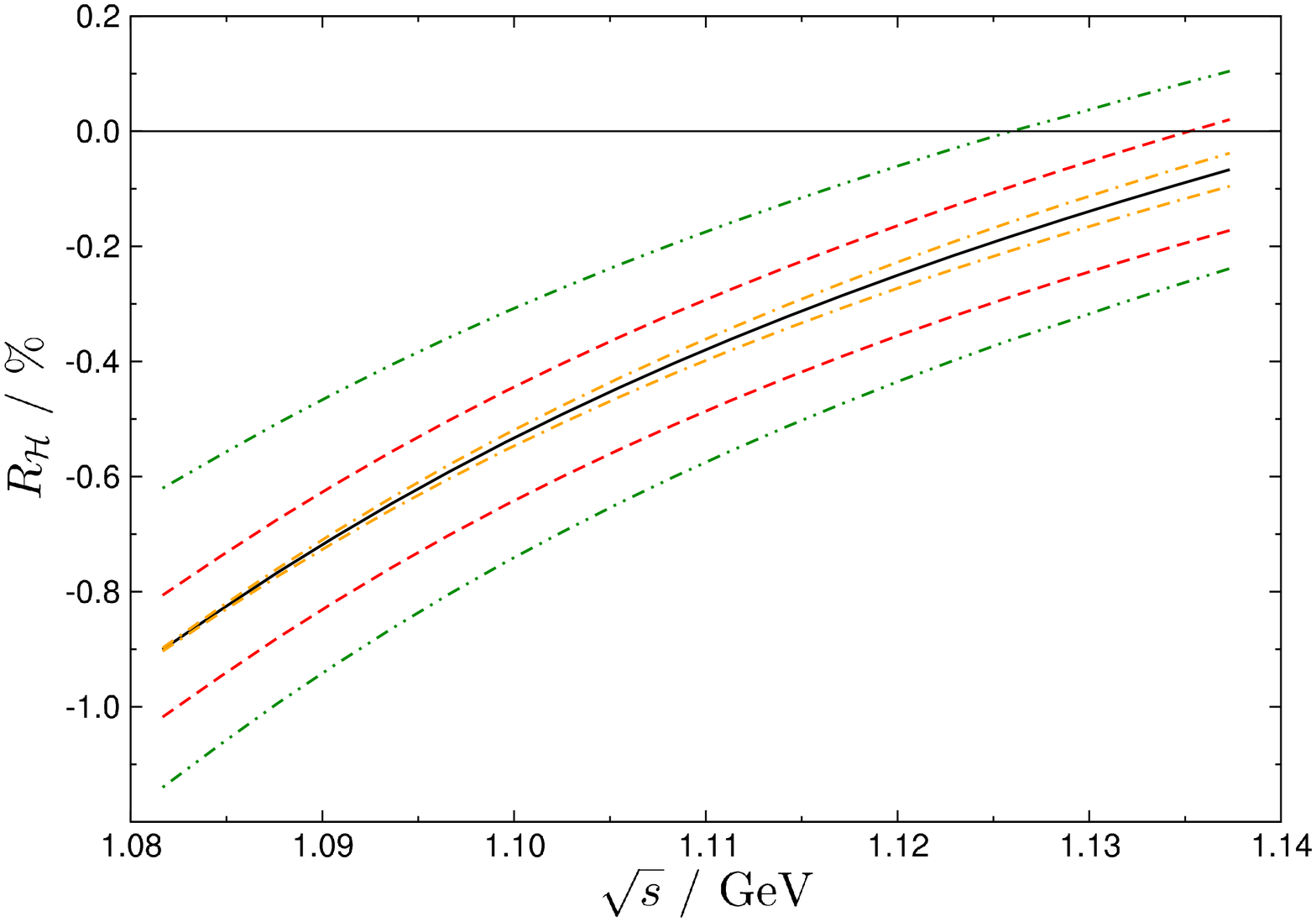}
\end{center}
\caption{Triangle relation in the $\mathcal{H}$-projection.
For notations, see Fig.~\ref{fig:swave}.
}
\label{fig:hwave}
\end{figure}

\begin{figure}
\begin{center}

\vspace{-23pt}

\includegraphics[width=0.55\linewidth,clip]{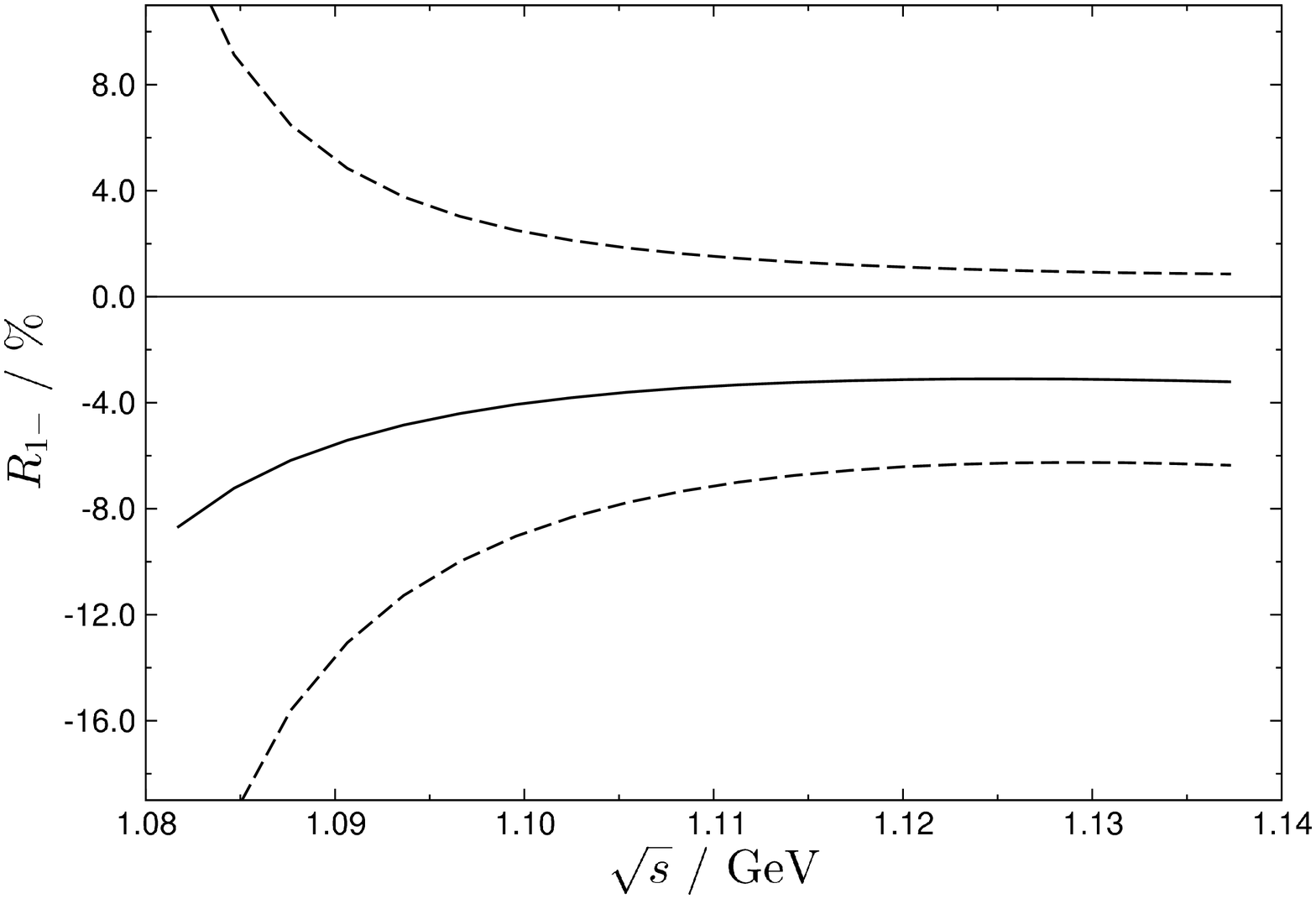}

\vspace{0.3cm}

\includegraphics[width=0.55\linewidth,clip]{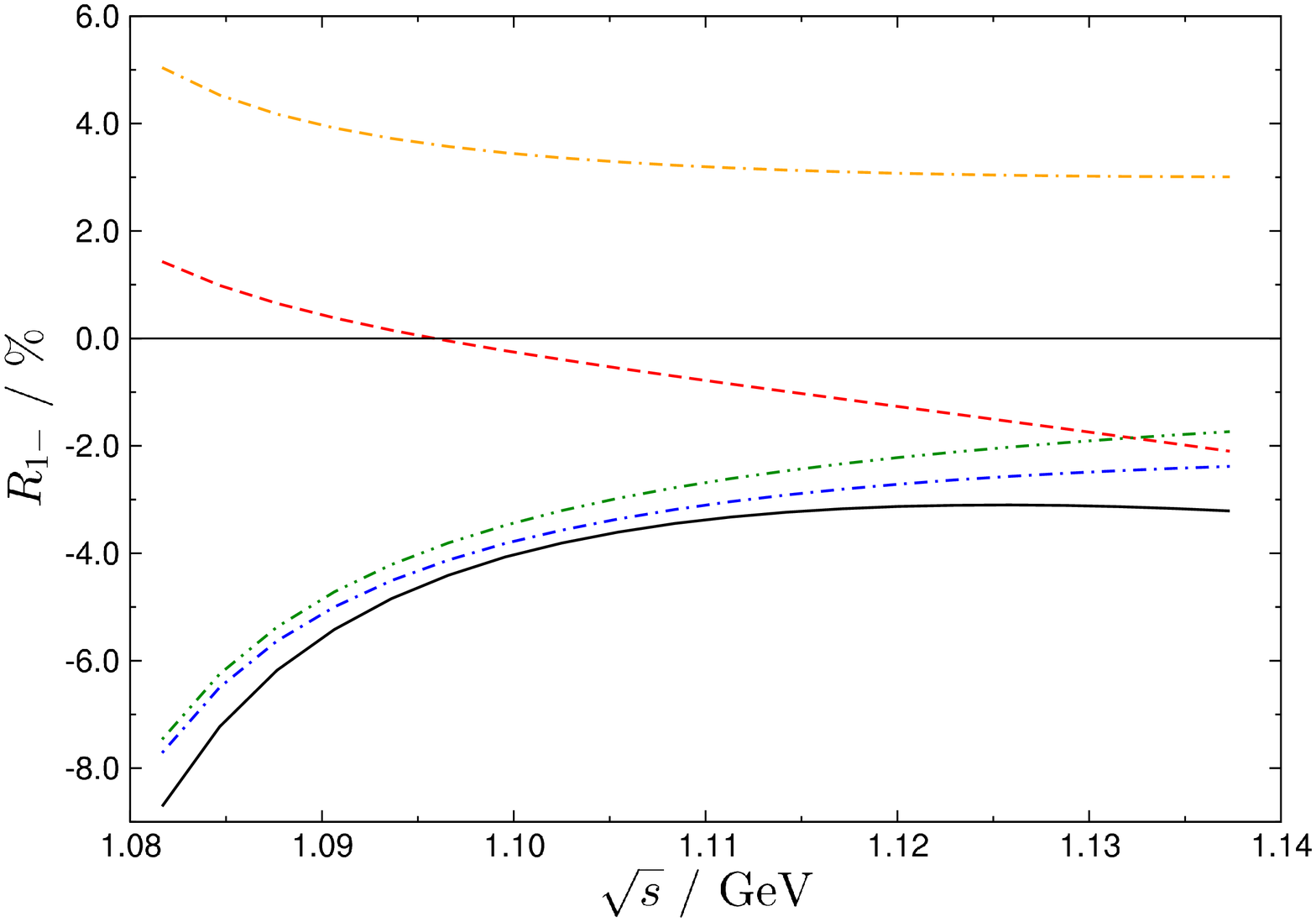}

\vspace{0.3cm}

\includegraphics[width=0.55\linewidth,clip]{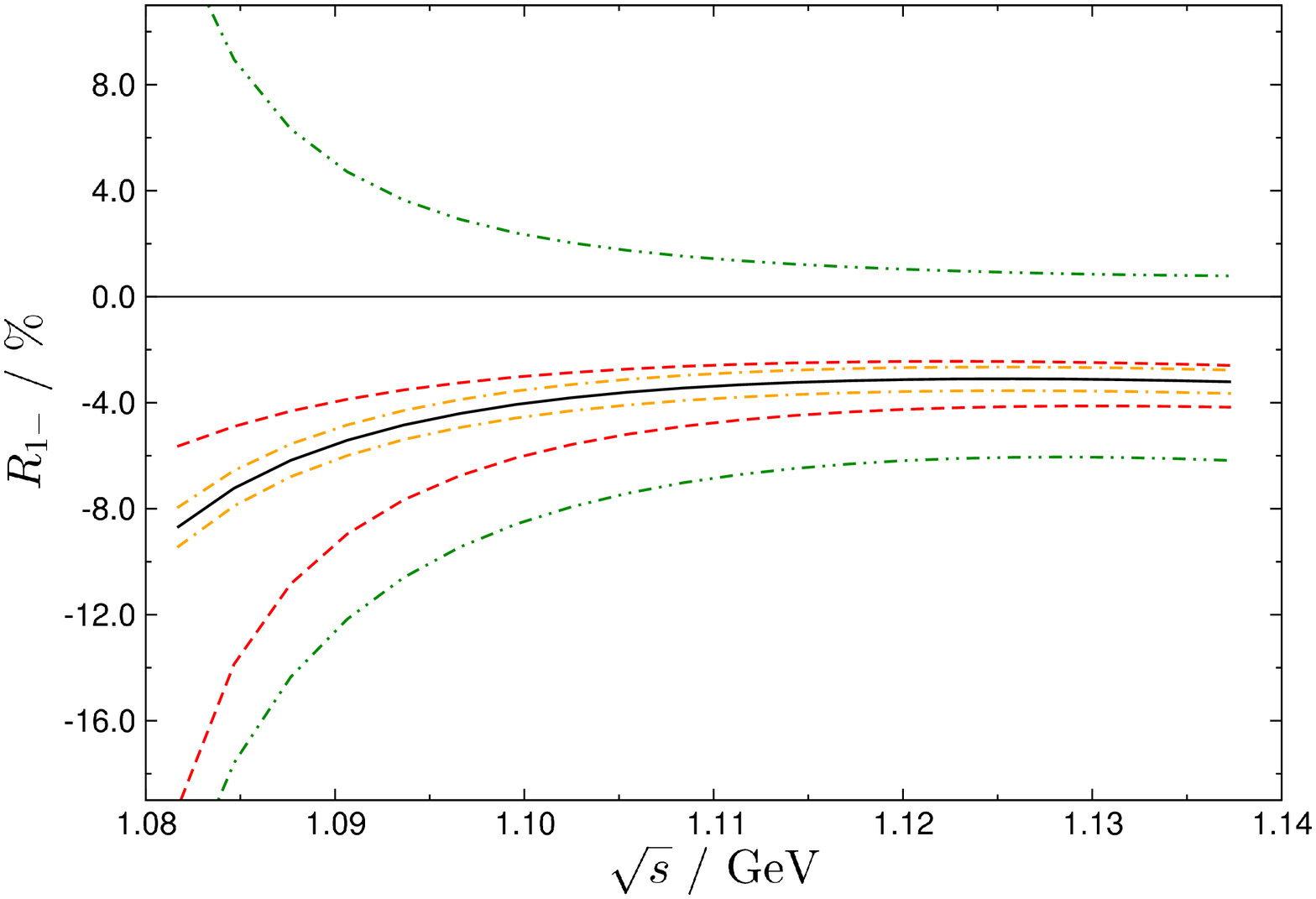}
\end{center}
\caption{Triangle relation in the $P_1$-wave.
For notations, see Fig.~\ref{fig:swave}.
}
\label{fig:p1wave}
\end{figure}

\begin{figure}
\begin{center}

\vspace{-23pt}

\includegraphics[width=0.55\linewidth,clip]{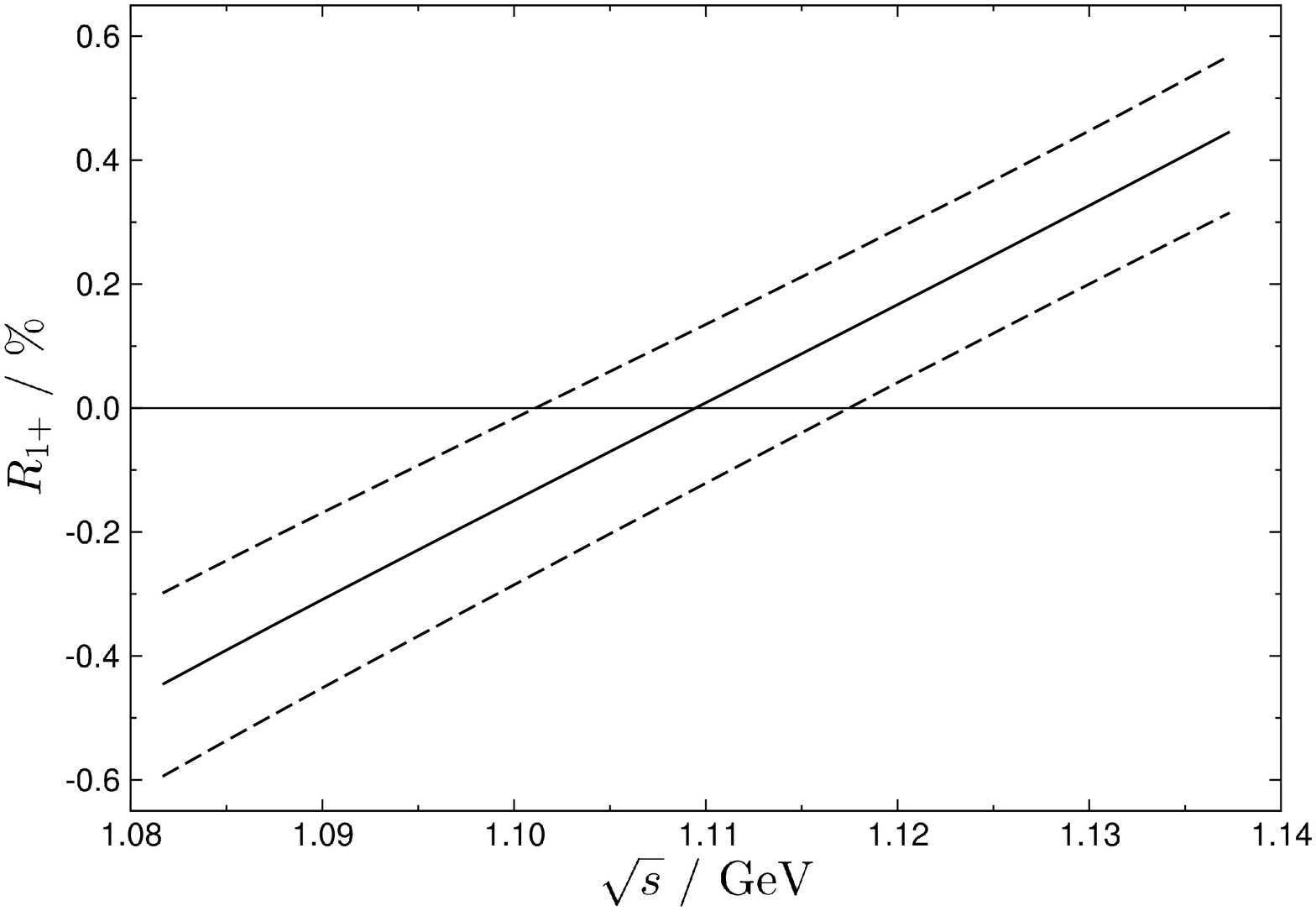}

\vspace{0.3cm}

\includegraphics[width=0.55\linewidth,clip]{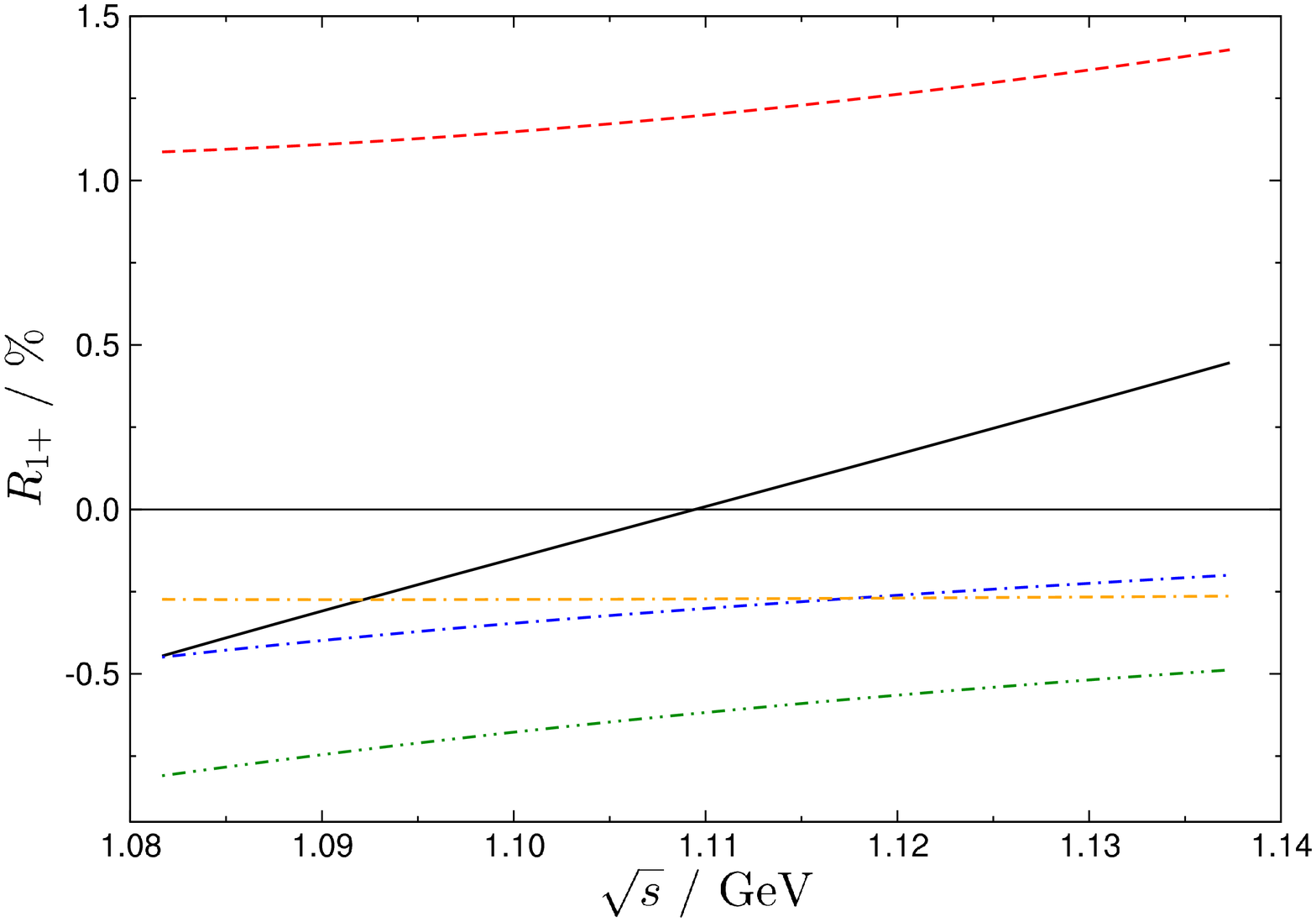}

\vspace{0.3cm}

\includegraphics[width=0.55\linewidth,clip]{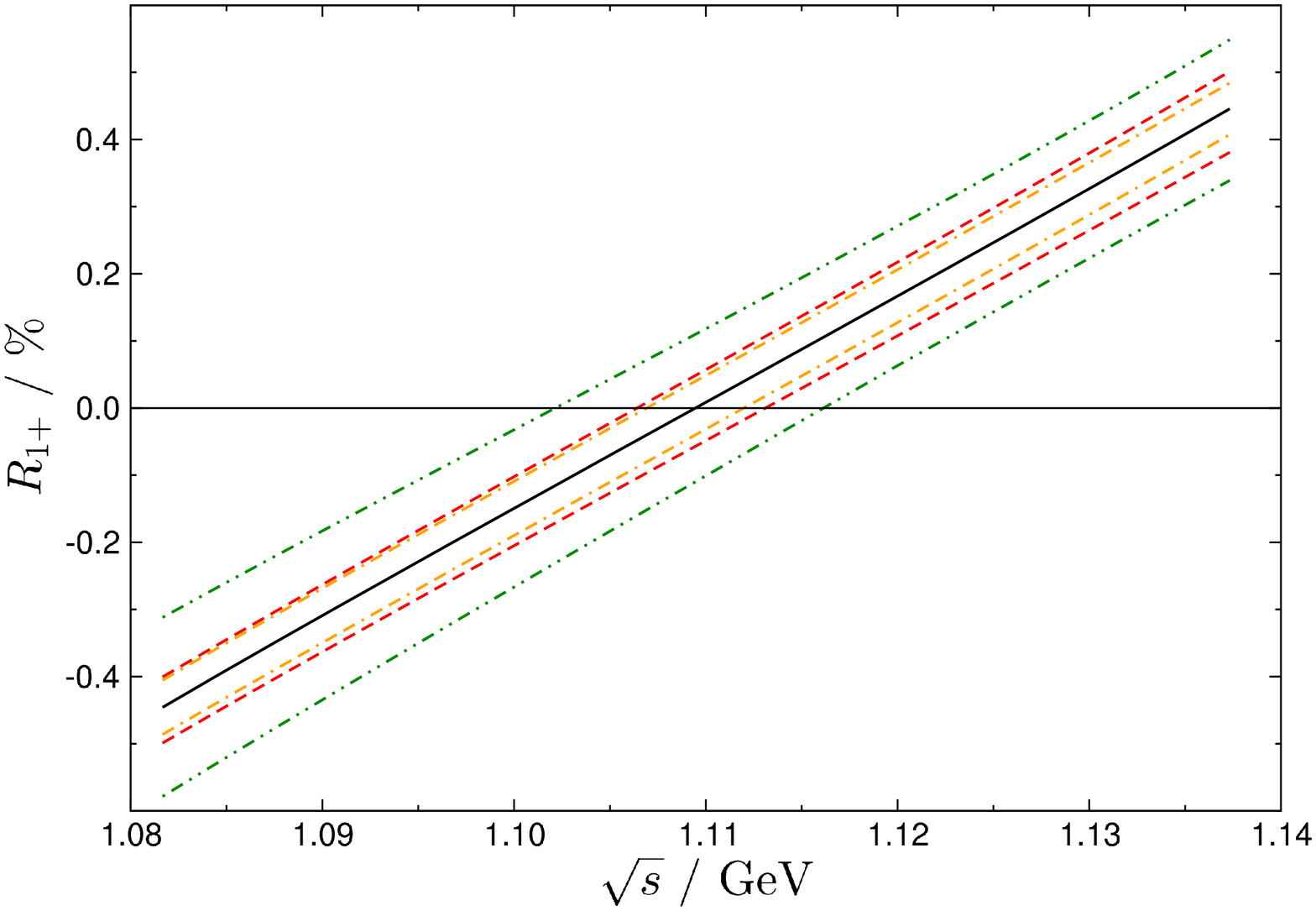}
\end{center}
\caption{Triangle relation in the $P_3$-wave.
For notations, see Fig.~\ref{fig:swave}.
}
\label{fig:p3wave}
\end{figure}

The results for the triangle relation above threshold in the $S$-wave ($R$), the $P_1$- and $P_3$-waves ($R_{1-}$ and $R_{1+}$), and the $\mathcal{G}$- and $\mathcal{H}$-projections ($R_\mathcal{G}$ and $R_\mathcal{H}$), are depicted in Figs.~\ref{fig:swave}--\ref{fig:p3wave}. The central values correspond to the LECs compiled in Sect.~\ref{sec:LEC_triangle}, a renormalization scale $\mu= 1\,{\rm GeV}$ and a detector resolution $E_{\rm max}=10\,{\rm MeV}$. The scale is varied between the mass of the $\rho$ meson $M_\rho=775.49\,{\rm MeV}$ and the $\Delta$ resonance ${m_\Delta=1.232\,{\rm GeV}}$, the detector resolution between $5\,{\rm MeV}$ and $20\,{\rm MeV}$. The errors induced by the LECs are calculated via Gaussian error propagation, where the errors of the $d_i$ are directly traced back to those of the corresponding threshold parameters. This ensures that the uncertainties in the different LECs are essentially uncorrelated. Finally, all three contributions are added in quadrature. We display the individual contributions to the triangle relation by tree graphs, strong loops, virtual photons, and bremsstrahlung, as well as the individual contributions to the final uncertainty. Due to the IR divergences, the separation of virtual photons and bremsstrahlung is scale dependent: by changing the scale $\mu_{\rm IR}$ in $\log m_\gamma / \mu_{\rm IR} $, contributions can be shifted between virtual photons and bremsstrahlung. We choose $\mu_{\rm IR}=\mpp$, which emerges naturally in the calculation.

Let us first consider the $S$-wave. At threshold, the triangle relation is
violated by about $2.5\,\%$, which is significantly larger than the $1.5\,\%$
at threshold obtained in \cite{HKM09} (but within the uncertainty given there). The reason is that both tree and loop contributions have not been chirally expanded; in particular, the diagrams which at threshold only start at $\Order(p^4)$ are included. In this way, the difference between both values may be taken as an estimate of the potential impact of higher chiral orders. From this point of view, the error bands in Fig.~\ref{fig:swave} dominated by the uncertainty in the LECs would therefore not be modified substantially if the uncertainty due to higher orders were included. Remarkably, the residual dependence on the renormalization scale as well as on the detector resolution turns out to be negligible in comparison.
The net effect of tree graphs and strong loops is rather small and practically independent of $\sqrt{s}$. In contrast, virtual photons yield the dominant contribution, which grows significantly with increasing energy. For higher energies, bremsstrahlung becomes quite important: it counteracts the rise induced by virtual photons. The size of the finite pieces of bremsstrahlung apart from the leading terms $\propto\log \mpp^2/4E_{\rm max}^2$ amounts to about $20\, \%$ of the total bremsstrahlung contribution.

For the $P_1$-wave, we observe the same problem as in \cite{FM00_above}: the denominator itself becomes very small, such that especially at threshold $R_{1-}$ is very sensitive to the precise values of the LECs. We conclude that for this projection no meaningful statement can be made. In the other $P$-waves isospin breaking is very small; the triangle relation is violated by $1\,\%$ at most. The observation that the variation of $E_{\rm max}$ yields a shift comparable to the uncertainty due to the LECs could be taken as a hint that the total error might be underestimated. Higher chiral orders and the $\Delta$ resonance might lead to larger shifts than indicated by the error bands in Figs.~\ref{fig:gwave}, \ref{fig:hwave}, and \ref{fig:p3wave}. In particular the $\Delta$ resonance, which is only parametrized in the LECs of the theory and not included as explicit degree of freedom in the present work, could become important in the $P$-wave. Similarly to the $S$-wave we find that virtual photons and bremsstrahlung tend to evolve in the opposite direction for increasing energy, but at least for $R_\mathcal{H}$ and $R_{1+}$ these contributions no longer outweigh tree graphs and strong loops. Since the $P$-wave projections themselves vanish at threshold, bremsstrahlung can generate finite contributions in the full energy range, which is indeed observed in the case of $R_\mathcal{G}$ and $R_{1+}$.

\subsection{Comparison to earlier work}

Isospin violation above threshold has been the subject of earlier studies in the framework of heavy-baryon ChPT \cite{FM00_above,FM01}. The findings in \cite{FM00_above} differ quite substantially from our results: in the $S$-wave, the triangle relation is found to be violated by about $-2.5\,\%$ instead of $+2.5\,\%$,\footnote{Note that the scattering lengths given in the same paper yield $R\sim 1\,\%$, cf.\ \cite{FMS99,HKM09}. The discrepancy is related to the question which threshold energies are inserted and again illustrates nicely that the final result for $R$ is quite sensitive to mass differences.} and also for the $P$-waves (especially for the $\mathcal{H}$-projection) significantly larger negative values for $R$ are obtained. However, this analysis is incomplete in the sense that virtual photons were neglected. Indeed, we observe that in all cases their contributions shift the curves upwards, which partly explains the discrepancy. Further differences concern the definition of the isospin limit, the values of the LECs (the $c_i$ and $d_i$ are taken from \cite{FMS98}), and the treatment of higher chiral orders. The philosophy in \cite{FM00_above} is to expand the amplitude strictly at $\Order(p^3)$ and only to include some higher-order terms whenever necessary to ensure the correct threshold behavior. As we have seen in the previous section, such higher-order effects can indeed lead to differences of $\sim 1\,\%$. Moreover, the denominator is not reduced to its isospin limit, which would generate in our work a dependence on some new combinations of electromagnetic LECs, involving e.g.\ $k_1$ and $k_2$. As for the $P_1$-wave, the problem of the small denominator manifests itself in \cite{FM00_above}  in a different way: the curve is regular at threshold but diverges around $\sqrt{s}=1.115\,{\rm GeV}$. Since we can reproduce a similar picture by varying the LECs in a broader range than described in Sect.~\ref{sec:LEC_triangle}, this does not point to a fundamental difference, but rather to the difficulties in the $P_1$-wave.
Finally, we remark that our results agree with \cite{FM00_above} in the qualitative energy dependence of the triangle relation in so far as $R$ is found to be a monotonically increasing function of $\sqrt{s}$. 

In \cite{FM01} virtual photons are taken into account, but this work eventually focuses on strong isospin breaking, such that in particular electromagnetic contributions to the mass differences are switched off. A consequence of this point of view is that the mass of the $\pi^0 n$ state is larger than of $\pi^- p$, since all pion masses may be identified with the mass of the neutral pion once the tiny strong contribution to $\Delta_\pi$ is neglected. Therefore, the three amplitudes entering the triangle relation no longer possess the same threshold. Nevertheless, we observe a similar discrepancy as in the case of \cite{FM00_above}: the values for the triangle relation in the $S$-wave (about $-0.7\,\%$), and in the $\mathcal{G}$- and especially $\mathcal{H}$-projection, lie significantly lower than the present results. However, virtual photons (and higher-order contributions) again tend to shift the curves into the right direction. Although no reliable error bands are provided,\footnote{In \cite{FM00_above} only shifts due to the ignorance of $f_1$ and $f_2$ are considered, whereas the error bands displayed in \cite{FM01} refer to the outcome of the MINUIT minimization routine of the CERN library, which, as pointed out by the authors, by no means reflect the total uncertainty.} we therefore consider the origin of the deviations from former studies in ChPT to be essentially understood.

It is very difficult to compare our results to phenomenological models \cite{Gibbs95,Matsinos97}, since in contrast to the analyses in ChPT  an unambiguous separation of the different sources of isospin breaking cannot be achieved. In \cite{Gibbs95}, a coupled-channel potential model \cite{Siegel86} is used to remove Coulomb interactions and hadronic mass differences from the data. The triangle relation is found to be violated by about $-7\,\%$ in the $S$-wave at $\sqrt{s}=1.10\,{\rm GeV}$. This value is practically independent of $\sqrt{s}$ (cf.\ Fig.~1 in~\cite{Gibbs95}), such that a similar number could be deduced for the threshold extrapolation of the amplitude. In the $\mathcal{H}$-projection no isospin violation is found, whereas $R_\mathcal{G}$ displays an effect comparable to the $S$-wave.\footnote{About $-7\,\%$ at $\sqrt{s}=1.10\,{\rm GeV}$. However, the situation is less clear at higher energies, cf.\ Fig.~2 in~\cite{Gibbs95}.}   

The investigation in \cite{Matsinos97} is based on a tree-level meson-exchange model \cite{Mat94}; electromagnetic effects and hadronic mass differences are separated relying on the NORDITA method \cite{Tromborg73,Tromborg74,Tromborg77}. In the $S$-wave, $R$ is found to be nearly independent of $\sqrt{s}$ in the energy range ${\sqrt{s}=(1.10-1.14) \,{\rm GeV}}$. However, the final value $R=(-6.4 \pm 1.4)\,\%$ differs from the triangle relation for the scattering lengths obtained by extrapolating the amplitudes to threshold, which is violated by about $-3.5\,\%$. 
The central values for $R_\mathcal{G}$ and $R_\mathcal{H}$ can be extracted from Fig.\ 5 in \cite{Matsinos97} and amount to $(3-4)\,\%$ and $-(5-6)\,\%$ in the considered energy range, respectively. However, the quoted uncertainties are quite large, such that at least in the $\mathcal{H}$-projection the results are compatible with isospin symmetry.

A similar analysis is performed in \cite{GHBS06}. The corrections due to hadronic mass differences are calculated within the $K$-matrix formalism and found to be in reasonable agreement with the NORDITA results, while the electromagnetic effects are taken from NORDITA directly. For the $S$-wave, isospin violation is found to be smaller than $1\,\%$ in magnitude. The discrepancy to \cite{Matsinos97} is traced back to the fact that a coupled-channel approach was chosen, arguing that elastic $\pi^-p$ scattering and  the charge exchange are coupled channels even if isospin is not conserved, whereas in \cite{Matsinos97} each reaction was analyzed separately. Isospin violation in the $P$-waves is not considered.

We conclude that the phenomenological models do not provide a consistent picture. Moreover, the systematic uncertainties are very difficult to assess.  As already remarked in \cite{FM00_above}, in principle a meaningful comparison between the model calculations and ChPT would require a detailed confrontation of the NORDITA method with our treatment of isospin-violating effects, such that the outcome of both methods cannot be naively compared. However, we have seen in the course of the numerical evaluation of the triangle relation that the final result (and especially the threshold behavior) is quite sensitive to the correct incorporation of mass differences and can in principle generate large effects. Certainly, these fine
effects are not sufficiently well under control in the model calculations
because they do not provide a systematic and consistent framework to
analyze isospin breaking.

\subsection*{Acknowledgements}

We acknowledge the support of the European 
Community-Research Infrastructure Integrating Activity ``Study of 
Strongly Interacting Matter'' (acronym HadronPhysics2, Grant Agreement 
n. 227431) under the Seventh Framework Programme of the EU.
Work supported in part by DFG (SFB/TR 16, ``Subnuclear Structure of Matter''),
by the Helmholtz Association through funds provided to the virtual 
institute``Spin and strong QCD'' (VH-VI-231), and by the Bonn-Cologne Graduate School of Physics and Astronomy.

\appendix

\settocdepth{2}

\section{Effective Lagrangians}
\def\theequation{\Alph{section}.\arabic{equation}}
\setcounter{equation}{0}
\label{app:Leff}
Our investigation of isospin-breaking effects relies on the effective Lagrangian for nucleons, pions, and virtual photons, as constructed in \cite{GR02}
\beq
	\Lagr_{\rm eff}=\Lagr_\pi^{(p^2)}+\Lagr_\pi^{(e^2)}+\Lagr_\pi^{(p^4)}+\Lagr_\pi^{(e^2p^2)}+\Lagr_{\rm N}^{(p)}+\Lagr_{\rm N}^{(p^2)}+\Lagr_{\rm N}^{(e^2)}+\Lagr_{\rm N}^{(p^3)}+\Lagr_{\rm N}^{(e^2p)}+\Lagr_\gamma\label{effective_Lagr}.
\eeq
Apart from the leading piece
\begin{align}
\Lagr_\pi^{(p^2)}+\Lagr_\pi^{(e^2)}+\Lagr_\gamma&=\frac{F^2}{4}\langle d^\mu U^\dagger d_\mu U+\chi^\dagger U+U^\dagger \chi\rangle+Z F^4\langle \mathcal{Q}U\mathcal{Q}U^\dagger\rangle-\frac{1}{4}F_{\mu\nu}F^{\mu\nu}-\frac{1}{2}\big(\partial_\mu A^\mu\big)^2,\notag\\
\Lagr_{\rm N}^{(p)}&=\bar{\Psi}\Big\{i\slashed{D}-m+\frac{1}{2}g \slashed{u}\gamma_5\Big\}\Psi, \label{lead_lagr}
\end{align}
we actually need the following terms:
\begin{align}
\Lagr_\pi^{(p^4)}&=\frac{l_4}{4}\langle d^\mu U^\dagger d_\mu \chi+d^\mu \chi^\dagger d_\mu U \rangle+l_5\langle \mathcal{F}^{\mu\nu}_{\rm R}U \mathcal{F}^{\rm L}_{\mu\nu} U^\dagger\rangle+l_6\frac{i}{2}\langle \mathcal{F}^{\mu\nu}_{\rm R}d_\mu U d_\nu U^\dagger+\mathcal{F}^{\mu\nu}_{\rm L}d_\mu U^\dagger d_\nu U\rangle,\notag\\
\Lagr_\pi^{(e^2p^2)}&=F^2 \Bigl\{\langle d^\mu U^\dagger d_\mu U \rangle \bigl(k_1\langle \mathcal{Q}^2 \rangle
+k_2 \langle \mathcal{Q}U\mathcal{Q}U^\dagger \rangle\bigr)\notag\\
&+k_3\bigl(\langle d^\mu U^\dagger \mathcal{Q} U \rangle\langle d_\mu U^\dagger \mathcal{Q} U \rangle+
 \langle d^\mu U \mathcal{Q} U^\dagger\rangle\langle d_\mu U \mathcal{Q} U^\dagger\rangle\bigr)
 +k_4\langle d^\mu U^\dagger \mathcal{Q} U \rangle \langle d_\mu U \mathcal{Q} U^\dagger \rangle\Bigr\} ,\notag\\
\Lagr_{\rm N}^{(p^2)}&=\bar{\Psi}\Big\{c_1 \langle\chi_+\rangle -\frac{c_2}{4m^2}\langle u_\mu u_\nu\rangle \left(D^\mu D^\nu + {\rm h.c.}\right)+\frac{c_3}{2}\langle u_\mu u^\mu\rangle+\frac{i}{4}c_4 \sigma^{\mu\nu}[u_\mu,u_\nu]+c_5 \hat{\chi}_+\notag\\
&+\frac{c_6}{8m}\sigma^{\mu\nu}F_{\mu\nu}^++\frac{c_7}{8m}\sigma^{\mu\nu}\langle F_{\mu\nu}^+\rangle\Big\}\Psi\notag,\\
\Lagr_{\rm N}^{(e^2)}&=F^2\bar{\Psi}\Big\{f_{1} \langle \hat{Q}^2_+ - Q_-^2\rangle+ f_2\langle Q_+\rangle \hat{Q}_++f_{3} \langle \hat{Q}^2_+ + Q_-^2\rangle\Big\}\Psi,\notag\\
\Lagr_{\rm N}^{(p^3)}&=\bar{\Psi}\Big\{-\frac{d_1}{2m}([u_\mu,[D_\nu,u^\mu]]D^\nu+{\rm h.c.})
-\frac{d_2}{2m}([u_\mu,[D^\mu,u_\nu]]D^\nu+{\rm h.c.})\notag\\
&+\frac{d_3}{12m^3}([u_\mu,[D_\nu,u_\lambda]](D^\mu D^\nu D^\lambda+{\rm sym})+{\rm h.c.})+d_5\frac{i}{2m}([\chi_-,u_\mu]D^\mu-{\rm h.c.})\notag\\
&+d_6\frac{i}{2m}([D^\mu,\hat{F}_{\mu\nu}^+]D^\nu-{\rm h.c.})+d_7\frac{i}{2m}([D^\mu,\langle F_{\mu\nu}^+\rangle]D^\nu-{\rm h.c.})\notag\\
&+d_{14}\frac{i}{4m}(\sigma^{\mu\nu}\langle u_\nu[D_\lambda,u_\mu]\rangle D^\lambda-{\rm h.c.})
+d_{15}\frac{i}{4m}(\sigma^{\mu\nu}\langle u_\mu[D_\nu,u_\lambda]\rangle D^\lambda-{\rm h.c.})\notag\\
&+\frac{d_{16}}{2}\gamma^\mu\gamma_5\langle\chi_+\rangle u_\mu+\frac{d_{17}}{2}\gamma^\mu\gamma_5\langle\chi_+ u_\mu\rangle+d_{18}\frac{i}{2}\gamma^\mu\gamma_5[D_\mu,\chi_-]+d_{19}\frac{i}{2}\gamma^\mu\gamma_5[D_\mu,\langle\chi_-\rangle]\Big\}\Psi,\notag\\
\Lagr_{\rm N}^{(e^2p)}&=F^2\bar{\Psi}\Big\{\frac{g_1}{2}\langle Q_+^2-Q_-^2\rangle \gamma^\mu\gamma_5u_\mu+\frac{g_2}{2}\langle Q_+\rangle^2\gamma^\mu\gamma_5u_\mu+\frac{g_3}{2}\gamma^\mu\gamma_5\langle Q_+\rangle\langle Q_+u_\mu\rangle\notag\\
&+\frac{g_4}{2}\gamma^\mu\gamma_5 Q_+\langle Q_+u_\mu\rangle+g_6\frac{i}{2m}(\langle Q_+\rangle \langle Q_-u_\mu\rangle D^\mu-{\rm h.c.})
+g_{7}\frac{i}{2m}(Q_-\langle Q_+ u_\mu\rangle  D^\mu-{\rm h.c.})\notag\\
&+g_{8}\frac{i}{2m}(Q_+\langle Q_- u_\mu\rangle  D^\mu-{\rm h.c.})\Big\}\Psi\label{eff_Lagr},
\end{align}
where $\langle A\rangle$ denotes the trace of a matrix $A$, $\hat{A}=A-\langle A\rangle/2$ its traceless part,
$
\bar{\Psi}(\mathcal{O}+{\rm h.c.})\Psi\equiv \bar{\Psi}\mathcal{O}\Psi+{\rm h.c.}
$
for an operator $\mathcal{O}$ and
\begin{align}
 d_\mu U&=\partial_\mu U-i A_\mu[\mathcal{Q},U], \quad \chi=2 B\,\text{diag}(\muu,\md), \quad U=u^2,\notag\\
F_{\mu\nu}&=\partial_\mu A_\nu-\partial_\nu A_\mu, \quad \mathcal{Q}=\frac{e}{3}\,\text{diag}(2,-1),\quad \mathcal{F}_{\mu\nu}^{\rm L}=\mathcal{F}_{\mu\nu}^{\rm R}=\mathcal{Q}F_{\mu\nu}, \notag\\
Q&=e\,\text{diag}(1,0),\quad F_{\mu\nu}^\pm=F_{\mu\nu}\Big(u^\dagger Q u\pm u Q u^\dagger\Big),\notag\\
D_\mu &=\partial_\mu + \Gamma_\mu, \quad  \Gamma_\mu= \frac{1}{2}\Big(u^\dagger(\partial_\mu-i Q A_\mu)u+u(\partial_\mu-i Q A_\mu)u^\dagger\Big),\notag\\
\chi_\pm&=u^\dagger \chi u^\dagger\pm u \chi^\dagger u,\quad  u_\mu= i\Big(u^\dagger(\partial_\mu-i Q A_\mu)u-u(\partial_\mu-i Q A_\mu)u^\dagger\Big),\notag\\
Q_\pm&=\frac{1}{2}(u Q u^\dagger \pm u^\dagger Q u), \quad [D_\mu,u_\nu]=\partial_\mu u_\nu+[\Gamma_\mu,u_\nu].
\end{align} 
$\Psi=(p,n)^T$ contains the nucleon fields and the pion matrix $U$ in the $\sigma$ representation reads
\beq
U(\boldsymbol\pi)=\sqrt{1-\frac{\boldsymbol\pi^2}{F^2}}+i\frac{\boldsymbol\tau\cdot\boldsymbol\pi}{F},\quad 
\boldsymbol\tau\cdot\boldsymbol\pi=\begin{pmatrix}\pi^0& \sqrt{2}\, \pi^+\\ \sqrt{2}\,\pi^-&-\pi^0\end{pmatrix}\label{sigma_param}.
\eeq
The renormalization of the LECs proceeds along the following pattern (the $\beta$-functions are given in \cite{GR02})
\beq
l_i=\gamma_i\lambda+l_i^{\rm r}(\mu),\quad k_i=\sigma_i\lambda+k_i^{\rm r}(\mu),\quad d_i=\frac{\beta_i}{F^2}\lambda+d_i^{\rm r}(\mu),\quad g_i=\frac{\eta_i}{F^2}\lambda+g_i^{\rm r}(\mu).
\eeq
Note that the choice of operator basis in $\Lagr_\pi^{(p^4)}$ corresponds to \cite{GL84}. As discussed in detail in \cite{GR02}, changing this Lagrangian to the version used in \cite{GSS88} results in a redefinition of several LECs and their $\beta$-functions. In particular, some $d_i$ are replaced by $\tilde{d}_i$ according to
\beq
\tilde{d}_5^{\rm r}(\mu)=d_5^{\rm r}(\mu)+\frac{l_4^{\rm r}(\mu)}{8F^2},\quad \tilde{d}_{18}^{\rm r}(\mu)=d_{18}^{\rm r}(\mu)-\frac{g}{2F^2}l_4^{\rm r}(\mu),\quad \tilde{d}_{19}^{\rm r}(\mu)=d_{19}^{\rm r}(\mu)+\frac{g}{4F^2}l_4^{\rm r}(\mu).
\eeq

Counting $e\sim\Order(p)$, the full calculation at order $\Order(p^3)$ involves terms of order $\Order(e^4p^{-1})$, which are numerically tiny and were therefore dropped in \cite{GR02}. Although we neglect these effects in the numerical analysis too, which reduces the number of LECs considerably, the amplitude shall be given in its full generality for the following reason: the cancelation of UV and IR divergences serves as a powerful check on the calculation, and the consistent treatment of $\Order(e^4)$ terms ensures that this cancelation works for {\it all} contributions at order $\Order(p^3)$. However, for this purpose, there are further contributions to the effective Lagrangian to be taken into account: first, a Lagrangian of the type $\Lagr_\pi^{(e^4)}$ should be added to \eqref{effective_Lagr}, and second we need some terms which do not contain pion fields (``contact terms''), but are indispensable in the context of the renormalization of the one photon exchange. Both pieces can be taken from \cite{KnechtUrech}, since there the same basis for $\Lagr_\pi^{(p^4)}$ is used, such that the additional Lagrangians are consistent with \cite{GR02}. Eventually, it turns out that we only need the counterterms encoded in
\beq
\Lagr_\pi^{\rm contact}=-h_2\langle \mathcal{F}^{\mu\nu}_{\rm R} \mathcal{F}_{\mu\nu}^{\rm R}+\mathcal{F}^{\mu\nu}_{\rm L} \mathcal{F}_{\mu\nu}^{\rm L}\rangle+h_4\langle \mathcal{Q}^2\rangle F_{\mu\nu}F^{\mu\nu}.
\eeq
The first part is already given in \cite{GL84}, whereas the second one is specific to the inclusion of photons, since the electromagnetic field strength tensor appears separately. As in \cite{KnechtUrech} no counterterm is specified to absorb the pertinent UV divergence as occurring in the renormalization of the generating functional, we extend the operator list of terms without pion fields as set up in \cite{GL84} by a forth term $\propto h_4$. The corresponding $\beta$-functions read
\beq
h_i=\delta_i\lambda+h_i^{\rm r}(\mu), \quad \delta_2=\frac{1}{12}, \quad \delta_4=\frac{1}{30}.
\eeq

\section{Loop functions}
\def\theequation{\Alph{section}.\arabic{equation}}
\setcounter{equation}{0}

\subsection{Definitions}
\label{app:def}

In this chapter, we will give explicit expressions for the loop functions originating from the integration of scalar propagators and for the basis functions of the tensor decomposition. Working in $d<4$ dimensions, the UV divergences are captured in the pole term
\beq
\lambda=\frac{\mu^{d-4}}{16\pi^2}\left\{\frac{1}{d-4}-\frac{1}{2}\left(\log 4\pi -\gamma_{\rm E}+1\right)\right\},
\eeq
where $\mu$ is the renormalization scale and $\gamma_{\rm E}=-\Gamma'(1)=0.5772156649\ldots$ denotes the Euler-Mascheroni constant.
To ease notation, also the variants
\begin{align}
C&=2m^{d-4}\left\{\frac{1}{d-4}-\frac{1}{2}\left(\log 4\pi -\gamma_{\rm E}+1\right)\right\}=32\pi^2\lambda+\log\frac{m^2}{\mu^2},\\
C_i&=2M_i^{d-4}\left\{\frac{1}{d-4}-\frac{1}{2}\left(\log 4\pi -\gamma_{\rm E}+1\right)\right\}=32\pi^2\lambda+\log\frac{M_i^2}{\mu^2},\quad i\in\{\pi,\pi^0\},\notag
\end{align}
are used. In contrast, IR divergences are regularized by a finite photon mass $m_\gamma$ and expressed by
\beq
L=\log\frac{m_\gamma^2}{m^2},\quad L_i=\log\frac{m_\gamma^2}{M_i^2}=L+\log\frac{m^2}{M_i^2}, \quad i\in\{\pi,\pi^0\}.
\eeq
Our results may be compared to calculations taming both UV and IR divergences in dimensional regularization by the replacement
\beq
C_{\rm IR}=2m^{d-4}\left\{\left.\frac{1}{d-4}\right|_{\rm IR}-\frac{1}{2}\left(\log 4\pi -\gamma_{\rm E}+1\right)\right\} \rightarrow -L-1.
\eeq
The propagators are implicitly understood to carry a small imaginary part according to
\beq
m^2-p^2\rightarrow m^2-p^2-i\epsilon,\quad \epsilon\rightarrow 0^+.
\eeq 
We will make use of the so-called dilogarithm or Spence's function 
\beq
\Li(x)=-\int\limits_0^x\diff t\frac{\log(1-t)}{t},
\eeq
in terms of which some integrals involving virtual photons can be done semi-analytically. 
Furthermore, the Feynman parameter integrals 
\begin{align}
f(t,m)&=\int\limits_0^1\frac{\diff x}{1-x(1-x)\frac{t}{m^2}}=\frac{4 m^2}{\sqrt{-t(4 m^2-t)}}\arsinh\frac{\sqrt{-t}}{2 m},\\
h\left(t,m,M_i\right)&\equiv h_1^{ii}(t)=\int\limits^1_0\diff x\frac{\log\left(1- x(1-x)\frac{t}{M_i^2}\right)}{1-x(1-x)\frac{t}{m^2}},\notag\\
f_{11}\left(s,m,M_i\right)&=\int\limits_0^1\frac{\diff x}{x+(1-x)\frac{s}{m^2}-x(1-x)\frac{M_i^2}{m^2}}\notag\\
&=\left\{\begin{array}{cl}
&\frac{m^2 }{\lambda^{1/2} \left(s, m^2, M_i^2\right)}\log\left(\frac{s+m^2-M_i^2+\lambda^{1/2} \left(s,m^2,M_i^2\right)}{2 m\sqrt{s} }\right)^2 \quad {\rm if} \, \lambda\left(s,m^2,M_i^2\right)\geq 0, \\
&\frac{2 m^2}{\sqrt{-\lambda \left(s, m^2, M_i^2\right)}}\arctan\frac{\sqrt{-\lambda \left(s, m^2, M_i^2\right)}}{s+m^2-M_i^2} \quad\hspace{50pt} {\rm otherwise},
\end{array}\right.\notag\\
g_{11}\left(s,m,M_i\right)&=\int\limits_0^1\frac{\diff x}{x+(1-x)\frac{M_i^2}{m^2}-x(1-x)\frac{s}{m^2}}\notag\\
&=\left\{\begin{array}{cl}\
&\frac{m^2 }{\lambda^{1/2} \left(s, m^2, M_i^2\right)}\log\left(\frac{s-m^2-M_i^2-\lambda^{1/2} \left(s,m^2,M_i^2\right)}{2 m M_i }\right)^2 \quad  {\rm if} \, \lambda\left(s,m^2,M_i^2\right)\geq 0,\\
&\frac{2 m^2}{\sqrt{-\lambda \left(s, m^2, M_i^2\right)}}\left(\arctan\frac{\sqrt{-\lambda \left(s, m^2, M_i^2\right)}}{m^2+M_i^2-s}+\pi\,\theta\left(s-m^2-M_i^2\right)\right) \quad {\rm otherwise},\\
\end{array}\right.\notag
\end{align}
appear in several loop functions.\footnote{The imaginary parts are not shown, since they are treated separately in the following sections.} 
Some of the loop integrals involving virtual photons can be obtained from integrals which are already present in the isospin limit by sending the appropriate meson mass to zero. Therefore, $i$ and  $j$ run over $\{\pi,\pi^0,\gamma\}$ in Sect.~\ref{Scal_loop_func} (with $M_\gamma\rightarrow m_\gamma$).

Whenever possible, i.e.\ if no infrared divergences prevent the dispersion integral from converging in the limit of vanishing photon mass, we give the loop functions both in terms of Feynman parameter integrals and based on dispersion relations. For this purpose, we also need the imaginary parts of the $t$-channel diagrams, which only emerge for positive $t$, while the formulae for the real part in Feynman parameter representation are only valid in the given form for physical values $t\leq 0$. The dispersion integrals in some cases require one subtraction, which is specified by the subtraction point $s_0$.  In general, the expressions for $u$-channel diagrams follow by substituting $s\leftrightarrow u$ and $\Sigma\leftrightarrow \Lambda$, such that they are given explicitly only if modifications are necessary.

As far as the charge exchange reaction is concerned, $\Delta_\pi$ can appear both via pion propagators and via the external kinematics. To obtain representations valid for all channels in question, we will make use of
\beq
\Delta_{\rm cex}=\left\{\begin{array}{ll}
\Delta_\pi \quad {\rm for } \ \pi^- p\rightarrow \pi^0 n, \\
\, 0 \quad \ \ {\rm otherwise}, \, 
\end{array}\right. \quad \mx=\sqrt{\mpi^2-\Deltax}.
\eeq
As remarked in \ref{sec:thres_div}, we will also take care of $\Delta_\pi$ in scalar loop functions involving photons, though this is an $\Order(e^4)$ effect. 

\subsection{Scalar loop functions}
\label{Scal_loop_func}

We adopt the following notation: the infrared (regular) part of any integral over $m$ meson and $n$ baryon propagators is in general denoted by $I_{mn}$ ($R_{mn}$), the finite pieces as $\bar{I}_{mn}=I_{mn}|_{\lambda\rightarrow 0}$ ($\bar{R}_{mn}=R_{mn}|_{\lambda\rightarrow 0}$).

\subsubsection{Meson integrals}

The result for the tadpole reads
\beq
\Delta_i\equiv I_{10}^i=\frac{1}{i}\int_{\rm I} \frac{\diff^d k}{(2\pi)^d}\frac{1}{M_i^2-k^2}=\frac{M_i^2}{16\pi^2}C_i\label{tadpole}.
\eeq

For the integral over two meson propagators we find ($J_{ij}(t)\equiv I_{20}^{ij}(t)$)
\begin{align}
J_{ij}(t)&=\frac{1}{i}\int_{\rm I} \frac{\diff^d k}{(2\pi)^d}\frac{1}{\left(M_i^2-k^2\right)\left(M_j^2-(k-\Delta)^2\right)}
=J_{ij}(t_0)+\frac{t-t_0}{\pi}\hspace{-18pt}\dashint{29pt}\limits_{\left(M_i+M_j\right)^2}^{\infty}\hspace{-6pt}\frac{\diff t' \text{Im}\,J_{ij}(t')}{(t'-t)(t'-t_0)},\notag\\
J_{ij}(t)&=-\frac{C_i+C_j}{32 \pi ^2}-\frac{1}{16 \pi ^2}\bigg\{1+k_{ij}(t)\bigg\}, \quad
J_{ii}(t)=-\frac{C_i}{16 \pi ^2}-\frac{1}{16 \pi ^2}\bigg\{1+k_{ii}(t)\bigg\}, \notag\\
k_{ij}(t)&=\frac{1}{2}\int\limits^1_0\diff x\log\frac{M_{ij}^2(x)- x(1-x)t}{M_i^2}+(i\leftrightarrow j),
\quad M_{ij}^2(x)=x M_i^2+(1-x) M_j^2,\notag\\
&=-2+\frac{M_i^2-M_j^2}{t}\log\frac{M_i}{M_j}+\frac{ \lambda^{1/2} \left(M_i^2,M_j^2,t\right)}{t}
\log\frac{M_i^2+M_j^2-t-\lambda^{1/2}\left(M_i^2,M_j^2,t\right)}{2 M_i M_j} ,\notag\\
k_{ii}(t)&=\int\limits^1_0\diff x\log\left(1-\frac{t}{M_i^2}x(1-x)\right)=-2+2\sqrt{\frac{4M_i^2-t}{-t}}\arsinh\frac{\sqrt{-t}}{2M_i}, \notag\\
\text{Im}\,J_{ij}(t)&=\frac{\theta\left(t-\left(M_i+M_j\right)^2\right)}{16\pi t}\lambda^{1/2}\left(M_i^2,M_j^2,t\right),\quad
\text{Im}\,J_{ii}(t)=\frac{\theta\left(t-4M_i^2\right)}{16\pi}\sqrt{1-\frac{4M_i^2}{t}},\notag\\
J_{\gamma \gamma}(t)&=-2\lambda+\frac{1}{16 \pi ^2}\bigg\{1-\log\frac{-t}{\mu ^2}\bigg\},\quad \text{Im}\,J_{\gamma\gamma}(t)=\frac{\theta(t)}{16\pi}.
\end{align}

\subsubsection{1 meson, 1 nucleon}

\begin{align}
I_i(s)&=\frac{1}{i}\int_{\rm I} \frac{\diff^d k}{(2\pi)^d}\frac{1}{\left(M_i^2-k^2\right)\left(m^2-(\Sigma-k)^2\right)}
=-\frac{s-m^2+M_i^2}{32\pi^2 s}\left(C_\pi-1\right)+\frac{d_i(s)}{32\pi^2},\notag\\
d_i&=\left\{\begin{array}{cl}
&\frac{\lambda^{1/2}\left(s,m^2,M_i^2\right)}{s}\log\left(\frac{s-m^2+M_i^2-\lambda^{1/2}\left(s,m^2,M_i^2\right)}{2M_i\sqrt{s}}\right)^2 \quad 
{\rm if} \, \lambda\left(s,m^2,M_i^2\right)\geq 0, \\
&-\frac{2 \sqrt{-\lambda\left(s,m^2,M_i^2\right)} }{s}\arccos \frac{m^2-s-M_i^2}{2 M_i \sqrt{s}} \quad\hspace{57pt} {\rm otherwise}, 
\end{array}\right.\notag\\
I_\gamma(s)&=-\frac{s-m^2}{32 \pi ^2 s}(C-1)-\frac{s-m^2 }{32 \pi ^2 s}\log\left( \frac{s-m^2}{m \sqrt{s}}\right)^2,\quad I_\gamma(m^2)=0,\notag\\
\text{Im}\,I_i(s)&=\frac{ \theta\left(s-\left(m+M_i\right)^2\right)}{16 \pi  s}\lambda^{1/2}\left(s,m^2,M_i^2\right),\quad
\text{Im}\,I_\gamma(s)=\theta(s-m^2)\frac{s-m^2}{16 \pi  s},\notag\\
I_i(s)&=I_i(s_0)+R_i(s_0)+\frac{s-s_0}{\pi}\hspace{-18pt}\dashint{24pt}\limits_{\left(m+M_i\right)^2}^{\infty}\hspace{-5pt}\frac{\diff s' \text{Im}\,I_i(s')}{(s'-s)(s'-s_0)}-R_i(s),\notag\\
R_i(s)&=-\frac{s+m^2-M_i^2}{32\pi^2 s}(C-1)+\frac{\tilde{d}_i(s)}{16 \pi^2},\notag\\
\tilde{d}_i&=\left\{\begin{array}{cl}
&-\frac{\lambda^{1/2}\left(s,m^2,M_i^2\right)}{s}  \arsinh\frac{\lambda^{1/2}\left(s,m^2,M_i^2\right)}{2 m \sqrt{s}} \quad {\rm if} \, \lambda\left(s,m^2,M_i^2\right)\geq 0, \\
&\frac{\sqrt{-\lambda\left(s,m^2,M_i^2\right)}}{s}  \arcsin\frac{\sqrt{-\lambda\left(s,m^2,M_i^2\right)}}{2 m \sqrt{s}} \quad\hspace{4pt}  {\rm otherwise}, 
\end{array}\right.\notag\\
R_\gamma(s)&=-\frac{s+m^2}{32 \pi ^2 s}(C-1)-\frac{s-m^2 }{16 \pi ^2 s}\log\frac{\sqrt{s}}{m}.
\end{align}

\subsubsection{2 mesons, 1 nucleon}

$I_{21}^{ij}(t)$ is symmetric in $i$ and $j$. The regular part will be denoted by $R_{21}^{ij}(t)$.
\begin{align}
I_{21}^{ij}(t)&=\frac{1}{i}\int_{\rm I} \frac{\diff^d k}{(2\pi)^d}\frac{1}{\left(M_i^2-k^2\right)\left(M_j^2-(k-\Delta)^2\right)
\left(m^2-(p-k)^2\right)},\notag\\
I_{21}^{ij}(t)&=\frac{1}{64\pi^2 m^2}\left(C_i+C_j+2\right)f(t,m)+\frac{1}{32\pi^2 m^2}\left(h_1^{ij}(t)+h_2^{ij}(t)\right),\notag\\
I_{21}^{ii}(t)&=\frac{1}{32\pi^2 m^2}\left(C_i+1\right)f(t,m)+\frac{1}{32\pi^2 m^2}\left(h\left(t,m,M_i\right)+h_2^{ii}(t)\right),\notag\\
I_{21}^{\gamma\gamma}(t)&=\frac{1}{32\pi^2 m^2}\left(C+\log\frac{-t}{m^2}+1\right)f(t,m) +\frac{1}{32\pi^2 m^2}\left(h^{\gamma \gamma }_1(t)+h^{ \gamma \gamma }_2(t)\right),\notag\\
h_1^{ij}(t)&=\frac{1}{2}\int\limits^1_0\diff x\frac{\log\frac{M_{ij}^2(x)- x(1-x)t}{M_i^2}}{1-x(1-x)\frac{t}{m^2}}+(i\leftrightarrow j),
\quad M_{ij}^2(x)=x M_i^2+(1-x) M_j^2,\notag\\
h_1^{\gamma\gamma}(t)&=\frac{2m^2}{\sqrt{-t\left(4 m^2-t\right)}}\Bigg\{\Li\left(\frac{t-\sqrt{-t\left(4m^2-t\right)}}{2m^2}\right)-\Li\left(\frac{t+\sqrt{-t\left(4m^2-t\right)}}{2m^2}\right)\Bigg\},\notag\\
h_2^{ij}(t)&=2\int\limits^1_0\diff x\frac{\left(2 m^2-M_{ij}^2(x)\right) \arccos\frac{-M_{ij}^2(x) +2x(1-x)t}{2 \sqrt{\left(m^2-x (1-x) t\right) \left(M_{ij}^2(x)-x(1-x) t\right)}}}{\left(1-x(1-x) \frac{t}{m^2}\right) \sqrt{M_{ij}^2(x)\left(4 m^2-M_{ij}^2(x)\right)-4m^2 x(1-x) t }},\notag\\
h_2^{ii}(t)&=2\int\limits^1_0\diff x\frac{\left(\frac{m}{M_i}-\frac{M_i}{2m}\right) \arccos\frac{-M_i^2 +2x(1-x)t}{2 \sqrt{\left(m^2-x (1-x) t\right) \left(M_i^2-x(1-x) t\right)}}}{\left(1-x(1-x) \frac{t}{m^2}\right) \sqrt{1-\frac{M_i^2}{4m^2}-x(1-x) \frac{t}{M_i^2} }},\notag\\
h_2^{\gamma\gamma}(t)&=2\int \limits_0^1\diff x\frac{m\arccos\frac{-\sqrt{-x(1-x)t}}{\sqrt{m^2-x(1-x) t}}}{\left(1-x(1-x)\frac{t}{m^2}\right)\sqrt{-x(1-x)t}} ,\notag\\
\text{Im}\,I^{ij}_{21}(t)&=\frac{ \theta\left(t-(M_i+M_j)^2\right)\theta\left(4m^2-t\right)}{8 \pi  \sqrt{\left(4 m^2-t\right) t}}\arctan\frac{\sqrt{\left(4 m^2-t\right) \lambda\left(M_i^2,M_j^2,t\right)}}{\sqrt{t} \left(t-M_i^2-M_j^2\right)}\notag\\
&+\frac{\theta\left(t-4m^2\right)}{8 \pi  \sqrt{t \left(t-4 m^2\right)}}\log\frac{\sqrt{t} \left(t-M_i^2-M_j^2\right)+\sqrt{\left(t-4 m^2\right)\lambda\left(M_i^2,M_j^2,t\right)}}{2 \sqrt{m^2 \lambda\left(M_i^2,M_j^2,t\right)+M_i^2M_j^2 t}},\notag\\
\text{Im}\,I^{ii}_{21}(t)&=\frac{ \theta\left(t-4 M_i^2\right)\theta\left(4m^2-t\right)}{8 \pi  \sqrt{\left(4 m^2-t\right) t}}\arctan\frac{\sqrt{\left(4 m^2-t\right) \left(t-4 M_i^2\right)}}{t-2 M_i^2}\notag\\
&+\frac{\theta\left(t-4m^2\right)}{8 \pi  \sqrt{t \left(t-4 m^2\right)}}\log\frac{t-2 M_i^2+\sqrt{\left(t-4 m^2\right)\left(t-4 M_i^2\right) }}{2 \sqrt{m^2 \left(t-4 M_i^2\right)+M_i^4}},\notag\\
\text{Im}\,I^{\gamma\gamma}_{21}(t)&=\frac{ \theta(t) \theta\left(4 m^2-t\right)}{8 \pi  \sqrt{\left(4 m^2-t\right) t}}\arctan\sqrt{\frac{4 m^2}{t}-1}+\frac{\theta\left(t-4 m^2\right)}{8 \pi  \sqrt{t\left(t-4 m^2\right)}}\arcosh \frac{\sqrt{t}}{2m},\notag\\
I_{21}^{ij}(t)&=\frac{1}{\pi}\hspace{-18pt}\dashint{29pt}\limits_{\left(M_i+M_j\right)^2}^{\infty}\hspace{-5pt}\frac{\diff t' \text{Im}\,I_{21}^{ij}(t')}{t'-t}-R_{21}^{ij}(t),\quad R_{21}^{ij}(t)=-\frac{C+1}{32\pi^2 m^2}f(t,m)-\frac{h_3^{ij}(t)}{32\pi^2 m^2},\notag\\
h_3^{ij}(t)&=2\int\limits^1_0\diff x\frac{\left(2 m^2-M_{ij}^2(x)\right) \arccos\frac{2m^2-M_{ij}^2(x)}{2m \sqrt{m^2-x (1-x) t }}}{\left(1-x(1-x) \frac{t}{m^2}\right) \sqrt{M_{ij}^2(x)\left(4 m^2-M_{ij}^2(x)\right)-4m^2 x(1-x) t }},\notag\\
h_3^{ii}(t)&=2\int\limits^1_0\diff x\frac{\left(\frac{m}{M_i}-\frac{M_i}{2m}\right) \arccos\frac{2m^2-M_i^2}{2m\sqrt{m^2-x (1-x) t }}}{\left(1-x(1-x) \frac{t}{m^2}\right) \sqrt{1-\frac{M_i^2}{4m^2}-x(1-x) \frac{t}{M_i^2} }},\notag\\
h_3^{\gamma\gamma}(t)&=2\int \limits_0^1\diff x\frac{m\arccos\frac{m}{\sqrt{m^2-x(1-x) t}}}{\left(1-x(1-x)\frac{t}{m^2}\right)\sqrt{-x(1-x)t}}.
\end{align}

\subsubsection{1 meson, 2 nucleons}
\label{sec:1m2n}

The scalar loop integral over 1 meson and 2 nucleon propagators in its general form reads 
\beq
I_{12}^{i}=\frac{1}{i}\int_{\rm I} \frac{\diff^d k}{(2\pi)^d}\frac{1}{\left(M_i^2-k^2\right)\left(m^2-(p_1-k)^2\right)
\left(m^2-(p_2-k)^2\right)}.
\eeq
However, we will only consider two special cases relevant for our calculation, in which either both $p_1$ and $p_2$ are on-shell or at least one of them. Inspired by \cite{BL01}, we write $I_{12}\equiv I_A$ for $p_1^2=p_2^2=m^2$, $(p_1-p_2)^2=t$ and $I_{12}\equiv I_B$ for e.g.\ $p_1^2=s$, $p_2^2=m^2$, $(p_1-p_2)^2=\mpi^2$. In the first case, we find
\begin{align}
I_{A}^{i}(t)&=\frac{1}{i}\int_{\rm I} \frac{\diff^d k}{(2\pi)^d}\frac{1}{\left(M_i^2-k^2\right)\left(m^2-(p-k)^2\right)
\left(m^2-(p'-k)^2\right)},\notag\\
I_{A}^{i}(t)&=-\frac{1}{32\pi^2 m^2}\left(C_i+1\right)f(t,m)+\frac{M_i}{32\pi^2 m^3}g_i(t),
\quad I_{A}^{\gamma}(t)=-\frac{f(t,m)}{32\pi^2 m^2}\left(C+L+1\right),\notag\\
g_i(t)&=\int\limits_0^1\diff x\frac{\arccos\frac{-M_i}{2 \sqrt{m^2-x(1-x)t}}}{\left(1- x(1-x)\frac{t}{m^2}\right) \sqrt{1-\frac{M_i^2}{4 m^2}-x(1-x)\frac{t}{m^2}}},\notag\\
\text{Im}\,I^{i}_{A}(t)&=\frac{ \theta\left(t-4 m^2\right)}{16 \pi  \sqrt{t\left(t-4 m^2\right)}}\log\frac{t-4m^2+M_i^2}{M_i^2},
\quad I_{A}^{i}(t)=\frac{1}{\pi}\dashint{4.5pt}\limits_{4m^2}^{\infty}\frac{\diff t' \text{Im}\,I_{A}^{i}(t')}{t'-t}-R_{A}^{i}(t),\notag\\
\text{Im}\,I^{\gamma}_{A}(t)&=\frac{ \theta\left(t-4 m^2\right)}{16 \pi  \sqrt{t\left(t-4 m^2\right)}}\left(\log\frac{t-4m^2}{m^2}-L\right),\notag\\
R_{A}^{i}(t)&=\frac{1}{32\pi^2 m^2}\left(C+1\right)f(t,m)+\frac{1}{32\pi^2 m^2}\left(h(t,m,m)-\frac{M_i}{m}\tilde{g}_i(t)\right),\notag\\
\tilde{g}_i(t)&=\int\limits_0^1\diff x\frac{\arccos\left(1-\frac{M_i^2}{2 \left(m^2-x(1-x)t\right)}\right)}{\left(1- x(1-x)\frac{t}{m^2}\right) \sqrt{1-\frac{M_i^2}{4 m^2}-x(1-x)\frac{t}{m^2}}}.
\end{align}
The definition
\beq
I_{B}^{ij}(s)=\frac{1}{i}\int_{\rm I} \frac{\diff^d k}{(2\pi)^d}\frac{1}{\left(M_i^2-k^2\right)\left(m^2-(p_1-k)^2\right)
\left(m^2-(p_2-k)^2\right)}\label{IB},
\eeq
for the second case needs some comment: the momenta are assumed to fulfil the on-shell conditions $p_1^2=m^2$ and $p_2^2=s$ (or $p_2^2=u$ for the $u$-channel). Apart from $M_i$ the masses of the external pions, which may enter through the kinematics, can contribute to isospin breaking. To account for this effect, we introduce a pion mass $M_j$ that always corresponds to $\mpi$ for the elastic channels, but has to be modified for the charge exchange reaction according to
\beq
M_j=\left\{\begin{array}{cl}
\mpi \quad {\rm if} \, p_1=p \\
\mpii \quad {\rm if} \, p_1=p'
\end{array}\right. , \qquad p_2=\Sigma,
\eeq
for $s$-channel diagrams and 
\beq
M_j=\left\{\begin{array}{cl}
\mpi \quad {\rm if} \, p_1=p' \\
\mpii \quad {\rm if} \, p_1=p
\end{array}\right. , \qquad p_2=\Lambda,
\eeq
for the $u$-channel case. $M_j$ enters \eqref{IB} by
\beq
\left(p_2-p_1\right)^2=M_j^2.
\eeq
The result for $I_{B}^{ij}(s)$ reads
\begin{align}
I_{B}^{ij}(s)&=-\frac{1}{32\pi^2m^2}\left(C_i+1\right)f_{11}(s,m,M_j)+\frac{1}{32\pi^2m^2}g_{ij}(s),\notag\\
I_{B}^{\gamma j}(s)&=-\frac{1}{32\pi^2m^2}\left(C+\log\left(\frac{s-m^2}{m^2}\right)^2+1\right)f_{11}(s,m,M_j)+\frac{1}{32\pi^2m^2}g_{\gamma j}(s),\notag\\
g_{ij}(s)&=\left\{\begin{array}{cl}
&\int_0^1\diff x g_{ij}^{(1)}(s,x) \quad \hspace{115pt} {\rm if} \, \lambda\left(s,m^2,M_i^2\right)\leq 0, \\
&\int_{x_B^{ij}(s)}^1\diff x g_{ij}^{(1)}(s,x)+\int_0^{x_B^{ij}(s)}\diff x g_{ij}^{(2)}(s,x) \quad {\rm otherwise}, \, 
\end{array}\right.\notag\\
g_{ij}^{(1)}(s,x)&=\frac{2 m^2\left(M_i^2+(1-x)\left(s-m^2\right)\right)\arccos\frac{-M_i^2-(1-x)\left(s-m^2\right)}{2 M_i \sqrt{(1-x)s+x\left(m^2-(1-x) M_j^2\right)}}}{\left\{(1-x)s +x\left(m^2-(1-x)M_j^2\right)\right\}\sqrt{-s^2_{ij}(x)}},\notag\\
g_{ij}^{(2)}(s,x)&=\frac{2 m^2\left(M_i^2+(1-x)\left(s-m^2\right)\right)\log\frac{M_i^2+(1-x)\left(s-m^2\right)-\sqrt{s^2_{ij}(x)}}{2 M_i \sqrt{(1-x)s+x\left(m^2-(1-x)M_j^2\right)}}}{\left\{(1-x)s +x\left(m^2-(1-x)M_j^2\right)\right\}\sqrt{s_{ij}^2(x)}},\notag\\
s^2_{ij}(x)&=\left((1-x)\left(s-m^2\right)-M_i^2\right)^2-4 M_i^2 \left(m^2-x(1-x)M_j^2\right),\notag\\
x_B^{ij}(s)&=\frac{1}{\left(s-m^2\right)^2-4M_i^2M_j^2}\bigg\{\left(s-m^2\right)^2-M_i^2 \left(s-m^2+2M_j^2\right)\notag\\
&-2M_i \sqrt{m^2\left(s-m^2\right)^2-M_i^2M_j^2\left(s+3m^2-M_i^2-M_j^2\right)}\bigg\},\notag\\
g_{\gamma j}(s)&=\int\limits_0^1\diff x\frac{m^2\log\frac{(1-x)s+x \left(m^2-(1-x)M_j^2\right) }{m^2 (1-x)^2}}{(1-x)s+ x\left(m^2-(1-x)M_j^2\right)},\notag\\
I_B^{ij}(s)&=I_B^{ij}(s_0)+R_B^{ij}(s_0)+\frac{s-s_0}{\pi}\hspace{-18pt}\dashint{24pt}\limits_{\left(m+M_i\right)^2}^{\infty}\hspace{-5pt}\frac{\diff s' \text{Im}\,I_B^{ij}(s')}{(s'-s)(s'-s_0)}-R_B^{ij}(s),\notag\\
\text{Im}\,I^{ij}_{B}(s)&=\frac{\theta\left(s-\left(m+M_i\right)^2\right)}{16 \pi \lambda^{1/2} \left(s,m^2,M_j^2\right)}\notag\\
\times\log & \frac{\left\{\left(s-m^2+M_i^2\right) \left(s+m^2-M_j^2\right)-2 s M_i^2 +\lambda^{1/2} \left(s,m^2,M_i^2\right)\lambda^{1/2} \left(s,m^2,M_j^2\right)\right\}^2}{4 s \left\{m^2 \left(s-m^2\right)^2-M_i^2 M_j^2 \left(s+3m^2-M_i^2-M_j^2\right)\right\}},\notag\\
\text{Im}\,I^{ii}_{B}(s)&=\frac{\theta\left(s-\left(m+M_i\right)^2\right)}{16 \pi \lambda^{1/2} \left(s,m^2,M_i^2\right)}\log\frac{s\left(s-m^2-2M_i^2\right)}{m^2 s-\left(m^2-M_i^2\right)},\notag\\
\text{Im}\,I^{\gamma j}_{B}(s)&=\frac{ \theta\left(s-m^2\right)}{8 \pi  \lambda^{1/2}\left(s,m^2,M_j^2\right)}\log\frac{s+m^2-M_j^2+\lambda^{1/2}\left(s,m^2,M_j^2\right)}{2 m \sqrt{s}},\notag\\
R_B^{ij}(s)&=\frac{1}{32\pi^2m^2}\left(C+1\right)f_{11}(s,m,M_j)+\frac{1}{32\pi^2m^2}\left(h_{ij}(s)-\tilde{g}_{ij}(s)\right),\notag\\
h_{ij}(s)&=h_{\gamma j}(s)=\int\limits_0^1\diff x\frac{m^2\log\left(1-x(1-x)\frac{M_j^2}{m^2}\right)}{(1-x)s+x\left(m^2-(1-x)M_j^2\right)},\notag\\
\tilde{g}_{ij}(s)&=\left\{\begin{array}{cl}
&\int_0^1\diff x \tilde{g}_{ij}^{(1)}(s,x) \quad \hspace{115pt} {\rm if} \, \lambda\left(s,m^2,M_i^2\right)\leq 0, \\
&\int_{x_B^{ij}(s)}^1\diff x \tilde{g}_{ij}^{(1)}(s,x)+\int_0^{x_B^{ij}(s)}\diff x \tilde{g}_{ij}^{(2)}(s,x) \quad {\rm otherwise}, \, 
\end{array}\right.\notag\\
\tilde{g}_{ij}^{(1)}(s,x)&=\frac{2 m^2\left(M_i^2+(1-x)\left(s-m^2\right)\right)\arccos\frac{m^2 (1+x)+(1-x) \left(s-2x M_j^2 \right)-M_i^2}{2\sqrt{\left((1-x)s+x\left(m^2-(1-x)M_j^2\right) \right) \left(m^2-x(1-x)M_j^2  \right)}}}{\left\{(1-x)s +x\left(m^2-(1-x)M_j^2\right)\right\}\sqrt{-s^2_{ij}(x)}},\notag\\
\tilde{g}_{ij}^{(2)}(s,x)&=\frac{2 m^2\left(M_i^2+(1-x)\left(s-m^2\right)\right)\log\frac{m^2 (1+x)+(1-x) \left(s-2x M_j^2 \right)-M_i^2+\sqrt{s_{ij}^2(x)}}{2\sqrt{\left((1-x)s+x\left(m^2-(1-x)M_j^2 \right)\right) \left(m^2-x(1-x)M_j^2\right)}}}{\left\{(1-x)s +x\left(m^2-(1-x)M_j^2\right)\right\}\sqrt{s_{ij}^2(x)}},\notag\\
\tilde{g}_{\gamma j}(s)&=\int\limits_0^1\diff x\frac{m^2\log\frac{(1-x)s+x\left(m^2-(1-x)M_j^2\right) }{m^2-x(1-x)M_j^2}}{(1-x)s +x\left(m^2-(1-x)M_j^2 \right) }.
\end{align}

\subsubsection{1 meson, 3 nucleons}
\label{sec:scal_1m3n}

The so-called box graph leads to the scalar integral
\beq
I_{13}^i(s,t)=\frac{1}{i}\int_{\rm I} \frac{\diff^d k}{(2\pi)^d}\frac{1}{\left(M_i^2-k^2\right)\left(m^2-(p-k)^2\right)
\left(m^2-(\Sigma-k)^2\right)\left(m^2-(p'-k)^2\right)}\label{box}.
\eeq
At least the imaginary part can be given in a rather compact form
\begin{align}
\text{Im}\,I_{13}^i(s,t)&=\frac{ \theta\left(s-\left(m+M_i\right)^2\right)}{8 \pi  \sqrt{-t}\sqrt{\zeta_3(s)}}\notag\\
&\times\log\frac{2\zeta _1(s)-t \lambda \left(s,m^2,M_i^2\right)+ \Deltax  \zeta _2(s)+\sqrt{-t \lambda\left(s,m^2,M_i^2\right)}\sqrt{\zeta_3(s)}}{2\sqrt{\zeta _1(s)|_{\mpi\rightarrow \mx}\zeta _1(s)}},\notag\\
\text{Im}\,I_{13}^i(s,t)\Big|_{\Deltax=0}&=\frac{ \theta\left(s-\left(m+M_i\right)^2\right)}{4 \pi  \sqrt{-t}\sqrt{4\zeta_1(s)-t \lambda\left(s,m^2,M_i^2\right)}}\arsinh\frac{\sqrt{-t\lambda \left(s,m^2,M_i^2\right)}}{2\sqrt{\zeta_1(s)}},\notag\\
\zeta_1(s)&=m^2 \left(s-m^2\right)^2-M_i^2 \mpi^2\left(s+3m^3-M_i^2-\mpi^2\right),\notag\\
\zeta_2(s)&=M_i^2 \left(s+3m^2-M_i^2-2 \mpi^2\right),\notag\\
\zeta_3(s)&=4\zeta _1(s)-t \lambda \left(s,m^2,M_i^2\right)+2\Deltax  \zeta _2(s)+M_i^2\left(4m^2-M_i^2\right)\frac{\Deltax^2 }{t},\notag\\
\text{Im}\,I_{13}^\gamma(s,t)&=\frac{\theta\left(s-m^2\right)}{4\pi \left(s-m^2\right)\sqrt{-t\left(4m^2-t\right)}}\arsinh\frac{\sqrt{-t}}{2m}.
\end{align}
After performing the integration corresponding to the extraction of the infrared part, we are left with an integral over two ordinary Feynman parameters. However, the representation
\begin{align}
I_{13}^i(s,t)&=\int\limits_0^1\diff x \int\limits_0^1\diff y \frac{y}{16\pi^2} f_{13}^i(s,t,x,y),\notag\\ f_{13}^i(s,t,x,y)&=\frac{M_i^2+(1-y)\left(s-m^2\right)}{s_{13}^i(s,t,x,y)\left\{4M_i^2s_{13}^i(s,t,x,y)-\left(M_i^2+(1-y)\left(s-m^2\right)\right)^2\right\}}\notag\\
&+\frac{4M_i^2 \arccos\frac{-M_i^2-(1-y)\left(s-m^2\right) }{2M_i\sqrt{s_{13}^i(s,t,x,y)}}}{ \left\{4M_i^2s_{13}^i(s,t,x,y)-\left(M_i^2+(1-y)\left(s-m^2\right)\right)^2\right\}^{3/2}},\notag\\
s_{13}^i(s,t,x,y)&=(1-y) s +y m^2-x(1-x) y^2t-y(1-y)\left(\mpi^2-(1-x) \Deltax \right),\label{box_feyn}
\end{align}
is numerically only viable below threshold. One possibility to obtain a representation valid above threshold is to employ dispersion relations
\begin{align}
I_{13}^{i}(s,t)&=\frac{1}{\pi}\hspace{-18pt}\dashint{24pt}\limits_{\left(m+M_i\right)^2}^{\infty}\hspace{-5pt}\frac{\diff s' \text{Im}\,I_{13}^{i}(s',t)}{s'-s}-R_{13}^{i}(s,t),
\quad  R_{13}^i(s,t)=\int\limits_0^1\diff x \int\limits_0^1\diff y \frac{y}{16\pi^2} \tilde{f}_{13}^i(s,t,x,y),\notag\\
\tilde{f}_{13}^i(s,t,x,y)&=-\frac{4 M_i^2\arccos\frac{2s_{13}^i(s,t,x,y)-M_i^2-(1-y)\left(s-m^2\right) }{2 \sqrt{s_{13}^i(s,t,x,y)-(1-y)\left(s-m^2\right)}\sqrt{s_{13}^i(s,t,x,y)}}}{ \left\{4M_i^2s_{13}^i(s,t,x,y)-\left((1-y)\left(s-m^2\right)+M_i^2\right)^2 \right\}^{3/2}}\notag\\
&+\frac{(1-y)^2\left(s-m^2\right)^2 +M_i^2 (1-y)\left(s-m^2\right)-2M_i^2s_{13}^i(s,t,x,y)}{s_{13}^i(s,t,x,y)\left\{s_{13}^i(s,t,x,y)-(1-y)\left(s-m^2\right)\right\}} \notag\\
&\times\frac{1}{4M_i^2s_{13}^i(s,t,x,y)-\left((1-y)\left(s-m^2\right)+M_i^2\right)^2 },\label{box_disp}
\end{align}
where we have discarded the $t$-channel cut, which lies outside the low-energy region. The regular part $R_{13}^i(s,t)$ is numerically uncritical. Unfortunately, \eqref{box_disp} does not apply to $I_{13}^{\gamma}(s,t)$, since $\text{Im}\,I_{13}^\gamma(s,t)\propto 1/\left(s-m^2\right)$, such that there is a non-integrable singularity at $s'=m^2$. However, this divergence should be captured by a finite $m_\gamma$ leading to a term proportional to $L$. Thus, the dispersion integral is certainly not suited to extract the infrared divergence. As described in Sect.~\ref{sec:box}, a representation appropriate for this purpose can be found by performing the $y$ integration in \eqref{box_feyn}. Below threshold, the result is given by
\begin{align}
I_{13}^i(s,t)&=\int\limits_0^1\diff x g_{13}^{i}(s,t,x),\notag\\
g_{13}^{i}(s,t,x)&=\frac{1}{c_1c_2}\Bigg\{\frac{s-m^2}{y_2-y_1}\log\frac{y_1\left(1-y_2\right)}{-y_2\left(y_1-1\right)}-\frac{A_1+y_0A_2}{\sqrt{c_3}}\arctan\frac{c_1\sqrt{c_3}}{s-c_1y_0}\notag\\
&-\frac{A_2}{2}\log\frac{\left(1-y_0\right)^2+c_3}{y_0^2+c_3}-\frac{B_1-A_2y_2}{y_2-y_1}\log\frac{1-y_2}{-y_2}+\frac{B_1-A_2y_1}{y_2-y_1}\log\frac{y_1-1}{y_1}\Bigg\}\notag\\
&+\frac{8 M_i^2}{c_2^{3/2}\left(y_1-y_2\right)^2}\Bigg\{\frac{y_1+y_2-2y_1y_2}{\sqrt{\left(y_1-1\right)\left(1-y_2\right)}}\arccos\frac{-M_i}{2\sqrt{c_1}\sqrt{\left(1-y_0\right)^2+c_3}}\notag\\
&-2\sqrt{-y_1y_2}\arccos\frac{m^2-s-M_i^2}{2M_i \sqrt{s}}\Bigg\}\notag\\
&+\frac{4 M_i^2}{c_2^{3/2}\left(y_1-y_2\right)^2\sqrt{d_3}}\Bigg\{2d_1\frac{y_1+y_2}{y_1-y_2}\log\frac{y_1\left(1-y_2\right)}{-y_2\left(y_1-1\right)}+\frac{A_3+y_0A_4}{\sqrt{c_3}}\arctan\frac{c_1\sqrt{c_3}}{s-c_1y_0}\notag\\
&+\frac{A_4}{2}\log\frac{\left(1-y_0\right)^2+c_3}{y_0^2+c_3}+\frac{B_2+y_1A_4}{y_2-y_1}\log\frac{y_1-1}{y_1}-\frac{B_2+y_2A_4}{y_2-y_1}\log\frac{1-y_2}{-y_2}\Bigg\},\notag\\
A_1(s,t,x)&=-\frac{c_3+y_0^2}{N}\Big\{\left(s-m^2\right) \left(c_3+y_0^2- y_1 y_2\right) +\left(s-m^2+M_i^2\right)\left( y_1+ y_2-2y_0\right)\Big\},\notag\\
A_2(s,t,x)&=\frac{1}{N}\Big\{\left(s-m^2\right)\left(c_3+y_0^2\right)\left( y_1+y_2\right) -2 \left(s-m^2\right) y_0 y_1 y_2\notag\\
&-\left(s-m^2+M_i^2\right)\left( c_3 +y_0^2-y_1 y_2\right)\Big\},\notag\\
A_3(s,t,x)&=-\frac{2}{N}\Big\{d_1 \Big(\left(c_3+y_0{}^2\right)^2\left(y_1+y_2\right)+\left(c_3+y_0^2\right)y_1 y_2 \left(y_1+y_2-4y_0\Big)\right)\notag\\
&+d_2 \Big(\left(c_3+ y_0^2\right) \left(2y_0\left(y_1+y_2\right)-y_1^2-y_2^2\right)+4 y_0 y_1 y_2 \left(y_1+y_2-2y_0\right)-2 y_1^2 y_2^2\Big)\Big\},\notag\\
A_4(s,t,x)&=\frac{2}{N}\Big\{d_1 \Big(\left(c_3+y_0{}^2\right) \left(y_1{}^2+y_2{}^2\right)+2 y_1 y_2\left(y_1y_2-y_0 \left(y_1+y_2\right)\right)\Big)\notag\\
&+d_2 \Big(\left(c_3+y_0^2\right) \left(y_1+y_2\right)+y_1 y_2 \left(y_1+y_2-4y_0\right)\Big)\Big\},\notag\\
B_1(s,t,x)&=\frac{1}{N}\Big\{\left(s-m^2+M_i^2\right) y_1 y_2 \left(y_1+y_2-2 y_0\right)+\left(s-m^2\right)\Big(\left(c_3+y_0{}^2\right)^2\notag\\
&+\left(c_3+y_0^2\right)\left(y_1^2+y_1 y_2+y_2^2\right)-2\left(c_3+y_0^2\right) y_0 \left(y_1+y_2\right)-2y_0 y_1 y_2 \left(y_1+y_2-2y_0 \right)\Big)\Big\},\notag\\
B_2(s,t,x)&=-\frac{2}{N}\Big\{d_1 \Big(\left(c_3+y_0^2\right)^2\left(y_1+y_2\right)+\left(c_3+y_0^2\right)\left(y_1+y_2\right)^3-2\left(c_3+y_0^2\right)y_0\left(y_1+y_2\right)^2\notag\\
&-\left(c_3+y_0^2\right)\left(y_1+y_2\right)y_1y_2+2 y_1y_2\left(y_1y_2-y_0\left(y_1+y_2\right)\right) \left(y_1+y_2-2y_0\right)\Big)\notag\\
&+d_2 y_1 y_2 \Big(2 \left(c_3+y_0^2\right)+y_1^2+y_2^2-2 y_0 \left(y_1+y_2\right)\Big)\Big\},\notag\\
N(s,t,x)&=\left(c_3+\left(y_0-y_1\right)^2\right) \left(c_3+\left(y_0-y_2\right)^2\right),\notag\\
d_1(s,t,x)&=M_i^2+\left(s-m^2\right)\left(1-y_0\right), \quad 
d_2(s,t,x)=\left(s-m^2\right)c_3-y_0 d_1, \notag\\
d_3(s,t,x)&=\left(s-m^2\right)^2-4M_i^2 c_1,\notag\\
y_{1/2}(s,t,x)&=\frac{c_4\pm \sqrt{c_4^2-c_2 \lambda\left(s,m^2,M_i^2\right)}}{c_2},\quad y_0(s,t,x)=\frac{s-m^2+\mpi^2-(1-x)\Deltax }{2\left(\mpi^2- x(1-x)t-(1-x)\Deltax \right)},\notag\\
c_1(t,x)&=\mpi^2- x(1-x)t-(1-x)\Deltax, \quad c_3(s,t,x)=\frac{s}{c_1}-y_0^2,\notag\\
c_2(s,t,x)&=\left(s-m^2\right)^2-4 M_i^2 \mpi^2+4x (1-x) M_i^2t+4 M_i^2(1-x)\Deltax,\notag\\
c_4(s,t,x)&=\left(s-m^2\right)\left(s-m^2-M_i^2\right)-2 M_i^2 \mpi^2+2 M_i^2(1-x)\Deltax \label{box_one_int},  
\end{align}
where, for convenience, we have omitted the arguments on the right hand side of the equations and suppressed the index $i$ in the case of the auxiliary functions. Note that below threshold, the kinematical region where
\vspace{-1pt} 
\beq
m^2-M_i\sqrt{4\mpi^2-\frac{\left(t-\Deltax\right)^2}{t}}\leq s \leq m^2+M_i\sqrt{4\mpi^2-\frac{\left(t-\Deltax\right)^2}{t}}
\eeq
has to be be paid attention to, since $c_2(s,t,x)$ can become negative and $y_{1/2}(s,t,x)$ complex. However, the representation \eqref{box_one_int} can still be used in this range if the analytic continuation is done properly, which can be achieved by evaluating the amplitude at $s\rightarrow s+i\epsilon$. As we are mainly interested in the continuation above threshold, we will not pursue this any further.
While $y_2(s,t,x)\leq 0$ for $\lambda\left(s,m^2,M_i^2\right)\leq 0$, we have $0\leq y_2(s,t,x)\leq 1$ for $\lambda\left(s,m^2,M_i^2\right)\geq 0$ (and $y_1(s,t,x)\geq 1$ in both cases). Taking this into account, the continuation of \eqref{box_one_int} above threshold reads
\begin{align}
I_{13}^i(s,t)&=\int\limits_0^1\diff x g_{13}^{i}(s,t,x),\label{box_above}\\
g_{13}^{i}(s,t,x)&=\frac{1}{c_1c_2}\Bigg\{\frac{s-m^2}{y_2-y_1}\log\frac{y_1\left(1-y_2\right)}{y_2\left(y_1-1\right)}-\frac{A_1+y_0A_2}{\sqrt{c_3}}\arctan\frac{c_1\sqrt{c_3}}{s-c_1y_0}\notag\\
&-\frac{A_2}{2}\log\frac{\left(1-y_0\right)^2+c_3}{y_0^2+c_3}-\frac{B_1-A_2y_2}{y_2-y_1}\log\frac{1-y_2}{y_2}+\frac{B_1-A_2y_1}{y_2-y_1}\log\frac{y_1-1}{y_1}\Bigg\}\notag\\
&+\frac{8 M_i^2}{c_2^{3/2}\left(y_1-y_2\right)^2}\Bigg\{\frac{y_1+y_2-2y_1y_2}{\sqrt{\left(y_1-1\right)\left(1-y_2\right)}}\arccos\frac{-M_i}{2\sqrt{c_1}\sqrt{\left(1-y_0\right)^2+c_3}}\notag\\
&+\sqrt{y_1y_2}\log\left(\frac{s-m^2+M_i^2-\sqrt{\lambda \left(s,m^2,M_i^2\right)}}{2M_i \sqrt{s}}\right)^2\Bigg\}\notag\\
&+\frac{4 M_i^2}{c_2^{3/2}\left(y_1-y_2\right)^2\sqrt{d_3}}\Bigg\{2d_1\frac{y_1+y_2}{y_1-y_2}\log\frac{y_1\left(1-y_2\right)}{y_2\left(y_1-1\right)}+\frac{A_3+y_0A_4}{\sqrt{c_3}}\arctan\frac{c_1\sqrt{c_3}}{s-c_1y_0}\notag\\
&+\frac{A_4}{2}\log\frac{\left(1-y_0\right)^2+c_3}{y_0^2+c_3}+\frac{B_2+y_1A_4}{y_2-y_1}\log\frac{y_1-1}{y_1}-\frac{B_2+y_2A_4}{y_2-y_1}\log\frac{1-y_2}{y_2}\Bigg\}.\notag  
\end{align}
Based on \eqref{box_above}, it is now straightforward to perform the limit $m_\gamma\rightarrow 0$
\begin{align}
I_{13}^\gamma(s,t)&=\frac{\arsinh\frac{\sqrt{-t}}{2m}}{8\pi ^2\left(s-m^2\right)\sqrt{-t\left(4m^2-t\right)}}\left\{L+\log\left(\frac{ m\sqrt{s}}{s-m^2}\right)^2\right\}+\frac{l_{\gamma }(s,t)}{16\pi ^2},\notag\\
l_{\gamma }(s,t)&=\int\limits_0^1\diff x\frac{\left(s+m^2-\mpi^2+(1-x)\Deltax\right)\arctan\frac{\sqrt{s_{13}^\gamma}}{s+m^2-\mpi^2+(1-x)\Deltax } }{\left(s-m^2\right) \left(m^2-x(1-x)t \right) \sqrt{s_{13}^\gamma}},\\
s_{13}^\gamma&=4s\left(\mpi^2-x(1-x)t-(1-x)\Deltax \right)-\left(s-m^2+\mpi^2-(1-x)\Deltax \right)^2.\notag
\end{align}

\subsubsection{1 meson, 1 photon}

\beq
V_{10}=\frac{1}{i}\int_{\rm I} \frac{\diff^d k}{(2\pi)^d}\frac{1}{\left(m_\gamma^2-k^2\right)\left(\mpi^2-(q-k)^2\right)}=-\frac{C_\pi-1}{16\pi^2}.
\eeq

\subsubsection{2 mesons, 1 photon}

\begin{align}
V_{20}(t)&=\frac{1}{i}\int_{\rm I} \frac{\diff^d k}{(2\pi)^d}\frac{1}{\left(m_\gamma^2-k^2\right)\left(\mpi^2-(q-k)^2\right)\left(\mpi^2-(q'-k)^2\right)}
=\frac{1}{\pi}\dashint{5.8pt}\limits_{4\mpi^2}^{\infty}\frac{\diff t' \text{Im}\,V_{20}(t')}{t'-t}\notag\\
&=-\frac{L_\pi}{32\pi^2\mpi^2}f\left(t,\mpi\right)+\frac{h\left(t,\mpi,\mpi\right)}{32\pi^2\mpi^2},\notag\\
\text{Im}\,V_{20}(t)&=\frac{\theta\left(t-4 \mpi^2\right)}{16\pi \sqrt{t\left(t-4\mpi^2\right)}}\left(\log\frac{t-4\mpi^2}{\mpi^2}-L_{\pi }\right).
\end{align}

\subsubsection{1 meson, 2 photons}

\begin{align}
P_{10}(t)&=\frac{1}{i}\int_{\rm I} \frac{\diff^d k}{(2\pi)^d}\frac{1}{\left(m_\gamma^2-k^2\right)\left(m_\gamma^2-(k-\Delta)^2\right)\left(\mpi^2-(q+k)^2\right)}=(q\leftrightarrow -q')
\notag\\
&=\frac{1}{i}\int_{\rm I} \frac{\diff^d k}{(2\pi)^d}\frac{1}{\left(m_\gamma^2-(q-k)^2\right)\left(m_\gamma^2-(q'-k)^2\right)\left(\mpi^2-k^2\right)}
=\frac{1}{\pi}\dashint{0pt}\limits_{0}^{\infty}\frac{\diff t' \text{Im}\,P_{10}(t')}{t'-t}\notag\\
&=\frac{f\left(t,\mpi\right)}{32\pi^2\mpi^2}\log\frac{-t}{\mpi^2}+\frac{g_{10}(t)}{16\pi^2\mpi^2}+h_{10}(t),\notag\\
g_{10}(t)&=\int\limits_0^1\frac{\diff x\log x}{1-x(1-x)\frac{t}{\mpi^2}},\notag\\
&=\frac{\mpi^2}{\sqrt{-t\left(4 \mpi^2-t\right)}}\Bigg\{\Li\left(\frac{t-\sqrt{-t\left(4\mpi^2-t\right)}}{2\mpi^2}\right)-\Li\left(\frac{t+\sqrt{-t\left(4\mpi^2-t\right)}}{2\mpi^2}\right)\Bigg\},\notag\\
h_{10}(t)&=\frac{\mpi}{32\pi}\int\limits_0^1\frac{\diff x}{\left(\mpi^2-x(1-x)t\right)\sqrt{-x(1-x)t}}=\frac{1}{16\sqrt{-t\left(4\mpi^2-t\right)}},\notag\\
\text{Im}\,P_{10}(t)&=\frac{\theta(t)\theta\left(4\mpi^2-t\right)}{8\pi \sqrt{\left(4\mpi^2-t\right)t}}\arctan\sqrt{\frac{4\mpi^2}{t}-1}+\frac{\theta\left(t-4\mpi^2\right)}{8\pi \sqrt{t\left(t-4\mpi^2\right)}}\arcosh\frac{\sqrt{t}}{2\mpi}.
\end{align}
$P_{10}(t)$ is merely needed in the tensor decomposition of $P_{11}(s,t)$.

\subsubsection{1 meson, 1 nucleon, 1 photon}
\label{sec:1m1n1p}

\begin{align}
V_{11}(s)&=\frac{1}{i}\int_{\rm I} \frac{\diff^d k}{(2\pi)^d}\frac{1}{\left(m_\gamma^2-k^2\right)\left(m^2-(p-k)^2\right)\left(\mpi^2-(q+k)^2\right)}=(q\rightarrow q',\ p\rightarrow p')\notag\\
&=\frac{1}{32\pi^2m^2}\left\{h_{11}(s)-L g_{11}\left(s,m,\mpi\right)\right\}-R_{11}(s),\notag\\ 
R_{11}(s)&=-\frac{C+1}{32\pi ^2m^2}f_{11}\left(s,m,M_i\right)-\frac{1}{16\pi^2m^2} k_{11}(s),\notag\\
h_{11}(s)&=\int\limits_0^1\diff x \frac{\log\left(x+(1-x)\frac{\mpi^2}{m^2}-x(1-x)\frac{s}{m^2}\right)}{x+(1-x)\frac{\mpi^2}{m^2}-x(1-x)\frac{s}{m^2}},\notag\\
h_{11}(s)&=\frac{m^2}{\lambda^{1/2}\left(s,m^2,\mpi^2\right)}\Bigg\{2\Li\left(\frac{2s-s_{11}^+}{2s-s_{11}^-}\right)-2\Li\left(\frac{s_{11}^+}{s_{11}^-}\right)+\frac{1}{2} \log^2\frac{2s-s_{11}^+}{2s-s_{11}^-}\notag\\
&-\frac{1}{2} \log^2\frac{s_{11}^+}{s_{11}^-}+\log\frac{\lambda \left(s,m^2,\mpi^2\right)}{m^2s}\log\frac{\left(2s-s_{11}^+\right)s_{11}^-}{\left(2s-s_{11}^-\right)s_{11}^+}\Bigg\}\quad \text{if} \ s\leq \left(m-\mpi\right)^2,\notag\\
h_{11}(s)&=\frac{2\pi^2m^2}{\lambda^{1/2}\left(s,m^2,\mpi^2\right)}+\frac{m^2}{\lambda^{1/2}\left(s,m^2,\mpi^2\right)}\Bigg\{2\Li\left(\frac{2s-s_{11}^+}{2s-s_{11}^-}\right)+2\Li\left(\frac{s_{11}^-}{s_{11}^+}\right)-\frac{2\pi^2}{3}\notag\\
&+\frac{1}{2} \log^2\frac{2s-s_{11}^+}{2s-s_{11}^-}+\frac{1}{2} \log^2\frac{s_{11}^+}{s_{11}^-}+\log\frac{\lambda \left(s,m^2,\mpi^2\right)}{m^2s}\log\frac{\left(2s-s_{11}^+\right)s_{11}^-}{\left(2s-s_{11}^-\right)s_{11}^+}\Bigg\}\notag\\
& \text{if} \ s\geq \left(m+\mpi\right)^2,\hspace{50pt}
s_{11}^\pm(s)=s-m^2+\mpi^2\pm\lambda^{1/2} \left(s,m^2,\mpi^2\right),\notag\\
k_{11}(s)&=\int\limits_0^1\diff x\frac{s_{11}(s,x)\log\frac{s_{11}(s,x)+x \lambda^{1/2} \left(s,m^2,\mpi^2\right)}{2m \sqrt{(1-x)m^2+x s-x(1-x) \mpi^2}}}{x\lambda^{1/2} \left(s,m^2,\mpi^2\right)\left(1-x+x \frac{s}{m^2}-x(1-x) \frac{\mpi^2}{m^2}\right)}\quad \text{if} \ \lambda\left(s,m^2,\mpi^2\right)\geq 0,\notag\\
k_{11}(s)&=\int\limits_0^1\diff x\frac{s_{11}(s,x)\arccos\frac{s_{11}(s,x)}{2m\sqrt{(1-x)m^2+x s-x(1-x)\mpi^2}}}{x\sqrt{-\lambda\left(s,m^2,\mpi^2\right)}\left(1-x+x \frac{s}{m^2}-x(1-x) \frac{\mpi^2}{m^2}\right)}\quad \text{if} \ \lambda\left(s,m^2,\mpi^2\right)\leq 0,\notag\\
s_{11}(s,x)&=2m^2+x\left(s-m^2-\mpi^2\right),\notag\\
V_{11}(s)&=V_{11}(s_0)+R_{11}(s_0)+\frac{s-s_0}{\pi}\hspace{-18pt}\dashint{26.5pt}\limits_{\left(m+\mpi\right)^2}^{\infty}\hspace{-5pt}\frac{\diff s' \text{Im}\,V_{11}(s')}{(s'-s)(s'-s_0)}-R_{11}(s),\notag\\
\text{Im}\,V_{11}(s)&=\frac{\theta\left(s-\left(m+\mpi\right)^2\right)}{16\pi \sqrt{\lambda \left(s,m^2,\mpi^2\right)}}\left(\log\frac{\lambda \left(s,m^2,\mpi^2\right)}{m^2s}-L\right).
\end{align}
For $s\geq \left(m+\mpi\right)^2$, $V_{11}(s)$ contributes to the Coulomb pole at threshold. The divergence is contained in $h_{11}(s)$ and solely due to the term $\propto 2\pi^2$, while the remaining bracket is regular at threshold. Besides $V_{11}(s)$, which emerges from vector-type diagrams, there is a second kind of loop integrals including one meson, nucleon, and photon propagator, arising from axial-type topologies
\begin{align}
A_{11}(s)&=\frac{1}{i}\int_{\rm I} \frac{\diff^d k}{(2\pi)^d}\frac{1}{\left(m_\gamma^2-k^2\right)\left(m^2-(\Sigma-k)^2\right)\left(\mpi^2-(q-k)^2\right)}=(q\rightarrow q')\notag\\
&=\frac{C_\pi+1}{32\pi^2m^2}f_{11}\left(s,m,\mpi\right)+\frac{1}{16\pi ^2m^2}\left(g_{11}^{\rm ax}(s)+h_{11}^{\rm ax}(s)\right),\notag\\ 
g_{11}^{\rm ax}(s)&=\int\limits_0^1\diff x\frac{\log x}{x+(1-x) \frac{s}{m^2}-x(1-x)\frac{\mpi^2}{m^2}}\notag\\
&=\frac{m^2}{\lambda^{1/2} \left(s,m^2,\mpi^2\right)}\Bigg\{\Li\left(\frac{s-m^2+\mpi^2-\lambda^{1/2} \left(s,m^2,\mpi^2\right)}{2s}\right)\notag\\
&-\Li\left(\frac{s-m^2+\mpi^2+\lambda^{1/2} \left(s,m^2,\mpi^2\right)}{2s}\right)\Bigg\}\quad \text{if} \ \lambda\left(s,m^2,\mpi^2\right)\geq 0,\notag\\
h_{11}^{\rm ax}(s)&=\int\limits_0^{x_{11}^{\rm ax}(s)}\diff x\frac{\left\{(1-x)s+(1+x)m^2-x \mpi^2\right\}\log\frac{\pm(1-x)\left(s-m^2\right)\pm(2x-1)x \mpi^2\mp\sqrt{s_{11\,\rm ax}^{2}(s,x)}}{2x \mpi\sqrt{(1-x)s+x m^2-x(1-x)\mpi^2}}}{\left(x+(1-x) \frac{s}{m^2}-x(1-x)\frac{\mpi^2}{m^2}\right)\sqrt{s_{11\,\rm ax}^{2}(s,x)}}\notag\\
&+\int\limits_{x_{11}^{\rm ax}(s)}^1\diff x\frac{\left\{(1-x)s+(1+x)m^2-x \mpi^2\right\}\arccos\frac{-(1-x)\left(s-m^2\right)-(2x-1)x \mpi^2}{2x \mpi\sqrt{(1-x)s+x m^2 -x(1-x)\mpi^2}}}{\left(x+(1-x) \frac{s}{m^2}-x(1-x)\frac{\mpi^2}{m^2}\right)\sqrt{-s_{11\,\rm ax}^{2}(s,x)}},\notag\\
x_{11}^{\rm ax}(s)&=\frac{s-m^2}{s-m^2+\mpi^2\pm2m\mpi} \quad \text{if} \ s\gtrless m^2,\notag\\
s_{11\,\rm ax}^{2}(s,x)&=(1-x)^2\left(s-m^2\right)^2+x^2\mpi^4-2x \mpi^2\left((1-x) s+(3x-1)m^2\right),\notag\\
A_{11}(s)&=A_{11}(s_0)+R_{11}^{\rm ax}(s_0)+\frac{s-s_0}{\pi}\dashint{0pt}\limits_{m^2}^{\infty}\frac{\diff s' \text{Im}\,A_{11}(s')}{(s'-s)(s'-s_0)}-R_{11}^{\rm ax}(s),\notag\\
\text{Im}\,A_{11}(s)&=\frac{\theta\left(s-m^2\right)}{8\pi \lambda^{1/2} \left(s,m^2,\mpi^2\right)}\log\frac{s-m^2+\mpi^2+\lambda^{1/2} \left(s,m^2,\mpi^2\right)}{2\mpi\sqrt{s}},\notag\\
R_{11}^{\rm ax}(s)&=-\frac{C+1}{32\pi ^2m^2}f_{11}\left(s,m,\mpi\right)-\frac{1}{16\pi ^2m^2}\tilde{h}_{11}^{\rm ax}(s),\\
\tilde{h}_{11}^{\rm ax}(s)&=\int\limits_0^{x_{11}^{\rm ax}(s)}\diff x\frac{\left\{(1-x)s+(1+x)m^2-x \mpi^2\right\}\log\frac{(1-x)s+(1+x)m^2-x \mpi^2+\sqrt{s_{11\,\rm ax}^{2}(s,x)}}{2m\sqrt{(1-x)s+x m^2 -x(1-x)\mpi^2}}}{\left(x+(1-x) \frac{s}{m^2}-x(1-x)\frac{\mpi^2}{m^2}\right)\sqrt{s_{11\,\rm ax}^{2}(s,x)}}\notag\\
&+\int\limits_{x_{11}^{\rm ax}(s)}^1\diff x\frac{\left\{(1-x)s+(1+x)m^2-x \mpi^2\right\}\arccos\frac{(1-x)s+(1+x)m^2-x \mpi^2}{2m\sqrt{(1-x)s+x m^2 -x(1-x)\mpi^2}}}{\left(x+(1-x) \frac{s}{m^2}-x(1-x)\frac{\mpi^2}{m^2}\right)\sqrt{-s_{11\,\rm ax}^{2}(s,x)}}.\notag
\end{align}

\subsubsection{2 mesons, 1 nucleon, 1 photon}

\begin{align}
A_{21}(s,t)&=\frac{1}{i}\int_{\rm I} \frac{\diff^d k}{(2\pi)^d}\frac{1}{\left(m_\gamma^2-k^2\right)\left(m^2-(\Sigma-k)^2\right)\left(\mpi^2-(q-k)^2\right)\left(\mpi^2-(q'-k)^2\right)}\notag\\
&=\frac{f\left(t,\mpi\right)}{32\pi ^2\mpi^2\left(s-m^2\right)}\Bigg\{L+\log\left(\frac{m^2}{s-m^2}\right)^2\Bigg\}+\frac{g_{21}(t)}{16\pi ^2\left(s-m^2\right)}-R_{21}(s,t),\notag\\
g_{21}(t)&=\int\limits_0^1\diff x\frac{\arctan\frac{\sqrt{4m^2\left(\mpi^2-x(1-x)t\right)-\mpi^4}}{\mpi^2}}{\left(1-x(1-x)\frac{t}{\mpi^2}\right)\sqrt{4m^2\left(\mpi^2-x(1-x)t\right)-\mpi^4}},\notag\\
R_{21}(s,t)&=\int\limits_0^1\diff x\int\limits_0^1\diff y\frac{y}{16\pi ^2}\Bigg\{-\frac{4m^2\arccos\frac{ (1-y)s+(1+y)m^2 -y \mpi^2 }{2m \sqrt{s_{21}^\pi(s,t,x,y)}}}{\left(s_{21}^2(s,t,x,y)\right)^{3/2}}\notag\\
&+\frac{ (1-y)s+(1+y)m^2 -y \mpi^2 }{s_{21}^2(s,t,x,y)s_{21}^\pi(s,t,x,y)}\Bigg\},\quad
s_{21}^\pi(s,t,x,y)\equiv s_{13}^\pi(s,t,x,y)|_{\Deltax=0},\notag\\
s_{21}^2(s,t,x,y)&=4y^2m^2\left(\mpi^2-x(1-x)t\right)-\left((1-y)\left(s-m^2\right)-y \mpi^2\right)^2,\notag\\
\text{Im}\,A_{21}(s,t)&=\frac{\theta\left(s-m^2\right)}{4\pi \left(s-m^2\right)\sqrt{-t\left(4\mpi^2-t\right)}}\arsinh\frac{\sqrt{-t}}{2\mpi}.
\end{align}
We cannot write down the usual dispersive representation due to the non-integrable singularity at $s=m^2$ of the imaginary part, which, once regularized by a finite photon mass, generates the term proportional to $L$.

\subsubsection{1 meson, 2 nucleons, 1 photon}
\label{sec:1m2n1p}

$A_{12}(s,t)$ possesses both $s$- and $u$-channel cuts, is infrared divergent and contributes to the Coulomb pole at threshold. Note that we still consider $s$ and $t$ as independent variables and $u$ to be fixed by the Mandelstam relation $s+t+u=2m^2+2\mpi^2-\Deltax$. 
The $s$-channel integrals
\begin{align}
A_{12}(s,t)&=\frac{1}{i}\int_{\rm I} \frac{\diff^d k}{(2\pi)^d}\frac{1}{\left(m_\gamma^2-k^2\right)\left(m^2-(p-k)^2\right)\left(m^2-(\Sigma-k)^2\right)\left(\mpi^2-(q'-k)^2\right)}\notag\\
&=\frac{1}{i}\int_{\rm I} \frac{\diff^d k}{(2\pi)^d}\frac{1}{\left(m_\gamma^2-k^2\right)\left(m^2-(p'-k)^2\right)\left(m^2-(\Sigma-k)^2\right)\left(\mpi^2-(q-k)^2\right)}
\end{align}
as well as
\beq
A_{12}(u,t)=\frac{1}{i}\int_{\rm I} \frac{\diff^d k}{(2\pi)^d}\frac{1}{\left(m_\gamma^2-k^2\right)\left(m^2-(p'-k)^2\right)\left(m^2-(\Lambda-k)^2\right)\left(\mpi^2-(q'+k)^2\right)}
\eeq
are only relevant for the elastic channels, while 
\beq
A_{12}(u,t)=\frac{1}{i}\int_{\rm I} \frac{\diff^d k}{(2\pi)^d}\frac{1}{\left(m_\gamma^2-k^2\right)\left(m^2-(p-k)^2\right)\left(m^2-(\Lambda-k)^2\right)\left(\mpi^2-(q+k)^2\right)}\label{1m2ncex}
\eeq
does contribute to the charge exchange reaction. We will give explicit expressions for the $s$-channel case, where $\Deltax$ is incorporated such that \eqref{1m2ncex} is reproduced correctly when deducing  the $u$-channel via $s\leftrightarrow u$. The result for the imaginary part reads
\begin{align}
\text{Im}\,A_{12}(s,t)&=\text{Im}\,A_{12}^s(s,t)+\text{Im}\,A_{12}^u(s,t),\notag\\
\text{Im}\,A_{12}^s(s,t)&=\frac{\theta\left(s-m^2\right)\theta\left(u-\left(m+\mpi\right)^2\right)}{8\pi \left(s-m^2\right)\sqrt{-\lambda \left(u,m^2,\mpi^2\right)}}\arccos\frac{m^2+\mpi^2-u}{2m \mpi}\notag\\
&+\frac{\theta\left(\left(m+\mpi\right)^2-u\right)}{8\pi \left(s-m^2\right)\sqrt{\lambda \left(u,m^2,\mpi^2\right)}}\arcosh\frac{m^2+\mpi^2-u}{2m \mpi},\notag\\
\text{Im}\,A_{12}^u(s,t)&=\frac{\theta\left(u-\left(m+\mpi\right)^2\right)}{16\pi \left(m^2-s\right)\lambda^{1/2} \left(u,m^2,\mpi^2\right)}\Bigg\{\log\frac{\left(m^2-s\right)^2\lambda^2 \left(u,m^2,\mpi^2\right)}{m^2\left(\lambda \left(u,m^2,\mpi^2\right)-\mpi^2\Deltax\right)\xi(s,t)}-L\Bigg\}\notag\\
&\overset{\hspace{-10pt}\Deltax=0}{\hspace{-10pt}=}\frac{\theta\left(u-\left(m+\mpi\right)^2\right)}{16\pi \left(m^2-s\right)\lambda^{1/2} \left(u,m^2,\mpi^2\right)}\Bigg\{\log\frac{\left(m^2-s\right)^2\lambda \left(u,m^2,\mpi^2\right)}{m^2\xi(s,t)}-L\Bigg\},\notag\\
\xi(s,t)&=2 \mpi^2\left(\mpi^4-\Deltax ^2\right)-\mpi^2\left(\mpi^2-\Deltax \right) \left(3 m^2+u+3 \Deltax \right)+m^2\left(u-m^2\right)^2\notag\\
&\overset{\hspace{-10pt}\Deltax=0}{\hspace{-10pt}=}2 \mpi^6-\mpi^4 \left(3 m^2+u\right)+m^2\left(u-m^2\right)^2.
\end{align}
Again, $\text{Im}\,A_{12}^s(s,t)\propto 1/(s-m^2)$ prevents the direct application of dispersion relations. The real part is found to be
\begin{align}
A_{12}(s,t)&=\frac{g_{11}\left(u,m,\mpi\right)}{32\pi ^2m^2\left(s-m^2\right)}\Bigg\{L+\log\left(\frac{m^2}{s-m^2}\right)^2\Bigg\}+\frac{g_{12}(s,t)}{32\pi ^2\left(s-m^2\right)}-R_{12}(s,t),\notag\\
R_{12}(s,t)&=\int\limits_0^1\diff x\int\limits_0^1\diff y\frac{1}{16\pi^2}\Bigg\{\frac{2x^2\mpi^2-\tilde{s}_{12}(x,y)}{s_{12}^2(x,y)\left(m^2-y(1-y)\left(\mpi^2-\Deltax \right)+x^2\mpi^2-\tilde{s}_{12}(x,y)\right)}\notag\\
&+\frac{2\tilde{s}_{12}(x,y)\arccos\frac{2\left(m^2-y(1-y)\left(\mpi^2-\Deltax \right)\right)-\tilde{s}_{12}(x,y)}{2\sqrt{m^2-y(1-y)\left(\mpi^2-\Deltax \right)}\sqrt{m^2-y(1-y)\left(\mpi^2-\Deltax \right)+x^2\mpi^2-\tilde{s}_{12}(x,y)}}}{\left(s_{12}^2(x,y)\right)^{3/2}}\Bigg\},\notag\\
s^2_{12}(s,t,x,y)&=4x^2\mpi^2\left(m^2-y(1-y)\left(\mpi^2-\Deltax \right)\right)-\tilde{s}^2_{12}(x,y),\notag\\
\tilde{s}_{12}(s,t,x,y)&=(x-y)\left(s-m^2-x \mpi^2\right)+x(1-y)\left(t-\mpi^2+\Deltax \right)+x^2\mpi^2.\notag
\end{align}
For $u\leq\left(m+\mpi\right)^2$, $g_{12}(s,t)$ is given by
\begin{align}
g_{12}(s,t)&=\left\{\begin{array}{cl}
&\hspace{-10pt}\int_0^1\diff x g_{12}^{(1)}(s,t,x) \quad \hspace{251pt}{\rm if} \, s\geq s^{\rm min}_{12},  \\
&\hspace{-10pt}\int_0^{x_{12}^-(s,t)}\diff x g_{12}^{(1)}(s,t,x)+\int_{x_{12}^-(s,t)}^{x_{12}^+(s,t)}\diff x g_{12}^{(2)}(s,t,x)+
\int_{x_{12}^+(s,t)}^1\diff x g_{12}^{(1)}(s,t,x) \quad {\rm otherwise}, \, 
\end{array}\right.\hspace{-5pt}\notag\\
s^{\rm min}_{12}&=\frac{1}{2m^2}\Big\{\mpi^2\left(4m^2+\Deltax \right)-\mpi^4+2m^2\left(m^2-t-\Deltax \right)\notag\\
&-\mpi \sqrt{\left(\mpi^2-\Deltax \right)\left(4m^2-\mpi^2+\Deltax \right)\left(4m^2-\mpi^2\right)}\Big\}
\overset{\Deltax=0}{=}m^2-t,\notag\\
x_{12}^\pm(s,t)&=\frac{1}{2}\pm\frac{4m\sqrt{\xi(s,t)}\mp2\left(2m^2-\mpi^2\right)\Deltax \mp\Deltax^2}{8 \left(m^2 u-\left(m^2-\mpi^2\right)^2\right)-8\Deltax \left(m^2-\mpi^2\right)-2\Deltax^2},\notag\\
g_{12}^{(1)}(s,t,x)&=\frac{2\left((1-x)\left(2m^2+\Deltax \right)+(2x-1)\mpi^2\right) \arctan\frac{\sqrt{-u_{12}^{(1)}(s,t,x)}}{(1-x) \left(2 m^2+\Deltax \right)+(2 x-1)\mpi^2 }}{ \Big\{x \mpi^2+(1-x)m^2 -x(1-x)u\Big\}\sqrt{-u_{12}^{(1)}(s,t,x)}},\notag\\
g_{12}^{(2)}(s,t,x)&=\frac{\left((1-x)\left(2m^2+\Deltax \right)+(2x-1)\mpi^2\right)\log\frac{\left\{(1-x)\left(2m^2+\Deltax \right)+\mpi^2(2x-1)+\sqrt{u_{12}^{(1)}(s,t,x)}\right\}^2}{4m^2\left(x \mpi^2+(1-x)m^2 -x(1-x)u\right)}}{\Big\{x \mpi^2+(1-x)m^2 -x(1-x)u\Big\}\sqrt{u_{12}^{(1)}(s,t,x)}},\notag\\
u_{12}^{(1)}(s,t,x)&=4m^2x(1-x)\left(u-m^2-2\mpi^2+\Deltax \right)-\left(4m^2-\mpi^2\right)\mpi^2(1-2x)^2\notag\\
&+2\Deltax  \left(2m^2-\mpi^2\right)(1-x)(1-2x)+(1-x)^2\Deltax^2.
\end{align}
Above the $u$-channel threshold, we perform an integration by parts to improve the numerical properties of $g_{12}(s,t)$. In particular, the contribution to the Coulomb pole is separated into the boundary terms. It is solely given by the term proportional to $\pi^2/\sqrt{\lambda\left(u,m^2,\mpi^2\right)}$. Suppressing the dependence on $s$ and $t$ of the auxiliary functions on the right hand side of the equations, we arrive at

\begin{align}
g_{12}(s,t)&=\int\limits_0^{x_{12}^-}\diff x g_{12}^{(1)}(x)+\int\limits_{x_{12}^+}^1\diff x g_{12}^{(1)}(x)+\int\limits_{x_{12}^-}^{x_{12}^{(1)}}\diff x g_{12}^{(2)}(x)+\int\limits_{x_{12}^{(2)}}^{x_{12}^+}\diff x g_{12}^{(2)}(x)\notag\\
&-\int\limits_{x_{12}^{(1)}}^{x_{12}^{(2)}}\diff x\left(k_{12}^{(1)}(x)l_{12}^{(1)}(x)-k_{12}^{(2)}(x)l_{12}^{(2)}(x)\right)
-\frac{\pi^2u\left\{h_{12}^{(2)}\left(\frac{u_{12}^-}{2u}\right)+h_{12}^{(2)}\left(\frac{u_{12}^+}{2u}\right)\right\}}{\lambda^{1/2}\left(u,m^2,\mpi^2\right)}\notag\\
&+k_{12}^{(1)}(x)h_{12}^{(1)}(x)-k_{12}^{(2)}(x)h_{12}^{(2)}(x)\bigg|^{x_{12}^{(2)}}_{x_{12}^{(1)}},\notag\\
x_{12}^{(1)}(s,t)&=\frac{1}{2}\left(x_{12}^-+\frac{u_{12}^-}{2u}\right), \quad x_{12}^{(2)}(s,t)=\frac{1}{2}\left(x_{12}^++\frac{u_{12}^+}{2u}\right),\notag\\
u_{12}^\pm(s,t)&=u+m^2-\mpi^2\pm\lambda^{1/2}\left(u,m^2,\mpi^2\right),\quad x_{12}^-<\frac{u_{12}^-}{2u}<\frac{u_{12}^+}{2u}<x_{12}^+,\notag\\
h_{12}^{(1)}(s,t,x)&=\frac{2\left((1-x)\left(2m^2+\Deltax \right)+(2x-1)\mpi^2\right)\log\frac{(1-x)\left(2m^2+\Deltax \right)+(2x-1)\mpi^2+\sqrt{u_{12}^{(1)}(x)}}{2m \sqrt{u}}}{u\sqrt{u_{12}^{(1)}(x)}},\notag\\
h_{12}^{(2)}(s,t,x)&=\frac{(1-x)\left(2m^2+\Deltax \right)+(2x-1)\mpi^2}{u\sqrt{u_{12}^{(1)}(x)}},\notag\\
l_{12}^{(1)}(s,t,x)&=\frac{\diff h_{12}^{(1)}(x)}{\diff x}=\frac{4m^2 u_{12}^{(2)}(x)\log\frac{(1-x)\left(2m^2+\Deltax \right)+(2x-1)\mpi^2+\sqrt{u_{12}^{(1)}(x)}}{2m \sqrt{u}}}{u\left(u_{12}^{(1)}(x)\right)^{3/2} }\notag\\
&-\hspace{-0.6pt}\frac{(1-x) \left(2 m^2+\Deltax \right)+(2 x-1)\mpi^2 }{u \left\{x \mpi^2+(1-x)m^2 -x(1-x) u\right\}u_{12}^{(1)}(x)}
\Big(\!\left(m^2 -\mpi^2 +(1-2 x) u\right)\sqrt{u_{12}^{(1)}(x)}+u_{12}^{(2)}(x)\Big),\notag\\
l_{12}^{(2)}(s,t,x)&=\frac{\diff h_{12}^{(2)}(x)}{\diff x}=\frac{2 m^2 u_{12}^{(2)}(x)}{u \left(u_{12}^{(1)}(x)\right)^{3/2}},\notag\\
k_{12}^{(1)}(s,t,x)&=\frac{u \log\left(\frac{2x u-u_{12}^+}{2x u-u_{12}^-}\right)^2}{2\lambda^{1/2} \left(u,m^2,\mpi^2\right)},
\quad \frac{\diff k_{12}^{(1)}(x)}{\diff x}=\frac{u}{x \mpi^2+(1-x) m^2-x(1-x)u},\notag\\
k_{12}^{(2)}(s,t,x)&=\frac{u}{\lambda^{1/2} \left(u,m^2,\mpi^2\right)}\Bigg\{\log^2\left(\frac{u_{12}^+}{2u }-x\right)-\log^2\left(\frac{u_{12}^-}{2u}-x\right)+\frac{1}{2}\log^2\frac{u_{12}^--2x u}{2\lambda^{1/2} \left(u,m^2,\mpi^2\right)}\notag\\
&+\frac{2\pi^2}{3}+\Li\left(\frac{2\lambda^{1/2} \left(u,m^2,\mpi^2\right)}{u_{12}^+-2x u}\right)+\Li\left(\frac{2x u-u_{12}^-}{2\lambda^{1/2} \left(u,m^2,\mpi^2\right)}\right)\Bigg\}\quad \text{if}\ x<\frac{u_{12}^-}{2u},\notag\\
k_{12}^{(2)}(s,t,x)&=\frac{u}{\lambda^{1/2} \left(u,m^2,\mpi^2\right)}\Bigg\{\frac{1}{2}\log^2\left(\frac{u_{12}^+}{2u}-x\right)
-\frac{1}{2}\log^2\left(x-\frac{u_{12}^-}{2u }\right)-\frac{\pi ^2}{2}\notag\\
&+\log\frac{\lambda^{1/2} \left(u,m^2,\mpi^2\right)}{u}\log\frac{u_{12}^+-2x u}{2x u-u_{12}^-}-\Li\left(\frac{u_{12}^+-2x u }{2\lambda^{1/2} \left(u,m^2,\mpi^2\right)}\right)\notag\\
&+\Li\left(\frac{2x u-u_{12}^-}{2\lambda^{1/2} \left(u,m^2,\mpi^2\right)}\right)\Bigg\}\quad \text{if}\ \frac{u_{12}^-}{2u}<x<\frac{u_{12}^+}{2u},\notag\\
k_{12}^{(2)}(s,t,x)&=\frac{u}{\lambda^{1/2} \left(u,m^2,\mpi^2\right)}\Bigg\{\log^2\left(x-\frac{u_{12}^+}{2u}\right)-\log^2\left(x-\frac{u_{12}^-}{2u}\right)
-\frac{1}{2}\log^2\frac{2x u-u_{12}^+}{2\lambda^{1/2} \left(u,m^2,\mpi^2\right)}\notag\\
&+\frac{\pi^2}{3}-\Li\left(\frac{u_{12}^+-2xu}{2\lambda^{1/2} \left(u,m^2,\mpi^2\right)}\right)-\Li\left(\frac{2\lambda^{1/2} \left(u,m^2,\mpi^2\right)}{2x u-u_{12}^-}\right)\Bigg\}\quad \text{if}\ x>\frac{u_{12}^+}{2u},\notag\\
u_{12}^{(2)}(s,t,x)&=4x m^2\mpi^2-\mpi^4 (1+2 x)-\left(2(1-x)m^2-\mpi^2\right) \left(u-m^2\right) \notag\\
&+\Deltax \left((1+x)\mpi^2-(1-x)\left(u-m^2\right)\right).
\end{align}

\subsubsection{1 meson, 1 nucleon, 2 photons}

\begin{align}
P_{11}(s,t)&=\frac{1}{i}\int_{\rm I} \frac{\diff^d k}{(2\pi)^d}\frac{1}{\left(m_\gamma^2-k^2\right)\left(m_\gamma^2-(k-\Delta)^2\right)\left(\mpi^2-(q+k)^2\right)\left(m^2-(p-k)^2\right)}\notag\\
&=\frac{1}{i}\int_{\rm I} \frac{\diff^d k}{(2\pi)^d}\frac{1}{\left(m_\gamma^2-(k-q)^2\right)\left(m_\gamma^2-(k-q')^2\right)\left(\mpi^2-k^2\right)\left(m^2-(\Sigma-k)^2\right)}\notag\\
&=-\frac{1}{16\pi ^2m^2t}\left(\log\frac{-t}{m^2}-L\right)g_{11}\left(s,m,\mpi\right)-R_{11}^{\rm p}(s,t),\notag\\
R_{11}^{\rm p}(s,t)&=\int\limits_0^1\diff x\int\limits_0^1\diff y\frac{1-y}{16\pi ^2}\Bigg\{-\frac{4m^2\arccos\frac{s_{11}(s,y)}{2m \sqrt{s_{11}^{\rm p}(s,t,x,y)}}}{\left\{- y^2\lambda \left(s,m^2,\mpi^2\right)-4x(1-x)(1-y)^2m^2t\right\}^{3/2}}\notag\\
&+\frac{s_{11}(s,y)}{s_{11}^{\rm p}(s,t,x,y)\left\{- y^2\lambda \left(s,m^2,\mpi^2\right)-4x(1-x)(1-y)^2m^2t\right\}}\Bigg\},\notag\\
s_{11}^{\rm  p}(s,t,x,y)&=s-(1-y)\left(s-m^2+\mpi^2\right)+(1-y)^2\left(\mpi^2- x(1-x)t\right),\notag\\
P_{11}(s,t)&=P_{11}(s_0,t)+R_{11}^{\rm p}(s_0,t)+\frac{s-s_0}{\pi}\hspace{-18pt}\dashint{26.5pt}\limits_{\left(m+\mpi\right)^2}^{\infty}\hspace{-5pt}\frac{\diff s' \text{Im}\,P_{11}(s',t)}{(s'-s)(s'-s_0)}-R_{11}^{\rm p}(s,t),\notag\\
\text{Im}\,P_{11}(s,t)&=-\frac{\theta\left(s-\left(m+\mpi\right)^2\right)}{8\pi t\lambda^{1/2} \left(s,m^2,\mpi^2\right)}\left(\log\frac{-t}{m^2}-L\right).
\end{align}

\subsection{Tensor decomposition}
\label{tensor_dec}

\subsubsection{Meson integrals}

\begin{align}
\Delta_i^\mu&=\frac{1}{i}\int_{\rm I} \frac{\diff^d k}{(2\pi)^d}\frac{k^\mu}{M_i^2-k^2}=0,\notag\\
J_{ij}^\mu(t)&=\frac{1}{i}\int_{\rm I} \frac{\diff^d k}{(2\pi)^d}\frac{k^\mu}{\left(M_i^2-k^2\right)\left(M_j^2-(k-\Delta)^2\right)}\notag\\
&=\frac{\Delta ^{\mu }}{2}J_{ij}(t)+\frac{\Delta ^{\mu }}{2t}\left(\left(M_i^2-M_j^2\right)J_{ij}(t)+\Delta _i-\Delta _j\right), \quad
J_{ii}^\mu(t)=\frac{\Delta ^{\mu }}{2}J_{ii}(t),\notag\\
J_{ij}^{\mu\nu}(t)&=\frac{1}{i}\int_{\rm I} \frac{\diff^d k}{(2\pi)^d}\frac{k^\mu k^\nu}{\left(M_i^2-k^2\right)\left(M_j^2-(k-\Delta)^2\right)}=\left(\Delta ^{\mu }\Delta ^{\nu }-g^{\mu \nu } t\right)J_{ij}^{(1)}(t)+\Delta ^{\mu }\Delta ^{\nu }J_{ij}^{(2)}(t),\notag\\
J_{ij}^{(1)}(t)&=\frac{1}{4(d-1)t}\bigg\{\Delta_i+\Delta _j+\frac{\lambda \left(M_i^2,M_j^2,t\right)}{t} J_{ij}(t)+\frac{\Delta_i-\Delta_j}{t}\left(M_i^2-M_j^2\right)\bigg\}\notag\\
&=\frac{1}{12t}\bigg\{\Delta_i+\Delta _j+\frac{\lambda \left(M_i^2,M_j^2,t\right)}{t} J_{ij}(t)+\frac{\Delta_i-\Delta_j}{t}\left(M_i^2-M_j^2\right)\bigg\}-\frac{3M_i^2+3M_j^2-t}{288\pi^2 t},\notag\\
J_{ii}^{(1)}(t)&=\frac{1}{4(d-1)t}\bigg\{2\Delta_i-\left(4M_i^2-t\right) J_{ii}(t)\bigg\}\notag\\
&=\frac{1}{12t}\bigg\{2\Delta_i-\left(4M_i^2-t\right) J_{ii}(t)\bigg\}-\frac{6M_i^2-t}{288\pi^2 t},\notag\\
J_{ij}^{(2)}(t)&=-\frac{\Delta _j}{2t}+\frac{\left(t+M_i^2-M_j^2\right)^2}{4t^2}J_{ij}(t)+\frac{\left(t+M_i^2-M_j^2\right)}{4t^2}\left(\Delta_i-\Delta_j\right),\notag\\
J_{ii}^{(2)}(t)&=-\frac{\Delta _i}{2t}+\frac{1}{4}J_{ii}(t).
\end{align}

\subsubsection{1 meson, 1 nucleon}

\begin{align}
I_i^\mu(s)&=\frac{1}{i}\int_{\rm I} \frac{\diff^d k}{(2\pi)^d}\frac{k^\mu}{\left(M_i^2-k^2\right)\left(m^2-(\Sigma-k)^2\right)}=\Sigma^{\mu }I_i^{(1)}(s),\notag\\
I_i^{(1)}(s)&=\frac{1}{2s}\left(\Delta_i+\left(s-m^2+M_i^2\right)I_i(s)\right),\quad
I_\gamma^{(1)}(s)=\frac{s-m^2}{2s}I_\gamma(s).
\end{align}

\subsubsection{2 mesons, 1 nucleon}

\begin{align}
I_{21}^{ij\,\mu}(t)&=\frac{1}{i}\int_{\rm I} \frac{\diff^d k}{(2\pi)^d}\frac{k^\mu}{\left(M_i^2-k^2\right)\left(M_j^2-(k-\Delta)^2\right)
\left(m^2-(p-k)^2\right)}\notag\\
&=Q^\mu I_{21}^{ij(1)}(t)+\frac{\Delta^\mu}{2}I_{21}^{ij}(t)+\frac{\Delta^\mu }{2t}\Big\{I_i\left(m^2\right)-I_j\left(m^2\right)+\left(M_i^2-M_j^2\right)I_{21}^{ij}(t)\Big\},\notag\\
I_{21}^{ii\,\mu}(t)&=Q^\mu I_{21}^{ii(1)}(t)+\frac{\Delta^\mu}{2}I_{21}^{ii}(t),\notag\\
I_{21}^{ij(1)}(t)&=\frac{1}{4m^2-t}\bigg\{J_{ij}(t)-\frac{1}{2}\left(I_i\left(m^2\right)+I_j\left(m^2\right)\right)+\frac{M_i^2+M_j^2-t}{2}I_{21}^{ij}(t)\bigg\},\notag\\
I_{21}^{ii(1)}(t)&=\frac{1}{4m^2-t}\bigg\{J_{ii}(t)-I_i\left(m^2\right)+\frac{2M_i^2-t}{2}I_{21}^{ii}(t)\bigg\},\notag\\
I_{21}^{\gamma\gamma(1)}(t)&=\frac{1}{4m^2-t}\bigg\{J_{\gamma\gamma}(t)-\frac{t}{2}I_{21}^{\gamma\gamma}(t)\bigg\},\notag\\
I_{21}^{ij\,\mu\nu}(t)&=\frac{1}{i}\int_{\rm I} \frac{\diff^d k}{(2\pi)^d}\frac{k^\mu k^\nu}{\left(M_i^2-k^2\right)\left(M_j^2-(k-\Delta)^2\right)
\left(m^2-(p-k)^2\right)}\notag\\
&=g^{\mu \nu } I_{21}^{ij(2)}(t)+Q^{\mu }Q^{\nu }I_{21}^{ij(3)}(t)+\Delta ^{\mu }\Delta ^{\nu }I_{21}^{ij(4)}(t)+\frac{1}{2}\left(\Delta ^{\mu }Q^{\nu }+Q^{\mu }\Delta ^{\nu }\right)I_{21}^{ij(5)}(t),\notag\\
I_{21}^{ij(2)}(t)&=\frac{1}{4(2-d)}\bigg\{I_i\left(m^2\right)+I_j\left(m^2\right)+\frac{\lambda\left(M_i^2,M_j^2,t\right)}{t} I_{21}^{ij}(t)
+2\left(M_i^2+M_j^2-t\right)I_{21}^{ij(1)}(t)\notag\\
&+\frac{M_i^2-M_j^2}{t}\left(I_i\left(m^2\right)-I_j\left(m^2\right)\right)\bigg\}\notag\\
&=-\frac{1}{8}\bigg\{I_i\left(m^2\right)+I_j\left(m^2\right)+\frac{\lambda\left(M_i^2,M_j^2,t\right)}{t} I_{21}^{ij}(t)
+2\left(M_i^2+M_j^2-t\right)I_{21}^{ij(1)}(t)\notag\\
&+\frac{M_i^2-M_j^2}{t}\left(I_i\left(m^2\right)-I_j\left(m^2\right)\right)\bigg\}\notag\\
&+\frac{2(1+f(t,m))\left(m^2\lambda \left(M_i^2,M_j^2,t\right)+t M_i^2M_j^2\right)-\left(4m^2-t\right)\left(M_i^2-M_j^2\right)^2}{128\pi ^2m^2\left(4m^2-t\right)t},\notag\\
I_{21}^{ii(2)}(t)&=-\frac{1}{4(d-2)}\bigg\{2I_i\left(m^2\right)-\left(4M_i^2-t\right) I_{21}^{ii}(t)
+2\left(2M_i^2-t\right)I_{21}^{ii(1)}(t)\bigg\}\notag\\
&=-\frac{1}{8}\bigg\{2I_i\left(m^2\right)-\left(4M_i^2-t\right) I_{21}^{ii}(t)
+2\left(2M_i^2-t\right)I_{21}^{ii(1)}(t)\bigg\}\notag\\
&+\frac{(1+f(t,m))\left(M_i^4-m^2(4M_i^2-t)\right)}{64\pi ^2m^2\left(4m^2-t\right)},\notag\\
I_{21}^{ij(3)}(t)&=\frac{1}{4(d-2)\left(4m^2-t\right)}\bigg\{\frac{\lambda \left(M_i^2,M_j^2,t\right)}{t}I_{21}^{ij}(t)-(d-2)\left(I_{i}^{(1)}\left(m^2\right)+I_j^{(1)}\left(m^2\right)\right)\notag\\
&+I_i\left(m^2\right)+I_j\left(m^2\right)+2(d-1)\left(M_i^2+M_j^2-t\right)I_{21}^{ij(1)}(t)+\frac{M_i^2-M_j^2}{t} \left(I_i\left(m^2\right)-I_j\left(m^2\right)\right)\bigg\}\notag\\
&=\frac{1}{8\left(4m^2-t\right)}\bigg\{I_i\left(m^2\right)+I_j\left(m^2\right)-2\left(I_{i}^{(1)}\left(m^2\right)+I_j^{(1)}\left(m^2\right)\right)+\frac{\lambda \left(M_i^2,M_j^2,t\right)}{t}I_{21}^{ij}(t)\notag\\
&+6\left(M_i^2+M_j^2-t\right)I_{21}^{ij(1)}(t)+\frac{M_i^2-M_j^2}{t} \left(I_i\left(m^2\right)-I_j\left(m^2\right)\right)\bigg\}\notag\\
&-\frac{2(1+f(t,m))\left(m^2\lambda \left(M_i^2,M_j^2,t\right)+t M_i^2M_j^2\right)-\left(4m^2-t\right)\left(M_i^2-M_j^2\right)^2}{128\pi ^2m^2\left(4m^2-t\right)^2t},\notag\\
I_{21}^{ii(3)}(t)&=\frac{1}{4(d-2)\left(4m^2-t\right)}\bigg\{2I_i\left(m^2\right)-2(d-2)I_{i}^{(1)}\left(m^2\right)-\left(4M_i^2-t\right)I_{21}^{ii}(t)\notag\\
&+2(d-1)\left(2M_i^2-t\right)I_{21}^{ii(1)}(t)\bigg\}\notag\\
&=\frac{1}{8\left(4m^2-t\right)}\bigg\{2I_i\left(m^2\right)-4I_{i}^{(1)}\left(m^2\right)-\left(4M_i^2-t\right)I_{21}^{ii}(t)+6\left(2M_i^2-t\right)I_{21}^{ii(1)}(t)\bigg\}\notag\\
&-\frac{(1+f(t,m))\left(M_i^4-m^2(4M_i^2-t)\right)}{64\pi ^2m^2\left(4m^2-t\right)^2},\notag\\
I_{21}^{ij(4)}(t)&=\frac{1}{4(d-2)t}\bigg\{-2(d-3)I_j\left(m^2\right)+(d-2)\left(I_{i}^{(1)}\left(m^2\right)+I_{j}^{(1)}\left(m^2\right)\right)\notag\\
&+2\left(M_i^2+M_j^2-t\right)I_{21}^{ij(1)}(t)+(d-1)\frac{t+M_i^2-M_j^2}{t} \left(I_i\left(m^2\right)-I_j\left(m^2\right)\right)\notag\\
&+\frac{1}{t}\left\{(d-1)\lambda \left(M_i^2,M_j^2,t\right)+4(d-2) t M_i^2\right\}I_{21}^{ij}(t)\bigg\}\notag\\
&=\frac{1}{8t}\bigg\{-2I_j\left(m^2\right)+2\left(I_{i}^{(1)}\left(m^2\right)+I_{j}^{(1)}\left(m^2\right)\right)+2\left(M_i^2+M_j^2-t\right)I_{21}^{ij(1)}(t)\notag\\
&+3\frac{t+M_i^2-M_j^2}{t} \left(I_i\left(m^2\right)-I_j\left(m^2\right)\right)+\frac{1}{t}\left\{3\lambda \left(M_i^2,M_j^2,t\right)+8 t M_i^2\right\}I_{21}^{ij}(t)\bigg\}\notag\\
&-\frac{2(1+f(t,m))\left(m^2\lambda \left(M_i^2,M_j^2,t\right)+t M_i^2M_j^2\right)-\left(4m^2-t\right)\left(M_i^2-M_j^2\right)^2}{128\pi ^2m^2\left(4m^2-t\right)t^2},\notag\\
I_{21}^{ii(4)}(t)&=\frac{1}{4(d-2)t}\bigg\{-2(d-3)I_i\left(m^2\right)+2(d-2)I_{i}^{(1)}\left(m^2\right)+2\left(2M_i^2-t\right)I_{21}^{ii(1)}(t)\notag\\
&-\left(4M_i^2-(d-1)t\right)I_{21}^{ii}(t)\bigg\}\notag\\
&=\frac{1}{8t}\bigg\{-2I_i\left(m^2\right)+4I_{i}^{(1)}\left(m^2\right)+2\left(2M_i^2-t\right)I_{21}^{ii(1)}(t)-\left(4M_i^2-3t\right)I_{21}^{ii}(t)\bigg\}\notag\\
&-\frac{(1+f(t,m))\left(M_i^4-m^2(4M_i^2-t)\right)}{64\pi ^2m^2\left(4m^2-t\right)t},\\
I_{21}^{ij(5)}(t)&=\frac{1}{2t}\left(I_{i}^{(1)}\left(m^2\right)-I_{j}^{(1)}\left(m^2\right)\right)+\frac{1}{t}\left(t+M_i^2-M_j^2\right)I_{21}^{ij(1)}(t),\quad I_{21}^{ii(5)}(t)=I_{21}^{ii(1)}(t).\notag
\end{align}
Note that $I_{21}^{ij(1)}(t)$, $I_{21}^{ij(2)}(t)$ and  $I_{21}^{ij(3)}(t)$ are symmetric in $i$ and $j$, but that $I_{21}^{ij(4)}(t)$ and $I_{21}^{ij(5)}(t)$ are not.

\subsubsection{1 meson, 2 nucleons}

\begin{align}
I_{A}^{i\,\mu}(t)&=\frac{1}{i}\int_{\rm I} \frac{\diff^d k}{(2\pi)^d}\frac{k^\mu}{\left(M_i^2-k^2\right)\left(m^2-(p-k)^2\right)
\left(m^2-(p'-k)^2\right)}=Q^{\mu }I_{A}^{i(1)}(t),\notag\\
I_{A}^{i(1)}(t)&=\frac{1}{4m^2-t}\left(I_i\left(m^2\right)+M_i^2I_{A}^{i}(t)\right),\notag\\
I_{A}^{i\,\mu\nu}(t)&=\frac{1}{i}\int_{\rm I} \frac{\diff^d k}{(2\pi)^d}\frac{k^\mu k^\nu}{\left(M_i^2-k^2\right)\left(m^2-(p-k)^2\right)
\left(m^2-(p'-k)^2\right)}\notag\\
&=g^{\mu \nu }I_{A}^{i(2)}(t)+Q^{\mu }Q^{\nu }I_{A}^{i(3)}(t)+\Delta^{\mu}\Delta^{\nu }I_{A}^{i(4)}(t),\notag\\
I_{A}^{i(2)}(t)&=\frac{M_i^2}{d-2}\left(I_{A}^i(t)-I_{A}^{i(1)}(t)\right)\notag\\
&=\frac{M_i^2}{2}\left(I_{A}^i(t)-I_{A}^{i(1)}(t)\right)-\frac{M_i^2 \left\{M_i^2+\left(M_i^2-4m^2+t\right) f(t,m)\right\}}{64\pi ^2m^2\left(4 m^2-t\right)},\notag\\
I_{A}^{i(3)}(t)&=\frac{1}{\left(4m^2-t\right)(d-2)}\bigg\{M_i^2\left((d-1)I_{A}^{i(1)}(t)-I_{A}^{i}(t)\right)+\frac{d-2}{2} I_{i}^{(1)}\left(m^2\right)\bigg\}\notag\\
&=\frac{1}{2\left(4m^2-t\right)}\bigg\{M_i^2\left(3I_{A}^{i(1)}(t)-I_{A}^{i}(t)\right)+I_{i}^{(1)}\left(m^2\right)\bigg\}\notag\\
&+\frac{M_i^2 \left\{M_i^2+\left(M_i^2-4m^2+t\right) f(t,m)\right\}}{64\pi ^2m^2\left(4 m^2-t\right)^2},\notag\\
I_{A}^{i(4)}(t)&=\frac{1}{t(d-2)}\bigg\{M_i^2\left(I_{A}^{i(1)}(t)-I_{A}^i(t)\right)-\frac{d-2}{2} I_{i}^{(1)}\left(m^2\right)\bigg\}\notag\\
&=\frac{1}{2t}\bigg\{M_i^2\left(I_{A}^{i(1)}(t)-I_{A}^i(t)\right)-I_{i}^{(1)}\left(m^2\right)\bigg\}
+\frac{M_i^2 \left\{M_i^2+\left(M_i^2-4m^2+t\right) f(t,m)\right\}}{64\pi ^2m^2\left(4 m^2-t\right)t},\notag\\
I_{A}^{\gamma(1)}(t)&=I_{A}^{\gamma(2)}(t)=I_{A}^{\gamma(3)}(t)=I_{A}^{\gamma(4)}(t)=0.
\end{align}
\begin{align}
I_{B}^{ij\,\mu}(s)&=\frac{1}{i}\int_{\rm I} \frac{\diff^d k}{(2\pi)^d}\frac{k^\mu}{\left(M_i^2-k^2\right)\left(m^2-(p_1-k)^2\right)
\left(m^2-(\Sigma-k)^2\right)}=\tilde{Q}^{\mu }I_{B}^{ij(1)}(s)+\tilde{\Delta }^{\mu }I_{B}^{ij(2)}(s),\notag\\
\tilde{Q}^{\mu }&=p_1^{\mu }+\Sigma ^{\mu }, \quad \tilde{\Delta }^{\mu }=\Sigma ^{\mu }-p_1^{\mu },\quad
\tilde{Q}^2=2 m^2+2s-M_j^2,\quad \tilde{\Delta }^2=M_j^2,\quad \tilde{Q}\cdot\tilde{\Delta }=s-m^2,\notag\\
I_{B}^{ij(1)}(s)&=\frac{1}{2\lambda \left(s,m^2,M_j^2\right)}\Big\{\left(s-m^2-M_j^2\right)I_i\left(m^2\right)-\left(s-m^2+M_j^2\right)I_i(s)\notag\\
&+\left(\left(s-m^2\right)^2-M_j^2\left(s-m^2+2M_i^2\right)\right)I_{B}^{ij}(s)\Big\},\notag\\
I_{B}^{ij(2)}(s)&=\frac{1}{2\lambda \left(s,m^2,M_j^2\right)}\Big\{\left(3s+m^2-M_j^2\right)I_i(s)-\left(s+3m^2-M_j^2\right)I_i\left(m^2\right)\notag\\
&-\left(s-m^2\right)\left(s+3m^2-2M_i^2-M_j^2\right)I_{B}^{ij}(s)\Big\}.
\end{align}
For the definition of $p_1$, $p_2$, and $M_j$, see Sect.~\ref{sec:1m2n}. In these conventions (in particular for $M_j$), the $u$-channel case follows by $\Sigma\rightarrow\Lambda$, $s\rightarrow u$. 

\subsubsection{1 meson, 3 nucleons}

\begin{align}
I_{13}^{i\,\mu}(s,t)&=\frac{1}{i}\int_{\rm I} \frac{\diff^d k}{(2\pi)^d}\frac{k^\mu}{\left(M_i^2-k^2\right)\left(m^2-(p-k)^2\right)
\left(m^2-(\Sigma-k)^2\right)\left(m^2-(p'-k)^2\right)}\notag\\
&=Q^{\mu }I_{13}^{i(1)}(s,t)+(\Delta +2q)^{\mu }I_{13}^{i(2)}(s,t)+\Delta ^{\mu }I_{13}^{i(3)}(s,t),\notag\\
I_{13}^{i(1)}(s,t)&=\frac{1}{4N_{13}(s,t)}\Big\{t(s-u)I_{A}^i(t)-\left(2t\left(s-m^2+\mpi^2\right)-(t+\Deltax )\Deltax \right)I_{B}^{i\pi}(s)\notag\\
&+\left(t\left(s-m^2\right)(s-u)-4t M_i^2\mpi^2+M_i^2(t+\Deltax )^2\right)I_{13}^i(s,t)\notag\\
&+2 \delta_{13}^i(s)  \left(\left(s-m^2\right)(t-\Deltax )+(t+\Deltax )\left(\mpi^2-\Deltax \right)\right)\Big\}\notag\\
&\overset{\hspace{-25pt}\Deltax=0,\ \delta_{13}^i=0}{\hspace{-25pt}=}\frac{1}{4\left(\lambda \left(s,m^2,\mpi^2\right)+s t\right)}\Big\{(s-u)I_{A}^i(t)-2\left(s-m^2+\mpi^2\right)I_{B}^{i\pi}(s)\notag\\
&+\left(\left(s-m^2\right)(s-u)-M_i^2(4\mpi^2-t)\right)I_{13}^i(s,t)\Big\},\notag\\
I_{13}^{\gamma(1)}(s,t)&=\frac{1}{4N_{13}(s,t)}\Big\{t(s-u)\left(I_{A}^\gamma(t)+\left(s-m^2\right)I_{13}^\gamma(s,t)\right)\notag\\
&-\left(2t\left(s-m^2+\mpi^2\right)-(t+\Deltax )\Deltax \right)I_{B}^{\gamma\pi}(s)\notag\\
&+2 \delta_{13}^\gamma(s)  \left(\left(s-m^2\right)(t-\Deltax )+(t+\Deltax )\left(\mpi^2-\Deltax \right)\right)\Big\}\notag\\
&\overset{\hspace{-25pt}\Deltax=0,\ \delta_{13}^\gamma=0}{\hspace{-25pt}=}\frac{(s-u)\left(I_{A}^\gamma(t)+\left(s-m^2\right)I_{13}^\gamma(s,t)\right)-2\left(s-m^2+\mpi^2\right)I_{B}^{\gamma\pi}(s)}{4\left(\lambda \left(s,m^2,\mpi^2\right)+s t\right)},\notag\\
I_{13}^{i(2)}(s,t)&=-\frac{1}{4N_{13}(s,t)}\Big\{t\left(4m^2-t\right)I_{A}^i(t)-t\left(2\left(s+m^2-\mpi^2\right)+\Deltax \right)I_{B}^{i\pi}(s)\notag\\
&+t\left(\left(s-m^2\right)\left(4m^2-t\right)-M_i^2(s-u)\right)I_{13}^i(s,t)\notag\\
&+2\delta_{13}^i(s) \left(\left(s+m^2-\mpi^2\right)t- \Deltax \left(2m^2-t\right)\right)\Big\}\notag\\
&\overset{\hspace{-25pt}\Deltax=0,\ \delta_{13}^i=0}{\hspace{-25pt}=}-\frac{1}{4\left(\lambda \left(s,m^2,\mpi^2\right)+s t\right)}\Big\{\left(4m^2-t\right)I_{A}^i(t)-2\left(s+m^2-\mpi^2\right)I_{B}^{i\pi}(s)\notag\\
&+\left(\left(s-m^2\right)\left(4m^2-t\right)-M_i^2(s-u)\right)I_{13}^i(s,t)\Big\},\notag\\
I_{13}^{\gamma(2)}(s,t)&=-\frac{1}{4N_{13}(s,t)}\Big\{t\left(4m^2-t\right)I_{A}^\gamma(t)-t\left(2\left(s+m^2-\mpi^2\right)+\Deltax \right)I_{B}^{\gamma\pi}(s)\notag\\
&+t\left(s-m^2\right)\left(4m^2-t\right)I_{13}^\gamma(s,t)+2\delta_{13}^\gamma(s) \left(\left(s+m^2-\mpi^2\right)t- \Deltax \left(2m^2-t\right)\right)\Big\}\notag\\
&\overset{\hspace{-25pt}\Deltax=0,\ \delta_{13}^\gamma=0}{\hspace{-25pt}=}-\frac{\left(4m^2-t\right)\left(I_{A}^\gamma(t)+\left(s-m^2\right)I_{13}^\gamma(s,t)\right)-2\left(s+m^2-\mpi^2\right)I_{B}^{\gamma\pi}(s)}{4\left(\lambda \left(s,m^2,\mpi^2\right)+s t\right)},\notag\\
I_{13}^{i(3)}(s,t)&=\frac{\Deltax}{4N_{13}(s,t)} \Big\{\left(2\left(s+m^2-\mpi^2\right)+\Deltax \right)I_{B}^{i\pi}(s)-\left(4m^2-t\right)I_{A}^i(t)\notag\\
&- \left(\left(4m^2-t\right)\left(s-m^2\right)-M_i^2\left(s-u+\mpi^2\right)\right)I_{13}^i(s,t)\Big\}\notag\\
&+\frac{\delta_{13}^i(s,t)}{4N_{13}(s,t)} \Big\{\left(4m^2-t\right)\left(4\mpi^2-t-2\Deltax \right)-(s-u)^2\notag\\
&-\Deltax \left(2\left(s+m^2-\mpi^2\right)+\Deltax \right)\Big\}
\overset{\Deltax=0,\ \delta_{13}^i=0}{=}0,\notag\\
I_{13}^{\gamma(3)}(s,t)&=\frac{\Deltax}{4N_{13}(s,t)} \Big\{\left(2\left(s+m^2-\mpi^2\right)+\Deltax \right)I_{B}^{\gamma\pi}(s)-\left(4m^2-t\right)I_{A}^\gamma(t)\notag\\
&- \left(4m^2-t\right)\left(s-m^2\right)I_{13}^\gamma(s,t)\Big\}+\frac{\delta_{13}^\gamma(s,t)}{4N_{13}(s,t)} \Big\{\left(4m^2-t\right)\left(4\mpi^2-t-2\Deltax \right)\notag\\
&-(s-u)^2-\Deltax \left(2\left(s+m^2-\mpi^2\right)+\Deltax \right)\Big\}\overset{\Deltax=0,\ \delta_{13}^\gamma=0}{=}0,\notag\\
N_{13}(s,t)&=t\left(\lambda \left(s,m^2,\mpi^2\right)+s t\right)+ t \Deltax \left(s+m^2-\mpi^2\right)+ m^2 \Deltax ^2,\notag\\
\delta_{13}^i(s)&=\left\{\begin{array}{ll}
\frac{1}{2}\left(I_{B}^{i\pi}(s)-I_{B}^{i\pi^0}(s)\right) \quad {\rm for} \, \pi^- p\rightarrow \pi^0 n, \\
0 \hspace{102pt} {\rm otherwise}. 
\end{array}\right. \label{box_s}
\end{align}
The corresponding decomposition for the $u$-channel reads
\begin{align}
I_{13}^{i\,\mu}(u,t)&=\frac{1}{i}\int_{\rm I} \frac{\diff^d k}{(2\pi)^d}\frac{k^\mu}{\left(M_i^2-k^2\right)\left(m^2-(p-k)^2\right)
\left(m^2-(\Lambda-k)^2\right)\left(m^2-(p'-k)^2\right)}\notag\\
&=Q^{\mu }I_{13}^{i(1)}(u,t)-(\Delta +2q)^{\mu }I_{13}^{i(2)}(u,t)-\Delta ^{\mu }I_{13}^{i(3)}(u,t),\label{box_u}
\end{align}
where the basis functions as defined in \eqref{box_s} may be used. The minus in front of $I_{13}^{i(2)}(u,t)$ is due to the replacement $q\leftrightarrow -q'$ to arrive at the $u$-channel, whereas the second minus is caused by our parametrization of the pion mass difference occurring in the kinematics of the charge exchange reaction: contracting \eqref{box_s} with $Q_\mu$, $(\Delta +2q)_{\mu }$, and $\Delta_{\mu }$, yields
\begin{align}
&\begin{pmatrix}
Q^2 & Q\cdot(\Delta+2q)& 0\\
Q\cdot (\Delta+2q) & (\Delta+2q)^2 & -\Deltax\\
0 & -\Deltax & t \\
\end{pmatrix}
\begin{pmatrix}
I_{13}^{i(1)}(s,t)\\
I_{13}^{i(2)}(s,t)\\
I_{13}^{i(3)}(s,t)
\end{pmatrix}\notag\\
&=\begin{pmatrix}
I_B^{i\pi}(s)-\delta_{13}^{i}(s)+M_i^2 I_{13}^i(s,t)\\
I_A^i(t)+\left(s-m^2\right)I_{13}^i(s,t)-I_B^{i\pi}(s)+\delta_{13}^{i}(s)\\
-\delta_{13}^{i}(s)
\end{pmatrix},
\end{align}
where we have used that $M_j=\mpi$ for $p_1=p$ in the $s$-channel. Supposing that the sign of $I_{13}^{i(3)}(u,t)$ is not changed as compared to $I_{13}^{i(3)}(s,t)$, we obtain
\begin{align}
&\begin{pmatrix}
Q^2 & -Q\cdot(\Delta+2q)& 0\\
-Q\cdot (\Delta+2q) & (\Delta+2q)^2 & \Deltax\\
0 & \Deltax & t \\
\end{pmatrix}
\begin{pmatrix}
I_{13}^{i(1)}(u,t)\\
I_{13}^{i(2)}(u,t)\\
I_{13}^{i(3)}(u,t)
\end{pmatrix}\notag\\
&=\begin{pmatrix}
I_B^{i\pi}(u)-\delta_{13}^{i}(u)+M_i^2 I_{13}^i(u,t)\\
I_A^i(t)+\left(u-m^2\right)I_{13}^i(u,t)-I_B^{i\pi}(u)+\delta_{13}^{i}(u)\\
\delta_{13}^{i}(u)
\end{pmatrix}.
\end{align}
Changing the sign of $I_{13}^{i(3)}(u,t)$ and multiplying the third equation by $(-1)$ ensures that both systems of equations are identical up to $s\leftrightarrow u$.

\subsubsection{1 meson, 1 photon}

\beq
V_{10}^\mu=\frac{1}{i}\int_{\rm I} \frac{\diff^d k}{(2\pi)^d}\frac{k^\mu}{\left(m_\gamma^2-k^2\right)\left(\mpi^2-(q-k)^2\right)}
=q^\mu V_{10}^{(1)}, \quad V_{10}^{(1)}=-\frac{\Delta_\pi}{2\mpi^2}.
\eeq

\subsubsection{2 mesons, 1 photon}

\begin{align}
V_{20}^\mu(t)&=\frac{1}{i}\int_{\rm I} \frac{\diff^d k}{(2\pi)^d}\frac{k^\mu}{\left(m_\gamma^2-k^2\right)\left(\mpi^2-(q-k)^2\right)\left(\mpi^2-(q'-k)^2\right)}
=(\Delta +2q)^{\mu }V_{20}^{(1)}(t),\notag\\
V_{20}^{(1)}(t)&=\frac{V_{10}-J_{\pi\pi}(t)}{4\mpi^2-t}.
\end{align}

\subsubsection{1 meson, 1 nucleon, 1 photon}

\begin{align}
&\frac{1}{i}\int_{\rm I} \frac{\diff^d k}{(2\pi)^d}\frac{k^\mu}{\left(m_\gamma^2-k^2\right)\left(m^2-(p-k)^2\right)\left(\mpi^2-(q+k)^2\right)}
=\Sigma ^{\mu }V_{11}^{(1)}(s)+p^{\mu }V_{11}^{(2)}(s),\notag\\
&\frac{1}{i}\int_{\rm I} \frac{\diff^d k}{(2\pi)^d}\frac{k^\mu}{\left(m_\gamma^2-k^2\right)\left(m^2-(p'-k)^2\right)\left(\mpi^2-(q'+k)^2\right)}
=\Sigma ^{\mu }V_{11}^{(1)}(s)+p'^{\mu }V_{11}^{(2)}(s),\notag\\
&\frac{1}{i}\int_{\rm I} \frac{\diff^d k}{(2\pi)^d}\frac{k^\mu}{\left(m_\gamma^2-k^2\right)\left(m^2-(p-k)^2\right)\left(\mpi^2-(q'-k)^2\right)}
=\Lambda ^{\mu }V_{11}^{(1)}(u)+p^{\mu }V_{11}^{(2)}(u),\notag\\
&\frac{1}{i}\int_{\rm I} \frac{\diff^d k}{(2\pi)^d}\frac{k^\mu}{\left(m_\gamma^2-k^2\right)\left(m^2-(p'-k)^2\right)\left(\mpi^2-(q-k)^2\right)}
=\Lambda ^{\mu }V_{11}^{(1)}(u)+p'^{\mu }V_{11}^{(2)}(u),\notag\\
V_{11}^{(1)}(s)&=\frac{\left(s-m^2-\mpi^2\right)V_{10}-\left(s+m^2-\mpi^2\right)I_\pi(s)}{\lambda \left(s,m^2,\mpi^2\right)},\notag\\
V_{11}^{(2)}(s)&=\frac{2s I_\pi(s)-\left(s-m^2+\mpi^2\right)V_{10}}{\lambda \left(s,m^2,\mpi^2\right)},\\
&\frac{1}{i}\int_{\rm I} \frac{\diff^d k}{(2\pi)^d}\frac{1}{\left(m_\gamma^2-k^2\right)\left(m^2-(\Sigma-k)^2\right)\left(\mpi^2-(q-k)^2\right)}=\Sigma ^{\mu }A_{11}^{(1)}(s)+p^{\mu }A_{11}^{(2)}(s),\notag\\
&\frac{1}{i}\int_{\rm I} \frac{\diff^d k}{(2\pi)^d}\frac{1}{\left(m_\gamma^2-k^2\right)\left(m^2-(\Sigma-k)^2\right)\left(\mpi^2-(q'-k)^2\right)}=\Sigma ^{\mu }A_{11}^{(1)}(s)+p'^{\mu }A_{11}^{(2)}(s),\notag\\
&\frac{1}{i}\int_{\rm I} \frac{\diff^d k}{(2\pi)^d}\frac{1}{\left(m_\gamma^2-k^2\right)\left(m^2-(\Lambda-k)^2\right)\left(\mpi^2-(q'+k)^2\right)}=\Lambda ^{\mu }A_{11}^{(1)}(u)+p^{\mu }A_{11}^{(2)}(u),\notag\\
&\frac{1}{i}\int_{\rm I} \frac{\diff^d k}{(2\pi)^d}\frac{1}{\left(m_\gamma^2-k^2\right)\left(m^2-(\Lambda-k)^2\right)\left(\mpi^2-(q+k)^2\right)}=\Lambda ^{\mu }A_{11}^{(1)}(u)+p'^{\mu }A_{11}^{(2)}(u),\notag\\
A_{11}^{(1)}(s)&=\frac{\left(s-m^2-\mpi^2\right)\left(V_{10}+\left(s-m^2\right)A_{11}(s)\right)-\left(s+m^2-\mpi^2\right)I_{\gamma }(s)+2m^2 I_\pi\left(m^2\right)}{\lambda \left(s,m^2,\mpi^2\right)},\notag\\
A_{11}^{(2)}(s)&=-\frac{\left(s-m^2+\mpi^2\right)\left(V_{10}+\left(s-m^2\right)A_{11}(s)\right)+\left(s+m^2-\mpi^2\right)I_\pi\left(m^2\right)-2s I_{\gamma }(s)}{\lambda \left(s,m^2,\mpi^2\right)}.
\end{align}

\subsubsection{2 mesons, 1 nucleon, 1 photon}

\begin{align}
A_{21}^\mu(s,t)&=\frac{1}{i}\int_{\rm I} \frac{\diff^d k}{(2\pi)^d}\frac{k^\mu}{\left(m_\gamma^2-k^2\right)\left(m^2-(\Sigma-k)^2\right)\left(\mpi^2-(q-k)^2\right)\left(\mpi^2-(q'-k)^2\right)}\notag\\
&=Q^{\mu }A_{21}^{(1)}(s,t)+(\Delta +2q)^{\mu }A_{21}^{(2)}(s,t),\notag\\
A_{21}^\mu(u,t)&=\frac{1}{i}\int_{\rm I} \frac{\diff^d k}{(2\pi)^d}\frac{k^\mu}{\left(m_\gamma^2-k^2\right)\left(m^2-(\Lambda-k)^2\right)\left(\mpi^2-(q+k)^2\right)\left(\mpi^2-(q'+k)^2\right)}\notag\\
&=Q^{\mu }A_{21}^{(1)}(u,t)-(\Delta +2q)^{\mu }A_{21}^{(2)}(u,t),\notag\\
A_{21}^{(1)}(s,t)&=\frac{1}{4\left(\lambda \left(s,m^2,\mpi^2\right)+s t\right)}\Big\{2\left(s-m^2+\mpi^2\right)A_{11}(s)-(s-u)I_{21}^{\pi\pi}(t)\notag\\
&-\left(4\mpi^2-t\right)\left(\left(s-m^2\right)A_{21}(s,t)+V_{20}(t)\right)\Big\},\notag\\
A_{21}^{(2)}(s,t)&=\frac{1}{4\left(\lambda \left(s,m^2,\mpi^2\right)+s t\right)}\Big\{\left(4m^2-t\right)I_{21}^{\pi\pi}(t)-2\left(s+m^2-\mpi^2\right)A_{11}(s)\notag\\
&+(s-u)\left(\left(s-m^2\right)A_{21}(s,t)+V_{20}(t)\right)\Big\}.
\end{align}

\subsubsection{1 meson, 2 nucleons, 1 photon}

The tensor decomposition for the $s$-channel may be chosen as
\begin{align}
&\frac{1}{i}\int_{\rm I} \frac{\diff^d k}{(2\pi)^d}\frac{k^\mu}{\left(m_\gamma^2-k^2\right)\left(m^2-(p-k)^2\right)\left(m^2-(\Sigma-k)^2\right)\left(\mpi^2-(q'-k)^2\right)}\notag\\
&=Q^{\mu }A_{12}^{(1)}(s,t)+(\Delta +2q)^{\mu }A_{12}^{(2)}(s,t)+\Delta ^{\mu }A_{12}^{(3)}(s,t),\notag\\
&\frac{1}{i}\int_{\rm I} \frac{\diff^d k}{(2\pi)^d}\frac{k^\mu}{\left(m_\gamma^2-k^2\right)\left(m^2-(p'-k)^2\right)\left(m^2-(\Sigma-k)^2\right)\left(\mpi^2-(q-k)^2\right)}\notag\\
&=Q^{\mu }A_{12}^{(1)}(s,t)+(\Delta +2q)^{\mu }A_{12}^{(2)}(s,t)-\Delta ^{\mu }A_{12}^{(3)}(s,t),
\end{align}
which for the $u$-channel becomes
\begin{align}
&\frac{1}{i}\int_{\rm I} \frac{\diff^d k}{(2\pi)^d}\frac{k^\mu}{\left(m_\gamma^2-k^2\right)\left(m^2-(p-k)^2\right)\left(m^2-(\Lambda-k)^2\right)\left(\mpi^2-(q+k)^2\right)}\notag\\
&=Q^{\mu }A_{12}^{(1)}(u,t)-(\Delta +2q)^{\mu }A_{12}^{(2)}(u,t)+\Delta ^{\mu }A_{12}^{(3)}(u,t),\notag\\
&\frac{1}{i}\int_{\rm I} \frac{\diff^d k}{(2\pi)^d}\frac{k^\mu}{\left(m_\gamma^2-k^2\right)\left(m^2-(p'-k)^2\right)\left(m^2-(\Lambda-k)^2\right)\left(\mpi^2-(q'+k)^2\right)}\notag\\
&=Q^{\mu }A_{12}^{(1)}(u,t)-(\Delta +2q)^{\mu }A_{12}^{(2)}(u,t)-\Delta ^{\mu }A_{12}^{(3)}(u,t)\label{2n_box_u}.
\end{align}
Since only the first version of \eqref{2n_box_u} contributes to the charge exchange reaction, $\Deltax$ is incorporated according to the prescription in Sect.~\ref{sec:1m2n1p}. The pion mass which has to be inserted into the definition of $I_{B}^{\gamma j}(s)$ is $M_j=\sqrt{\mpi^2-\Deltax}$. In the present context, we thus deviate from the original definition in Sect.~\ref{sec:1m2n}.
\begin{align}
A_{12}^{(1)}(s,t)&=\frac{1}{4N_{12}(s,t)}\Big\{ \left(t \left(u-m^2+\mpi^2\right)+\Deltax \left(s-m^2-\mpi^2\right)\right) I_{B}^{\pi j}(u)\notag\\
&+ \left(t\left(s-m^2+\mpi^2\right)+\Deltax\left(u-m^2-\mpi^2\right)\right) I_{B}^{\gamma j}(s)\notag\\
&-\left( t\left(s-m^2+\mpi^2\right) +\Deltax  \left(s-m^2-\mpi^2\right)\right) A_{11}(s)\notag\\
&-\left(t \left(u-m^2+\mpi^2\right)+\Deltax \left(u-m^2-\mpi^2\right)\right) \left(\left(s-m^2\right) A_{12}(s,t)+V_{11}(u)\right)\Big\}\notag\\
&\overset{\hspace{-10pt}\Deltax=0}{\hspace{-10pt}=}\frac{1}{4\left(\lambda \left(s,m^2,\mpi^2\right)+s t\right)}\Big\{  \left(u-m^2+\mpi^2\right) I_{B}^{\pi\pi}(u)+ \left(s-m^2+\mpi^2\right) I_{B}^{\gamma \pi}(s)\notag\\
&-\left(s-m^2+\mpi^2\right) A_{11}(s)-\left(u-m^2+\mpi^2\right) \left(\left(s-m^2\right) A_{12}(s,t)+V_{11}(u)\right)\Big\},\notag\\
A_{12}^{(2)}(s,t)&=\frac{1}{4N_{12}(s,t)}\Big\{\left(t \left(u+m^2-\mpi^2\right)+\Deltax  \left(t-2m^2\right)\right) I_{B}^{\pi j}(u)\notag\\
&-\left(t\left(s+m^2-\mpi^2\right)+\Deltax  \left(t-2m^2\right)\right) I_{B}^{\gamma j}(s)+\left(t \left(s+m^2-\mpi^2\right)+2m^2\Deltax \right) A_{11}(s)\notag\\
&-\left(t \left(u+m^2-\mpi^2\right)+2m^2\Deltax \right) \left(\left(s-m^2\right) A_{12}(s,t)+V_{11}(u)\right)\Big\}\notag\\
&\overset{\hspace{-10pt}\Deltax=0}{\hspace{-10pt}=}\frac{1}{4\left(\lambda \left(s,m^2,\mpi^2\right)+s t\right)}\Big\{ \left(u+m^2-\mpi^2\right)I_{B}^{\pi\pi}(u)-\left(s+m^2-\mpi^2\right)I_{B}^{\gamma \pi}(s)\notag\\
&+\left(s+m^2-\mpi^2\right) A_{11}(s)-\left(u+m^2-\mpi^2\right) \left(\left(s-m^2\right) A_{12}(s,t)+V_{11}(u)\right)\Big\},\notag\\
A_{12}^{(3)}(s,t)&=-\frac{1}{2 t}\Big\{I_{B}^{\pi j}(u)-I_{B}^{\gamma j}(s)-A_{11}(s)+\left(s-m^2\right) A_{12}(s,t)+V_{11}(u)\Big\}-\frac{\Deltax}{t}A_{12}^{(2)}(s,t)\notag\\
&\overset{\hspace{-10pt}\Deltax=0}{\hspace{-10pt}=}-\frac{1}{2 t}\Big\{I_{B}^{\pi\pi}(u)-I_{B}^{\gamma \pi}(s)-A_{11}(s)+\left(s-m^2\right) A_{12}(s,t)+V_{11}(u)\Big\},\notag\\
N_{12}(s,t)&=t \left(\lambda \left(s,m^2,\mpi^2\right)+s t\right) +\Deltax \left(\left(s+m^2-\mpi^2\right)t+m^2\Deltax \right).
\end{align}

\subsubsection{1 meson, 1 nucleon, 2 photons}

\begin{align}
P_{11}^\mu(s,t)&=\frac{1}{i}\int_{\rm I} \frac{\diff^d k}{(2\pi)^d}\frac{k^\mu}{\left(m_\gamma^2-k^2\right)\left(m_\gamma^2-(k-\Delta)^2\right)\left(\mpi^2-(q+k)^2\right)\left(m^2-(p-k)^2\right)}\notag\\
&=-q^{\mu }P_{11}(s,t)+Q^{\mu }P_{11}^{(1)}(s,t)+(\Delta +2q)^{\mu }P_{11}^{(2)}(s,t),\notag\\
P_{11}^\mu(u,t)&=\frac{1}{i}\int_{\rm I} \frac{\diff^d k}{(2\pi)^d}\frac{k^\mu}{\left(m_\gamma^2-k^2\right)\left(m_\gamma^2-(k-\Delta)^2\right)\left(\mpi^2-(q'-k)^2\right)\left(m^2-(p-k)^2\right)}\notag\\
&=q'^{\mu }P_{11}(u,t)+Q^{\mu }P_{11}^{(1)}(u,t)-(\Delta +2q)^{\mu }P_{11}^{(2)}(u,t),\notag\\
P_{11}^{(1)}(s,t)&=-\frac{1}{4\left(\lambda \left(s,m^2,\mpi^2\right)+s t\right)}\Big\{(s-u)I_{21}^{\gamma\gamma} (t)+\left(4\mpi^2-t\right)P_{10}(t)\notag\\
&-\left(s-m^2+\mpi^2\right)\left(t P_{11}(s,t)+2V_{11}(s)\right)\Big\},\notag\\
P_{11}^{(2)}(s,t)&=\frac{P_{11}(s,t)}{2}+\frac{1}{4\left(\lambda \left(s,m^2,\mpi^2\right)+s t\right)}\Big\{(4m^2-t)I_{21}^{\gamma \gamma }(t)+(s-u)P_{10}(t)\notag\\
&-\left(s+m^2-\mpi^2\right)\left(t P_{11}(s,t)+2V_{11}(s)\right)\Big\}.
\end{align}

\subsection{Bremsstrahlung}
\label{app:brems_func}

The following representation of the integrals appearing in \eqref{brems_ampl_result} is valid for $s\geq(m+\mpi)^2$:
\begin{align}
g_{p_1p_2}&=\int\limits_0^1\frac{\diff x}{\tilde{p}^2}\frac{E_{\tilde{p}}}{|\mathbf{\tilde{p}}|}\log\frac{E_{\tilde{p}}+|\mathbf{\tilde{p}}|}{E_{\tilde{p}}-|\mathbf{\tilde{p}}|},\quad
\tilde{p}=x p_1+(1-x) p_2=\left(E_{\tilde{p}},\mathbf{\tilde{p}}\right),\notag\\
g_{pp}(s)&=\frac{2}{m^2}\frac{s+m^2-\mpi^2}{\lambda^{1/2}\big(s,m^2,\mpi^2\big)}\log\frac{s+m^2-\mpi^2+\lambda^{1/2}\big(s,m^2,\mpi^2\big)}{2m\sqrt{s}},\notag\\
g_{qq}(s)&=\frac{2}{\mpi^2}\frac{s-m^2+\mpi^2}{\lambda^{1/2}\big(s,m^2,\mpi^2\big)}\log\frac{s-m^2+\mpi^2+\lambda^{1/2}\big(s,m^2,\mpi^2\big)}{2\mpi\sqrt{s}},\notag\\
g_{pq}(s)&=2\int\limits_0^1\frac{\text{d}x}{s_{pq}(x)}\frac{s+(2x-1)\big(m^2-\mpi^2\big)}{\sqrt{\left((s+(2x-1)\big(m^2-\mpi^2\big)\right)^2-4s s_{pq}(x)}}\notag\\
&\times\log\frac{s+(2x-1)\big(m^2-\mpi^2\big)+\sqrt{\left(s+(2x-1)\big(m^2-\mpi^2\big)\right)^2-4 s s_{pq}(s)}}{2\sqrt{s s_{pq}(x)}},\notag\\
s_{pq}(x)&=x(1-x)s+(2x-1)\big(x m^2-(1-x)\mpi^2\big)\notag\\
&=-x(1-x)(u+t)+x m^2+(1-x)\mpi^2,\notag\\
g_{pp'}(s,t)&=2\int\limits_0^1\frac{\text{d}x}{m^2- x(1-x)t}\frac{s+m^2-\mpi^2}{\sqrt{s^2_{pp'}(x)}}
\log\frac{s+m^2-\mpi^2+\sqrt{s^2_{pp'}(x)}}{2\sqrt{s}\sqrt{m^2- x(1-x)t}},\notag\\
g_{qq'}(s,t)&=2\int\limits_0^1\frac{\text{d}x}{\mpi^2- x(1-x)t}\frac{s-m^2+\mpi^2}{\sqrt{s^2_{qq'}(x)}}
\log\frac{s-m^2+\mpi^2+\sqrt{s^2_{qq'}(x)}}{2\sqrt{s}\sqrt{\mpi^2- x(1-x)t}},\notag\\
s^2_{pp'}&=s^2_{qq'}=\lambda\big(s,m^2,\mpi^2\big)+4s t x(1-x),\notag\\
g_{pq'}(s,t)&=2\int\limits_0^1\frac{\text{d}x}{s_{pq'}(x)}\frac{s+(2x-1)\big(m^2-\mpi^2\big)}{\sqrt{\left(s+(2x-1)\big(m^2-\mpi^2\big)\right)^2-4s s_{pq'}(x)}}\notag\\
&\times\log\frac{s+(2x-1)\big(m^2-\mpi^2\big)+\sqrt{\left(s+(2x-1)\big(m^2-\mpi^2\big)\right)^2-4 s s_{pq'}(x)}}{2\sqrt{s s_{pq'}(x)}},\notag\\
s_{pq'}(x)&=-x(1-x)u+x m^2+(1-x)\mpi^2\notag\\
&=x(1-x)(s+t)+(2x-1)\big(x m^2-(1-x)\mpi^2\big).
\end{align}

\section{Contributions of individual diagrams}
\def\theequation{\Alph{section}.\arabic{equation}}
\setcounter{equation}{0}
\label{app:ind_diagrams}

\subsection{Strong diagrams}

\subsubsection*{$(s_1)$}

\begin{align}
A_{s_1}^{i}(s)&=4M_i^2\frac{s+m^2}{s-m^2}I_i(s)-\left(3s+m^2\right)I_i^{(1)}(s),\notag\\
B_{s_1}^{i}(s)&=\frac{s^2+10 s m^2 +5 m^4}{\left(s-m^2\right)^2}M_i^2I_i(s)-\frac{s^2+6s m^2 +m^4}{s-m^2}I_i^{(1)}(s),\\
A_{s_1}^{\pi^- p}(s)&=\frac{m g^4}{8F^4}\left(2A_{s_1}^{\pi}(s)+A_{s_1}^{\npi}(s)\right),\quad 
B_{s_1}^{\pi^- p}(s)=-\frac{g^4}{8F^4}\left(2B_{s_1}^{\pi}(s)+B_{s_1}^{\npi}(s)\right),\notag\\
A_{s_1}^{\pi^+ p}(s,t)&=\frac{m g^4}{8F^4}\left(2A_{s_1}^{\pi}(u)+A_{s_1}^{\npi}(u)\right),\quad 
B_{s_1}^{\pi^+ p}(s,t)=\frac{ g^4}{8F^4}\left(2B_{s_1}^{\pi}(u)+B_{s_1}^{\npi}(u)\right),\notag\\
A_{s_1}^{{\rm cex}}(s,t)&=\frac{\sqrt{2}\,m g^4}{16F^4}\left(2A_{s_1}^{\pi}(u)+A_{s_1}^{\npi}(u)-2A_{s_1}^{\pi}(s)-A_{s_1}^{\npi}(s)\right),\notag\\
B_{s_1}^{{\rm cex}}(s,t)&=
\frac{\sqrt{2}\,g^4}{8F^4}\left(2B_{s_1}^{\pi}(u)+B_{s_1}^{\npi}(u)
+2B_{s_1}^{\pi}(s)+B_{s_1}^{\npi}(s)\right).\notag
\end{align}

\subsubsection*{$(s_2)$}

\begin{align}
A_{s_2}^{ij}(s,t)&=4\left(s+m^2\right)I_i(s)-\left(3s+m^2\right)I_i^{(1)}(s)+4m^2I_i^{(1)}\left(m^2 \right)\notag\\
&+8m^2\Big[2M_i^2I_{B}^{ij}(s)-4\left(s+m^2\right)I_{B}^{ij(1)}(s)-2\left(s-m^2\right)I_{B}^{ij(2)}(s)\notag\\
&-M_i^2I_{A}^i(t)+4m^2 I_{A}^{i(1)}(t)-\left(s-u\right)I_{A}^{i(3)}(t)\Big]+32 m^4\left(s-m^2\right)I_{13}^{i(1)}(s,t),\notag\\
B_{s_2}^{ij}(s,t)&=\left(s+3m^2\right)I_{i}(s)+4m^2I_{i}^{(1)}\left(m^2\right)-\left(s+m^2\right)I_{i}^{(1)}(s)
\notag\\
&+4m^2\Big[M_i^2I_{A}^{i}(t)-2 I_{A}^{i(2)}(t)+2M_i^2 I_{B}^{ij}(s)-2\left(s+3m^2\right)I_B^{ij(1)}(s)\notag\\ 
&-2\left(s-m^2\right)I_B^{ij(2)}(s)\Big]-16m^4\left(M_i^2I_{13}^i(s,t)-2\left(s-m^2\right)I_{13}^{i(2)}(s,t)\right),\notag\\
A_{s_2}^{\pi^- p}(s,t)&=\frac{m g^4}{8F^4}A_{s_2}^{\npi\pi}(s,t)+\frac{m g^4}{4F^4}A_{s_2}^{\pi\pi}(u,t),\quad
B_{s_2}^{\pi^- p}(s,t)=\frac{g^4}{8F^4}B_{s_2}^{\npi\pi}(s,t)-\frac{g^4}{4F^4}B_{s_2}^{\pi\pi}(u,t),\notag\\
A_{s_2}^{\pi^+ p}(s,t)&=\frac{m g^4}{4F^4}A_{s_2}^{\pi\pi}(s,t)+\frac{m g^4}{8F^4}A_{s_2}^{\npi\pi}(u,t),\quad
B_{s_2}^{\pi^+ p}(s,t)=\frac{g^4}{4F^4}B_{s_2}^{\pi\pi}(s,t)-\frac{g^4}{8F^4}B_{s_2}^{\npi\pi}(u,t),\notag\\
A_{s_2}^{\rm cex}(s,t)&=\frac{\sqrt{2} \, m g^4}{32F^4}\left(A_{s_2}^{\npi\pi}(s,t)+A_{s_2}^{\npi\npi}(s,t)\right)-\frac{\sqrt{2} \, m g^4}{32F^4}\left(A_{s_2}^{\npi\pi}(u,t)+A_{s_2}^{\npi\npi}(u,t)\right),\notag\\
B_{s_2}^{\rm cex}(s,t)&=\frac{\sqrt{2} \, g^4}{32F^4}\left(B_{s_2}^{\npi\pi}(s,t)+B_{s_2}^{\npi\npi}(s,t)\right)+\frac{\sqrt{2}\,  g^4}{32F^4}\left(B_{s_2}^{\npi\pi}(u,t)+B_{s_2}^{\npi\npi}(u,t)\right).
\end{align}

\subsubsection*{$(s_3)$}

\begin{align}
A_{s_3}^{ij}(s)&=\left(s-m^2\right)I_i^{(1)}(s)-2 \Delta_i-8m^2\left(M_i^2 I_{B}^{ij}(s)-\left(s-m^2\right)I_{B}^{ij(2)}(s)\right),\notag\\
B_{s_3}^{ij}(s)&=-M_i^2I_i(s)+\left(s-m^2\right)\left(I_i^{(1)}(s)-4m^2 I_{B}^{ij(1)}(s)\right)\notag\\
&+\frac{4m^2}{s-m^2}\left(\Delta_i+\left(s+3m^2\right)\left(M_i^2 I_{B}^{ij}(s)-\left(s-m^2\right)I_B^{ij(2)}(s)\right)\right),\notag\\
A_{s_3}^{\pi^- p}(s)&=\frac{m g^4}{4F^4}A_{s_3}^{\npi\pi}(s),\quad B_{s_3}^{\pi^- p}(s)=\frac{g^4}{4F^4}B_{s_3}^{\npi\pi}(s),\notag\\
A_{s_3}^{\pi^+ p}(s,t)&=\frac{m g^4}{4F^4}A_{s_3}^{\npi\pi}(u),\quad B_{s_3}^{\pi^+ p}(s,t)=-\frac{g^4}{4F^4}B_{s_3}^{\npi\pi}(u),\notag\\
A_{s_3}^{\rm cex}(s,t)&=-\frac{\sqrt{2} \, m g^4}{16F^4}\left(2A_{s_3}^{\pi\npi}(s)+A_{s_3}^{\npi\pi}(s)-A_{s_3}^{\npi\npi}(s)\right)\notag\\
&+\frac{\sqrt{2} \, m g^4}{16F^4}\left(2A_{s_3}^{\pi\npi}(u)+A_{s_3}^{\npi\pi}(u)-A_{s_3}^{\npi\npi}(u)\right),\notag\\
B_{s_3}^{\rm cex}(s,t)&=-\frac{\sqrt{2} \, g^4}{16F^4}\left(2B_{s_3}^{\pi\npi}(s)+B_{s_3}^{\npi\pi}(s)-B_{s_3}^{\npi\npi}(s)\right)\notag\\
&-\frac{\sqrt{2} \, g^4}{16F^4}\left(2B_{s_3}^{\pi\npi}(u)+B_{s_3}^{\npi\pi}(u)-B_{s_3}^{\npi\npi}(u)\right).
\end{align}

\subsubsection*{$(s_4)$}

\begin{align}
A_{s_4}^{i}(s)&=-2M_i^2 I_i(s)+\left(s-m^2\right)I_{i}^{(1)}(s),\notag\\
B_{s_4}^{i}(s)&=-\left(s+m^2\right)I_{i}^{(1)}(s)+M_i^2\frac{s+3m^2}{s-m^2}I_i(s),\notag\\
A_{s_4}^{\pi^- p}(s)&=\frac{m g^2}{2F^4}\left(A_{s_4}^{\pi}(s)+A_{s_4}^{\npi}(s)\right),\quad 
B_{s_4}^{\pi^- p}(s)=\frac{g^2}{2F^4}\left(B_{s_4}^{\pi}(s)+B_{s_4}^{\npi}(s)\right),\notag\\
A_{s_4}^{\pi^+ p}(s,t)&=\frac{m g^2}{2F^4}\left(A_{s_4}^{\pi}(u)+A_{s_4}^{\npi}(u)\right),\quad 
B_{s_4}^{\pi^+ p}(s,t)=-\frac{g^2}{2F^4}\left(B_{s_4}^{\pi}(u)+B_{s_4}^{\npi}(u)\right),\notag\\
A_{s_4}^{\rm cex}(s,t)&=-\frac{\sqrt{2} \, m g^2}{8F^4}\left(3A_{s_4}^{\pi}(s)+A_{s_4}^{\npi}(s)\right)+\frac{\sqrt{2} \, m g^2}{8F^4}\left(3A_{s_4}^{\pi}(u)+A_{s_4}^{\npi}(u)\right),\notag\\
B_{s_4}^{\rm cex}(s,t)&=-\frac{\sqrt{2} \, g^2}{8F^4}\left(3B_{s_4}^{\pi}(s)+B_{s_4}^{\npi}(s)\right)-\frac{\sqrt{2} \, g^2}{8F^4}\left(3B_{s_4}^{\pi}(u)+B_{s_4}^{\npi}(u)\right).
\end{align}

\subsubsection*{$(s_5)$}

\begin{align}
A_{s_5}^{i}&=-\frac{m g^2M_i^2}{F^4} I_i\left(m^2 \right),\quad 
B_{s_5}^{i}(s)=\frac{g^2M_i^2}{4F^4}\frac{s+7m^2}{s-m^2}I_i\left(m^2\right),\notag\\
A_{s_5}^{\pi^- p}&=A_{s_5}^{\pi^+ p}=A_{s_5}^{\pi}+A_{s_5}^{\npi},\quad B_{s_5}^{\pi^- p}(s)=B_{s_5}^{\pi}(s)+B_{s_5}^{\npi}(s),
\quad B_{s_5}^{\pi^+ p}(s,t)=-B_{s_5}^{\pi}(u)-B_{s_5}^{\npi}(u),\notag\\
A_{s_5}^{\rm cex}&=0,\quad B_{s_5}^{\rm cex}(s,t)=-\frac{\sqrt{2}}{4}\left(3B_{s_5}^{\pi}(s)+B_{s_5}^{\npi}(s)+3B_{s_5}^{\pi}(u)+B_{s_5}^{\npi}(u)\right).
\end{align}

\subsubsection*{$(s_6)$}

\begin{align}
A_{s_6}^{ij}(s)&=\left(s-m^2\right)\left(-2I_i(s)+I_i^{(1)}(s)+8m^2I_{B}^{ij(1)}(s)\right),\notag\\
B_{s_6}^{ij}(s)&=-M_i^2I_i\left(m^2\right)-2\Delta_i-8m^2M_i^2I_{B}^{ij}(s)\notag\\
&+2\left(s-m^2\right)\left(I_{i}^{(1)}(s)-I_i(s)
+4m^2\left(I_{B}^{ij(1)}(s)+I_{B}^{ij(2)}(s)\right)\right),\notag\\
A_{s_6}^{\pi^- p}(s,t)&=\frac{m g^2}{2F^4}\left(A_{s_6}^{\npi\pi}(s)+A_{s_6}^{\pi\pi}(u)\right),
\quad B_{s_6}^{\pi^- p}(s,t)=\frac{g^2}{4F^4}\left(B_{s_6}^{\npi\pi}(s)-B_{s_6}^{\pi\pi}(u)\right),\notag\\
A_{s_6}^{\pi^+ p}(s,t)&=\frac{m g^2}{2F^4}\left(A_{s_6}^{\pi\pi}(s)+A_{s_6}^{\npi\pi}(u)\right),
\quad B_{s_6}^{\pi^+ p}(s,t)=\frac{g^2}{4F^4}\left(B_{s_6}^{\pi\pi}(s)-B_{s_6}^{\npi\pi}(u)\right),\notag\\
A_{s_6}^{\rm cex}(s,t)&=\frac{\sqrt{2} \, m g^2}{8F^4}\left(-A_{s_6}^{\pi\npi}(s)+A_{s_6}^{\npi\npi}(s)+A_{s_6}^{\pi\npi}(u)-A_{s_6}^{\npi\npi}(u)\right),\notag\\
B_{s_6}^{\rm cex}(s,t)&=\frac{\sqrt{2} \, g^2}{16F^4}\left(-B_{s_6}^{\pi\npi}(s)+B_{s_6}^{\npi\npi}(s)-B_{s_6}^{\pi\npi}(u)+B_{s_6}^{\npi\npi}(u)\right).
\end{align}

\subsubsection*{$(s_7)$}

\begin{align}
A_{s_7}^{i}(s,t)&=\frac{m^3g^2}{F^4}\left(s-u\right)I_{A}^{i(3)}(t),\notag\\ 
B_{s_7}^{i}(s,t)&=\frac{g^2}{8F^4}\left(\Delta_i-4m^2\left(I_{i}^{(1)}\left(m^2\right)+M_i^2 I_{A}^i(t)-2I_{A}^{i(2)}(t)\right)\right),\notag\\
A_{s_7}^{\pi^- p}(s,t)&=-A_{s_7}^{\pi^+ p}(s,t)=A_{s_7}^{\npi}(s,t)-2A_{s_7}^{\pi}(s,t),\notag\\ 
B_{s_7}^{\pi^- p}(s,t)&=-B_{s_7}^{\pi^+ p}(s,t)=B_{s_7}^{\npi}(s,t)-2B_{s_7}^{\pi}(s,t),\notag\\
A_{s_7}^{\rm cex}(s,t)&=\sqrt{2}A_{s_7}^{\npi}(s,t),\quad B_{s_7}^{\rm cex}(s,t)=\sqrt{2}B_{s_7}^{\npi}(s,t).
\end{align}

\subsubsection*{$(s_8)$}

\begin{align}
A_{s_8}^{\pi^- p}&=A_{s_8}^{\pi^+ p}=A_{s_8}^{\rm cex}=0,\notag\\
B_{s_8}^{\pi^- p}(t)&=\frac{t}{F^4}J_{\pi\pi}^{(1)}(t),\quad B_{s_8}^{\pi^+ p}(t)=-\frac{t}{F^4}J_{\pi\pi}^{(1)}(t),
\quad B_{s_8}^{\rm cex}(t)=-\frac{\sqrt{2}\,t}{F^4}J_{\pi\npi}^{(1)}(t).
\end{align}

\subsubsection*{$(s_9)$}

\begin{align}
A_{s_9}^{i}(s)&=\frac{m\left(s-m^2\right)}{4F^4}I_{i}^{(1)}(s),\quad
B_{s_9}^{i}(s)=\frac{1}{16F^4}\left(-4M_i^2I_i(s)-\Delta_i+4\left(s-m^2\right)I_{i}^{(1)}(s)\right),\notag\\
A_{s_9}^{\pi^- p}(s,t)&=2A_{s_9}^{\npi}(s)+A_{s_9}^{\pi}(s)+A_{s_9}^{\pi}(u),\quad
B_{s_9}^{\pi^- p}(s,t)=2B_{s_9}^{\npi}(s)+B_{s_9}^{\pi}(s)-B_{s_9}^{\pi}(u),\notag\\
A_{s_9}^{\pi^+ p}(s,t)&=A_{s_9}^{\pi}(s)+2A_{s_9}^{\npi}(u)+A_{s_9}^{\pi}(u),\quad
B_{s_9}^{\pi^+ p}(s,t)=B_{s_9}^{\pi}(s)-2B_{s_9}^{\npi}(u)-B_{s_9}^{\pi}(u),\notag\\
A_{s_9}^{\rm cex}(s,t)&=-\sqrt{2}\left(A_{s_9}^{\pi}(s)-A_{s_9}^{\pi}(u)\right),
\quad B_{s_9}^{\rm cex}(s,t)=-\sqrt{2}\left(B_{s_9}^{\pi}(s)+B_{s_9}^{\pi}(u)\right).
\end{align}

\subsubsection*{$(s_{10})$}

\begin{align}
A_{s_{10}}^{\pi^- p}&=A_{s_{10}}^{\pi^+ p}=\frac{m g^2 }{F^4}\Delta_\pi,\quad A_{s_{10}}^{\rm cex}=0,\quad
B_{s_{10}}^i(s)=\frac{g^2}{8F^4} \Delta _i \frac{s+3m^2}{s-m^2},\notag\\
B_{s_{10}}^{\pi^- p}(s)&=-4B_{s_{10}}^\pi(s),\quad B_{s_{10}}^{\pi^+ p}(s,t)=4B_{s_{10}}^\pi(u),\notag\\
B_{s_{10}}^{\rm cex}(s,t)&=\sqrt{2}\left(B_{s_{10}}^\pi(s)+B_{s_{10}}^\npi(s)+B_{s_{10}}^\pi(u)+B_{s_{10}}^\npi(u)\right).
\end{align}

\subsubsection*{$(s_{11})$}

\begin{align}
A_{s_{11}}^{\pi^- p}&=A_{s_{11}}^{\pi^+ p}=-\frac{m g^2\mpi^2}{F^4} I_\pi\left(m^2\right),\quad A_{s_{11}}^{\rm cex}=0,\notag\\
B_{s_{11}}^{\pi^- p}&=-B_{s_{11}}^{\pi^+ p}=-\frac{g^2 \mpi^2}{2F^4} I_\pi\left(m^2\right),\quad 
B_{s_{11}}^{\rm cex}=\frac{\sqrt{2} \, g^2}{4F^4}\left(\mpi^2I_\pi\left(m^2\right)+\mpii^2I_\npi\left(m^2\right)\right).
\end{align}

\subsubsection*{$(s_{12})$}

\begin{align}
A_{s_{12}}^{\pi^\pm p}(s,t)&=-\frac{m g^2}{2F^4}\left(t-\mpii^2\right)\Big\{4m^2I_{21}^{\npi\npi(1)}(t)-J_{\npi\npi}(t)\Big\}+\frac{m g^2 }{2F^4}\Big\{\pm 8m^2\left(s-u\right)I_{21}^{\pi\pi(3)}(t)\notag\\
&-\left(t+4\left(\mpi^2-\mpii^2\right)\right)\left(4m^2I_{21}^{\pi\pi(1)}(t)-J_{\pi\pi}(t)\right)+2\mpi^2I_\pi\left(m^2\right)\Big\},\notag\\
B_{s_{12}}^{\pi^- p}(t)&=-B_{s_{12}}^{\pi^+ p}(t)=-\frac{g^2}{F^4}\Big\{t J_{\pi\pi}^{(1)}(t)+4m^2 I_{21}^{\pi\pi(2)}(t)\Big\},\\
A_{s_{12}}^{\rm cex}(s,t)&=\frac{4\sqrt{2}\,m^3 g^2}{F^4}\left(s-u\right)I_{21}^{\pi\npi(3)}(t),\quad 
B_{s_{12}}^{\rm cex}(t)=\frac{\sqrt{2} \, g^2}{F^4}\Big\{t J_{\pi\npi}^{(1)}(t)+4m^2 I_{21}^{\pi\npi(2)}(t)\Big\}.\notag
\end{align}

\subsubsection*{$(s_{13})$}

\begin{align}
A_{s_{13}}^{\pi^- p}&=A_{s_{13}}^{\pi^+ p}=A_{s_{13}}^{\rm cex}=0,\quad B_{s_{13}}^{\pi^- p}&=-B_{s_{13}}^{\pi^+ p}=\frac{4 \Delta _\pi+\Delta _\npi}{8F^4},\quad
B_{s_{13}}^{\rm cex}=-\frac{\sqrt{2}\left(3 \Delta _\pi+2\Delta _\npi\right)}{8F^4}.
\end{align}

\subsubsection*{$(s_{14})$}

\begin{align}
A_{s_{14}}^{\pi^- p}&=A_{s_{14}}^{\pi^+ p}=A_{s_{14}}^{\rm cex}=B_{s_{14}}^{\pi^- p}=B_{s_{14}}^{\pi^+ p}=B_{s_{14}}^{\rm cex}=0.
\end{align}

\subsection{Vector-type electromagnetic diagrams}

\subsubsection*{$(v_{1})$}

\begin{align}
A_{v_{1}}^{\pi^- p}(s)&=A_{v_{1}}^{\pi^+ p}(s)=-\sqrt{2}A_{v_{1}}^{\rm cex}(s)=\frac{2m e^2}{F^2}\left(s-m^2\right)\Big\{V_{11}^{(1)}(s)+V_{11}^{(2)}(s)\Big\},\notag\\
B_{v_{1}}^{\pi^- p}(s)&=B_{v_{1}}^{\pi^+ p}(s)=-\sqrt{2}B_{v_{1}}^{\rm cex}(s)\notag\\
&=\frac{e^2}{2F^2}\Big\{-V_{10}^{(1)}+4\left(s-m^2-\mpi^2\right)V_{11}(s)+4\left(s-m^2\right) V_{11}^{(1)}(s)\Big\}.
\end{align}

\subsubsection*{$(v_{2})$}

\begin{align}
A_{v_{2}}^{\pi^- p}(s,t)&=A_{v_{2}}^{\pi^+ p}(s,t)=A_{v_{1}}^{\pi^- p}(u),
\quad B_{v_{2}}^{\pi^- p}(s,t)=B_{v_{2}}^{\pi^+ p}(s,t)=-B_{v_{1}}^{\pi^- p}(u),\notag\\
A_{v_{2}}^{\rm cex}&=B_{v_{2}}^{\rm cex}=0.
\end{align}

\subsubsection*{$(v_{3})$}

\begin{align}
A_{v_{3}}^{\pi^- p}&=A_{v_{3}}^{\pi^+ p}=A_{v_{3}}^{\rm cex}=B_{v_{3}}^{\rm cex}=0,\quad 
B_{v_{3}}^{\pi^- p}(t)=-B_{v_{3}}^{\pi^+ p}(t)=-\frac{e^2}{F^2}\left(2m^2-t\right)I_{A}^\gamma(t).
\end{align}

\subsubsection*{$(v_{4})$}

\begin{align}
A_{v_{4}}^{\pi^- p}&=A_{v_{4}}^{\pi^+ p}=A_{v_{4}}^{\rm cex}=B_{v_{4}}^{\rm cex}=0,\notag\\ 
B_{v_{4}}^{\pi^- p}(t)&=-B_{v_{4}}^{\pi^+ p}(t)=-\frac{e^2}{F^2}\Big\{\left(2\mpi^2-t\right)\left(V_{20}(t)-2V_{20}^{(1)}(t)\right)
+V_{10}^{(1)}-V_{10}\Big\}.
\end{align}

\subsubsection*{$(v_{5})$}

\begin{align}
A_{v_{5}}^{\pi^- p}&=A_{v_{5}}^{\pi^+ p}=A_{v_{5}}^{\rm cex}=B_{v_{5}}^{\pi^- p}=B_{v_{5}}^{\pi^+ p}=B_{v_{5}}^{\rm cex}=0.
\end{align}

\subsubsection*{$(v_{6})$}

\begin{align}
A_{v_{6}}^{\pi^- p}&=A_{v_{6}}^{\pi^+ p}=A_{v_{6}}^{\rm cex}=0,\quad 
B_{v_{6}}^{\pi^- p}=-B_{v_{6}}^{\pi^+ p}=-2\sqrt{2}B_{v_{6}}^{\rm cex}=-\frac{e^2}{F^2}\left(2V_{10}-V_{10}^{(1)}\right).
\end{align}

\subsection{Axial-type electromagnetic diagrams}

\subsubsection*{$(a_{1})$}

\begin{align}
A_{a_{1}}^{\pi^- p}&=A_{a_{1}}^{\pi^+ p}=A_{a_{1}}^{\rm cex}=B_{a_{1}}^{\pi^- p}=B_{a_{1}}^{\pi^+ p}=B_{a_{1}}^{\rm cex}=0.
\end{align}

\subsubsection*{$(a_{2})$}

\begin{align}
A_{a_{2}}^{\pi^- p}&=A_{a_{2}}^{\pi^+ p}=-\frac{2m e^2 g^2 }{F^2}\left(2V_{10}-V_{10}^{(1)}\right),\quad A_{a_{2}}^{\rm cex}=0,\notag\\
B_{a_{2}}^{\pi^- p}(s)&=\frac{e^2g^2}{F^2}\frac{s+3m^2}{s-m^2}\left(2V_{10}-V_{10}^{(1)}\right),\quad 
B_{a_{2}}^{\pi^+ p}(s,t)=-B_{a_{2}}^{\pi^- p}(u),\notag\\
B_{a_{2}}^{\rm cex}(s,t)&=-\frac{\sqrt{2}}{4}\left(B_{a_{2}}^{\pi^- p}(s)+B_{a_{2}}^{\pi^- p}(u)\right).
\end{align}

\subsubsection*{$(a_{3})$}

\begin{align}
A_{a_{3}}^{j}(s)&=-\frac{m e^2g^2}{F^2}\Big\{(d-2)\left(I_{\gamma }(s)-I_{\gamma}^{(1)}(s)\right)
+2(d-4)\left(s-m^2-M_j^2\right)I_{B}^{\gamma j(1)}(s)\notag\\
&-\left((d-4)\left(s-m^2\right)+4M_j^2\right)I_{B}^{\gamma j}(s)\Big\}=-\frac{m e^2g^2}{F^2}\Big\{2\left(I_{\gamma }(s)-I_{\gamma}^{(1)}(s)\right)-4M_j^2I_{B}^{\gamma j}(s)\Big\}\notag\\
&-\frac{m e^2 g^2 \left(s-m^2\right) ^2}{32 \pi ^2F^2 s^2}-\frac{m e^2 g^2 M_j^2\left(s-m^2\right) \left(s+m^2-M_j^2-2 sf_{11}\left(s,m,M_j\right) \right)}{8 \pi ^2F^2 s \lambda \left(s,m^2,M_j^2\right)},\notag\\
B_{a_{3}}^{j}(s)&=-\frac{e^2g^2}{F^2}\Big\{(d-2)I_{\gamma }(s)-4m^2I_{B}^{\gamma j}(s)-4m^2(d-4)I_{B}^{\gamma j(1)}(s)\Big\}\notag\\
&=-\frac{2e^2g^2}{F^2}\Big\{I_{\gamma }(s)-2m^2I_{B}^{\gamma j}(s)\Big\}\notag\\
&+\frac{e^2g^2\left(s-m^2\right)\left(s-m^2-M_j^2\right)\left(s+m^2-M_j^2-2s f_{11}\left(s,m,M_j\right)\right)}{16\pi ^2F^2s \lambda \left(s,m^2,M_j^2\right)},\notag\\
A_{a_{3}}^{\pi^-p}(s)&=A_{a_{3}}^{\pi}(s),\quad B_{a_{3}}^{\pi^-p}(s)=B_{a_{3}}^{\pi}(s),\quad 
A_{a_{3}}^{\pi^+p}(s,t)=A_{a_{3}}^{\pi}(u),\quad B_{a_{3}}^{\pi^+p}(s,t)=-B_{a_{3}}^{\pi}(u),\notag\\
A_{a_{3}}^{\rm cex}(s,t)&=-\frac{\sqrt{2}}{4}A_{a_{3}}^{\npi}(u),\quad B_{a_{3}}^{\rm cex}(s,t)=\frac{\sqrt{2}}{4}B_{a_{3}}^{\npi}(u).
\end{align}

\subsubsection*{$(a_{4})$}

\begin{align}
A_{a_{4}}^{\pi^-p}(s)&=\frac{m e^2g^2}{2F^2}\left\{(2d-2)I_{\gamma }(s)-(d-2)I_{\gamma}^{(1)}(s)\right\}\notag\\
&=\frac{m e^2g^2}{F^2}\left(3I_{\gamma }(s)-I_{\gamma}^{(1)}(s)\right)+\frac{m e^2 g^2 }{64\pi ^2 F^2s^2}\left(\left(s+m^2\right)^2-4 s^2\right),\notag\\
B_{a_{4}}^{\pi^-p}(s)&=\frac{e^2g^2}{2F^2}(d-2)\left(I_{\gamma }(s)-I_{\gamma}^{(1)}(s)\right)
=\frac{e^2g^2}{F^2}\left(I_{\gamma }(s)-I_{\gamma}^{(1)}(s)\right)-\frac{e^2 g^2 \left(s^2-m^4\right)}{64 \pi ^2F^2s^2},\notag\\
A_{a_{4}}^{\pi^+p}(s,t)&=A_{a_{4}}^{\pi^-p}(u),\quad B_{a_{4}}^{\pi^+p}(s,t)=-B_{a_{4}}^{\pi^-p}(u),\quad
A_{a_{4}}^{\rm cex}=B_{a_{4}}^{\rm cex}=0.
\end{align}

\subsubsection*{$(a_{5})$}

\begin{align}
A_{a_{5}}^{\pi^-p}(s)&=\frac{2m e^2g^2}{F^2}
\Big\{V_{10}-V_{10}^{(1)}+2\left(s-m^2-\mpi^2\right)V_{11}(s)+\left(s-m^2\right)\left(V_{11}^{(1)}(s)-V_{11}^{(2)}(s)\right)\Big\},\notag\\
B_{a_{5}}^{\pi^-p}(s)&=-\frac{e^2g^2}{F^2}\bigg\{2\left(s+m^2\right)V_{11}^{(1)}(s)-4m^2V_{11}^{(2)}(s)\notag\\
&+\frac{s+3m^2}{s-m^2}\left(V_{10}-V_{10}^{(1)}+2\left(s-m^2-\mpi^2\right)V_{11}(s)\right)\bigg\},
\quad A_{a_{5}}^{\pi^+p}(s,t)=A_{a_{5}}^{\pi^-p}(u),\notag\\
A_{a_{5}}^{\rm cex}(s)&=-\frac{\sqrt{2}}{4}A_{a_{5}}^{\pi^-p}(s),
\quad B_{a_{5}}^{\rm cex}(s)=-\frac{\sqrt{2}}{4}B_{a_{5}}^{\pi^-p}(s),\quad B_{a_{5}}^{\pi^+p}(s,t)=-B_{a_{5}}^{\pi^-p}(u).
\end{align}

\subsubsection*{$(a_{6})$}
\vspace{-3pt}

\begin{align}
A_{a_{6}}^{\pi^-p}(s,t)&=\frac{2m e^2g^2}{F^2}\Big\{8m^2\left(u-m^2\right)A_{12}^{(1)}(s,t)-2\left(u-m^2-\mpi^2\right)V_{11}(u)+I_\pi(u)\notag\\
&-\mpi^2\left(V_{11}^{(1)}(u)+V_{11}^{(2)}(u)\right)+\left(s-m^2-2\mpi^2\right)\left(A_{11}(s)-A_{11}^{(1)}(s)\right)\Big\},\notag\\
B_{a_{6}}^{\pi^-p}(s,t)&=\frac{e^2g^2}{F^2}\Big\{V_{10}-8m^2\left(u-m^2-\mpi^2\right)A_{12}(s,t)+16m^2\left(u-m^2\right)A_{12}^{(2)}(s,t)\notag\\
&-8m^2I_{B}^{\gamma\pi}(s)-2I_\pi(u)+2\left(u-m^2-\mpi^2\right)V_{11}(u)+2\mpi^2V_{11}^{(1)}(u)\notag\\
&-2m^2\left(2A_{11}(s)-3A_{11}^{(1)}(s)
-A_{11}^{(2)}(s)\right)\Big\},\notag\\
A_{a_{6}}^{\pi^+p}(s,t)&=A_{a_{6}}^{\pi^-p}(u,t),\quad B_{a_{6}}^{\pi^+p}(s,t)=-B_{a_{6}}^{\pi^-p}(u,t),\quad 
A_{a_{6}}^{\rm cex}(s,t)=-\frac{\sqrt{2}}{4}A_{a_{6}}^{\pi^-p}(u,t),\notag\\
B_{a_{6}}^{\rm cex}(s,t)&=\frac{\sqrt{2}}{4}B_{a_{6}}^{\pi^-p}(u,t)+\frac{2\sqrt{2}\,m^2e^2g^2}{F^2}\left(I_B^{\gamma \pi}(u)-I_B^{\gamma \npi}(u)\right).
\end{align}

\subsubsection*{$(a_{7})$}
\vspace{-3pt}

\begin{align}
A_{a_{7}}^{\pi^-p}(s,t)&=\frac{m e^2g^2}{F^2}\Big\{2t I_{A}^\gamma(t)-I_{\gamma}(s)
-\frac{d-2}{2}I_{\gamma}^{(1)}(s)-\left(4\mpi^2+(d-4)\left(s-m^2\right)\right)I_{B}^{\gamma\pi}(s)\notag\\
&+2(d-4)\left(s-m^2-\mpi^2\right)I_{B}^{\gamma\pi(1)}(s)+4m^2\left(s-m^2\right)I_{13}^\gamma(s,t)\notag\\
&+4m^2\left(2\left(u-m^2\right)+\left(s-m^2\right)(d-4)\right)I_{13}^{\gamma(1)}(s,t)\Big\}\notag\\
&=\frac{m e^2g^2}{F^2}\Big\{2t I_{A}^\gamma(t)-I_{\gamma}(s)
-I_{\gamma}^{(1)}(s)-4\mpi^2I_{B}^{\gamma\pi}(s)+4m^2\left(s-m^2\right)I_{13}^\gamma(s,t)\notag\\
&+8m^2\left(u-m^2\right)I_{13}^{\gamma(1)}(s,t)\Big\}\notag\\
&+\frac{m e^2g^2\left(s-m^2\right)}{16\pi ^2F^2}\bigg\{\frac{s-m^2}{4s^2}+\frac{\left(s-m^2\right)^2-\mpi^4-4s \mpi^2f_{11}\left(s,m,\mpi\right)}{s \lambda \left(s,m^2,\mpi^2\right)}\notag\\
&+\frac{2\left(s-m^2+\mpi^2\right)f_{11}\left(s,m,\mpi\right)-\left(s-u\right)f(t,m)}{\lambda \left(s,m^2,\mpi^2\right)+s t}\bigg\},\notag\\
B_{a_{7}}^{\pi^-p}(s,t)&=\frac{e^2g^2}{2F^2}\Big\{2\left(2m^2-t\right)I_{A}^\gamma(t)+(d-2)\left(I_{\gamma }(s)-I_{\gamma}^{(1)}(s)\right)\notag\\
&+4(d-4)m^2\left(I_{B}^{\gamma\pi}(s)-2I_B^{\gamma\pi(1)}(s)\right)
+8m^2\left(t-2m^2\right)I_{13 }^\gamma(s,t)\notag\\
&+4(6-d)m^2\left(4m^2-t\right)I_{13}^{\gamma(1)}(s,t)+4(d-2)m^2\left(2\mpi^2-t\right)I_{13}^{\gamma(2)}(s,t)\Big\}\notag\\
&=\frac{e^2g^2}{2F^2}\Big\{2\left(2m^2-t\right)I_{A}^\gamma(t)+2\left(I_{\gamma }(s)-I_{\gamma}^{(1)}(s)\right)\notag\\
&+8m^2\Big[\left(t-2m^2\right)I_{13 }^\gamma(s,t)+\left(4m^2-t\right)I_{13}^{\gamma(1)}(s,t)+\left(2\mpi^2-t\right)I_{13}^{\gamma(2)}(s,t)\Big]\Big\}\notag\\
&-\frac{e^2g^2\left(s-m^2\right)}{8\pi ^2F^2}\bigg\{\frac{s+m^2}{8s^2}+\frac{m^2\left(s-m^2+\mpi^2\right)-s\left(s-m^2-\mpi^2\right)f_{11}\left(s,m,\mpi\right)}{s \lambda \left(s,m^2,\mpi^2\right)}\notag\\
&+\frac{2\left(s+m^2-\mpi^2\right)f_{11}\left(s,m,\mpi\right)-\left(4m^2-t\right)f(t,m)}{2\left(\lambda \left(s,m^2,\mpi^2\right)+s t\right)}\bigg\},\notag\\
A_{a_{7}}^{\pi^+p}(s,t)&=A_{a_{7}}^{\pi^-p}(u,t),\quad B_{a_{7}}^{\pi^+p}(s,t)=-B_{a_{7}}^{\pi^-p}(u,t),\quad A_{a_{7}}^{\rm cex}=B_{a_{7}}^{\rm cex}=0.
\end{align}

\subsubsection*{$(a_{8})$}

\begin{align}
A_{a_{8}}^{\pi^-p}(s,t)&=-\frac{m e^2g^2}{F^2}\Big\{2\left(2\mpi^2-t\right)V_{20}(t)-2V_{10}+J_{\pi\pi}(t)-8m^2\left(2\mpi^2-t\right)A_{21}^{(1)}(s,t)\notag\\
&+4m^2\left(A_{11}^{(1)}(s)+A_{11}^{(2)}(s)\right)-4m^2I_{21}^{\pi\pi(1)}(t)\Big\},\notag\\
B_{a_{8}}^{\pi^-p}(s,t)&=\frac{e^2g^2}{F^2}\Big\{\left(2\mpi^2-t\right)\left(V_{20}(t)-2V_{20}^{(1)}(t)\right)-V_{10}+V_{10}^{(1)}
+4m^2\left(A_{11}(s)-A_{11}^{(1)}(s)\right)\notag\\
&-4m^2\left(2\mpi^2-t\right)\left(A_{21}(s,t)-2A_{21}^{(2)}(s,t)\right)\Big\},\notag\\
A_{a_{8}}^{\pi^+p}(s,t)&=A_{a_{8}}^{\pi^-p}(u,t),\quad B_{a_{8}}^{\pi^+p}(s,t)=-B_{a_{8}}^{\pi^-p}(u,t),\quad A_{a_{8}}^{\rm cex}=B_{a_{8}}^{\rm cex}=0.
\end{align}

\subsubsection*{$(a_{9})$}

\begin{align}
A_{a_{9}}^{\pi^-p}(s)&=\frac{2m e^2 g^2}{F^2}\left\{-I_{\gamma }(s)+\mpi^2\left(2A_{11}(s)-A_{11}^{(1)}(s)\right)\right\},\notag\\
B_{a_{9}}^{\pi^-p}(s)&=\frac{e^2g^2}{F^2}\left\{2V_{10}-V_{10}^{(1)}-2m^2\left(A_{11}^{(1)}(s)+A_{11}^{(2)}(s)\right)\right\},\notag\\
A_{a_{9}}^{\pi^+p}(s,t)&=A_{a_{9}}^{\pi^-p}(u),\quad B_{a_{9}}^{\pi^+p}(s,t)=-B_{a_{9}}^{\pi^-p}(u),\quad A_{a_{9}}^{\rm cex}=B_{a_{9}}^{\rm cex}=0.
\end{align}

\subsubsection*{$(a_{10})$}

\begin{align}
A_{a_{10}}^{\pi^-p}&=B_{a_{10}}^{\pi^-p}=A_{a_{10}}^{\pi^+p}=B_{a_{10}}^{\pi^+p}=0,\notag\\
A_{a_{10}}^{\rm cex}(s,t)&=-\frac{\sqrt{2}\,m e^2g^2}{2F^2\left(u-m^2\right)^2}\Big\{\left(u^2-6u m^2+5m^4+(d-4)\left(u-m^2\right)^2\right)I_{\gamma }(u)\notag\\
&+\frac{d-2}{2}\left(\left( u+m^2\right)^2-4u^2\right)I_{\gamma}^{(1)}(u)\Big\}=\frac{\sqrt{2}\,m e^2g^2\left(u-m^2\right)^2}{128\pi ^2F^2u^2}\notag\\
&-\frac{\sqrt{2}\,m e^2g^2}{2F^2\left(u-m^2\right)^2}\Big\{\left(u^2-6u m^2+5m^4\right)I_{\gamma }(u)+\left(\left( u+m^2\right)^2-4u^2\right)I_{\gamma}^{(1)}(u)\Big\},\notag\\
B_{a_{10}}^{\rm cex}(s,t)&=\frac{\sqrt{2}\,e^2g^2}{4F^2\left(u-m^2\right)^2}\Big\{\left(16m^4-2\left(u-m^2\right)^2+(d-4)\left(4m^4-\left(u+m^2\right)^2\right)\right)I_{\gamma }(u)\notag\\
&+(d-2)\left(\left(u+m^2\right)^2+4u m^2\right)I_{\gamma}^{(1)}(u)\Big\}=\frac{\sqrt{2}\,e^2g^2\left(u^2-m^4\right)}{128\pi ^2F^2u^2}\\
&+\frac{\sqrt{2}\,e^2g^2}{2F^2\left(u-m^2\right)^2}\Big\{\left(8m^4-\left(u-m^2\right)^2\right)I_{\gamma }(u)+\left(\left(u+m^2\right)^2+4u m^2\right)I_{\gamma}^{(1)}(u)\Big\}.\notag
\end{align}

\subsubsection*{$(a_{11})$}

\begin{align}
A_{a_{11}}^{\pi^-p}&=B_{a_{11}}^{\pi^-p}=A_{a_{11}}^{\pi^+p}=B_{a_{11}}^{\pi^+p}=0,\notag\\
A_{a_{11}}^{\rm cex}(s,t)&=\frac{\sqrt{2}\,m e^2g^2}{2F^2}\Big\{\frac{d-2}{2} I_{\gamma}^{(1)}(u)+\left(2\left(u-3m^2+\mpii^2\right)+(d-4)\left(u-m^2\right)\right)I_{B}^{\gamma\npi}(u)\notag\\
&-\left(4\left(u-m^2\right)+(d-4)\left(2u+2m^2-\mpii^2\right)\right)I_{B}^{\gamma\npi(1)}(u)-d\mpii^2 I_{B}^{\gamma\npi(2)}(u)-I_{\gamma }(u)\Big\}\notag\\
&=\frac{\sqrt{2}\,m e^2g^2}{2F^2}\Big\{ I_{\gamma}^{(1)}(u)-I_{\gamma }(u)+2\left(u-3m^2+\mpii^2\right)I_{B}^{\gamma\npi}(u)\notag\\
&-4\left(u-m^2\right)I_{B}^{\gamma\npi(1)}(u)-4\mpii^2 I_{B}^{\gamma\npi(2)}(u)\Big\}\notag\\
&-\frac{\sqrt{2}\,m e^2g^2\left(u-m^2\right)^2}{128\pi ^2F^2u^2}\bigg\{1+4u\frac{u+m^2-\mpii^2-2u f_{11}\left(u,m,\mpii\right)}{\lambda \left(u,m^2,\mpii^2\right)}\bigg\},\notag\\
B_{a_{11}}^{\rm cex}(s,t)&=\frac{\sqrt{2}\,e^2g^2}{4F^2\left(u-m^2\right)}\Big\{\left(2\left(u-3m^2\right)+(d-4)\left(u-m^2\right)\right)I_{\gamma }(u)+2(d-2)m^2I_{\gamma}^{(1)}(u)\notag\\
&+\left(4m^2\left(u-5m^2+2\mpii^2\right)+4(d-4)m^2\left(u-m^2\right)\right)I_{B}^{\gamma\npi}(u)\notag\\
&-4m^2\left(4\left(u-m^2\right)+(d-4)\left(3u+m^2-\mpii^2\right)\right)I_B^{\gamma\npi(1)}(u)-4d m^2\mpii^2 I_B^{\gamma\npi(2)}(u)\Big\}\notag\\
&=\frac{\sqrt{2}\,e^2g^2}{2F^2\left(u-m^2\right)}\Big\{\left(u-3m^2\right)I_{\gamma }(u)+2m^2I_{\gamma}^{(1)}(u)+2m^2\left(u-5m^2+2\mpii^2\right)I_{B}^{\gamma\npi}(u)\notag\\
&-8m^2\left(\left(u-m^2\right)I_B^{\gamma\npi(1)}(u)+\mpii^2 I_B^{\gamma\npi(2)}(u)\right)\Big\}-\frac{\sqrt{2}\,e^2g^2\left(u-m^2\right)}{64\pi ^2F^2u}\bigg\{\frac{u+m^2}{u}\notag\\
&+\frac{2m^2\left(3u+m^2-\mpii^2\right)-2u\left(u+3m^2-\mpii^2\right)f_{11}\left(u,m,\mpii\right)}{\lambda \left(u,m^2,\mpii^2\right)}\bigg\}.
\end{align}

\subsubsection*{$(a_{12})$}

\begin{align}
A_{a_{12}}^{\rm cex}(s,t)&=\frac{\sqrt{2}\,m e^2g^2}{2F^2}\Big\{V_{10}-V_{10}^{(1)}+2I_{\gamma }(u)-\left(u-m^2+2\mpi^2\right)A_{11}(u)+\left( u-m^2\right)A_{11}^{(1)}(u)\Big\},\notag\\
B_{a_{12}}^{\rm cex}(s,t)&=\frac{\sqrt{2}\,e^2g^2}{4F^2}\bigg\{2m^2\left(A_{11}^{(1)}(u)-A_{11}^{(2)}(u)-2A_{11}(u)\right)
+2V_{10}-V_{10}^{(1)}\\
&+\frac{4m^2}{u-m^2}\left(V_{10}-V_{10}^{(1)}+2I_{\gamma }(u)-2\mpi^2A_{11}(u)\right)\bigg\},\quad A_{a_{12}}^{\pi^-p}=B_{a_{12}}^{\pi^-p}=A_{a_{12}}^{\pi^+p}=B_{a_{12}}^{\pi^+p}=0.\notag
\end{align}

\subsubsection*{$(a_{13})$}

\begin{align}
A_{a_{13}}^{\pi^-p}&=B_{a_{13}}^{\pi^-p}=A_{a_{13}}^{\pi^+p}=B_{a_{13}}^{\pi^+p}=0,\notag\\
A_{a_{13}}^{\rm cex}(s,t)&=-\frac{\sqrt{2}\,m e^2g^2}{2F^2}\Big\{I_{\gamma }(u)+\frac{d-2}{2}I_{\gamma}^{(1)}(u)\Big\}\notag\\
&=-\frac{\sqrt{2}\,m e^2g^2}{2F^2}\left(I_{\gamma }(u)+I_{\gamma}^{(1)}(u)\right)+\frac{\sqrt{2}\,m e^2g^2\left(u-m^2\right)^2}{128\pi ^2F^2u^2},\\
B_{a_{13}}^{\rm cex}(s,t)&=\frac{\sqrt{2}\,e^2g^2}{4F^2\left(u-m^2\right)}\left\{\left(2\left(u-3m^2\right)+(d-4)\left(u-m^2\right)\right)I_{\gamma }(u)-(d-2)\left(u+m^2\right)I_{\gamma}^{(1)}(u)\right\}\notag\\
&=\frac{\sqrt{2}\,e^2g^2}{2F^2\left(u-m^2\right)}\left\{\left(u-3m^2\right)I_{\gamma }(u)-\left(u+m^2\right)I_{\gamma}^{(1)}(u)\right\}-\frac{\sqrt{2}\,e^2g^2\left(u-m^2\right)^2}{128\pi ^2F^2u^2}.\notag
\end{align}

\subsection{Photon exchange}

$A_{p_i}^{\rm cex}=B_{p_i}^{\rm cex}=0$ for $i=1,\ldots,14$.

\subsubsection*{$(p_{1})$}

\beq
A_{p_{1}}^{\pi^-p}=A_{p_{1}}^{\pi^+p}=0,\quad B_{p_{1}}^{\pi^-p}(t)=-B_{p_{1}}^{\pi^+p}(t)=-\frac{4e^2}{F^2} J_{\pi\pi}^{(1)}(t).
\eeq

\subsubsection*{$(p_{2})$}

\beq
A_{p_{2}}^{\pi^-p}=A_{p_{2}}^{\pi^+p}=B_{p_{2}}^{\pi^-p}=B_{p_{2}}^{\pi^+p}=0.
\eeq

\subsubsection*{$(p_{3})$}

\begin{align}
A_{p_{3}}^{\pi^-p}(s,t)&=-A_{p_{3}}^{\pi^+p}(s,t)=\frac{8m^3e^2g^2}{F^2t}\left(s-u\right)I_{21}^{\pi\pi(3)}(t),\notag\\
B_{p_{3}}^{\pi^-p}(t)&=-B_{p_{3}}^{\pi^+p}(t)=\frac{2e^2g^2}{F^2t}\left\{t J_{\pi\pi}^{(1)}(t)+4m^2I_{21}^{\pi\pi(2)}(t)\right\}.
\end{align}

\subsubsection*{$(p_{4})$}

\beq
A_{p_{4}}^{\pi^-p}=A_{p_{4}}^{\pi^+p}=0,\quad B_{p_{4}}^{\pi^-p}(t)=-B_{p_{4}}^{\pi^+p}(t)=-\frac{2e^2}{F^2} J_{\pi\pi}^{(1)}(t).
\eeq

\subsubsection*{$(p_{5})$}

\beq
A_{p_{5}}^{\pi^-p}=A_{p_{5}}^{\pi^+p}=0,\quad B_{p_{5}}^{\pi^-p}(t)=-B_{p_{5}}^{\pi^+p}(t)=\frac{e^2}{F^2t} \Delta_\pi.
\eeq

\subsubsection*{$(p_{6})$}

\begin{align}
A_{p_{6}}^{\pi^-p}(s,t)&=-A_{p_{6}}^{\pi^+p}(s,t)=-\frac{4m^3e^2g^2}{F^2t}\left(s-u\right)I_{A}^{\npi(3)}(t),\\
B_{p_{6}}^{\pi^-p}(t)&=-B_{p_{6}}^{\pi^+p}(t)=-\frac{e^2g^2}{2F^2t}\Big\{\Delta_\npi-4m^2\left(I_\npi^{(1)}\left(m^2\right)+\mpii^2I_{A}^\npi(t)-2I_{A}^{\npi(2)}(t)\right)\Big\}.\notag
\end{align}

\subsubsection*{$(p_{7})$}

\beq
A_{p_{7}}^{\pi^-p}=A_{p_{7}}^{\pi^+p}=0,\quad B_{p_{7}}^{\pi^-p}(t)=-B_{p_{7}}^{\pi^+p}(t)=\frac{2e^2g^2}{F^2t} \mpi^2I_\pi\left(m^2\right).
\eeq

\subsubsection*{$(p_{8})$}

\beq
A_{p_{8}}^{\pi^-p}=A_{p_{8}}^{\pi^+p}=0,\quad B_{p_{8}}^{\pi^-p}(t)=-B_{p_{8}}^{\pi^+p}(t)=\frac{8e^4}{t} J_{\pi\pi}^{(1)}(t).
\eeq

\subsubsection*{$(p_{9})$}

\begin{align}
A_{p_{9}}^{\pi^-p}(s,t)&=A_{p_{9}}^{\pi^+p}(s,t)=8m e^4\left\{\left(s-m^2\right)P_{11}^{(1)}(s,t)+\left(u-m^2\right)P_{11}^{(1)}(u,t)\right\},\notag\\
B_{p_{9}}^{\pi^-p}(s,t)&=B_{p_{9}}^{\pi^+p}(s,t)=-4e^4\Big\{\mpi^2P_{11}(s,t)-\mpi^2P_{11}(u,t)\notag\\
&-2\left(s-m^2\right)P_{11}^{(2)}(s,t)+2\left(u-m^2\right)P_{11}^{(2)}(u,t)\Big\}.
\end{align}

\subsubsection*{$(p_{10})$}

\begin{align}
A_{p_{10}}^{\pi^-p}&=A_{p_{10}}^{\pi^+p}=0,\notag\\
 B_{p_{10}}^{\pi^-p}(t)&=-B_{p_{10}}^{\pi^+p}(t)
=\frac{4e^4}{t}\left\{\left(2\mpi^2-t\right)\left(V_{20}(t)-2V_{20}^{(1)}(t)\right)-V_{10}+V_{10}^{(1)}\right\}.
\end{align}

\subsubsection*{$(p_{11})$}

\beq
A_{p_{11}}^{\pi^-p}=A_{p_{11}}^{\pi^+p}=0,\quad B_{p_{11}}^{\pi^-p}(t)=-B_{p_{11}}^{\pi^+p}(t)=\frac{4e^4}{t}\left\{2V_{10}-V_{10}^{(1)}\right\}.
\eeq

\subsubsection*{$(p_{12})$}

\beq
A_{p_{12}}^{\pi^-p}=A_{p_{12}}^{\pi^+p}=0,\quad B_{p_{12}}^{\pi^-p}(t)=-B_{p_{12}}^{\pi^+p}(t)=\frac{4e^4}{t}\left(2m^2-t\right)I_A^\gamma(t).
\eeq

\subsubsection*{$(p_{13})$}

\begin{align}
A_{p_{13}}^{\pi^-p}(t)&=A_{p_{13}}^{\pi^+p}(t)=4m e^4 \left\{I_{21}^{\gamma\gamma}(t)+(d-2)I_{21}^{\gamma\gamma(1)}(t)\right\}\notag\\
&=4m e^4 \left\{I_{21}^{\gamma\gamma}(t)+2I_{21}^{\gamma\gamma(1)}(t)\right\}-\frac{e^4\left(4m^2+t f(t,m)\right)}{8\pi ^2m\left(4m^2-t\right)},\quad 
 B_{p_{13}}^{\pi^-p}=B_{p_{13}}^{\pi^+p}=0.
\end{align}

\subsubsection*{$(p_{14})$}

\beq
A_{p_{14}}^{\pi^-p}=A_{p_{14}}^{\pi^+p}=0,\quad B_{p_{14}}^{\pi^-p}(t)=-B_{p_{14}}^{\pi^+p}(t)=-\frac{4e^4}{t^2}\Delta_\pi.
\eeq

\subsection{Bremsstrahlung}
\label{app:Bremsstrahlung}

Apart from $m_\gamma^2$ terms in the denominators, the following expressions for the bremsstrahlung diagrams are still exact. $T^\mu$ is defined according to \eqref{brems_ampl}. The amplitude for $\pi^+p\rightarrow \pi^+ p$ follows from
\beq
T^{\pi^+p}_\mu\left(p,q,p',q',k\right)=T^{\pi^-p}_\mu\left(p,-q',p',-q,k\right).
\eeq

\subsubsection{Vector-type}

\subsubsection*{$(v_{1})$}

\beq
T_{v_1}^{\pi^-p\,\mu}=\frac{e}{2F^2}\bigg\{\frac{\slashed{q}\left(2q'+k\right)^\mu}{2q'\cdot k}-\frac{\slashed{q}'\left(2q-k\right)^\mu}{2q\cdot k}\bigg\},\quad T_{v_1}^{{\rm cex}\,\mu}=\frac{\sqrt{2}\,e}{2F^2}\frac{\slashed{q}'\left(2q-k\right)^\mu}{2q\cdot k}.
\eeq

\subsubsection*{$(v_{2})$}

\begin{align}
T_{v_2}^{\pi^-p\,\mu}&=\frac{e}{2F^2}\bigg\{\gamma^\mu-\frac{\gamma^\mu\left(\slashed{p}'+\slashed{k}+m\right)\slashed{q}}{2p'\cdot k}+\frac{\slashed{q}'\left(\slashed{p}-\slashed{k}+m\right)\gamma^\mu}{2p\cdot k}\bigg\},\notag\\
T_{v_2}^{{\rm cex}\,\mu}&=-\frac{\sqrt{2}\,e}{4F^2}\bigg\{\gamma^\mu+\frac{2\slashed{q}'\left(\slashed{p}-\slashed{k}+m\right)\gamma^\mu}{2p\cdot k}\bigg\}.
\end{align}

\subsubsection*{$(v_{3})$}

\beq
T_{v_3}^{\pi^-p\,\mu}=-\frac{e}{2F^2}\gamma^\mu,\quad T_{v_3}^{{\rm cex}\,\mu}=\frac{\sqrt{2}\,e}{4F^2}\gamma^\mu.
\eeq

\subsubsection{Axial-type}

\subsubsection*{$(a_{1})$}

\begin{align}
T_{a_1}^{\pi^-p\,\mu}&=\frac{e g^2}{2F^2}\bigg\{-\frac{\slashed{q}\left(2q'+k\right)^\mu}{2q'\cdot k}\bigg(1+\frac{4m^2}{(p+q)^2-m^2}\bigg)+\frac{2m\left(2q'+k\right)^\mu}{2q'\cdot k}\notag\\
&+\frac{\slashed{q}'\left(2q-k\right)^\mu}{2q\cdot k}\bigg(1+\frac{4m^2}{(p'+q')^2-m^2}\bigg)-\frac{2m\left(2q-k\right)^\mu}{2q\cdot k}\bigg\},\\
T_{a_1}^{{\rm cex}\,\mu}&=-\frac{\sqrt{2}\,e g^2}{2F^2}\bigg\{\frac{\slashed{q}'(2q-k)^\mu}{2q\cdot k}+\frac{2m^2\slashed{q}'(2q-k)^\mu}{2q\cdot k\left(\left(p'+q'\right)^2-m^2\right)}+\frac{2m^2\slashed{q}'(2q-k)^\mu}{2q\cdot k\left(\left(p-q'\right)^2-m^2\right)}\bigg\}.\notag
\end{align}

\subsubsection*{$(a_{2})$}
\vspace{-5pt}

\begin{align}
T_{a_2}^{\pi^-p\,\mu}&=-\frac{e g^2}{2F^2}\bigg\{-\frac{\gamma^\mu\left(\slashed{p}'+\slashed{k}+m\right)\slashed{q}}{2p'\cdot k}+\frac{2m\gamma^\mu\left(\slashed{q}-\slashed{q}'\right)}{2p'\cdot k}-\frac{2m\gamma^\mu \slashed{q}}{(p+q)^2-m^2}\notag\\
&+\frac{4m^2\gamma^\mu\slashed{q}'\slashed{q}}{2p'\cdot k\left((p+q)^2-m^2\right)}+2\gamma^\mu+\frac{\slashed{q}'\left(\slashed{p}-\slashed{k}+m\right)\gamma^\mu}{2p\cdot k}+\frac{2m\left(\slashed{q}-\slashed{q}'\right)\gamma^\mu}{2p\cdot k}\notag\\
&-\frac{2m \slashed{q}'\gamma^\mu}{\left(p'+q'\right)^2-m^2}-\frac{4m^2\slashed{q}'\slashed{q}\gamma^\mu}{2p\cdot k\left(\left(p'+q'\right)^2-m^2\right)}\bigg\},\notag\\
T_{a_2}^{{\rm cex}\,\mu}&=-\frac{\sqrt{2}\,e g^2}{4F^2}\bigg\{-\frac{2\slashed{q}'\left(\slashed{p}-\slashed{k}+m\right)\gamma^\mu}{2p\cdot k}+\frac{2m\slashed{q}'\gamma^\mu}{\left(p'+q'\right)^2-m^2}+\frac{2m\slashed{q}\gamma^\mu}{\left(p'-q\right)^2-m^2}\notag\\
&-\gamma^\mu+\frac{4m^2\slashed{q}'\slashed{q}\gamma^\mu}{2p\cdot k\left(\left(p'+q'\right)^2-m^2\right)}-\frac{4m^2\slashed{q}\slashed{q}'\gamma^\mu}{2p\cdot k\left(\left(p'-q\right)^2-m^2\right)}\bigg\}.
\end{align}

\subsubsection*{$(a_{3})$}
\vspace{-5pt}

\begin{align}
T_{a_3}^{{\rm cex}\,\mu}&=\frac{\sqrt{2}\,e g^2}{4F^2}\bigg\{\gamma^\mu+\frac{2m\slashed{q}\gamma^\mu}{\left(p'-q\right)^2-m^2}+\frac{2m\gamma^\mu\slashed{q}'}{\left(p-q'\right)^2-m^2}\notag\\
&+\frac{4m^2\slashed{q}\gamma^\mu\slashed{q}'}{\left(\left(p'-q\right)^2-m^2\right)\left(\left(p-q'\right)^2-m^2\right)}\bigg\},\quad T_{a_3}^{\pi^-p\,\mu}=0.
\end{align}

\subsubsection*{$(a_{4})$}
\vspace{-5pt}

\begin{align}
T_{a_4}^{\pi^-p\,\mu}&=\frac{e g^2}{F^2}\bigg\{\gamma^\mu-\frac{m\gamma^\mu\slashed{q}}{\left(p+q\right)^2-m^2}-\frac{m\slashed{q}'\gamma^\mu}{\left(p'+q'\right)^2-m^2}\bigg\},\notag\\
T_{a_4}^{{\rm cex}\,\mu}&=-\frac{\sqrt{2}\,e g^2}{2F^2}\bigg\{\gamma^\mu+\frac{m\gamma^\mu\slashed{q}'}{\left(p-q'\right)^2-m^2}-\frac{m\slashed{q}'\gamma^\mu}{\left(p'+q'\right)^2-m^2}\bigg\}.
\end{align}

\subsubsection{Photon exchange}

\subsubsection*{$(p_{1})$}
\vspace{-6pt}

\beq
T_{p_1}^{\pi^-p\,\mu}=\frac{2e^3 }{\left(p-p'\right)^2}\bigg\{\frac{\slashed{q}'\left(2q-k\right)^\mu}{2q\cdot k}-\frac{\slashed{q}\left(2q'+k\right)^\mu}{2q'\cdot k}\bigg\},\quad T_{p_1}^{{\rm cex}\,\mu}=0.
\eeq

\subsubsection*{$(p_{2})$}
\vspace{-6pt}

\beq
T_{p_2}^{\pi^-p\,\mu}=\frac{2e^3 }{\left(p-p'-k\right)^2}\bigg\{\gamma^\mu-\frac{\slashed{q}\left(\slashed{p}-\slashed{k}+m\right)\gamma^\mu}{2p\cdot k}+\frac{\gamma^\mu\left(\slashed{p}'+\slashed{k}+m\right)\slashed{q}'}{2p'\cdot k}\bigg\},\quad T_{p_2}^{{\rm cex}\,\mu}=0.
\eeq

\subsubsection*{$(p_{3})$}
\vspace{-6pt}

\beq
T_{p_3}^{\pi^-p\,\mu}=\frac{2e^3\gamma^\mu }{\left(p-p'\right)^2},\quad T_{p_3}^{{\rm cex}\,\mu}=0.
\eeq


\end{document}